\documentclass[a4paper,12pt]{report}
\usepackage{times, amsfonts,graphicx,amssymb,amstext,amsmath,amsthm,amssymb,pifont,textgreek,mathtools,siunitx,caption,atbegshi}
\usepackage[superscript,biblabel]{cite}
\usepackage[official]{eurosym}
\usepackage[T1]{fontenc}
\usepackage[left=2.5cm, right=2.5cm, top=3cm, bottom=3cm]{geometry}
\usepackage[labelfont=bf]{caption}
\usepackage{multirow}
\usepackage[normalem]{ulem}
\newcommand{\mli}[1]{\mathit{#1}}
\newcommand{\dbar}{d\hspace*{-0.08em}\bar{}\hspace*{0.1em}}

\usepackage{xcolor}
\usepackage{titlesec}
\definecolor{gray75}{gray}{0.75}

\titleformat{\chapter}[display]{\filright\Huge\bfseries}{\fontsize{100}{100}\selectfont\textcolor{gray75}\thechapter}{1ex}{}[]%
\AtBeginDocument{\AtBeginShipoutNext{\AtBeginShipoutDiscard}}

\begin{document}
\title{Quantum Shape Effects}
\maketitle
\vspace*{-3.45cm}
\noindent\makebox[\textwidth]{\includegraphics[width=\paperwidth]{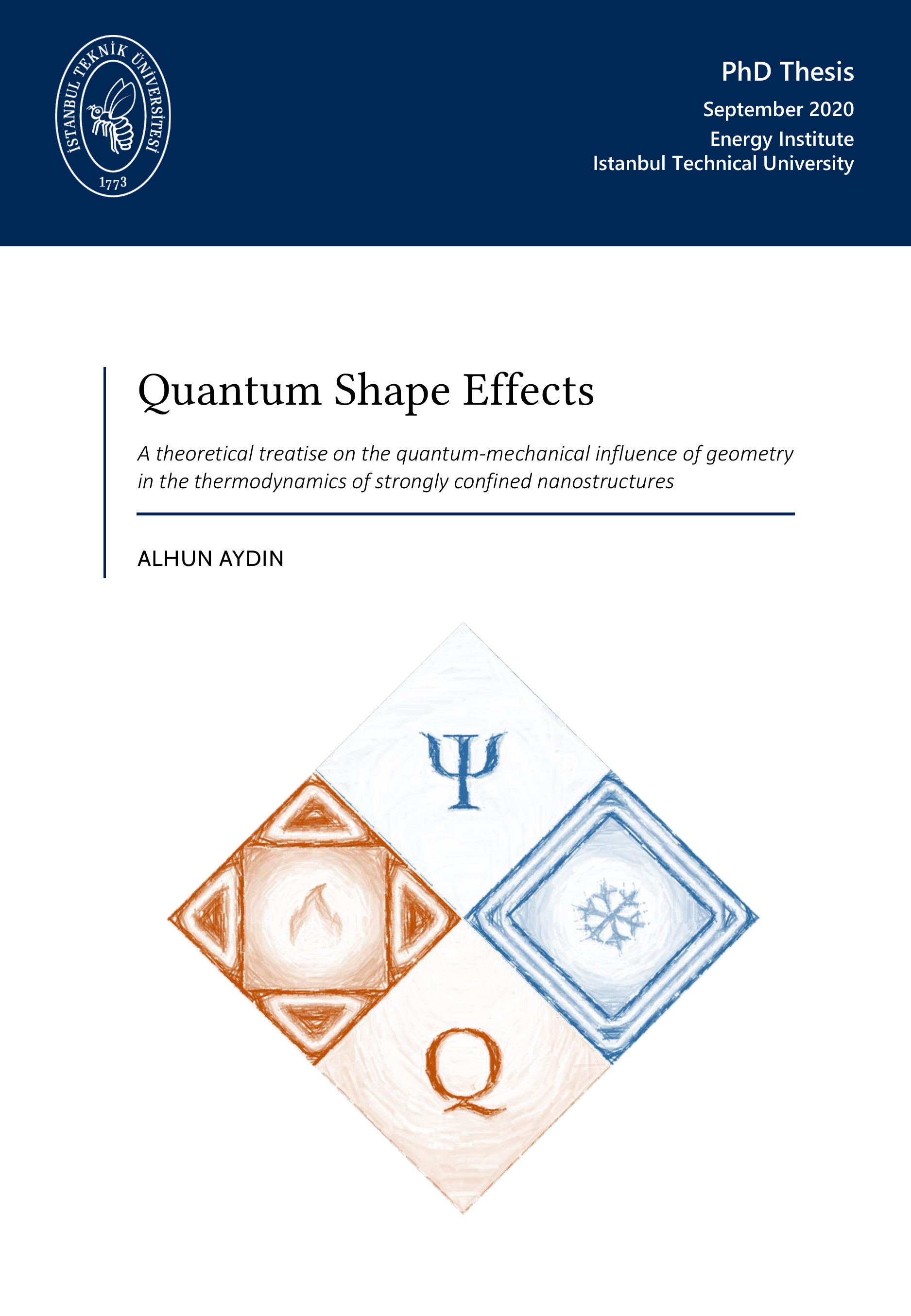}}
\thispagestyle{empty}
\newpage
\vspace*{40px}
\noindent
Alhun AYDIN, a Ph.D. student of ITU Energy Institute 301142001 successfully defended the thesis entitled "QUANTUM SHAPE EFFECTS", which he prepared after fulfilling the requirements specified in the associated legislations, before the jury whose signatures are below. \\
\vspace*{50px}

\def\arraystretch{2.5}
\begin{tabular}{clc}
\multirow[t]{2}{*}{\textbf{Thesis Advisor:}} & \textbf{Prof. Dr. Altu\u{g} \c{S}\.{I}\c{S}MAN}      & \multirow{2}{*}{..............................} \\[-20pt]
                                          & Istanbul Technical University &                                                 \\
\multirow[t]{2}{*}{\textbf{Jury Members:}}   & \textbf{Prof. Dr. \"{O}zg\"{u}r E. M\"{U}STECAPLIO\u{G}LU} & \multirow{2}{*}{..............................} \\[-20pt]
                                          &Ko\c{c} University&                                                 \\
\multirow{2}{*}{}                         &\textbf{Assoc. Prof. Dr. Z. Fatih \"{O}ZT\"{U}RK}& \multirow{2}{*}{..............................} \\[-20pt]
                                          &Istanbul Technical University&                                                 \\
\multirow{2}{*}{}                         &\textbf{Prof. Dr. Jonas FRANSSON}& \multirow{2}{*}{..............................} \\[-20pt]
                                          &Uppsala University&                                                 \\
\multirow{2}{*}{}                         &\textbf{Assoc. Prof. Dr. Ahmet Levent SUBAŞI}& \multirow{2}{*}{..............................} \\[-20pt]
                                          &Istanbul Technical University&                                                
\end{tabular}

\vspace*{\fill}
\begin{flushleft}
\textbf{Date of Submission : 26 June 2020} \\
\textbf{Date of Defense : 15 September 2020} \\
\end{flushleft}
\begin{flushright}
\textcopyright \; 2020 Alhun Aydın
\end{flushright}
\newpage
\vspace*{200px}

\begin{flushright}
\begin{large}\textit{To the love of wisdom,}\end{large}
\end{flushright}
\newpage
\vspace*{60px}
\begin{huge}\textbf{Foreword}\end{huge} \\
\vspace*{20px}

Isn't it ironic that forewords are actually written at the very last? I don't know where to start, so let me start from the very beginning. It was all gas and dust cloud... Well, okay maybe not from that much beginning… I had the passion for science, curiosity for nearly everything during my entire life. Still, I think I can name a few milestones on my path leading to science. I was greatly fascinated by physics, when I had watched the movie "Back to the Future" with tremendous excitement during my preschool childhood years. My encounter with Goldbach conjecture in the first year of high school was another defining point for my enthusiasm to explore and think upon the unsolved problems in mathematics and physics. Contrary to the mounting evidences on that direction, initially I was not thinking of choosing math or physics as a major. The chance was on my side and thanks to my "lower than expected score" on the national exam, I chose physics to study and I got enrolled to Koç University with a full merit-scholarship, which was a great decision as things stand. But there were two important turning points in my ever-changing career plan between music and science. The first one was the "Quantum teleportation" topic given to me by my then supervisor Tekin DERELİ (I owe him a great deal of gratitude for his guidance in science) for the investigation during the independent study course when I was a junior undergraduate. I remember that I’d worked day and night with an immense excitement and joy. I had the same feeling when I joined to my advisor Altuğ ŞİŞMAN’s Nano Energy Research Group and read their papers to work it out. Thanks to their approaches coupled with my own attitude, I have never considered research as a job or as a duty, rather I always felt that I am just doing what I enjoy. I’ll do my best to keep this amateur spirit.

The story of my thesis research started around August 2015, when Altug suddenly draw a square within a square (image below) and said "This thing must rotate!" by pointing the inner one. At the moment I saw it (and we discussed on it), I was literally amazed and in a sense I felt that it was destined to be my Ph.D. topic. Then its unofficial name became "d\"{o}nen \c{s}ey" (meaning "the rotating thing" in Turkish).

\begin{figure}[h]
\centering
\includegraphics[width=0.15\textwidth]{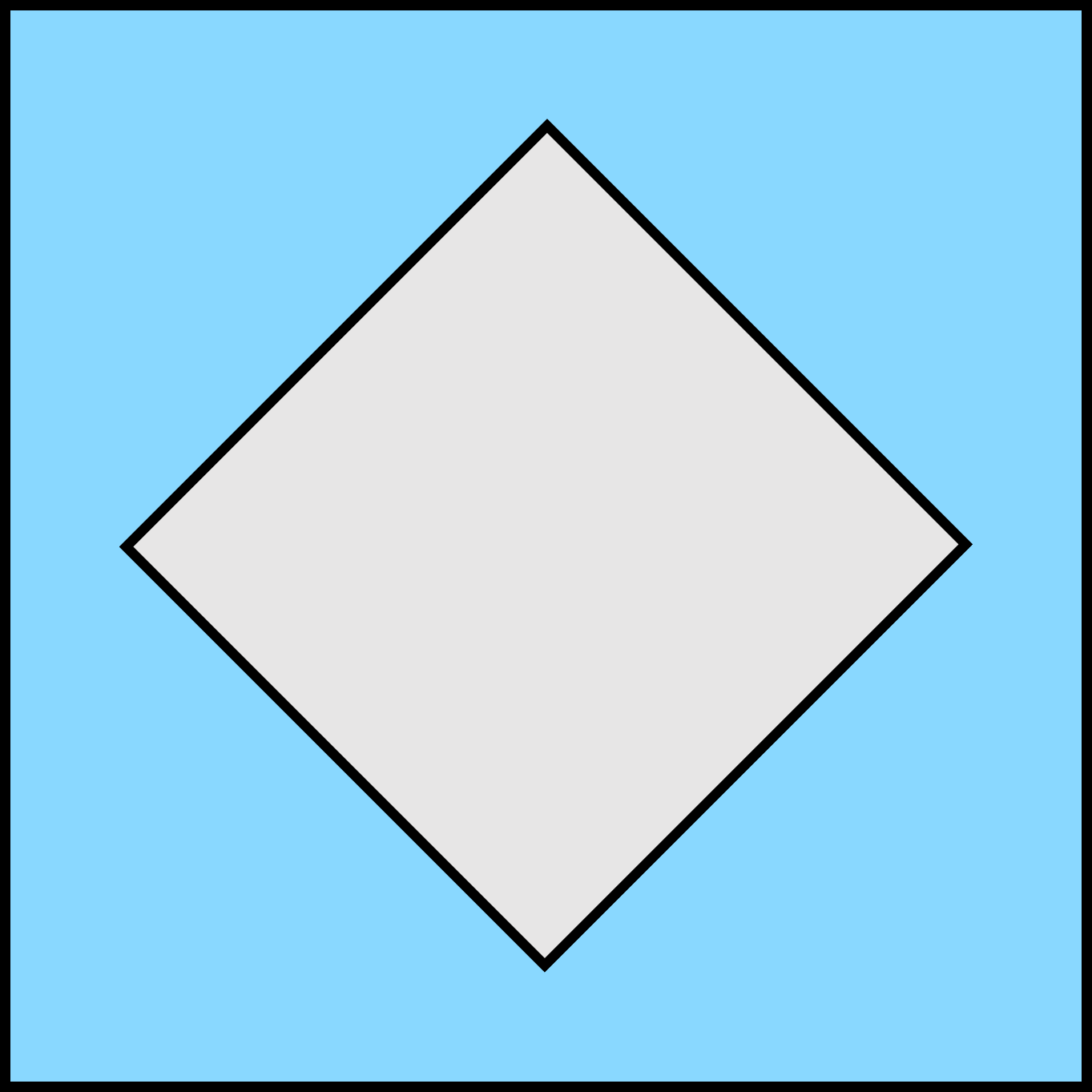}
\end{figure}

"Did you look at d\"{o}nen \c{s}ey?", he was constantly keep asking. We were discussing at the time whether it should rotate or not. Fatih was asking to me with his never-ending cheerful mood "Is it rotating or not huh?" (with a big smile on his face). From my initial calculations in September 2015, I noticed a free energy difference between two different angular configurations of the inner square, suggesting that it should somehow rotate. From September 2015 to the summer of 2016, I had to deal with courses and the infamous qualification exam. Finally, in August 2016, after I came back from California out of left field, I started to focus on "the rotating thing" and at that time we were sure that there is a finite amount of torque and so, under quasistatic process it must rotate!

The time I spent during Ph.D. helped me a lot to shape my thoughts about life, nature, human nature, universe, reality and everything. These years of doing Ph.D. was definitely highly crucial for my development both in terms of knowledge and experience not only in academia, but also in life.
%In the view of the things I learned, thought and questioned, I put quotes in the beginning of each chapter. They are not directly related to the thesis, but in fact they are related with everything.
Due to the interdisciplinary nature of our Institute, I tried my best to write the thesis understandable also by the non-experts of the field. Okay, now comes the thanks part.

Firstly, I’d like to express my great and warm gratitude to my advisor Altuğ ŞİŞMAN. To me, he has been much more than a scientific advisor, he is a very good friend. I’ve always felt his true care and sincere guidance. His dialogue with his students was exemplary. I’ve learnt a lot of things from him and not just knowledge but also principles and virtues. His wit, patience, kindness and work-ethic were invaluable for me. He once and for all revived my passion for science by his exceptional scientific approach. I know that whichever path I choose for my future, I’ll be doing scientific research at least in some part of the 24 hours. The unbreakable intellectual bond that we established between us constitutes a large part of the things that I gained during my Ph.D. time. Certainly, I’d like to thank also to his wife Ayşe KAŞLILAR ŞİŞMAN for her warm friendship, realistic advices and mutual sincerity. I am waiting our next dinner-table-conversations/discussions impatiently. 

Z. Fatih ÖZTÜRK was definitely the person who brings joy to my time during Ph.D. with his great humor that usually makes me rolling on the floor laughing. I’ll remember his conversations with my advisor on work agenda and memories, with a big smile on my face. I thank to him also for his tireless support, kindness and cheerfulness all the time. Likewise, I’d like to thank to the other members of Nano Energy Research Group; Sevan KARABETOĞLU, Gülru BABAÇ SCHÜBLER, Coşkun FIRAT and Türker ŞAHİN for their valuable supports and friendships. I appreciate all academic, administrative and employee staff of ITU Energy Institute for providing me a comfortable and beautiful workplace. For nearly 8 years, Energy Institute was like my true home, considering my usual late-evening/night, weekend and even holiday workings. 

I’d like to thank to my thesis steering committee member Özgür E. MÜSTECAPLIOĞLU (as well as his research group members) for his viewpoints, comments, advices and kindness. I’d like to express my gratitude to Jonas FRANSSON for inviting me to Uppsala University Physics \& Astronomy Department and for his sincere supports along with valuable advices and a fruitful collaboration. I am very grateful to Ronnie KOSLOFF for accepting me into his research group and transferring me a portion of his vast and valuable knowledge and experience along with our joyful conversations and moments in our picnics and gatherings. I'm thankful to A. Levent SUBAŞI for his detailed feedback on editing of the thesis. I thank to all jury members for their comments and suggestions on the final form of the thesis. I’d like to thank also to Nick TREFETHEN, İnanç ADAGİDELİ, Gökhan Barış BAĞCI, Mauro PATERNOSTRO, Obinna ABAH, Peter SALAMON and Raam UZDIN for our fruitful discussions.

I appreciate the hospitalities of Uppsala University, Department of Physics \& Astronomy and The Hebrew University of Jerusalem, Fritz Haber Research Center for Molecular Dynamics. I thank to "Uppsala Multidisciplinary Center for Advanced Computational Science" and "Computer Cluster Service of Fritz Haber Center for Molecular Dynamics" for providing me access to their computational resources. I thank to Israel Ministry of Foreign Affairs, Scientific and Technological Research Council of Turkey and AIM Energy Technologies for their partial supports during my Ph.D. work. 

Special thanks go to my former home (though I still feel at home when I'm there), Koç University, for preparing me to the academia, with its world-class education, academic and social environment. Most parts of this thesis are written in its lounges and terraces with spectacular forest and Black Sea views. I thank to Istanbul Technical University for providing a very nice academic environment, residential resources and financial supports on several conferences abroad. I really enjoyed my time in its beautiful campus at Ayazağa.

Life does not pass without good friends; a Ph.D. is never done. Luckily, I have many good ones. My close friend Serhat TETİKOL was the first person who listen to, comment on and criticize the core ideas of this thesis work. He helped me to look from different angles as he always blended the careful listening with his candid opinions. Our highly enjoyable intellectual discussions, accompanied by coffee and fresh air in the daytime and whisky and nuts in the night, helped a lot to polish my mind. Our deep talks are infamous for ending up with the most unexpected conclusions. I acknowledge all members of Papa John's Pizza Group (shortly Papa Johns or PJPG) for being such a coherent group, I’ll really miss our extremely funny and geeky parties at the New Energy Technologies Lab. Work hours (in fact after work hours as well) could not be without my coworker friends Murat Ferhat DOĞDU, Can AKSAKAL, Ahmet GÜLTEKİN, Osman ÜRPER, Ertuğrul DEMİR, Zeynep CAMTAKAN, Berker YURTSEVEN, Utku HARMANKAYA, Tuğçin KIRANT, Gökçen GÖKÇELİ, Uğur KAHVECİ, Neslihan KOYUNCU, EDAG friends as well as many other good friends who have worked with me in Energy Institute. Together we shared loads of nice memories, activities and joy. I’m grateful for all. My dear friends Erelcan YANIK and Burak ÖZAYDIN deserve also big credit, as members of our intriguing philosophical discussion group "Bitli Sinir", for our collective thinking without boundaries. I am still having a small hope for someday being able to regularly play basketball with my dear friend Erelcan. I especially thank to my childhood friend Çağın ÇİÇEK for our coherently absurd chit-chats, heart-to-heart talks, game-days and horror-nights with cool drinks, and to Yelda ÇİÇEK for enriching our meetings even more. I thank to Büşra TAYFUR and Berkay SAKİN for our laugh-filled lion milk table evenings, to my like-minded friend Luca TONDELLI for our enjoyable chats, activities and holidays, as well as to many other fantastic friends I’ve met during my business trip at Hawaii; a heaven on earth was the most entertaining with you all. Meetings with my undergrad circle, Samet ÖZCAN, Güven YILMAZ, Çığıl YILMAZ, Elif ÖZPAĞDA, Fatih ONUR, İrem LAÇİN, Gökçehan AKOĞUZ, Melek ARI and Atakan ARASAN, were occasional but enjoyably memorable. I love you all.

I spent one year of my Ph.D. at Uppsala, Sweden and at Jerusalem, Israel. My time at Uppsala was instructive and pleasant with Jonas, Henning, Juan David, Andreas, Johann, Emel, Paramita, Seif, Johannes, Oladunjoye, Francesco, Tomas, Anna and other nice people at the Villa and at the Angstrom Buildings. My Jerusalem times were far beyond my initial expectations, I never wanted my time there to end, thanks to good friendships of Ronnie, Marcel, Bar, Aviv, Efrat, Florian, Suleiman, Sayak, Roie, Laura, Estefania, Itay, Oren, both Ksenia’s, Han, Atılgan, Müslüm, Igor, Raam and other nice people at the Fritz Haber Center, also thanks to my roommates Sheng Fu, Ellen and dear Fransiscus Ismael with whom I feel we’ve developed a lifelong friendship bond. I’ll always miss our picnics, gatherings, trips and discussions. Beautiful people that I’ve met during conferences, seminars, meetings and workshops, I thank to luck for living those nice moments with you all which I won’t forget.

My cute, lovely sister Aslı, my sweet mother Hanife and my supportive father Öztürk, as well as my beloved aunts Nurten, Nurşen and Emine and my dear cousin Yasemin, I owe my greatest gratitude to all of you for the love and support that you have given to me.

Lastly, I thank to my friends who I couldn't mention here and are not necessarily related with the course of my Ph.D. thesis but surely important to me. I am glad to have/had you in my life.

Wow, wordy! Seems like I have been waiting for this... Here we go! \\

\begin{flushright}
Alhun AYDIN \\
Physicist \\
September 2020
\end{flushright}

%\vspace*{20px}
\vspace*{\fill}
\textit{Note: In this arXiv version of the thesis, several stylistic adaptations have been made from the original ITU format, for the sake of online reading compatibility.}

\newpage
\vspace*{60px}
\begin{huge}\textbf{Summary}\end{huge} \\
\vspace*{20px}

\begin{center}
\begin{large}\textbf{QUANTUM SHAPE EFFECTS}\end{large} \\
\end{center}
\vspace*{10px}

\noindent
How does geometry affect the physical properties of matter? Geometry has been considered as an important mathematical concept for understanding physical reality since the time of ancient Greek philosophers. The geometry of a physical object is associated with its sizes and overall shape. Sizes are characterized by volume, surface area, peripheral length and number of vertices of the object. With the development of quantum mechanics in the beginning of the last century, it is seen that nature has different appearances depending on the scale of physical systems. Physics of nanoscale (a billionth of a meter) is governed by quantum mechanics and nanomaterials have some superior properties in comparison with their macroscale counterparts.

Quantum mechanics taught us that matter exhibits both particle-like and wave-like characteristics, the so called wave-particle duality. Wave behavior is associated with the de Broglie wavelength, which is usually quite small for our macro world. Wave nature of particles become prominent when they are confined in domains with sizes that are comparable to their de Broglie wavelengths. In such a case, the physical properties of particles are affected by the confinement domain and so quantum size effects appear. Utilization of quantum size effects lead to tailoring and enhancing various properties of materials, constituting the backbone of modern nanoscience and nanotechnology.

Unlike size, shape is not so straightforward to define. In almost all systems size and shape effects coexist and interfere with each other. Is there any way to separate them? Can we change the shape of a domain without altering its sizes and focus on pure shape effects (that is completely overlooked)? 

This thesis shows the separation of quantum size and shape effects from each other. We propose the existence and explore the consequences of a new type of physical effect which we call quantum shape effect. We introduce a size-invariant shape transformation on nested domains which can be realized in core-shell nanostructures. Performing a rotation on the core structure causes a variation of the shape of the shell structure where the particles are confined. During this rotation all size parameters of the confined domain stay constant. By this way we perfectly separate quantum size and shape effects from each other and investigate quantum shape effects alone. Shape not only becomes a control parameter on the material properties, but also leads to novel physical behaviors which have never been seen before.

In the thesis, after the introduction to the topic, literature overview and a review of quantum size effects, we introduce quantum shape effects in the third chapter in detail. We solve time-independent Schr\"{o}dinger equation numerically for the confinement domains that are constituted by the nested structures with various geometries. Eigenvalue spectrum for each angular configuration is obtained and used to calculate partition function and all other thermodynamic quantities. It is demonstrated that the thermodynamic properties of non-interacting particles strongly confined in nested nanostructures significantly change with shape. Next, we develop an analytical method to predict the quantum shape dependence of thermodynamic state functions as well as to develop a physical insight to the quantum shape effect phenomenon. Other known methods such as Weyl density of states and first two terms of Poisson summation formula cannot predict any shape-dependence in the thermodynamic properties. Our analytical methodology is based on the quantum boundary layer approach. Considering the overlaps of quantum boundary layers forming in nested domains reveals the information about the shape dependence of the properties of confined particles. We call the analytical model as overlapped quantum boundary layer method and its accuracy is quite good at estimating the functional behaviors of shape-dependent thermodynamic properties. Influence of various boundary conditions and quantum size effects on quantum shape effects are also investigated.

Thermodynamic properties such as internal energy, free energy, entropy and specific heat of particles are examined under quantum shape effects for particles obeying Maxwell-Boltzmann and Fermi-Dirac statistics. Their behavior shows exotic characteristics that are previously unseen in the thermodynamics of confined non-interacting gases. Shape dependence of the chemical potential of electrons produces a novel kind of quantum oscillations which are intrinsically different than density- or size-dependent quantum oscillations.

Due to quantum shape effects, free energies of various angular configurations are different from each other. This suggests a spontaneous rotation of the core structure as a result of the torque generated by the particles confined within the shell structure to minimize their free energy. Formation of non-uniform and asymmetric pressure distribution even at thermodynamic equilibrium is the principal cause of this torque of quantum origin. 

From the application point of view, quantum shape effects lead to some novel heat engine and refrigeration cycles, opening up new possibilities in nanoscale thermodynamics. We propose the existence of a new thermodynamic process under constant shape, we call isoformal process. The thermodynamic cycles featuring isoformal process are examined and they show various novel properties that are not encountered in conventional thermodynamic cycles. It is also possible to design new nanoscale energy conversion devices based on quantum shape effects. A number of possible applications are presented in the fifth chapter. As a whole, this thesis constitutes a comprehensive investigation of the theory, methodology and applications of quantum shape effects in thermodynamics, which hopefully have a great potential to bring new ideas and advances to the field of nanoscale physics and energy. 

\newpage
\vspace*{60px}
\begin{huge}\textbf{Özet}\end{huge} \\
\vspace*{20px}

\begin{center}
\begin{large}\textbf{KUANTUM ŞEKİL ETKİLERİ}\end{large} \\
\end{center}
\vspace*{10px}

\noindent
Geometri maddenin fiziksel özelliklerini ne şekilde etkiler? Antik Yunan filozoflarının zamanından beri geometri fiziksel gerçekliği anlamada önemli bir matematiksel kavram olarak görülmüştür. Bir fiziksel nesnenin geometrisi genelde o nesnenin ölçeksel büyüklükleri (ebatları) ve bütünsel şekli ile ilişkilendirilir. Bir nesnenin ebatları Lebesgue ölçüsü altında o nesnenin hacmi, yüzey alanı, çevresel uzunluğu ve köşe sayıları ile belirlenir. Örneğin üç boyutlu bir nesne için esas ölçeksel büyüklük hacim iken, yüzey alanı, çevresel uzunluk ve barındırdığı köşe sayıları ise üç boyutlu nesnenin düşük boyutlu ölçeksel büyüklükleri olarak tanımlanır. Geçtiğimiz yüzyılın başında kuantum mekaniğinin keşfedilmesi ile birlikte doğanın fiziksel sistemlerin boyutuna ve ölçeğine göre farklı davranışları olduğu gözlemlendi. Metrenin milyarda birini ifade eden nano ölçek fiziği günümüzde kuantum mekaniği yasaları tarafından anlaşılabiliyor. Nano ölçekteki fizik bilim insanları açısından oldukça cazip bir konu zira hem teorik hem deneysel olarak gösterildiği üzere nano ölçek malzemeler birçok açıdan makro ölçekteki malzemelere göre üstün özelliklere sahip. Enerji bilim ve teknoloji alanında da nanoyapıların uygulaması gün geçtikçe artmakta ve bu malzemelerin fiziksel davranışlarının doğru ve detaylı bir biçimde anlaşılması önem arz etmektedir. Geçmişte teoride sınırlı kalan birçok fiziksel olgu, günümüzde laboratuvar olanaklarının hızlı gelişimi sayesinde deneysel olarak gösterilebilmekte, hatta bir kısmı ticari olarak uygulanabilmektedir.

Kuantum fiziği maddenin hem dalga hem de parçacık davranışları gösterdiğini ortaya koymuştur. Buna dalga-parçacık ikiliği diyoruz. Parçacıkların dalga karakteri de Broglie dalga boyları ile ölçülür ki bu genelde oldukça küçüktür. Parçacıkların içinde sınırlandığı domenin ebatları de Broglie dalga boyları ile karşılaştırılabilir bir mertebede ise parçacıkların dalga doğası önem kazanır. Böyle bir durumda, parçacıkların fiziksel özellikleri tutuklanma domeninden etkilenir ve kuantum ölçek etkileri ortaya çıkar. Kuantum ölçek etkileri malzemelerin çeşitli özelliklerini belirlenen amaca uygun ve daha iyi hale getirmeye yol açarak çağımız nanobilim ve nanoteknolojisinin temel taşını oluşturur.  

Bir nesnenin ebatlarının veya ölçeğinin aksine şeklini tanımlamak ve sayısallaştırmak çok daha zordur. Neredeyse tüm fiziksel sistemlerde ölçek ve şekil bir arada birbiri içine geçmiş bir şekilde bulunur. Bir nesnenin büyülüğünü ve şeklini ayırıp, ayrı ayrı inceleme altına almak olası mıdır? Sınırlandırılmış bir domenin ebatlarını değiştirmeden şeklini değiştirmek ve bu sayede yalnızca şekil etkilerini incelemek mümkün müdür? Sınırlandırılmış domenlerdeki parçacıklar üzerinde şekil etkilerini ölçek etkilerinden tamamen ayrı inceleme olasılığı literatürde şimdiye kadar göz ardı edilmiştir.

Bu tezde kuantum ölçek etkileri ve kuantum şekil etkileri birbirinden tamamen ayrılmıştır. Tezde kuantum şekil etkileri olarak adlandırdığımız yeni bir fiziksel etkinin varlığını ortaya koyuyor ve sonuçlarını tetkik ediyoruz. Birbirinin içine geçmiş domenlerde ölçekten bağımsız şekil değiştirimi tekniğini gösteriyoruz. Literatürde olduğu gibi bu tarz iç içe geçmiş domenler deneysel olarak çekirdek-kabuk nanoyapılarında gösterilebilir. Çekirdek nanoyapıda gerçekleştirilecek döndürme hareketi, kabuk ile çekirdek yapı arasında tutuklanmış parçacıkların sınırlandığı domenin şeklini değiştirir. Bu dönme hareketi esnasında parçacıkların bulunduğu domenin bütün ölçek değişkenleri aynı kalır. Bu sayede kuantum ölçek ve şekil etkileri birbirinden tamamen ayrılarak yalnızca kuantum şekil etkilerini inceleme olanağı oluşur. Şekil, malzeme özellikleri üzerinde bir kontrol değişkeni haline gelmekle kalmaz, aynı zamanda daha önce görülmemiş yepyeni fiziksel davranışların ortaya çıkmasına sebebiyet verir.

Tezin ilk bölümünde tezin motivasyonu ve tez konusu tanıtılarak, literatür incelemesi ile beraber tez çalışmasının ana çıktıları verilmiştir. İkinci bölümde kuantum ölçek etkilerinin istatistiksel termodinamik özelinde bir derlemesi yapılmıştır. Kuantum şekil etkilerinin nasıl ortaya çıktığı ve temelleri tezin üçüncü bölümünde detaylı bir biçimde incelenmiştir. Çeşitli geometrilerdeki iç içe geçmiş nanoyapılardan oluşan tutuklama domenleri için zamandan bağımsız Schrödinger denklemi sayısal olarak çözülmüştür. Çekirdek nanoyapı dediğimiz içteki objenin her bir açısal durumu için özdeğer görüngesi elde edilmiş ve bu özdeğerler kullanılarak bölüşüm fonksiyonu ile beraber diğer termodinamik büyüklükler hesaplanmıştır. İç içe geçmiş nanoyapılarda tutuklanmış etkileşmeyen parçacıkların termodinamik özelliklerinin şekil bağımlılığı ortaya konmuştur. Ardından bu şekil bağımlılığını öngörmek için analitik bir yöntem geliştirilmiştir. Geliştirilen yöntem tutuklanmış parçacıkların termodinamik özelliklerinin şekil bağımlılıklarının fonksiyonel davranışını doğru öngörmekle kalmayıp kuantum şekil etkisi olgusuna fiziksel bir kavrayış getirmeyi başarmıştır. Analitik yöntemimiz kuantum sınır tabaka yaklaşımı üzerine geliştirilmiştir. Kuantum sınır tabakaların üst üste bindiği (örtüştüğü) bölgelerin büyüklüğü parçacıkların bulunduğu domenin şekil bilgisini taşır. Bu örtüşen bölgelerin miktarları içteki nanoyapının dönüşü sırasında açıyla beraber değişir. Bu sayede domenin termodinamik özelliklerini bu örtüşme bölgelerini de göz önüne alan bir efektif hacim ile ilişkilendirmek mümkün olur. Örtüşen kuantum sınır tabaka modelimiz iç içe domenlerde güçlü bir şekilde tutuklanmış parçacıkların termodinamik özelliklerini oldukça iyi bir doğrulukla analitik olarak öngörmektedir. Kuantum şekil etkilerinin çeşitli sınır koşullarındaki davranışı ve kuantum ölçek etkileri sebebiyle değişimi de tezin bu bölümünde incelenmiştir.

Tezin dördüncü bölümünde parçacıkların iç enerji, Helmholtz serbest enerji, entropi ve özgül ısı gibi termodinamik özellikleri kuantum şekil etkileri altında Maxwell-Boltzmann ve Fermi-Dirac istatistikleri çerçevesinde ayrı ayrı incelenmiştir. Kuantum şekil etkileri sebebiyle bu termodinamik büyüklüklerin daha önce etkileşmeyen gazların termodinamiğinde görülmemiş ilginç fiziksel davranışlar gösterdiği görülmüş ve bu davranışların kökenleri ve mekanizmaları kurulan analitik modelin de yardımıyla açıklığa kavuşturulmuştur. Ayrıca elektronların kimyasal potansiyelinin şekil bağımlılığının yoğunluk veya ölçeğe bağlı kuantum salınımlardan temelde farklı olan başka bir tür kuantum salınımı gösterdiği ortaya konmuştur. Bu özgün kuantum salınımının özellikle özgül ısıda güncel deneysel olanaklarla gösterilebilecek büyüklükte değişimlere yol açtığı gözlenmiştir.

Kuantum şekil etkilerinden ötürü farklı açısal durumların serbest enerjileri birbirinden farklı olmaktadır. Bu da tutuklanmış parçacıkların serbest enerjilerini minimize etme amacıyla dönebilme serbestliği bulunan iç nanoyapı üzerinde tork uygulayacağını ve iç nanoyapının kendiliğinden dönerek serbest enerjinin minimum olduğu açıda duracağını işaret eder. Nano ölçekte tutuklanmış yapılarda termodinamik denge durumunda dahi geometrik simetrinin bozulduğu her durumda asimetrik ve düzensiz bir basınç dağılımı oluşur. İç içe geçmiş domenlerde bu tez çalışması kapsamında ortaya konan kuantum-mekaniksel torkun ortaya çıkmasının temel nedeni budur.

Kuantum şekil etkileri nano ölçek termodinamiğinde yeni uygulama olanakları da açar. Bu tezde sabit şekil durumunda izoformal proses olarak adlandırdığımız yeni bir termodinamik prosesin varlığını ortaya koyduk. İzoformal prosese dayalı özgün ısıtma ve soğutma çevrimleri önerdik. İki izotermal, iki izoformal prosesten oluşan bir termodinamik çevirimi ile iki izentropik, iki izoformal prosesten oluşan bir termodinamik çevrimin analizlerini yaptık ve alışılagelmiş termodinamik çevrimlerde karşılaşılmayan bazı özellikler içerdiğini gördük. Termodinamik çevrimlerin yanı sıra kuantum şekil etkilerine dayalı yeni nano ölçek enerji dönüşüm cihazlarının tasarlanması da mümkündür. Bu bağlamda muhtemel uygulamaların birkaçı tezin beşinci bölümünde sunulmuştur. Tek malzemeli tek kutuplu termoelektrik cihazlar ve kuantum Szilard ısı makineleri kuantum şekil etkilerinin uygulanabileceği birçok farklı alandan sadece birkaçıdır. Kuantum şekil etkilerinin termodinamikte ortaya çıkışı, teorisi, yöntemleri ve uygulamaları kapsamlı bir şekilde bu tezde incelenmiştir. Tezle ilişkili yapılan bazı çalışmaların da gösterdiği üzere, bu tezde ortaya konan yeni fiziksel etki ve uygulamaları nano enerji bilimi ve teknolojisinde yepyeni fikirlere ve gelişmelere yol açma potansiyeline sahiptir. 

\tableofcontents
\newpage
%%%
%1%
%%%

\chapter{Introduction}
\section{Motivation}
% Curiosity
We, humans, are curious animals. Science, and even technology, are still first and foremost driven by this unceasing curiosity, despite the gradual changes in the priorities of people during recent decades towards more economic and vanity-driven concerns. It's maybe not clear whether technology has made our lives easier or science has made us wiser, yet, one thing is clear that we are not just wondering things, but also understanding the nature of some things by using our intelligence, senses and the tools that we have created so far. The more we know things, the more we realize how large the lower part of the iceberg of knowledge might possibly be, and even more our intellectual curiosity grows.

% Quantum
Quantum theory, the physics of the tiny scales, is one of the biggest products in this quest of the roads leading to the true nature of things. Nature behaves surprisingly different at small scales. Many phenomena discovered at quantum realm are exceedingly counter-intuitive to us living in and experiencing a macroscopic world. But still we know by now that some things are not what they appear to be. For instance, it's nearly impossible to see the roundness of the Earth by the naked eye, when you look to the horizon. Similarly, we don't notice in our macro world the weird behaviors appearing at quantum scales. Even so, quantum effects actually play a significant role in our modern life, since many devices that we use today such as transistors, lasers, navigation devices and magnetic resonance imagers, directly rely on the principles of quantum mechanics.

% Nano
So, how is the physics of the small scales? First of all, by small we mostly mean nanoscale, which is sometimes also called quantum scale. Nanometer is one billionth of meter, and approximately one hundred thousandth of a human hair. We encounter with quantum phenomena at not only small scale, but also low temperature and low mass conditions. At least one of these conditions need to be satisfied in order for quantum effects to appear. Under these terms, nature exhibits some phenomena that cannot be explained by classical, pre-quantum, physics. Quantum mechanics has shown that matter has a probabilistic wave nature, see Fig. 1.1 where the famous double-slit experiment is illustrated. This concept underlies the roots of many different quantum phenomena such as Heisenberg uncertainty principle, coherence/decoherence, entanglement, superposition, tunneling, wavefunction collapse during a quantum measurement, zero-point energy, Casimir effect, discreteness of certain physical quantities, indistinguishability of identical particles and so on. Some weird consequences of these phenomena are: inherent uncertainty in position and velocity of particles (Heisenberg uncertainty principle), interconnectedness of particles that have a shared past (quantum entanglement) and inexistence of an objective reality before interacting with the particles (quantum probabilistic nature of wavefunction) \textit{et cetera}. All these quantum phenomena have surprised the world a lot and still have been continuing to surprise even scientists. We'll discuss more deeply on the wave nature of particles in the next chapter.

\begin{figure}[b]
\centering
\includegraphics[width=0.95\textwidth]{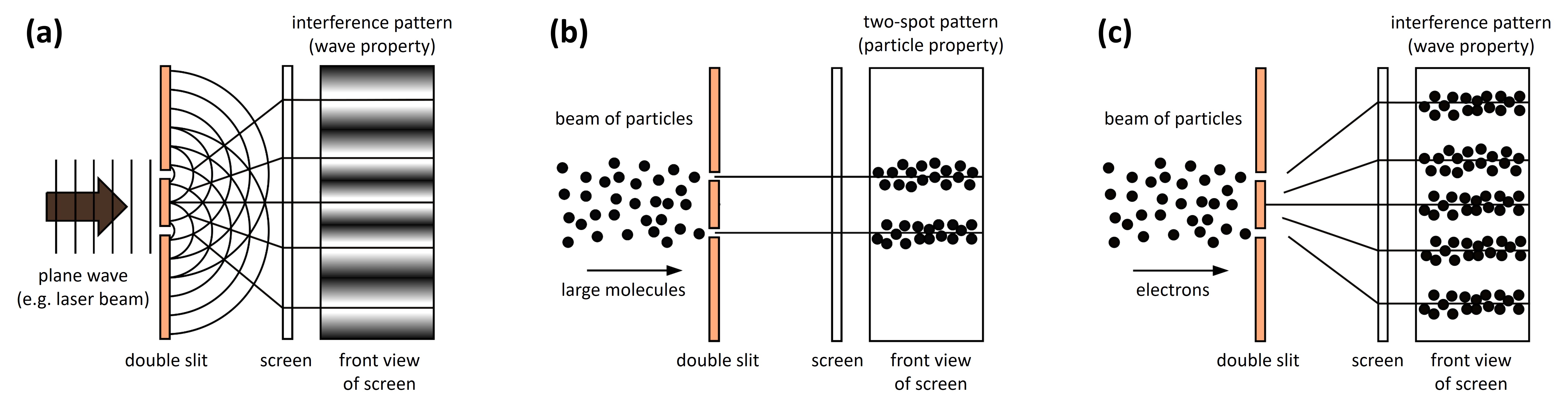}
\caption{Double-slit experiment with (a) photons, (b) very large particles, (c) electrons. Photons and electrons passing the double-slit form interference patterns on the screen, showing the wave nature of particles. Particles on (b) are too large/massive to exhibit their wave nature.}
\label{fig:c1f1}
\end{figure}

% What we will focus? (Thermodynamics)
What we will focus on in this thesis is the thermodynamics at quantum scales. Before that, let's briefly mention what thermodynamics deals with. Thermodynamics is the branch of physics that deals with the relationships between heat, work and other forms of energy. Let's concretize this over a simple example: Consider a gas confined in a macroscopic cylinder by a piston as shown in Fig. 1.2. For simplicity, we assume a weightless piston, depicted by the red bar that can move up or down without any friction. On top of the piston, there is a weight. The gas has a temperature and a pressure which are in equilibrium with the environment. In other words, both the temperature and the pressure of the gas are constant in the beginning. Now, let's give some heat to the system by bringing it in contact with a heat reservoir having a higher temperature than the gas. The temperature of the gas will rise and because of that the gas pressure will start to increase. However, remember the piston was free to move and the outside pressure is constant, it is just the atmospheric pressure plus the pressure exerted by the weight. Therefore, the system will try to keep the pressure in equilibrium with the outside pressure. To do that it will expand its volume and push the weight upwards, thereby doing work on the weight. This is one of the simplest heat engines. It converts heat into work or potential energy. Refrigerators, air conditioners, power plants, internal combustion transportation vehicles are all examples of thermodynamic machines driven by thermodynamic principles.

\begin{figure}[t]
\centering
\includegraphics[width=0.25\textwidth]{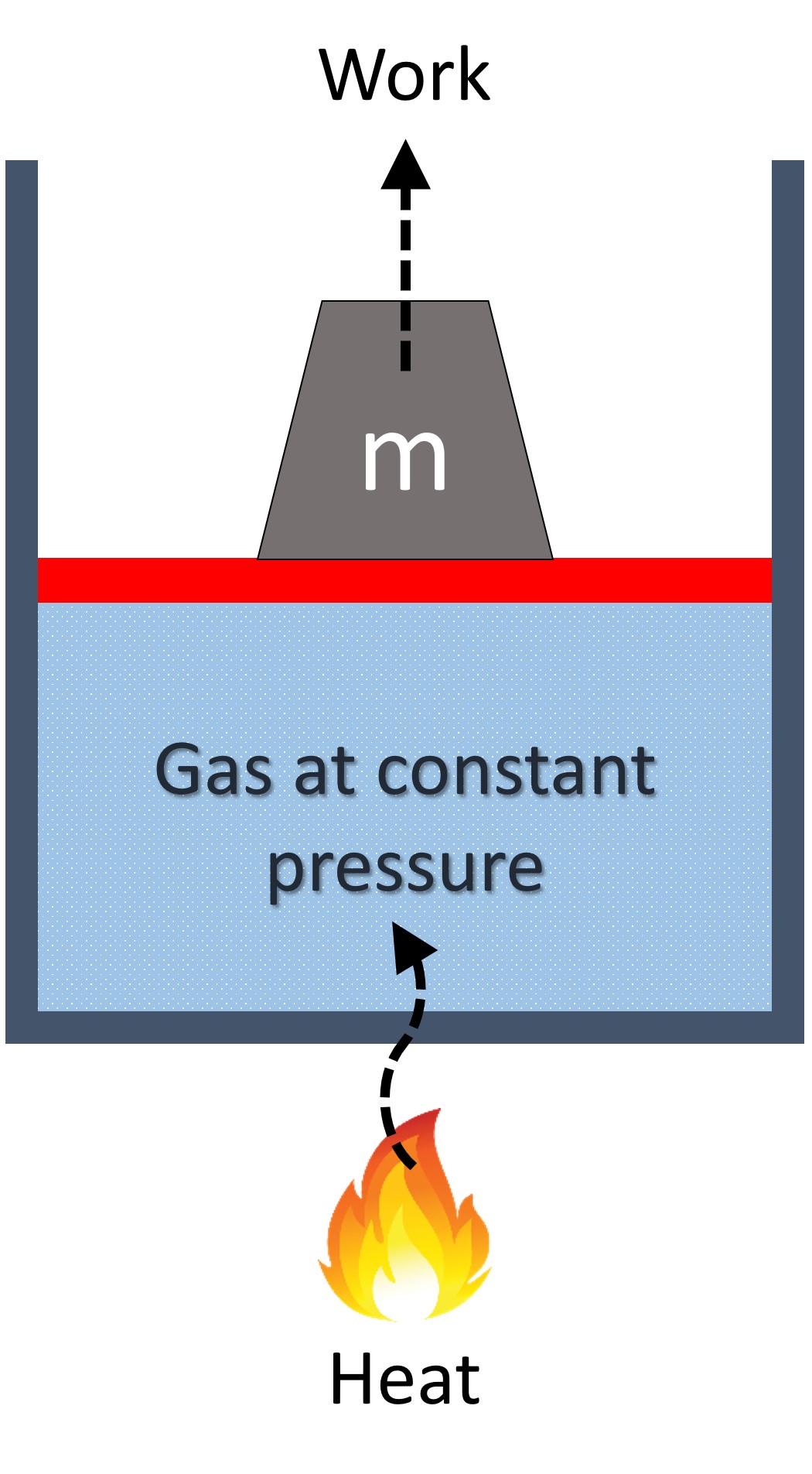}
\caption{A simple thermodynamic machine that converts heat into work. A gas at constant pressure held in a cylinder by a movable, frictionless piston, carrying a weight with mass $m$. Moving piston does work on the weight.}
\label{fig:c1f2}
\end{figure}

Going hand in hand with the above example, thermodynamics has started as a phenomenological theory during the beginning of the industrial revolution\cite{prigo1} (nowadays it's called the first industrial revolution by the trendy perspective). In the late 19th century, mainly with the efforts of James Clerk Maxwell, Ludwig Boltzmann and Josiah Willard Gibbs, it has developed into an analytical theory that is explainable by the statistical properties of a macroscopic physical system composed of many particles or states. Thus, statistical mechanics, one of the pillars of modern physics, was born. Statistical mechanics uses statistical methods and microscopic physical laws to explain the thermodynamic behaviors of macroscopic systems. It is also called statistical thermodynamics when it is applied to explain, in particular, thermodynamic properties of systems. 

Thermodynamics is usually considered as one of the most well-established disciplines of all science. It has very few (simple) premises, yet has a broad applicability with great success. But still, the statistical nature of its laws is both its strength and also its weakness. Despite its power of applicability, the quantum mechanical origins (which is different than classical microscopic origins) of thermodynamic laws are still a mystery. What is the meaning of temperature of a quantum object? What is entropy at quantum scale? How does heat behave in the quantum realm? Quantum thermodynamics has emerged as a stand-alone research field quite recently to answer all these questions and more\cite{qtbook09,qtbook17,qtbook18,qtbook19}. Before even coming to all these questions, we argue that even in equilibrium quantum statistical thermodynamics, there are interesting phenomena that yet to be discovered. How small can we reduce the sizes of our devices? What is the influence of system's geometry, on its physical properties? Nanoscale manipulation has reached a level so extreme that facing these questions are now inevitable. To this end, in this thesis, we will theoretically and computationally explore the quantum-mechanical influence of geometry in the thermodynamics of confined systems. A more detailed description of the thesis topic is given in next section.

% Future
From a broader perspective, utilization of quantum phenomena has a potential to revolutionize the current state of computing, communication, and energy technologies, to make them faster, safer and more efficient than the conventional ones\cite{PhysRevLett.117.100001}. In more detail, quantum computing can perform certain type of operations significantly faster by making use of quantum superposition of states. Quantum cryptography in principle provides a $100\%$ secure information transfer. Design and fabrication of new nanodevices and nanomaterials improves the efficiency (e.g in thermoelectrics)\cite{tebook2013,tebook2014}, stiffness (e.g graphene)\cite{graphene,stiff}, sensitivity (e.g quantum sensors)\cite{RevModPhys.89.035002} and capacity (e.g super/ultracapacitors)\cite{supercap} in energy generation, conversion and storage technologies. Nanotechnology offers to design materials with the desired mechanical, electrical, optical and thermal properties\cite{nanomat}. Nanostructured materials can make the solar cells cheaper and more efficient\cite{solarcell,solarcell2}, which may increase the usability of solar energy drastically. Quantum thermodynamics stands on the ground of all these seemingly distinct research fields. Substantial development can only be obtained when a solid quantum thermodynamic framework is established. As of 2020, today's conventional computers/smartphones use transistors with 7 nm semiconductor manufacturing processes, the scale of which was 32 nm ten years ago, 180 nm twenty years ago and 600 nm thirty years ago and 10{\textmu}m in 1971. Until recent years, we were still on the edge of the validity of classical thermodynamics. However, with the leap forwards in nanofabrication, we are now more aware of the major obstacles in the way of outstanding breakthroughs. Understanding and controlling the heat dissipation and enabling efficient cooling at quantum scales are still ongoing challenges. 

% Spendings
Due to the substantial influence on vastly different fields of modern life, countries spend huge resources on nanoscale/quantum research and development. For nanoscience and nanotechnology researches, the United States has invested nearly $\$27$ billion since 2001, and announced $\$1.4$ billion for the 2019 budget of the National Nanotechnology Initiative \cite{USNNIreport}, while the European Union spends around \euro{1.7} billion under Horizon 2020 Work Programme for 2018-2020 \cite{EU2020report}. Very recently, both the European Commission and the United States Congress announced their massive research programmes for quantum science and technologies, which are \euro{1} billion for the Quantum Flagship Programme and $\$1.3$ billion for the National Quantum Initiative respectively \cite{EUQFreport,USNQIreport}. Numerous projects on quantum and nanoscale thermodynamics field have been supported under these umbrella projects during recent years. Considering the trend and the direction of technology, it is almost certain that funding of these areas will continue increasingly. But regardless of the ongoing trend, nanoscale thermodynamics field provides me enough personal, scientific motivation to work on because it seeks for answers to deep and fundamental questions of physics. 

\section{Topic Statement}

% Discrete energy spectrum and summations
In statistical physics, physical properties of systems are calculated through a probability distribution function by summing over all possible values of quantum state variables (which are infinitely many in general) in each degree of freedom. A degree of freedom is any parameter that specifies a state. Quantum state variables determine the energy levels of a quantum system. Although energy levels are essentially discrete, it is customary to use continuum approximation and thereby replacing the infinite summations with integrals. Continuum approximation assumes the energy spectrum to be continuous rather than discrete. This assumption provides simpler, analytical expressions for physical quantities and make it easy to see and interpret the functional relations and the main physics.

Even though continuum approximation works well at macroscale, it fails to capture the characteristics of the systems confined at nanoscale, due to the reasons that we will explore in detail in the next chapter. Instead of replacing the summations directly with integrals, approaching them using better mathematical tools like Poisson summation formula (PSF) has been studied in the literature. First implementations of PSF have been done on Casimir effect and on lattice sums \cite{baltes,path1,path2} during 1960's and 70's. Later, this methodology has been extended to study the finite-size effects or quantum size effects in thermodynamic properties of confined systems \cite{baltes,rmpqse,molina,qsenat,pathria,dai1,dai2,sismanmuller,sisman,suprqsesc,PhysRevE.53.2360,pwallsis,qsewff,emequse,bcsbecqse,qsenjpg,qsescirp,qoscnat,RevModPhys.86.1127}. It has been realized that there is a connection between PSF and Weyl conjecture (which describes the asymptotic behavior of the eigenvalues of the Laplacian) \cite{weyl11,baltes}. Weyl conjecture has also been used to obtain the quantum size effect corrections to thermodynamic expressions \cite{baltes,qforce,aydin3}. Even a much more powerful concept called quantum boundary layer, allowing to get quantum size effects without solving the Schr\"{o}dinger equation explicitly, has been developed in 2006 by Sisman and his Nano Energy Research Group \cite{qbl,uqbl,nanocav}.

As a result of quantum size effects, considering the discreteness of the energy spectra by using the infinite summations leads to many interesting results in the thermodynamic and transport properties of confined systems. For example, extensivity rule breaks down at nanoscale and thermodynamic quantities become non-additive\cite{baltes,sismanmuller}. Pressure of a confined gas becomes a tensorial quantity even at thermodynamic equilibrium\cite{sismanmuller}. Although thermodynamics has always been shown as a theory of continuous variables, in 2014, intrinsic discrete nature in the thermodynamic properties of confined and degenerate Fermi gases has been shown\cite{msc,aydin1}. As manifestations of quantum size effects, dimensional transition points in thermodynamic properties of Maxwell-Boltzmann gases have been explored \cite{aydin2}. Discrete and Weyl density of states concepts are introduced to represent the thermodynamic properties of confined systems more accurately \cite{aydin3,aydin8}. The phase diagram of quantum oscillations in confined and degenerate Fermi gases has been constructed and the oscillations are successfully predicted by the half-vicinity model \cite{aydinhvm,aydin4,aydin5,aydin6}. In all these studies quantum size effects have been explored.

So far in literature, shape effects have never been investigated solely, because size and shape effects were inherently linked to each other. In this thesis, by building on the previously acquired knowledge on quantum confinement effects, we separate them from each other and can focus only on shape. Shape of an object is described as the geometric information which is invariant under Euclidean similarity transformations such as translation, rotation, reflection and uniform scaling. We propose and explore a new type of quantum effect, which we call the \textit{quantum shape effect}. The topic of this thesis is to introduce and examine quantum shape effects on the thermodynamic properties of nanoscale systems. We establish the theoretical background of this new effect and design novel nanoscale energy devices based on them. Fundamental questions we are going to answer (Spoiler: The answer is YES to all) in this thesis are listed below:
\begin{itemize}
\item Is there a way to change the shape of an object without changing its sizes?
\item Does shape enter as a separate control variable to the thermodynamic state space?
\item How do thermodynamic potentials and functions change solely with shape? 
\item Is it possible to analytically predict the shape-dependence of physical quantities?
\item Can we design novel nanoscale heat engines based on quantum shape effects?
\end{itemize}

We've introduced our first results on quantum shape effects in March 2017 at the 5th Quantum Thermodynamics Conference, held in Oxford, United Kingdom. Subsequently within the next years, we've presented our works on developing quantum shape effects in several other international conferences, workshops and seminars. We've published the fundamental results of this thesis in Ref. \cite{aydin7}. The thesis constitutes an introduction to and a comprehensive examination of quantum shape effects in the thermodynamics of confined systems along with a thorough review of its now established background.

\section{Literature Overview}

In this section, we will summarize the research that has been done in literature related to the subject of this thesis. Some concepts might appear to be mentioned without explicit definitions, however, they would hopefully be clear during the second chapter of the thesis.

% From general things to shape
Development of new techniques and technologies makes it possible to create and manipulate nanoscale systems much easier than before \cite{subasi11,class,qtech,Kurizki2015,dvira}. Many nanoscale systems exploiting quantum effects and having great application possibilities have been realized in recent years, such as nanomotors, single-atom heat engines etc.\cite{naturerot,naturerot2,PhysRevLett.102.230601,naturerot3,naturerot4,eplrot,goldmot,PhysRevLett.111.060802,Goswami_2013,PhysRevLett.112.030602,PhysRevLett.114.183602,atomhe,PhysRevE.98.052124,PhysRevE.92.032105,nanomotor5,PhysRevLett.116.163602,quamach,turkp,anrevw,neqqhm,PhysRevB.97.245414,gyrator,flywheel,entkos,PhysRevE.99.042121,PhysRevE.100.012109,patern1,thermcohe}. Study of confined systems and size-dependent phenomena are very active and promising research areas since they can revolutionize our understanding of thermodynamics at nanoscale as well as contributing to the development of nanoscale energy conversion and storage devices with excess properties \cite{rodun,mitchen,qbook}.

% QSE in general
Quantum size effects have shown to be of great importance in nanoscale thermodynamics, in fact it has been shown that they put some fundamental limitations on work extraction from non-equilibrium states\cite{limitsnano}. Quantum size effects are also very fundamental in nanoscale transport phenomena\cite{rodun,qsemet}. Indeed, they result to even some conceptual changes in physics such as questioning the meaning of "conductivity" at nanoscale and preference to use the word "conductance". Conductance quantization, De Haas-van Alphen effect, Shubnikov-De Haas effect are some of the important manifestations of quantum size effects which pretty much shaped the modern nanoscience and nanotechnology\cite{binek}. Quantum size effects are especially important for nanoscale energy conversion and storage technologies, in particular for thermoelectricity\cite{te1,te2,tebook2014}. Charge and heat transport even in a single molecular junction have attracted recent interest\cite{PhysRevB.94.054311,franss17,tejunc2018a}. As an energy application of quantum size effects, thermosize effects, which can be considered as a sub-branch of thermoelectric effects, proposed by Sisman and M\"{u}ller in 2004 and have been studied extensively during the last decade\cite{sismanmuller,tsef2008a,tsef2008b,tsef2010a,tsef2011a,tse2,tse3,tse4,tsef2012a,tsef2012b,tsef2013a,tsef2013b,tsef2014a,tse5,tsef2018a,tsef2018b}. Quantum size effects is a broad topic on its own, but to fully characterize the geometry of confined systems, we need more than just size.

% Topological
The role of geometry in physics is quite deep and diverse. Gravity, one of the fundamental forces in physics, is explained by the geometry of spacetime in general relativity. Along with topologically protected properties of matter, geometric properties like geometric phases and forces attract considerable attention nowadays. Berry phase induced forces \cite{PhysRevLett.97.190401,PhysRevA.76.042109} and using those to drive tiny nanomotors have been studied for that matter \cite{ruiteleiden}.

% Shape in spectral geometry, can one hear?
A more related problem to the topic of this thesis comes from the spectral geometry. In 1966, Polish mathematician Mark Kac popularized a fascinating problem in spectral geometry by posing an elegant question: "Can one hear the shape of a drum?" \cite{kac66}. Of course at first thought, it's reasonable that differently shaped drums will sound different which is supported by our daily-life experiences. For example, sound difference between high tom and low tom is due to their diameter difference. (Asking this question mathematically is something else of course.) What has been asked actually is "Are there \textit{differently} shaped drums that sound exactly \textit{the same}?" Kac didn't know the answer, however studies related to this problem actually go back to the end of the 17th century, when an English polymath Robert Hooke observed nodal patterns in vibrating plates. About a century later, a German physicist and musician Ernst Chladni made first systematic experiments on the phenomenon. When a plate fixed at the middle is set into vibration, excited modes of vibration can be visualized by sand grains poured on the plate as shown in Fig. 1.3. Sand grains accumulate towards the non-oscillating nodal lines of the plate and form patterns unique to each mode, which is called Chladni patterns. In 1821, a French mathematician Sophie Germain made a partial mathematical description. Membrane oscillations (also called cymatics) or in general the wave phenomenon, are mathematically formulated by Helmholtz partial differential equation. Solution of the Helmholtz equation for a domain gives eigenvalues and corresponding eigenfunctions, which define the vibrational and oscillational characteristics of the domain \cite{RevModPhys.82.2213}.
	
\begin{figure}[!b]
\centering
\includegraphics[width=0.85\textwidth]{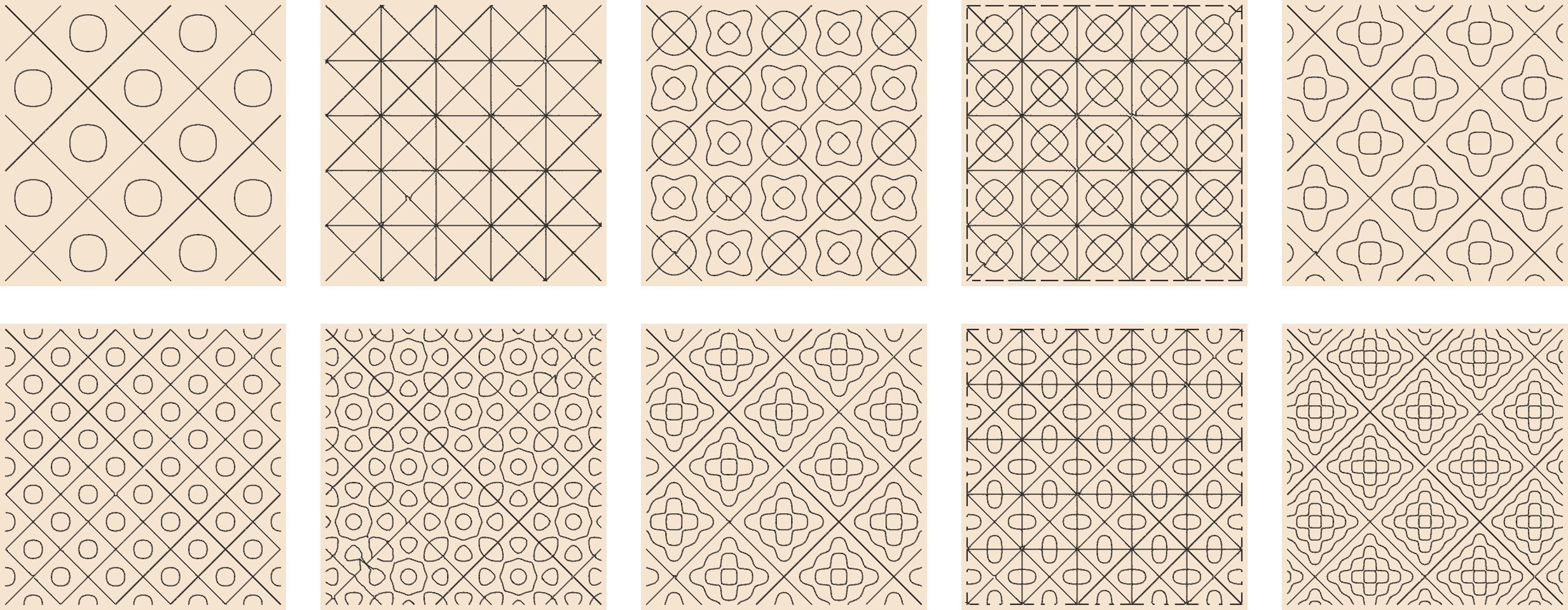}
\caption{Chladni patterns produced by scattered sand on an elastic, rectangular plate. When the plate is forced to vibrate (say through a loud speaker), grains end up in the places with zero vibration, thereby visualizing the modes of vibration on the plate surface.}
\label{fig:c1f3}
\end{figure}

While a result answering Kac's question has been published in 1964 by J. Milnor for 16-dimensional tori, it was not until 1992 that the first generalized mathematical proof answering Kac's question has been announced by Gordon, Webb and Wolpert \cite{tori64,cannothear,ansno}. They created two (pair of) domains having different shape but exactly the same eigenvalue spectrum (see Fig. 1.4). These kinds of domains are called isospectral. Their method of proof is based on Sunada's theory \cite{sunada}, though there is more than one way to show the existence of differently shaped yet isospectral domains \cite{RevModPhys.82.2213}. Besides showing it mathematically, there has been attempts to show isospectrality also experimentally, but of course it requires very precise equipments as well as ideal conditions \cite{PhysRevLett.72.2175,exphsdr,PhysRevE.72.056211,PhysRevLett.97.050404,manonat,hari,PhysRevE.85.036604}.

\begin{figure}
\centering
\includegraphics[width=0.5\textwidth]{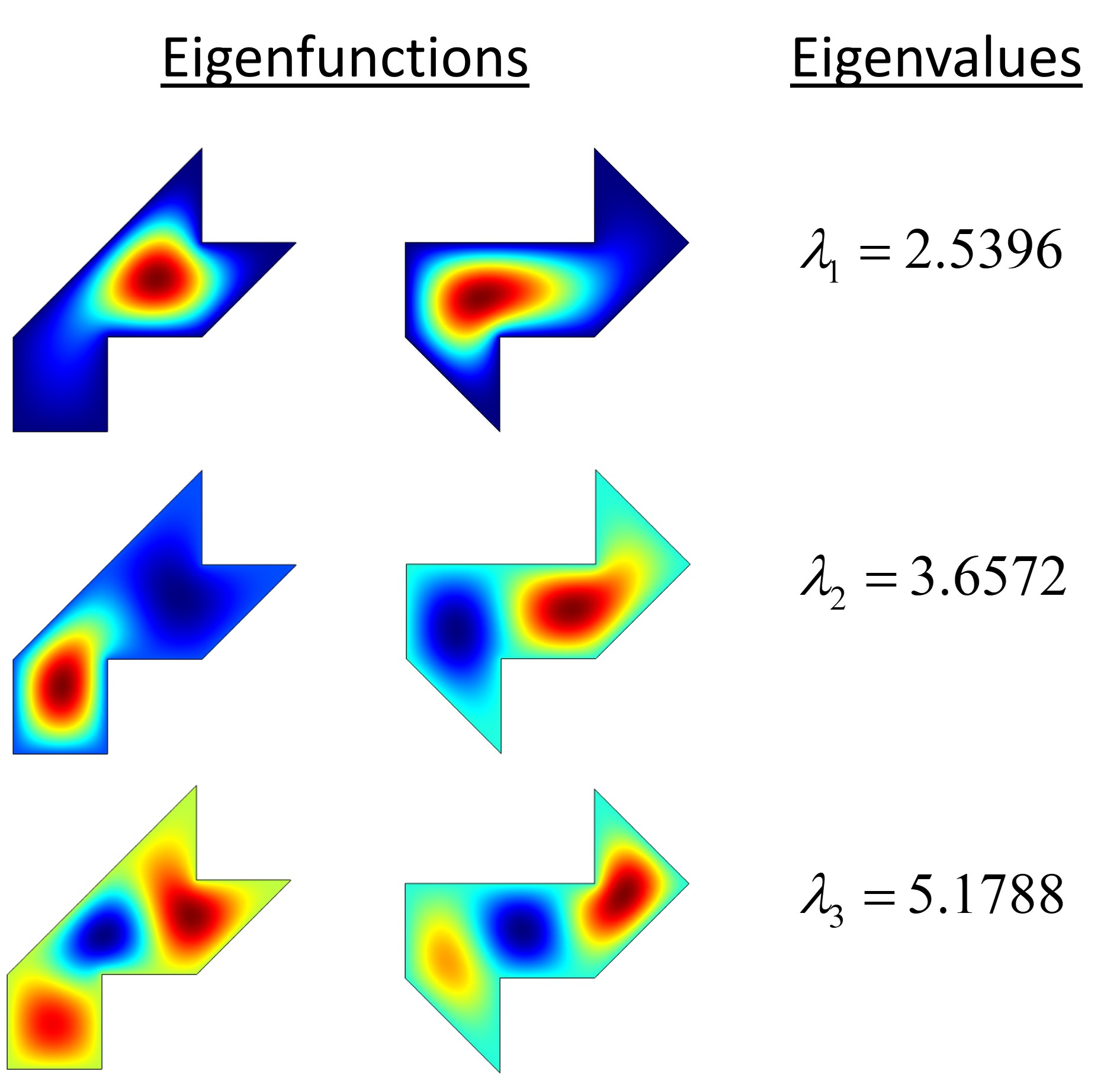}
\caption{An example of 2D isospectral domains. Left and right shapes have identical eigenvalue spectrum, therefore two drums made by these shapes sound exactly the same.}
\label{fig:c1f4}
\end{figure}

Hearing the shape is technically called an inverse problem. Shape determines the sound, but the catch is figuring out the shape from the sound. This inverse problem actually has been solving in everyday life of several living organisms, Fig. 1.5. Mammals like bats and dolphins constantly use echolocation techniques to navigate and communicate. They send sound waves into environment and from their reflections they determine the object's distance, shape and type. Bats actually "see" things by hearing them. It is remarkable that evolution results to such ingenious solutions for complicated problems.

\begin{figure}[!b]
\centering
\includegraphics[width=0.9\textwidth]{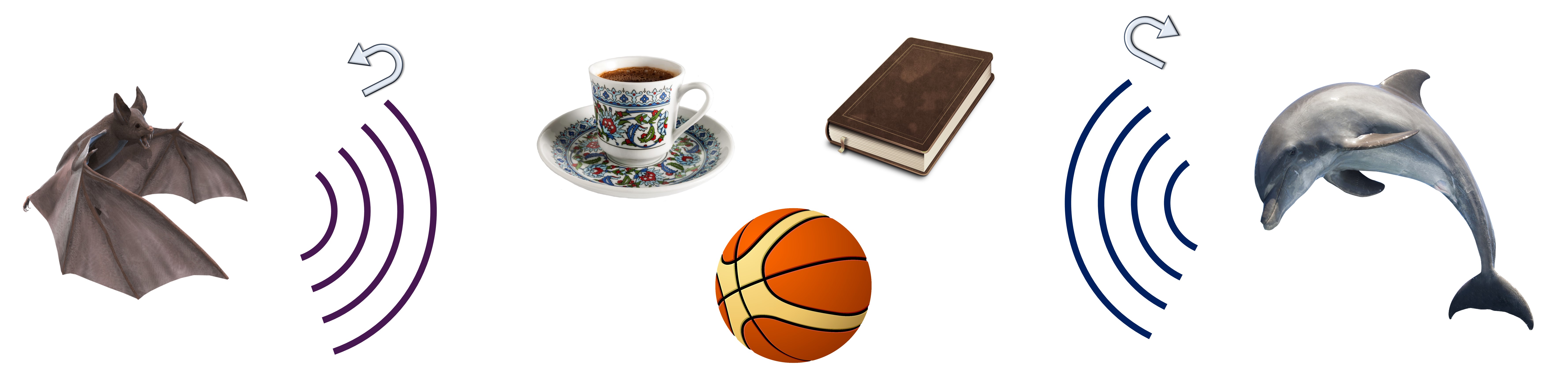}
\caption{An inverse problem of hearing the shapes in nature. Bats and dolphins use echolocation techniques to navigate, which relies on the solution of an inverse problem of "If an object sounds like that, where is it and what is its shape?".}
\label{fig:c1f5}
\end{figure}

In recent years, following the advancements in pattern recognition and numerical techniques, shape recognition based on eigenvalues of the Laplacian has become an attractive field \cite{bookshape,lozan15}. For non-rigid shape analysis a method called "Shape-DNA" based on Laplace-Beltrami spectra is introduced. Shape-DNA is basically a numerical fingerprint of any two- or three-dimensional manifold based on its Laplacian eigenvalue spectrum. Recognition of objects based on Shape-DNA can be done in digital environment. Uniform scaling of different objects is possible by the same method \cite{shapeDNA,prshape}. Although isospectral objects exist, their occurrence is extremely improbable in daily life. This makes methods of shape recognition based on eigenvalue spectrum attractive even for commercial applications.

Kac's question can also be generalized into the quantum mechanical applications. Since types of boundary conditions determine the solutions of differential equations, boundary conditions dramatically affect the energy spectra of the particles confined in a domain \cite{yalab,speedof,efspectra,M.Levitin2006}. Can a gas feel the size and shape of its confinement domain? This question has also been taken into consideration in several papers \cite{eskifeel,feelsize,qdrumtt}. The important thing to realize is that Weyl terms (surface/volume, periphery/volume and edge/volume) are not enough to represent full characteristics of the domain. Weyl conjecture is just another approximation (but a really good one) that only holds true in the asymptotic limit. Although there may be some special absolutely isospectral domains, there are infinitely many domains whose Weyl terms are completely identical, yet properties of gas confined in these domains can still be different. In an absolute mathematical sense, there are differently shaped domains that gas cannot feel on which one it's filled. However, these kinds of isospectral domains are extremely rare and very specially arranged domains \cite{RevModPhys.82.2213}.

All these works done in spectral geometry are profoundly related with the main theme of this thesis. Laplacian eigenvalue spectrum is equivalent to the solutions of time-independent Schr\"{o}dinger equation. While Weyl conjecture is giving some good information about the spectrum, it has limitations in strongly confined systems. The field is very rich both mathematically and physically, and research on open questions is still ongoing \cite{optdi,trishape,furtherhear,isomem,isodom,eimodiso,drumsame,revisit,frome,revisit2,PhysRevLett.97.090201,PhysRevE.58.610,densdrum,hdimfrac,fracdimdr,shapegr,pophis,room,triang,geop,PhysRevE.88.042915,PhysRevLett.66.1555,elec1,elec2,polyop,bells,10.1063/1.4733363,PhysRevE.87.062915,vfracd}. A very nice comprehensive review of the subject with its connections to physics can be found in Ref \cite{superdoku}.

% Matter wave pressure forces
Research on quantum forces and torques is also very much related with our thesis as we also introduce a new kind of quantum force and torque due to quantum shape effects. Some aspects of matter wave related forces such as quantum statistical forces due to boundary effects \cite{bermanq,bouncing,flopb1,flopb2,PhysRevE.83.041133,PhysRevLett.114.250601} and quantum-classical hybrid systems with matter wave pressure \cite{conpreq,PhysRevLett.97.190401,PhysRevA.82.062107,PhysRevA.83.062101} are investigated in literature. One of the remarkable consequences of the wave-like properties of quantum particles are quantum mirages at quantum corrals \cite{manoharan,RevModPhys.75.933,cohmaw}. They represent excellent examples of fluctuation-induced forces like Casimir forces which are also subject to size and shape related geometric effects \cite{RevModPhys.71.1233,PhysRevA.89.033835}.

Electromagnetic waves exert pressure. Several type of actuators \cite{georfafm,radprebook} and motors \cite{motpreraddm,lasmot,goldmot} have been proposed based on this radiation pressure. From the experience of these kinds of small pressures, measurement of as tiny as piconewton forces are possible \cite{natpos,PhysRevLett.116.147202,Joopti,PhysRevX.6.021001}. Like the radiation pressure produced by light, matter waves can also exert pressure. In literature, semi-classical approaches are used for the calculation of fluid-like properties of matter waves \cite{moddense,prestens,fluidm,mecmatt0,mecmatt,topre,PhysRevA.82.062107,PhysRevB.82.165406,PhysRevE.72.066704,localami}. Since semi-classical or quantum hydrodynamic approach to calculate properties of matter waves is quite adopted in literature \cite{moddense,prestens,topre,PhysRevA.82.062107,PhysRevB.82.165406,PhysRevE.72.066704,localami,fluidm,mecmatt0,mecmatt,conpreq,qtmwavedet,PhysRevA.90.012709,cif,aveave}, it may be proper to use similar kind of hydrodynamic transport approach in our calculations as well. Quantum kinetic theory and Wigner function methods are also used occasionally in literature \cite{ftffphase,phaesp,qhydo,boniqkt}.

Another closely related intriguing phenomenon is quantum backflow effect \cite{artime,qbackflow0,PhysRevA.86.042116,qbackflow1,PhysRevA.87.053618,qbackflow2,qbackflow3,qbackflow4} in which flow of negative probability, or in other words, negative current of particles with entirely positive momenta occurs. Although it is really a striking quantum phenomenon, the subject is largely overlooked and pretty much unexplained. One-sided momentum flux calculations for local pressures in this thesis will be done similar to the ones done in the calculation of the backflow effect.

Coherence of matter waves may also enhance matter wave related effects \cite{cohmaw}. Quantum forces that may appear even in superconductors are also proposed in literature \cite{PhysRevB.64.012505}. New methods for quantifying macroscopicity degrees of quantum phenomena \cite{outphd} may widen our view and further the enthusiasm beyond. Quantum gases confined under time dependent boundaries \cite{mov1,mov2} are also interesting to examine and may shed some light on the time-dependent studies.

\section{Thesis Structure}

We'll start with an overview of quantum size effects in statistical thermodynamics. In the following chapter, we first explain the quantum-mechanical origins of the size effects in confined nanostructures. We discuss how size effects can change the thermodynamic behaviors of systems. We cover some necessary mathematical tools to calculate quantum size effect corrections to the thermodynamic expressions. We review the quantum boundary layer method which is one of the most powerful methods for obtaining quantum size effect expressions as well as for understanding their underlying physical mechanisms. We are particularly interested in this methodology because we'll extend it to explain also quantum shape effects.

In the third chapter, we separate size and shape effects from each other completely and introduce the quantum shape effects. Signatures of shape effects in eigenspectrum will be discussed. We examine the shape dependence of partition function and extend quantum boundary layer methodology to analytically predict quantum shape effects as well. Different confinement domains and boundary conditions will be considered. Furthermore, the influence of quantum size effects on quantum shape effects will be discussed.

Investigation of change in thermodynamic properties due to shape effects is done in the fourth chapter. Shape enters as a new control variable on thermodynamic state functions and state space, opening up a whole new, before unexplored dimension in thermodynamics. Quantum shape effects on internal energy, free energy and entropy of confined systems are investigated. Non-uniform density distribution of particles causes a non-uniform pressure distribution in nested confinement domains. Due to quantum shape effects, an asymmetric non-uniform distribution occurs which induce a quantum torque. Examination of the pressure and torque is done from both local and global approaches. We further investigate quantum shape effects in electron gases using Fermi-Dirac statistics.

Several energy applications of quantum shape effects are proposed and explored during the fifth chapter. A new thermodynamic process called isoformal process is introduced and two new heat engine cycles are designed based on quantum shape effects. In this chapter we mention also a quantum Szilard engine variant and a single-material unipolar thermoelectric effect called thermoshape effect, which are not directly parts of the thesis but mentioned anyway as they are done by groups including the author of this thesis. Also, they constitute significant examples of the applications of quantum shape effects.

The main outputs and highlights of this thesis are itemized below:
\begin{itemize}
\item Quantum size and shape effects are completely separated from each other through a size-invariant shape transformation, which allows one to focus on purely shape-dependent physical properties of confined systems.
\item Shape variation alone is able to change the thermodynamic state functions while all other physical parameters and geometric size variables are constant.
\item Equilibrium statistical properties of the particles confined in an arbitrary domain, along with their size and shape dependence, are analytically estimated with a reasonable accuracy by extending the quantum boundary layer method.
\item The analytical methodology not only gives physical insights about the existence and underlying mechanisms of quantum shape effects, but also provides opportunity to predict them without doing cumbersome numerical calculations.
\item Existence of size effects increase the quantum confinement whereas the appearance of shape effects leads to a decrease in the effective confinement.
\item Quantum shape effects cause decrements in internal energy and Helmholtz free energy, whereas they have a more complicated effect on the entropy of the system.
\item Entropy and free energy of the confined system can decrease simultaneously and spontaneously due to quantum shape effects, which is a unique phenomenon in the thermodynamics of non-interacting gases.
\item Quantum shape effects cause a breakdown of extensivity of the thermodynamic quantities, just like size effects.
\item Quantum shape effects give rise to quasistatic spontaneous rotation and/or Casimir-like translational motion of the objects that are freely movable inside the confinement domain.
\item Thermodynamically stable configurations in nested confinement domains are determined by the symmetric periodicity points, whereas configurations other than that are unstable.
\item A quantum torque is induced due to the non-uniform pressure exerted by matter waves.
\item In the thermodynamics of electrons obeying Fermi-Dirac statistics, chemical potential oscillates with the varying shape parameter for fixed number of particles, temperature and sizes. This causes oscillatory behaviors in all thermodynamic properties of confined and degenerate Fermi gases regardless whether they intrinsically exhibit quantum oscillations or not.
\item A new type of thermodynamic process, called isoformal process, based on keeping the domain shape constant, is proposed and new thermodynamic cycles featuring the isoformal process are introduced and examined.
\item Quantum shape effects provide a novel way to design new type of nanoscale heat engines and energy harvesting devices.
\end{itemize}

The results that are found and the topics that are explored in this thesis are related to and can shed light into many diverse areas of physics and mathematics, such as spectral geometry, quantum thermodynamics, quantum billiards\cite{qubil}, quantum corrals, bound states in waveguides, localization, topological properties, dimensional transitions, local properties, quantum hydrodynamics, quantum transport, cross effects, quantum backflow, shape recognition, nanoscale energy technologies and so on.

%%%
%2%
%%%

%\chapter{A Review of Quantum Size Effects in Statistical Thermodynamics}
\chapter{A Review of Quantum Size Effects in Statistical Thermodynamics}
In this chapter, we'll introduce and review some primary concepts and methodologies that we've used in this thesis. 
\section{Quantum Confined Structures}

Our main purpose in this thesis is to develop a comprehensive understanding on the role of confinement geometry in the thermodynamic properties of physical systems at nanoscale. Therefore, in this review chapter we first mention what do we mean by confinement and how physical systems change behavior when confinement geometry is changed. Later on in this chapter, we'll discuss on how to quantify these changes and explore the influence of quantum size effects in the thermodynamics of confined systems.

\subsection{Wave nature of matter}

The foundations of quantum theory laid in 1900, when German physicist Max Planck (accidentally) discovered that the radiation is quantized. Yet the theory found a meaningful ground to develop, after French physicist Louis de Broglie had suggested the hypothesis of matter waves, \textit{id est} matter exhibit wave-like behavior, in 1924 in his PhD thesis. This behavior has been demonstrated many times experimentally and now sits on the central part of our current understanding of the universe. Before quantum mechanics, behaviors of particles are modeled as if they are points or hard spheres. Consider you have a box filled with point particles randomly moving around inside the box. Particles can only bounce from the walls of the box when they come infinitely close (touching basically) to the boundaries. On the other hand, if they are not point particles but waves, it would be more accurate to think of them occupying a finite amount of space without having a precise pointwise coordinate location. In that case, particles can feel the presence of boundaries without strictly touching them. In other words, they can reflect back from the walls without necessarily coming really close to them. 

Fundamentally, a quantum system is described by an abstract mathematical object called wavefunction to which all the quantum weirdness essentially boils down. The wavefunction is a square integrable complex-valued quantity that carries the possible outcomes of the measurements made on the system. For example, a position wavefunction of a single particle carries the information about the possible locations that particle can be found at a given momentum and time. The motion of the particle depends on the time evolution of its wavefunction, which is described by the Schr\"{o}dinger equation,
\begin{equation}
i\hbar\frac{\partial}{\partial t}\Psi(\textbf{r},t)=-\frac{\hbar^2}{2m}\nabla^2\Psi(\textbf{r},t)+V(\textbf{r},t)\Psi(\textbf{r},t),
\end{equation}
where $\hbar=h/(2\pi)$, $\nabla^2$ is the Laplace operator, $\textbf{r}$ is the position vector, $m$ is particle's mass, $V$ is the confinement potential, $t$ denotes time and $\Psi$ is the wavefunction in position basis. In thermodynamics, we deal with the equilibrium properties of systems and so we are interested in stationary solutions. To obtain the stationary states of a quantum system, we need to solve the time-independent Schr\"{o}dinger equation. By using the method of separation of variables, we reach an eigenvalue equation called the Helmholtz equation. Under zero potential, V(\textbf{r})=0, it is written as
\begin{equation}
\nabla^2\psi(\textbf{r})+k^2\psi(\textbf{r})=0
\end{equation}
This equation is a wave equation and it is the general form of time-independent Schr\"{o}dinger equation, where the wavenumber $k$ corresponds to $k=\sqrt{2mE}/\hbar$ and $E$ denotes the energy of the particle. 

Now we have an equation describing the wave nature of quantum particles and we'll use it in the next subsection. Before that, let's mention a bit more about the wave nature of particles and how do we quantify it. Essentially, what we are interested in is the position space representation, because we want to understand the geometry effects. In position space, the wave character of particles is quantified by their de Broglie wavelengths, which is defined as $\lambda_{dB}=h/p$, where $h$ is Planck's constant and $p$ is the momentum of the particle. For massive particles, momentum is the multiplication of particle's mass and velocity, $p=mv$. This means, larger the particle's mass, smaller its de Broglie wavelength. This is the reason why wave-like behavior is more apparent in subatomic particles but not in our macroscopic world.

In condensed matter physics, we mostly deal with a collection of particles rather than a single particle. Both the statistical behaviors and the wave nature of a collection of particles can be captured by the so called thermal de Broglie wavelength, which is given by the following expression, 
\begin{equation}
\lambda_{th}=\frac{h}{\sqrt{2\pi mk_BT}},
\end{equation}
where $k_B$ is Boltzmann's constant and $T$ is temperature. In Eq. (2.3), individual velocities of particles are replaced by the mean velocity corresponding to their average kinetic energies. By use of statistical mechanics, we don't have to deal with the individual behaviors of astronomic number of particles, instead we can capture their statistical behavior which provides a quite good representation even for dozens of particles.

In addition to mass of the particles, thermal de Broglie wavelength says that the temperature of the system is also important for the appearance of wave nature. Like mass, it is also inversely proportional with the wavelength. Despite all, $\lambda_{th}$ by itself is not enough to see the difference between macro world and nano world. What is important is its comparison with the system sizes. Wave nature does not disappear magically at macroscale. What happens is our macro world is too big in comparison with the thermal de Broglie wavelength. Recall the point particle vs wave-like particle example. Our macro world is on the order of meters, whereas the thermal de Broglie wavelength of a free electron gas at room temperature, for example, is around 4.3 nanometers. There is $10^9$ orders of magnitude difference. Compared to our macro world, electrons are like point particles, although their actual behavior is wave-like. Note that this analysis was for a subatomic particle electron, one of the lightest massive particles. For atoms and molecules, the order of magnitude difference is even larger. So the essential thing separating the nano world from macro one is the comparison of the thermal de Broglie wavelength of particles with the sizes of the domain where those particles are confined. In Fig. 2.1, such a comparison is given. When $L$ is much larger than $\lambda_{th}$, we can assume the particles to behave as point-like. When $L$ is comparable with $\lambda_{th}$, wave-like behavior of particles become apparent. Hence, electrons confined in domains with nanoscale dimensions exhibit its wave nature even at room temperature.

\begin{figure}
\centering
\includegraphics[width=0.5\textwidth]{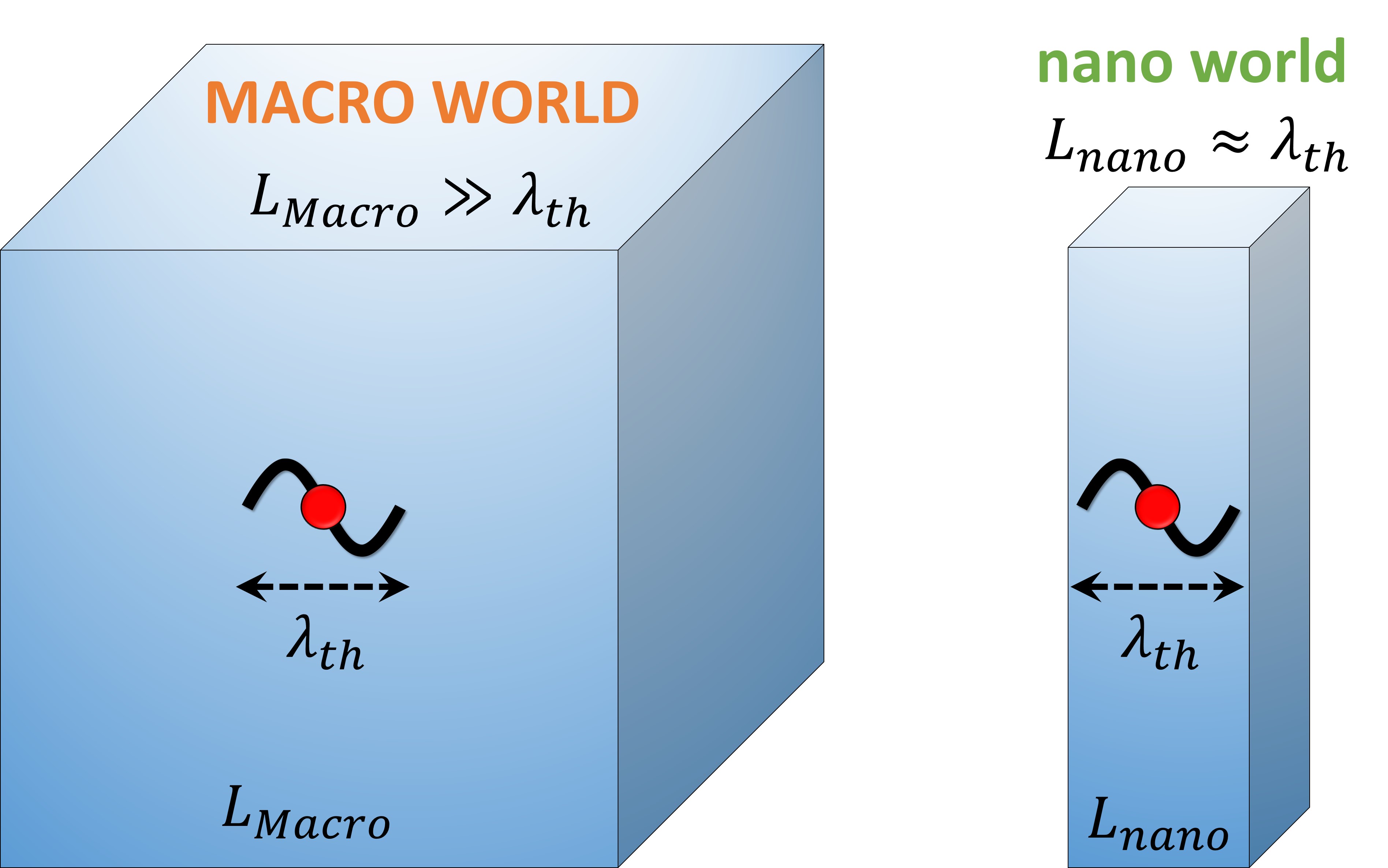}
\caption{Differentiation of macro and nano scales are coming from comparisons of the thermal de Broglie wavelength of particles and the sizes of the domain that the particles confined within. Depending on the nature of the system and particle statistics, one can replace the thermal de Broglie wavelength with the appropriate characteristic de Broglie wavelength.}
\label{fig:c2f1}
\end{figure}

It is necessary also to mention here that the statistical properties are not the same for all kinds of particles. In terms of statistical behavior, particles split up into two groups; Fermions and Bosons, satisfying the Fermi-Dirac statistics and Bose-Einstein statistics respectively. In quantum mechanics, particles are indistinguishable. Further to that, it is not just we cannot distinguish this electron from that electron, it is meaningless even to talk about such a thing\cite{introqm}. (So, it is like asking to draw a triangle having two sides. It's wrong by definition.) Degeneracy of particles comes into play when Fermions or Bosons are considered as confined particles. Quantum degeneracy brings a separate correction to the characteristic de Broglie wavelength (it can be chosen as thermal, mean or most probable de Broglie wavelengths). In such a case, rather than thermal de Broglie wavelength, it is more useful to consider Fermi wavelength for Fermions for example. Therefore, $\lambda_{c}$, the characteristic de Broglie wavelength, is a matter of choice which should be done according to the statistical nature of the particular system. We will mention more on this in section 4.9. When the indistinguishability of particles is ignored, we use the Maxwell-Boltzmann statistics, which gives accurate results for low density and high temperature systems. To keep the discussion simple, we keep using the thermal de Broglie wavelength and demonstrate our formalism considering the Maxwell-Boltzmann statistics.

\subsection{Quantum confinement and low-dimensionality}

Quantum confinement occurs when the motion of particles is restricted in at least one direction. For example, graphene is a 2D material (it consists of a single-layer of carbon atoms) in which electrons are free to move in two directions but restricted or confined in one direction. Accordingly, carbon nanotubes and quantum wires are considered as 1D structures that are confined in two directions. When the particles are confined in all directions, the structure is called a quantum dot, hence 0D. In Fig. 2.2, different structures of carbon-based materials confined in various directions can be seen as examples. The strength of confinement in a particular direction is determined by the confinement parameter of that direction,
\begin{equation}
\alpha=\frac{h}{\sqrt{8 mk_BT}L},
\end{equation}
where $L$ is the length in the confined direction. $\alpha$ is the ratio of the most probable de Broglie wavelength in Maxwell-Boltzmann statistics to the domain's length. In terms of thermal de Broglie wavelength, it is stated as $\alpha=(\sqrt{\pi}/2)*(\lambda_{th}/L)$.  Therefore, magnitude of the quantum confinement of a domain depends on the particles' mass, system's temperature and characteristic domain size.

\begin{figure}[!b]
\centering
\includegraphics[width=0.5\textwidth]{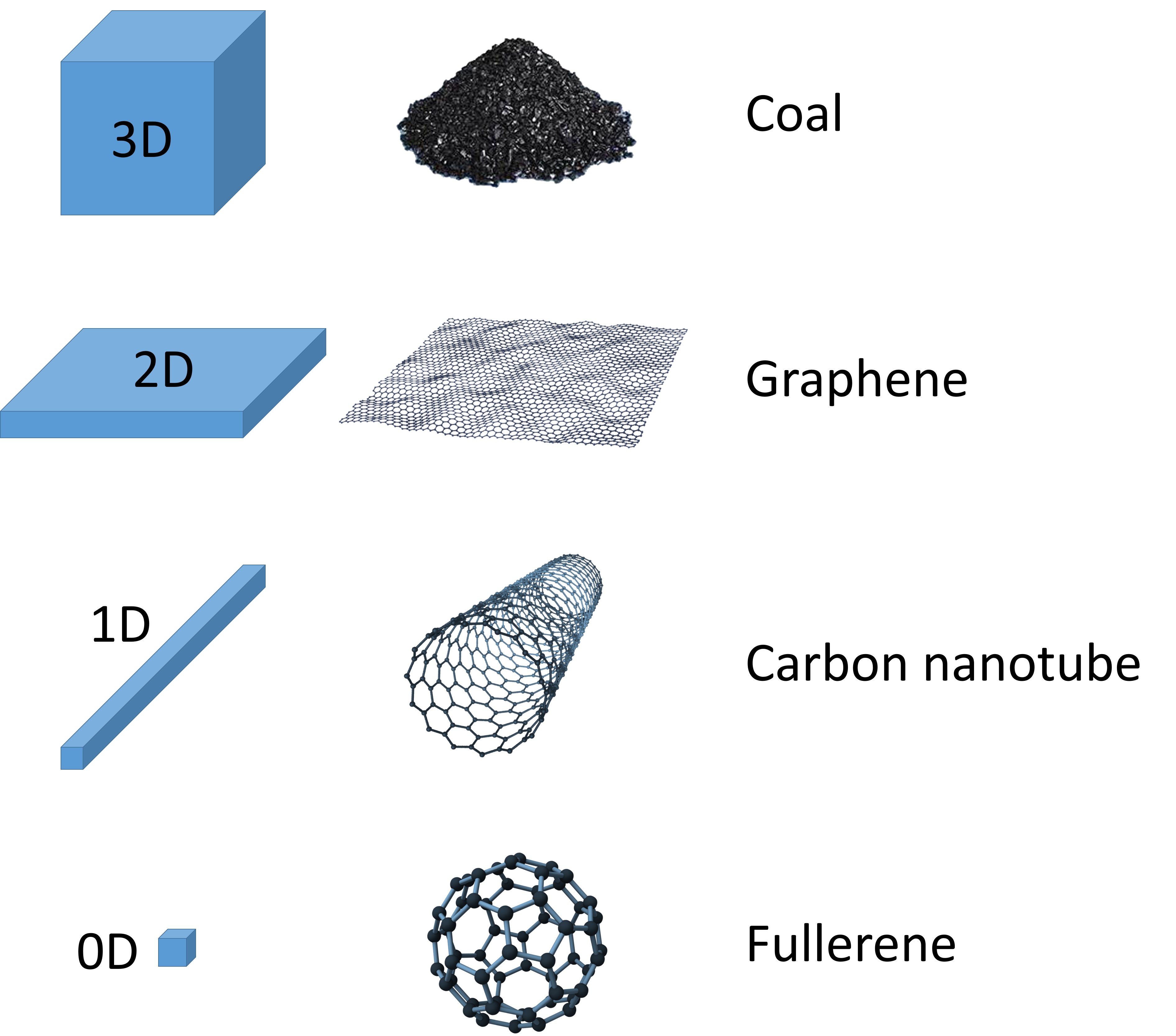}
\caption{Allotropes of carbon at various dimensions. Numbers describing the dimensions denote the number of free directions for charge carriers. Note that the particles are appeared to be confined on the surface of the fullerene, but its overall size is less than 1nm, which makes it 0D as it is isotropically and strongly confined from all directions.}
\label{fig:c2f2}
\end{figure}

Depending on the value of the confinement parameter, we can refer the domain as unconfined or free ($\alpha\approx 0$), weakly confined ($\alpha<0.1$), confined ($0.1<\alpha<1$) and strongly confined ($\alpha>1$). Note that $\alpha$ values give just a rough scale, since the transitions between these confinement regimes are not sharp but smooth. Mathematically, $\alpha$ enters as a proportionality constant of quantum states to statistical expressions of thermodynamic quantities and determines the essential discreteness in energy spectrum. Although the difference between each quantum state is always one, the number of quantum states within an energy interval is determined by $\alpha$. Thus, it scales the energy spectrum.

Quantum confinement plays a significant role in solid-state physics and material science. It paves a way not only to the discovery of materials with better properties in comparison to their conventional counterparts, but also to the development of nanoscale devices that are more efficient and better at certain tasks. Some well-known and well-studied examples of confined systems are the charge carriers (electrons and holes) and phonons in nanoscale metals and semiconductors in addition to ultracold atoms in optical traps. Theoretical study of these confined systems is usually done using the particle in a box model. Consider a non-relativistic single quantum particle confined in a 1D domain having length $L$, Fig. 2.3. The potential inside the well is zero and infinite at the outside, meaning the walls are impenetrable. Even a quantum particle cannot tunnel (leak) through the walls of an infinite well. This is a quite accurate model for the behaviors of conduction band electrons in metals for example. We will stick with the impenetrable boundaries throughout the thesis in order to maximize the effect of confinement geometry on the particles as quantum tunneling leads to leakages and reduces the geometry influence.

\begin{figure}[!b]
\centering
\includegraphics[width=0.95\textwidth]{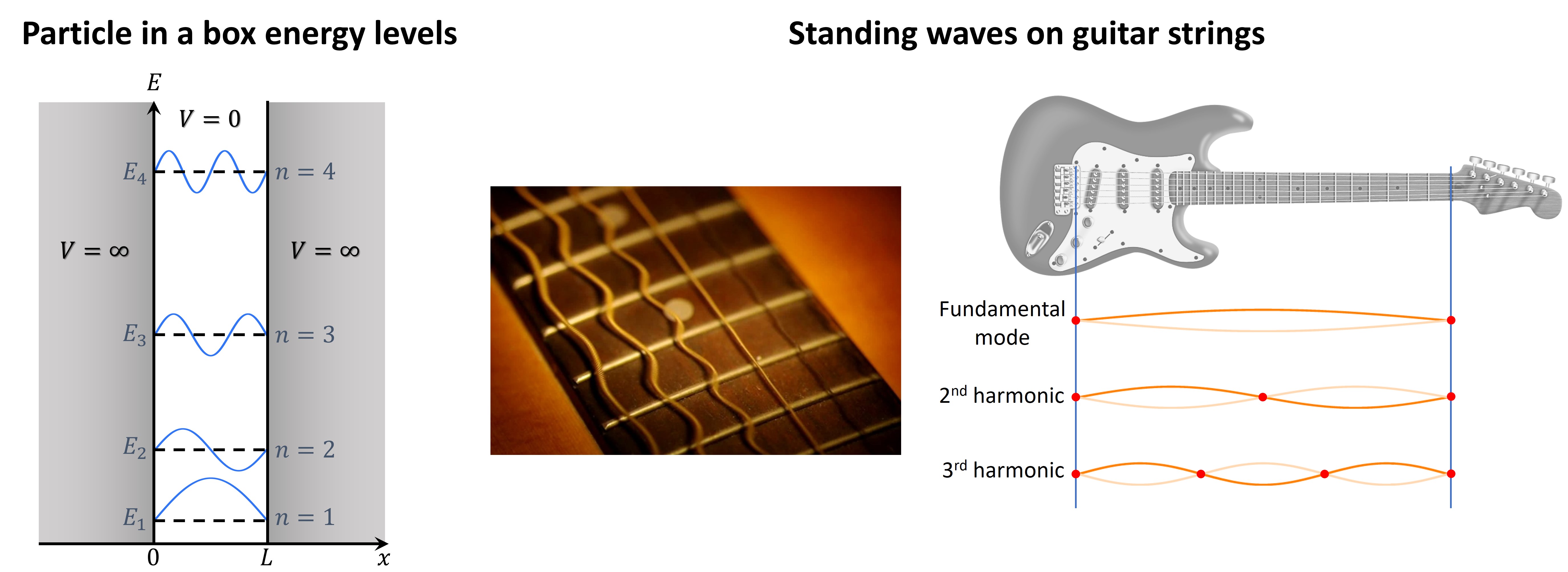}
\caption{Stationary states correspond to standing waves in systems exhibiting wave-like behavior, if the system has a fixed geometry. Leftmost subfigure shows the energy levels and the corresponding wavefunctions of a single particle confined in an infinite potential well. The figure on the center shows a picture of a guitar's vibrating strings. Rightmost subfigure shows the fixed nodes of the guitar strings leading to the formation of wave modes.}
\label{fig:c2f3}
\end{figure}

To solve the particle in a box model, we first solve the time-independent Schr\"{o}dinger equation, which is basically the Helmholtz equation in Eq. (2.2). Since the confinement potential is infinite outside the box and zero inside, the boundary conditions are Dirichlet so that $\psi(0)=0$ and $\psi(L)=0$, which means the wavefunction goes to zero at the fixed ends of the box. Under these boundary conditions, energy eigenvalues (corresponding to the energy levels of the system) of the particle confined in 1D domain with length $L$ can be obtained from the solution of Eq. (2.2) as 
 \begin{equation}
E_n=\frac{\hbar^2 \pi^2}{2m}\left(\frac{n}{L}\right)^2,
\end{equation}
where $n$ denotes the modes of the wave (also corresponds to the quantum states in this particular example), integers running from 1 to $\infty$. The important thing to notice here is that energy levels are not continuous but discrete. This is one of the properties of matter which was unexplainable by the classical physics. Discreteness of energies of confined particles is a direct result of their wave characteristic. Eigenfunctions (corresponding to the wavefunctions) describing the spatial behavior of the wave modes are also obtained as
 \begin{equation}
\psi_n=\sqrt{\frac{2}{L}}\sin\left(\frac{n\pi x}{L}\right),
\end{equation}
where $x$ is the position defined between the edges of the domain. The reason for the appearance of these modes is because the domain is fixed at both ends. The length of the domain is associated with the confinement of the particles. Smaller the length, higher the confinement and higher the energy of the particle by Eq. (2.5). In this regard, spatial confinement gives rise to the discreteness in momentum and energy space. Visualization of energy eigenfunctions can be seen in Fig. 2.3. In 1D systems, each mode corresponds to a different energy and a wavefunction. Modulus square of the wavefunction gives the probability of finding the particle for a given state at a certain location inside the box.

The resulting particle in a box eigenfunctions are extremely similar to the vibrating guitar strings. This is because, they are the results of the same phenomenon that is described by the same mathematical equation (differing only by constants); so they have the same physics. Just like in the particle in a box example, guitar strings are fixed at both ends and they can vibrate only in discrete set of modes. There is a fundamental mode, which has the lowest energy, lowest frequency and highest wavelength. Higher harmonics of guitar strings correspond to the excited states of a confined quantum particle.

Another important consequence of the wave nature of particles is the existence of the non-zero ground state (or zero point) energy. By their very nature, wave modes and quantum state variables start from their ground state value $n=1$, corresponding to their fundamental mode. In Fig. 2.4, approximations on the representation of energy spectrum of particles can be seen. Classically, energy spectrum is considered to be continuous and starts from zero. As we have seen from the particle in a box example, the true nature of the energy spectrum is discrete and it starts from a non-zero value called the ground state. At macroscale, continuum approximation works very well because the separation between the levels is inversely proportional to the size of the system. The larger the size, the higher the accuracy of the continuum approximation. For weakly confined systems, we use so called bounded continuum approximation, which considers the non-zero value of the ground state while still assuming a continuous spectrum for the rest. Despite taking a continuous spectrum, the bounded continuum approximation is very powerful. In addition to giving quite accurate results, it also properly captures the boundedness of the domain and generates all quantum size effect correction terms. For strongly confined systems, however, even this approximation fails and one needs to consider the discreteness of the spectrum. We'll discuss more on the use of bounded continuum approximation and quantum size effects in the next section. 

\begin{figure}[!b]
\centering
\includegraphics[width=0.7\textwidth]{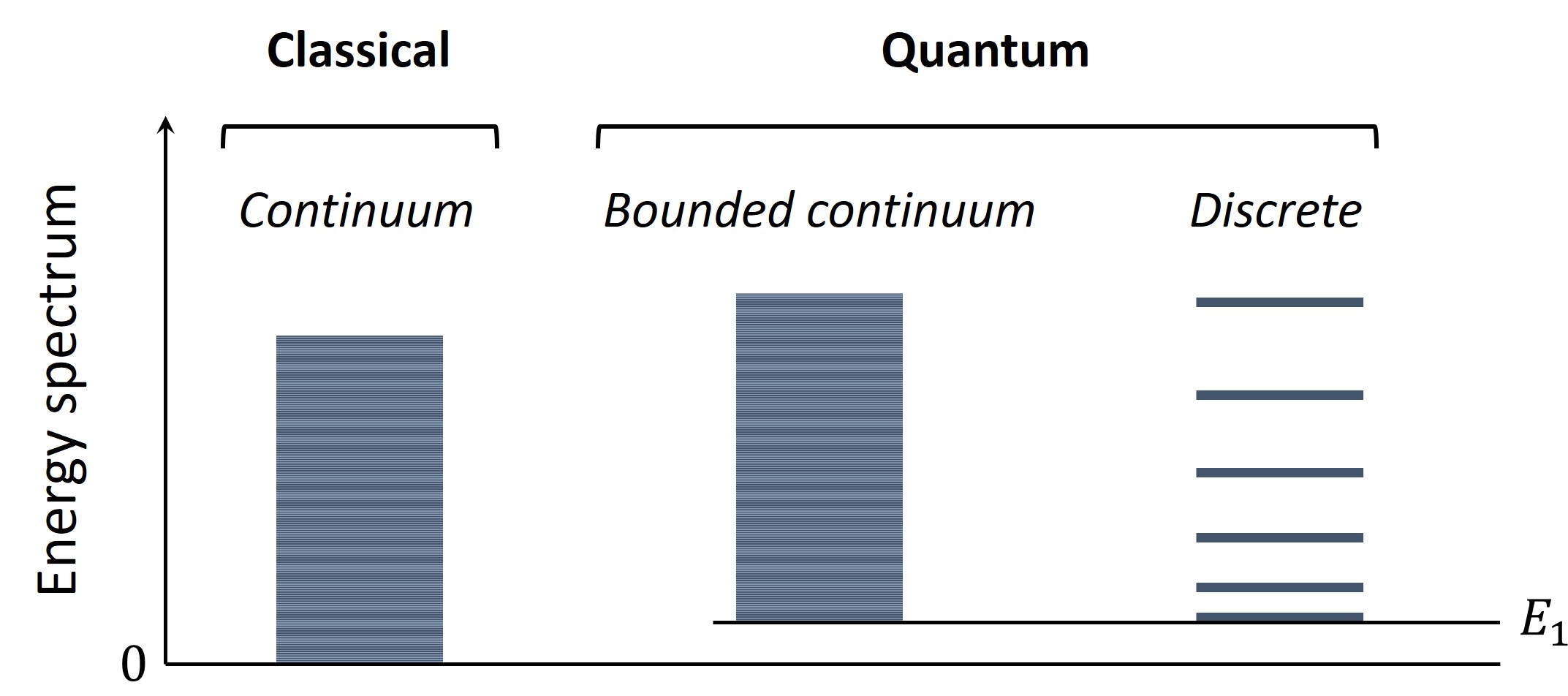}
\caption{Classical and quantum representations of the energy spectrum. Energy spectra are considered to be continuous in classical physics by the continuum approximation. For bounded systems, a better approximation called bounded continuum approximation considers the non-zero value of the ground state $E_1$. Without any approximation, actual energy spectra of confined systems are discrete.}
\label{fig:c2f4}
\end{figure}

\subsection{What is size?}

We talked about quantum confinement in a general way. But in fact, we only dealt with a 1D model so far. Size of a 1D domain is characterized by its length. Once you know the length, you have the energy levels. But what about the higher dimensions? How do we quantify the size in general? When we say the sizes of an object at macroscale, what does that even mean? To understand these, let's continue our discussion with the 2D model. Think of a simple 2D domain, for example a square with side length $L$. It has an area of $L^2$, a periphery of $4L$ and 4 vertices (cusps or corners). We can quantify the sizes of a square with three numbers, which means if we know the values of these size parameters, we can construct that particular square without having any additional knowledge. For 2D domains, area is defined as the bulk geometric size parameter, whereas the peripheral lengths and number of vertices are the lower dimensional geometric size parameters. Now you may have noticed that we actually missed something in our 1D domain's size analysis. What about the number of vertices in 1D domains? A 1D domain with length $L$ has 2 vertices, beginning and the end points of the domain. But isn't it trivial? The answer is no! Can we think of a 1D domain with different number of vertices? The answer is of course yes! Just add some new vertices to the domain or remove the ones on the edges. Consider for instance the keyboard of a guitar. Depending on the location that you press on the keyboard, it generates different sound. The reason is you are adding another fixed node when you press anywhere on the keyboard. Although the actual length of the string doesn't change, adding another node creates two 1D domains with different lengths and they sound according to their new lengths. An example of this can also be seen in Fig. 2.5 comparing columns II and III in the 1D domain row.

In Fig. 2.5 size characterization of domains with different dimensions is illustrated. For 3D objects, the sizes are volume, surface area, peripheral lengths and number of vertices. These are altogether named as geometric size variables. In measure theory, this definition coincides with the standard Lebesgue measure (or more generally the Hausdorff measure if one considers non-integer continuous dimensions). Depending on the dimension of the object, the bulk term becomes different (volume for 3D, area for 2D, length for 1D and vertices for 0D) while the remaining ones are named as lower dimensional geometric size variables. Sizes of a 3D object are characterized by these four variables. If all four of these variables are the same for two objects, they are considered to have the same sizes, but they don't necessarily have to have exactly the same shape, as we shall see later. This point is very crucial and actually constitutes the central point of the thesis. However, we'll continue to our review in this chapter and turn back to this point in Chapter 3.

\begin{figure}
\centering
\includegraphics[width=0.55\textwidth]{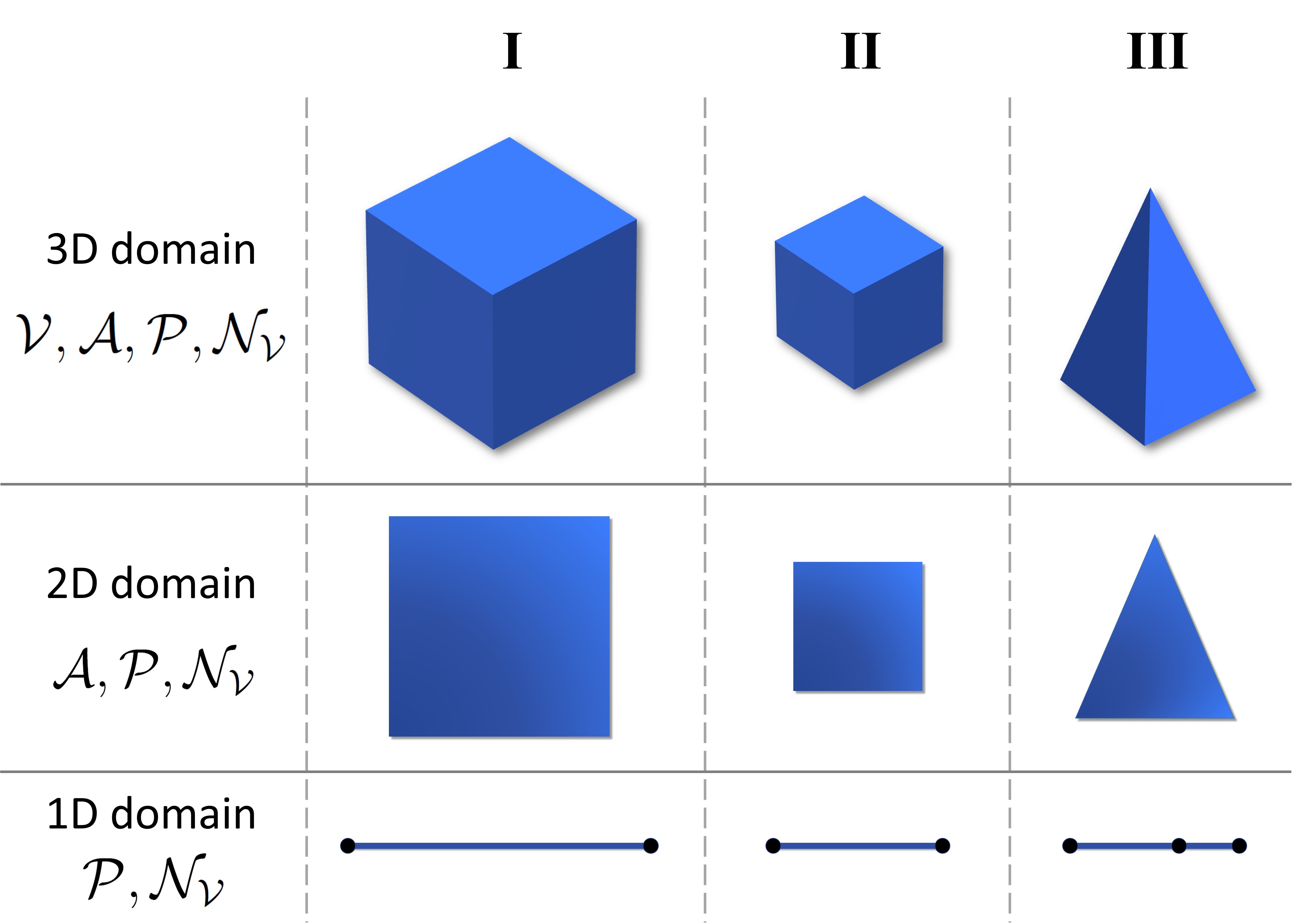}
\caption{A 3D domain is characterized by four different geometric size parameters: volume $\mathcal{V}$, surface area $\mathcal{A}$, peripheral lengths $\mathcal{P}$ and the number of vertices $\mathcal{N_V}$. Similarly, 2D domain is by $\mathcal{A,P,N_V}$ and 1D by $\mathcal{P,N_V}$. This is the standard Lebesgue measure. Sizes of an object can be changed while keeping its general shape constant by the process called uniform scaling (compare columns I and II). Between columns II and III, both size and shape of the objects are changed (except the last row where the size is unchanged).}
\label{fig:c2f5}
\end{figure}

As an explicit example, geometric size variables of a cube are shown in Fig. 2.6. Any 3D object has a single volume. A cube has a surface area consists of 6 squares with the side length $L$ of the cube. Surface area is a lower-dimensional (2D) property in this case. Periphery of the cube is the total lengths of the line segments that are present on the object which are the sides of squares. There are 12 of them. Note that it is not $6\times 4$ because all line segments are shared by 2 different squares. The lowest dimensional elements are the number of vertices of which the cube has 8. This analysis may look too simple, however, it is actually indispensable for the understanding of the quantification of sizes of a domain and they play a significant role on the physical properties of confined systems.

\begin{figure}
\centering
\includegraphics[width=0.7\textwidth]{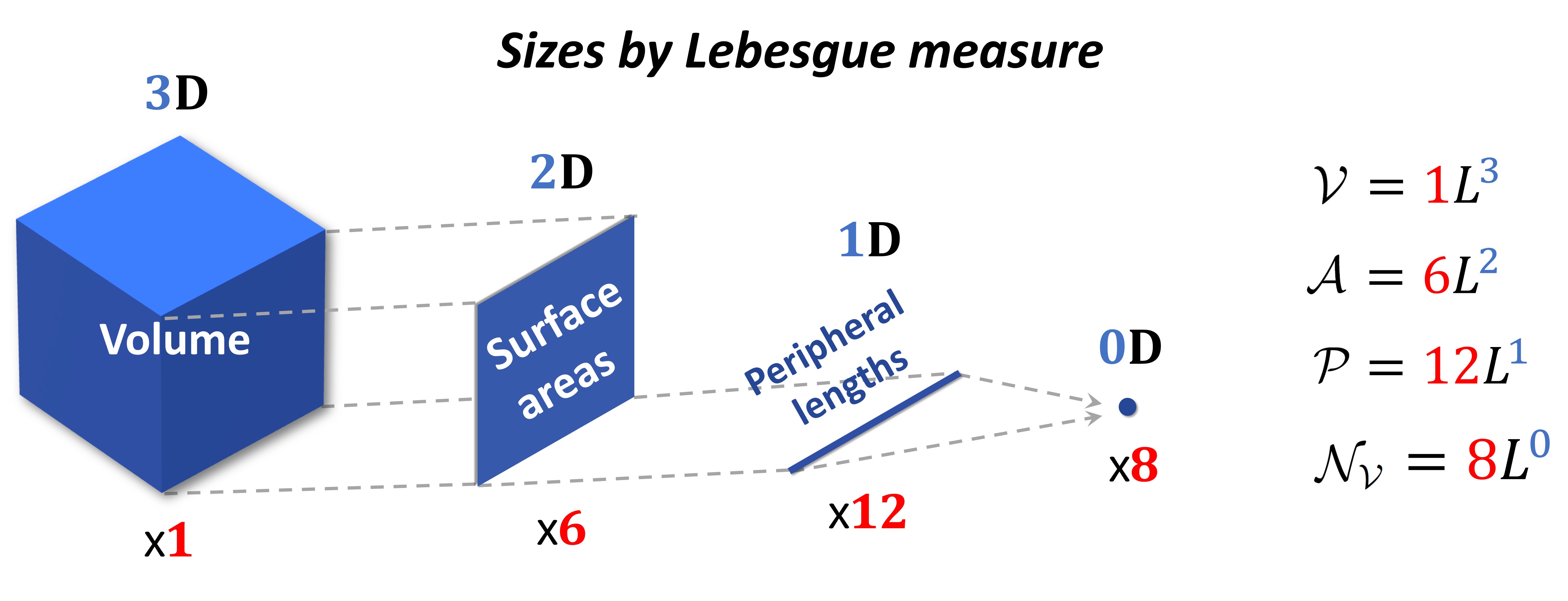}
\caption{Geometric size variables of a cube. It consists of 1 volume, 6 square surfaces, 12 lines and 8 points. Powers of the length $L$ generates the size variables of corresponding dimensions.}
\label{fig:c2f6}
\end{figure}

By definition, thermodynamics deals with the average values of physical quantities, as the principles of thermodynamics are statistical in origin. This fact is expressed by the concept of thermodynamic limit, which is defined as the limit of infinite number of particles $(N\rightarrow\infty)$ and infinite volume $(\mathcal{V}\rightarrow\infty)$, yet still finite density ($N/\mathcal{V}= $ constant). Consequently, in classical thermodynamics, there are no finite-size effects, because there is no finite size. In other words, thermodynamic properties classically independent of the scale. Nanoscale thermodynamics, on the other hand, challenges the thermodynamic limit by restricting finite number of particles into a finite volume. For a single or very few particles, both quantum and thermal fluctuations hinder the ability to determine the thermodynamic properties. Even if there are finite, but enough number of particles to be able to do statistics, then the methods of statistical thermodynamics can be extended into the nanoscale systems by proper methods which will be described in the following section.

\section{Mathematical Tools to Represent Infinite Sums and Their Usage}

In statistical physics, physical quantities are represented by infinite summations over quantum state variables of the microscopic quantities that are weighted over the relevant statistical distribution function:
 \begin{equation}
X=\sum_ix_if_i,
\end{equation}
where $X$ is a physical quantity, $i$ is the quantum state variable, $x$ is the microscopic quantity of X, $f$ is the relevant distribution function. Classically, the summations are replaced by the integrals under continuum approximation, which assumes a continuous spectrum of states. This approximation fails at nanoscale, because the separation between the energy levels are inversely proportional with the domain sizes. Namely, the higher the confinement, the more apparent the discrete nature of energy levels. Calculation of summations directly can be cumbersome in some cases. After all, they are infinite sums needing truncation operations during numerical calculations and they may depend on many degrees of freedom. Besides, most of the time it is very hard to predict the functional relations and dependencies of X on its control variables into the calculated quantities. To tackle with these problems, two distinct but interconnected mathematical tools are employed in the literature.

% Partition function
Before explaining the tools, in order to quantify our discussion and make it more physically intuitive, we'll first mention an important concept called partition function. The partition function is a powerful statistical concept that relates microscopic properties to macroscopic ones via the probability theory, Fig. 2.7. For a monatomic gas the partition function is written as
 \begin{equation}
\zeta=\sum_{\varepsilon}\exp\left(-\frac{\varepsilon}{k_BT}\right).
\end{equation}

Derivation of the partition function depends on the maximum entropy principle along with several constraints like fixed number of particles in a volume as well as their temperature.

\begin{figure}
\centering
\includegraphics[width=0.45\textwidth]{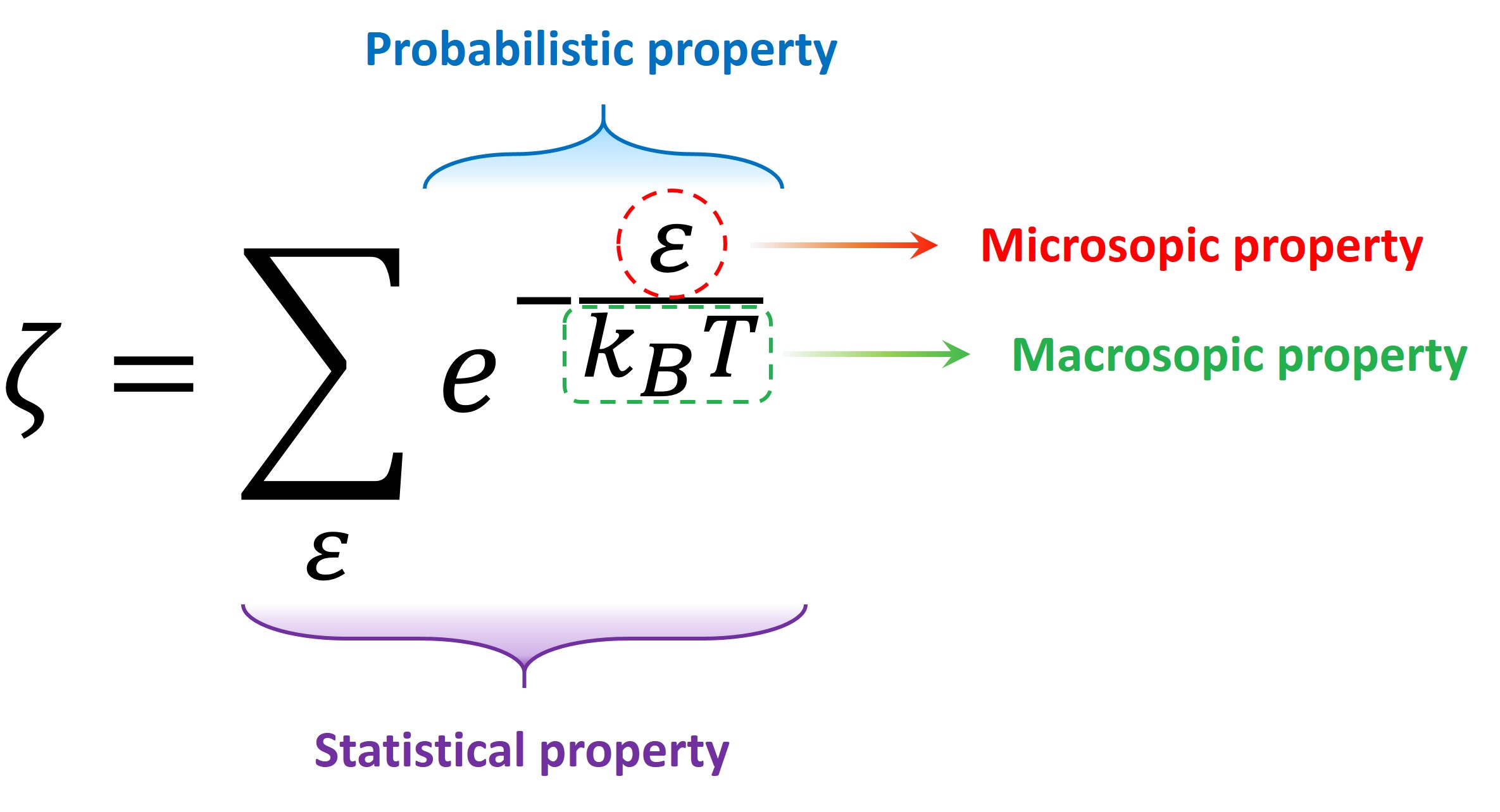}
\caption{Single-particle partition function with the Boltzmann statistical weight, providing a probabilistic connection between microscopic and macroscopic properties of a system.}
\label{fig:c2f7}
\end{figure}

$\varepsilon$ denotes the energy values obtained from the Schr\"{o}dinger equation, which is microsopic in origin. $k_BT$ contains the temperature which is an average, macroscopic quantity. They are brought together in a probabilistic weight factor, called Boltzmann factor, through the maximization of entropy subject to fixed number of particles and total energy. Summation over all states gives the statistical property called the single-particle partition function. Derivation of the partition function is done under thermodynamic equilibrium, that is the maximum entropy condition.

To show the application of the methods we'll use the partition function, since it captures the essence of the methods without loss of generality, as any other thermodynamic state function can be easily calculated from the partition function. However, it should be noted that partition function is not necessary at all to obtain thermodynamic properties (which can be obtained without using it but rather using the free energy), rather it is a convenient tool to derive them. Nevertheless, it also simplifies our investigations due to the fact it gives a self-contained quantity to explore and demonstrate the size and shape effects explicitly.

\subsection{Poisson summation formula}

Let's return to back to the mathematical tools that we refer. To calculate infinite sums, there are several approaches like Poisson summation formula and Abel-Plana formula as well as Euler-Maclaurin formula, which is for finite sums. The most convenient one for the evaluation of the sums encountered in statistical physics is the Poisson summation formula (PSF). PSF relates the original summation to the summation of its Fourier transform. The steps to get the reduced version of PSF for even functions has been given in the Appendix of my Master of Science thesis \cite{msc}. Here, we'll focus more on the understanding of PSF. For even functions (which applies to all thermodynamic state functions) the PSF can be written as
 \begin{equation}
\sum_{i=1}^{\infty}f(i)=\int_0^\infty f(i)di-\frac{f(0)}{2}+2\sum_{s=1}^{\infty}\int_0^\infty f(i)\cos(2\pi si)di.
\end{equation}
PSF consists of three terms. You may have noticed that the first term actually corresponds to the continuum approximation (i.e. replacement of the sum with integral). As it is shown in Fig. 2.8, it is a reasonable approximation for systems having less significance on near ground states which corresponds to the vanishing values of the confinement parameter $\alpha$. Look at the accuracy of the black curve representing the colored columns of the summations. It has been shown clearly in our other study that in order for continuum approximation to be used, occupation probabilities of the excited states need to be substantial \cite{aydin3}.

\begin{figure}
\centering
\includegraphics[width=0.95\textwidth]{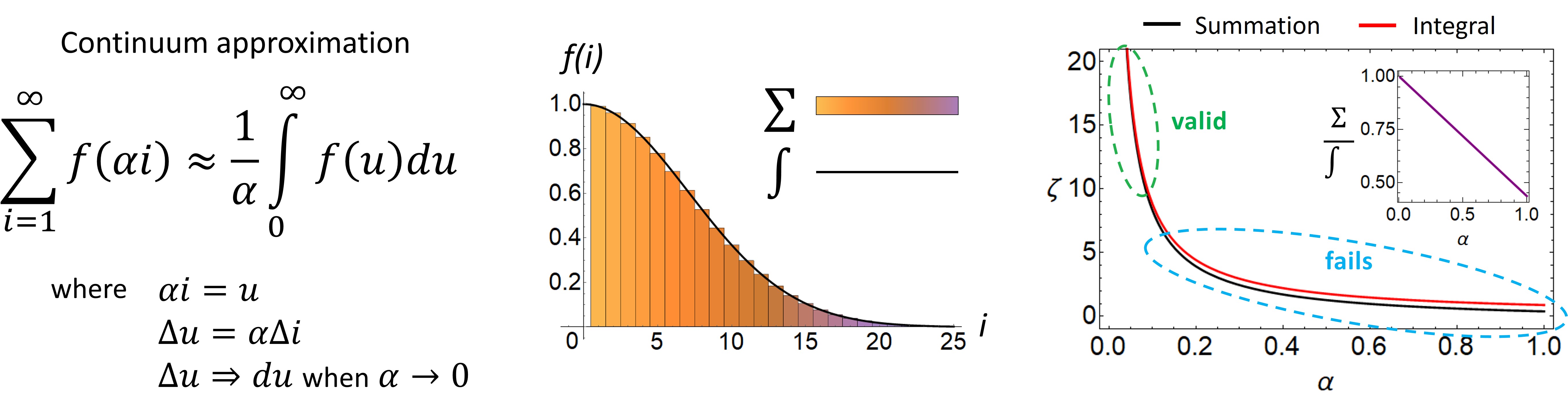}
\caption{The usual (unbounded) continuum approximation, which corresponds to the replacement of infinite sums by integrals. It is a good approximation for systems with a lesser degree of discreteness. Subfigure in the rightmost figure shows the ratio of the summation and integral representations. It fails for not very small $\alpha$ values.}
\label{fig:c2f8}
\end{figure}

\begin{figure}[!b]
\centering
\includegraphics[width=0.95\textwidth]{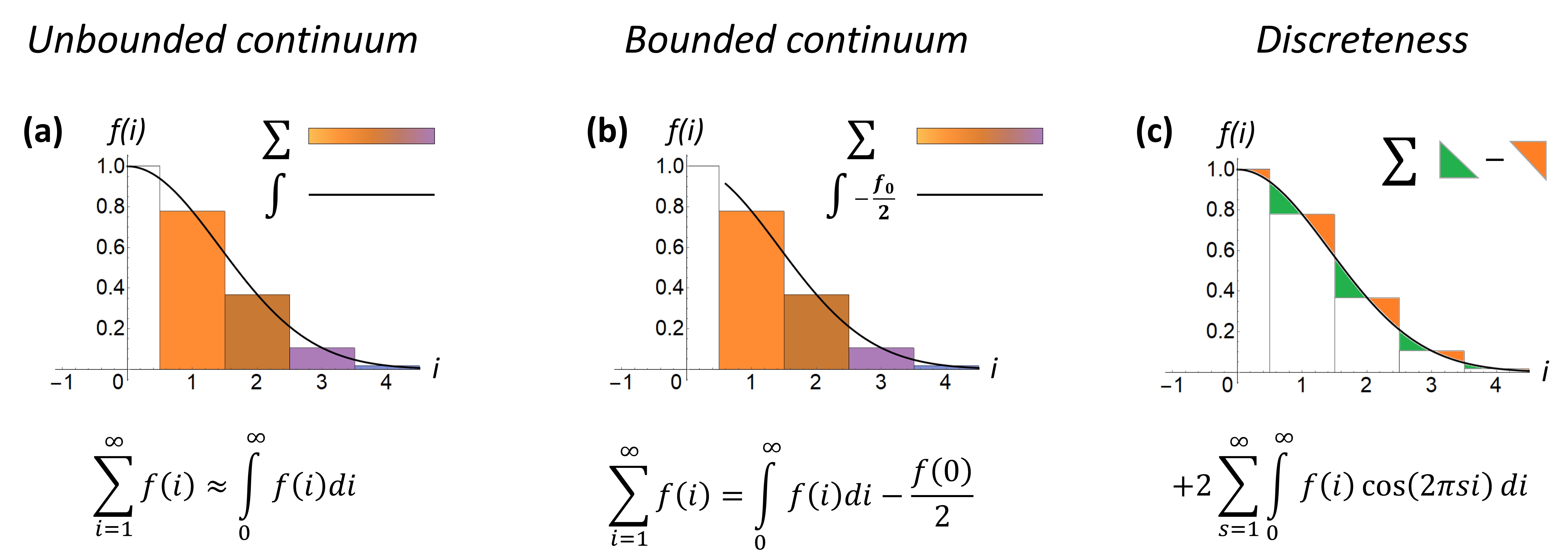}
\caption{Representations of (a) unbounded and (b) bounded continuum approximations via the Poisson summation formula (PSF). Consideration of the second term of the PSF enhances the usual continuum approximation into the bounded domains. (c) The third term of PSF corresponds to the discreteness correction and responsible from recovering the intrinsic discrete nature of the actual summation.}
\label{fig:c2f9}
\end{figure}

If the system is so confined that its energy is close to ground state, the first term of PSF fails to represent the summation, see Fig. 2.9a. Discreteness of the sum and the smoothness of the integral can be easily noticed in Fig. 2.9a. It should be noted that the summation starts not from zero but from unity, unlike the integral. Therefore, integral representation contains the half of the zeroth value incorrectly. Bounded continuum approximation corrects this improper calculation by removing the false contribution from the half of the function's zeroth value by the integral, Fig. 2.9b. The correction term is called the zero correction and corresponds to the second term of the PSF. Physically, zero correction excludes the false contribution of the zeroth quantum state, since there is no such state since all quantum states start from unity. All usual quantum size effect corrections to the thermodynamic properties come from this second term. 

For a system to be considered as confined, either the temperature is very low ($\lambda_{th}$ is large), or the sizes are very small so that the confinement parameter is near to or larger than unity. Even though bounded continuum approximation is accurate up to confinement values near unity, it considerably fails for $\alpha>1$. The third term of PSF becomes important for such strongly confined systems. Note that integral curve passes from the middle points of the summation columns. For $\left\{i-0.5\right\}$ integral representation mistakenly calculates over the top whereas for $\left\{i+0.5\right\}$ it falls short, Fig.2.9c. In total, this surplus (green triangles) and deficiency (orange triangles) roughly compensate each other, except for strong confinements. The third term of PSF is called the discreteness correction and corrects this total oscillatory difference between integral and the summation. 

Despite the power of PSF, it has some weaknesses as well. You can only use it if you can obtain the energy eigenvalues analytically, which can be done only for some limited number of geometries like rectangular, cylindrical and spherical. Even for cylindrical and spherical ones, we have to use some approximations to obtain analytical expressions for energy eigenvalues. If you want to explore quantum size effects for arbitrary geometries, PSF cannot do the job. Fortunately, there is another mathematical tool for that: the Weyl conjecture.

\subsection{Weyl density of states}

In 1911, German mathematician Hermann Weyl conjectured a law describing the asymptotic behavior of the Laplace-Beltrami operator. It is known as one of the earliest works on spectral geometry, the field of mathematics concerning with the relationships between spectrum of bounded differential operators and geometric structures of manifolds. Spectral geometry and Weyl's result had considerable influence on the mathematical formulation of quantum mechanics and the understanding of geometry effects in confined systems\cite{pathbook}. The question of "Can one hear the shape of a drum?" that we mentioned in the introduction chapter is related with the Weyl conjecture. This question is about an inverse problem which is determining the shape of a drum from its sound, see Fig. 2.10. On the other side, the Weyl's conjecture "solves" the direct problem, which is determining the behavior of the spectrum by using its geometric properties. Of course, Weyl's conjecture cannot be used to answer Kac's question, because it is valid only at asymptotics. But this result definitely has intrigued Kac's question. In particular, the Weyl's conjecture gives the asymptotic behavior of the number of eigenvalues less than $k$ for Helmholtz equation, which can be written in its general, d-dimensional, closed form as,
\begin{equation}
\Omega_d(k)= \sum_{n=0}^d\left(-\frac{1}{4}\right)^{d-n}\left(\frac{k}{2\sqrt{\pi}}\right)^n\frac{\mathcal{V}_n}{\Gamma\left[\frac{n+2}{2}\right]},
\end{equation}
where $\mathcal{V}_n$ is the $n$-dimensional volume according to Lebesgue measure ($\mathcal{V}_3\rightarrow \mathcal{V}$, $\mathcal{V}_2\rightarrow \mathcal{A}$, $\mathcal{V}_1\rightarrow \mathcal{P}$, $\mathcal{V}_0\rightarrow \mathcal{N_V}$). Note that Eq. (2.10) is analytical, as the summation over $n$ serves for generating the lower-dimensional correction terms. This result is known as the Weyl conjecture, but actually the term conjecture is a misnomer here, as it is proved and generalized by Victor Ivrii in 1980 \cite{ivri}. It can readily be seen that Eq. (2.10) contains all geometric size variables in separate terms, just like the first two terms of PSF gives. From the relation $p=\hbar k$ wavenumber equals to the momentum with the proportionality constant $\hbar$. Equations depending only on $k$ are said to be written in momentum space.

\begin{figure}
\centering
\includegraphics[width=0.9\textwidth]{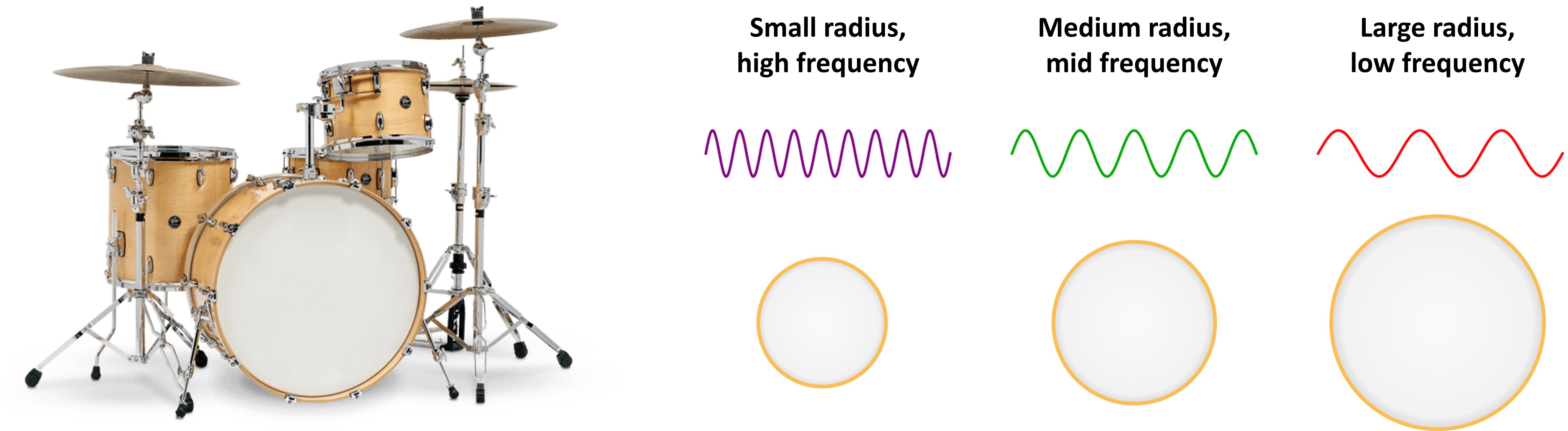}
\caption{An inverse problem of hearing the shapes. Surface area determines the pitch of the toms in a drum, physically showing the relationship between the size and the eigenvalue spectrum. Light waves are shown for analogy.}
\label{fig:c2f10}
\end{figure}

In statistical physics, by using a useful concept called density of states, integrals over quantum states can be replaced by integral over energy. Thereby, one can reduce multiple integrals into a single one. In other words, density of states represents the Jacobian when we change the coordinate systems from Cartesian to spherical one. Density of states describe the number of states within an infinitesimal momentum or energy interval. Note that Weyl conjecture given in Eq. (2.10), gives the number of states up to a wavenumber. Hence, the derivative of Eq. (2.10) with respect to wavenumber gives the density of states in momentum space. Likewise, density of states in energy space can be obtained by writing the Eq. (2.10) in energy space and differentiating it with respect to energy. 

Weyl density of states (WDOS) in $k$-space can easily be obtained by differentiating Eq. (2.10) with respect to $k$. For this reason, we won't write it explicitly, but instead below we write WDOS in energy space, which can also be used directly to calculate the thermodynamic properties.
\begin{equation}
WDOS_d(\varepsilon)= \sum_{n=0}^d\left(-\frac{1}{4}\right)^{d-n}\frac{n \mathcal{V}_n}{2\Gamma\left[\frac{n+2}{2}\right]\lambda_{th}^n}\left(\frac{\varepsilon}{k_BT}\right)^{\frac{n-2}{2}},
\end{equation}

% Weyl
Weyl conjecture gives identical results with the first two terms of PSF. Quantum size effects are studied in literature by using Weyl conjecture or first two terms of PSF in the calculation of infinite summations \cite{pathbook,molina,pathria,dai1,dai2,sismanmuller,sisman,dai3,tse4,qforce}. A comprehensive investigation of Weyl density of states has been done in Refs. \cite{aydin3,aydin8}.

% Casimir
Weyl conjecture and shape dependence is also significant when calculating forces that depend on the geometric structure of systems such as Casimir force \cite{PhysRevLett.87.260402,cass,PhysRevA.87.042519}. Geometric Weyl corrections shed light on the understanding of the ultraviolet divergences and cutoffs encountered in the calculation of Casimir force \cite{cass}. Weyl conjecture may be used also to obtain analytical expressions for such geometric dependencies \cite{PhysRevLett.87.260402}.

\subsection{Quantum size effect corrections to the partition function}

Now let's see our tools in action. A visual summary of the formalism of quantum statistical thermodynamics is presented in Fig. 2.11. Consider large number of non-interacting particles confined in a domain. First, the Schr\"{o}dinger equation is solved considering the boundary conditions exposed by the confinement domain, then the obtained eigenvalues are used in the appropriate partition function. In the figure, Maxwell-Boltzmann weight is used but the formalism can directly be applied to other statistics as well. There are two options. If one knows the eigenvalues analytically (which is only possible for limited number of geometries), the first two terms of PSF can be used to obtain the partition function under quantum size effects. Otherwise, WDOS concept can be used for any geometry. After obtaining the partition function, any thermodynamic state function can be easily calculated by standard thermodynamic relations. 

\begin{figure}
\centering
\includegraphics[width=0.99\textwidth]{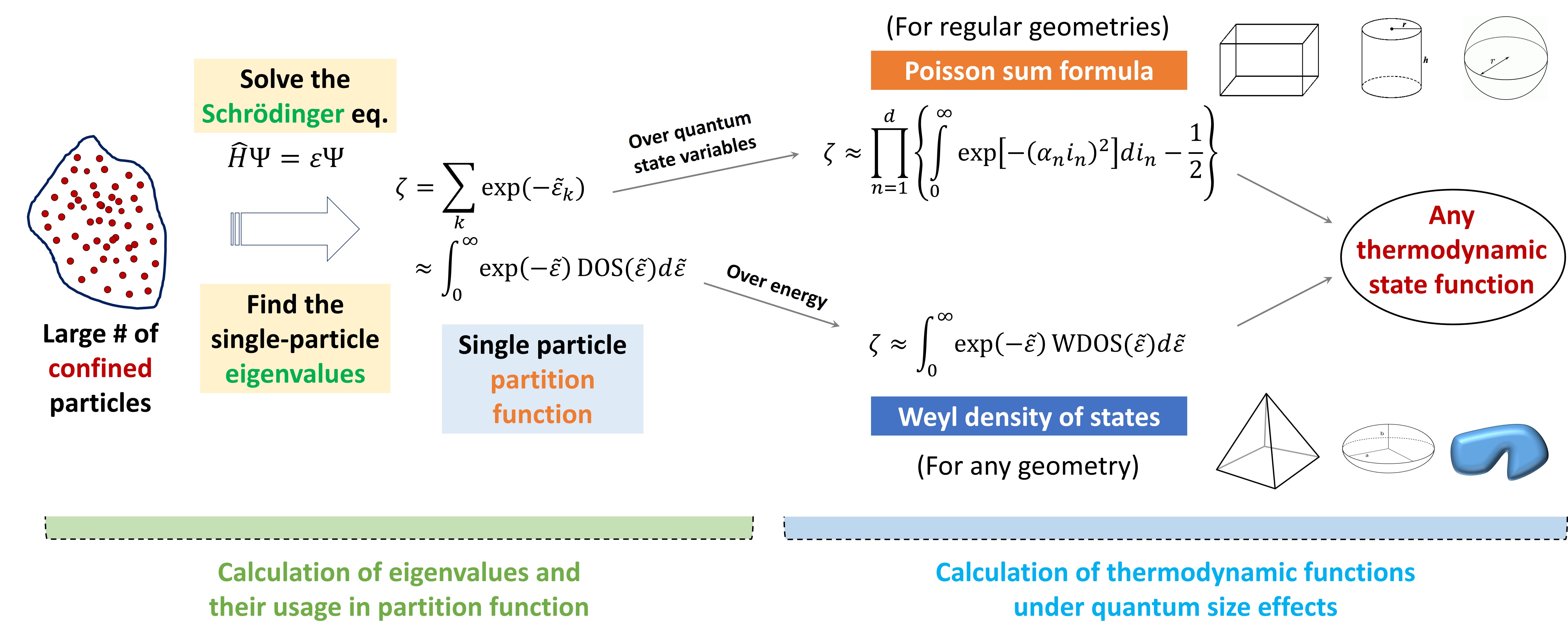}
\caption{Quantum statistical thermodynamics formalism for bounded continuum systems. PSF or WDOS concepts can be used to obtain quantum size effect corrections to the thermodynamic quantities.}
\label{fig:c2f11}
\end{figure}

For example, partition function by using either PSF or WDOS in three different dimensions along with quantum size effect corrections is written as
\begin{subequations}
\begin{align}
& \zeta_3=\frac{\mathcal{V}}{\lambda_{th}^3}-\frac{\mathcal{A}}{4\lambda_{th}^2}+\frac{\mathcal{P}}{4^2\lambda_{th}}-\frac{\mathcal{N_V}}{4^3}, \\
& \zeta_2=\frac{\mathcal{A}}{\lambda_{th}^2}-\frac{\mathcal{P}}{4\lambda_{th}}+\frac{\mathcal{N_V}}{4^2}, \\
& \zeta_1=\frac{\mathcal{P}}{\lambda_{th}}-\frac{\mathcal{N_V}}{4}.
\end{align}
\end{subequations}
Here $\lambda_{th}^d$ (dimensional powers of the thermal de Broglie wavelength) can be thought of the amount of $d$-dimensional volume roughly one particle occupies in an $d$-dimensional space. Then, we can define the de Broglie density $n_{dB}$, which is the reciprocal of $\lambda_{th}^d$. In $d$-dimensions it is defined as $n_{dB,d}=\lambda_{th}^{-d}$. This quantity is also called as quantum density or phase space density in literature, but to make it distinguishable from other different quantities with similar names, we call it de Broglie density, which is an appropriate terminology since it defines the density in the de Broglie volume. Additive nature of the lower dimensional terms is not due to an assumption, but directly comes from the exclusion of zero-energy states from the bulk term.

The partition function tells us very important things. If we pay attention to the functional dependencies, we see that the partition function is predominantly proportional with the bulk sizes of the particular dimension ($\zeta\propto\mathcal{V}$ in 3D, $\zeta\propto\mathcal{A}$ in 2D, $\zeta\propto\mathcal{P}$ in 1D). Also, the temperature dependency changes according to dimension as $\zeta\propto T^{3/2}$ in 3D, $\zeta\propto T$ in 2D and $\zeta\propto \sqrt{T}$ in 1D, so $\zeta\propto T^{n/2}$ in $n$D generally. In this regard, the partition function can also be interpreted as the number of small D-dimensional cubes with the size of thermal de Broglie wavelength that can fit into a confinement domain.

\section{Quantum Boundary Layer Method}

Both PSF and WDOS are useful mathematical tools to obtain quantum size effect corrections to the thermodynamic properties of particles at nanoscale. However, they won't provide deeper insights on how and why these correction terms appear. For example, why they add up with alternating sign? What is the meaning of the constants that multiply each individual term? These questions are answered in 2006, when the concept of quantum boundary layer has been developed and used as a new method for the calculation and understanding of the underlying mechanisms of quantum size effects \cite{qbl}. This method is crucial in the context of quantum statistical thermodynamics in the sense that it makes it possible to obtain the thermodynamical quantities with quantum confinement corrections without needing to explicitly solve the Schr\"{o}dinger equation or use Weyl conjecture, practically leaving out the occasionally burdensome statistical calculations. 

\subsection{Quantum thermal probability density distribution}

Quantum thermal probability density distribution, describing the number density distribution of quantum particles at thermal equilibrium, lies at the heart of the quantum boundary layer concept. Consider particles confined in a 1D domain. Due to their wave nature, particles have tendency to stay away from the boundaries of the domain, generating a non-uniform density distribution even in thermodynamic equilibrium. Quantum boundary layer (QBL) method approximates this non-uniform density distribution with a uniform one by introducing empty layers on boundaries, thereby constituting an effective size. How this is done can be seen in Fig. 2.12. Black curve represents the exact \textit{ensemble-averaged quantum-mechanical particle number density} distribution, in short we call \textit{quantum thermal density}, given by the following expression:
\begin{equation}
n(\textbf{x})=\left\langle\left|\Psi(\textbf{x})\right|^2\right\rangle_{\mli{ens}}=\frac{\sum_i{\exp(-\varepsilon_i\beta)}\left|\Psi_{i}(\textbf{x})\right|^2}{\sum_i{\exp(-\varepsilon_i\beta)}}
\end{equation}
where $\left|\Psi(\textbf{x})\right|^2$ is the quantum-mechanical probability density and $\beta=1/(k_BT)$. True density distribution of particles is non-uniform; forms a plateau at the center and gradually goes to zero to the boundaries. After around 2$\delta$, the density becomes uniform, but it is non-uniform within around 2$\delta$ thickness near to boundaries. This is a local information about the occupation probabilities of the confined particles. On the other hand, in thermodynamics, we deal with the global properties of matter. At this point, we can make an approximation by replacing the non-uniform distribution with a uniform one around the center and completely empty layer near to the boundaries, red-dashed curve in Fig. 2.12. This assumption gives us the possibility to still define global properties but also consider the wave behavior of particles. Two variables determine the thickness of the QBL, the height of the plateau (maximum density value) and the total area under the curves which has to equal to unity due to the law of conservation of probability. By these two constraints, the thickness of QBL is obtained as
\begin{equation}
\delta=\frac{h}{4\sqrt{2\pi mk_BT}}=\frac{\lambda_{th}}{4},
\end{equation}
which is not so surprisingly related to the thermal de Broglie wavelengths of particles, and it turns out to be conveniently quarter of it.

\begin{figure}
\centering
\includegraphics[width=0.95\textwidth]{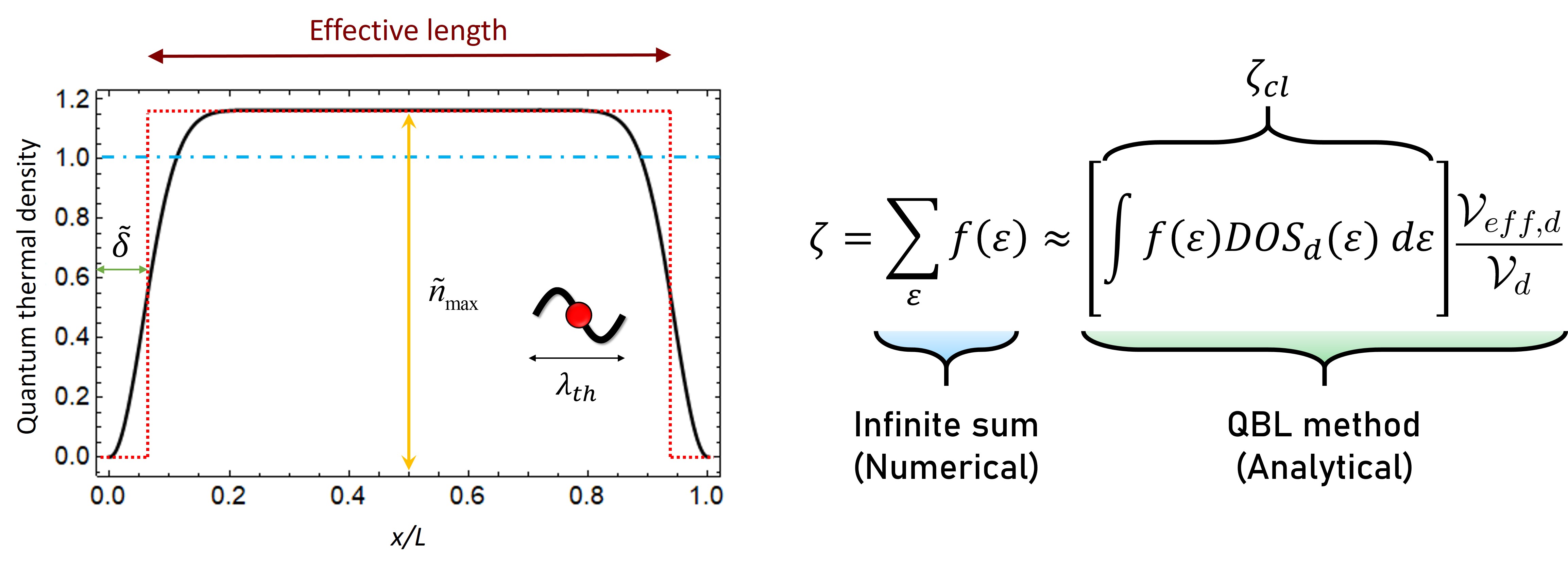}
\caption{Quantum boundary layer (QBL) methodology in a nutshell. Ensemble-averaged quantum-mechanical density distribution of confined particles is non-uniform (black curve) and it can be approximated by a uniform one (red-dashed curve) via the introduction of QBL $\delta$. QBL method can be applied to the calculation of any quantity $X$ by a summation over a distribution function $f$. Quantum size effect corrections on thermodynamic and transport properties can exactly and analytically be obtained from their conventional (bulk) expressions just by replacing the apparent sizes with the effective ones.}
\label{fig:c2f12}
\end{figure}

QBL contains the statistical and quantum nature of particles in an approximate and unified way, just like the effective mass concept, capturing the approximate influence of potential fields on particles, in solid state physics. After the thickness is determined (which turns out to be geometrically universal\cite{uqbl}), QBL method is basically based on the replacement of apparent geometric size variables with the effective ones. In the calculation of thermodynamic properties, this replacement allows to some extent eliminate the difference between the conventional integral approach and the exact summations over the eigenvalues of the Schr\"{o}dinger equation. In other words, QBL constitutes a bridge between classical and quantum pictures of thermodynamic and transport expressions. Even the Weyl conjecture is not needed.

Methodology of QBL is summarized in Fig. 2.12. The partition function consists of an infinite sum of which calculation might be numerically cumbersome. By using the standard formalism of statistical mechanics, sums can be replaced by integrals, with a minor trick where the actual volume is also replaced by the effective volume which can be found by QBL formalism. Rather than partition function, free energy could also very well be used for this transformation.

The methodology is pictured more visually in Fig. 2.13. Classically, particles are uniformly distributed inside the domain, whereas in confined domains at nanoscale, they form a non-uniform distribution due to the wave nature. Existence of a nearly empty layer near to boundaries can be explicitly seen also in the contour plot of the 2D domain's density. The approximation of the actual density with a uniform middle region and empty near-boundary region can be seen on the rightmost subfigure of Fig. 2.13. Although QBL is constructed by considering the 1D domain, it is straightforwardly applicable to the higher dimensional domains like the one shown in Fig. 2.13.

\begin{figure}[!b]
\centering
\includegraphics[width=0.95\textwidth]{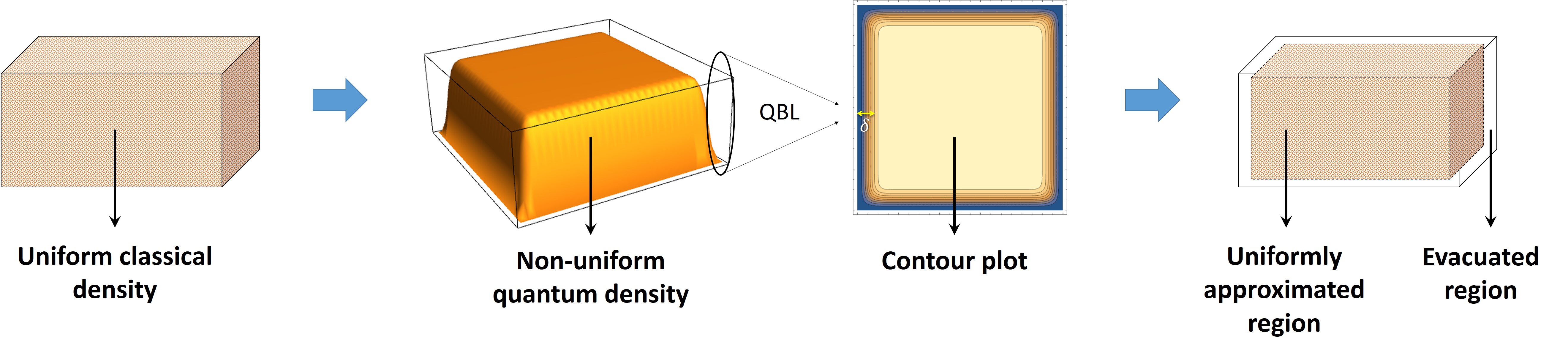}
\caption{Dressing QBL to a 2D square domain and constructing the effective area. Approximating the non-uniform quantum density by an effectively uniform one via the introduction of QBL.}
\label{fig:c2f13}
\end{figure}

In Fig. 2.14, calculation of the effective area in a 2D domain is shown. The 2D square domain has length $L$ and QBL dictates a reduction in length by $\delta$ from every outside boundary. This makes the effective square domain's length $L-2\delta$. If one calculates the effective area, it can be readily found that it is $\mathcal{A}_{\mli{eff}}=L^2-4L\delta+4\delta^2$. Note that $4L$ is the periphery of the domain and $4$ is the number of vertices, so $\mathcal{A}_{\mli{eff}}=\mathcal{A}-\delta\mathcal{P}+\delta^2\mathcal{N_V}$. This is the origin of the constants appearing in quantum size effect terms. To subtract the QBL area from the actual domain, we first subtract the peripheral QBL areas ($\delta$ times four side lengths of the square) but then the square areas at the corners are subtracted twice which needs to be added to correct the calculation. This is the origin of the alternating sign in the corrections.

\begin{figure}
\centering
\includegraphics[width=0.7\textwidth]{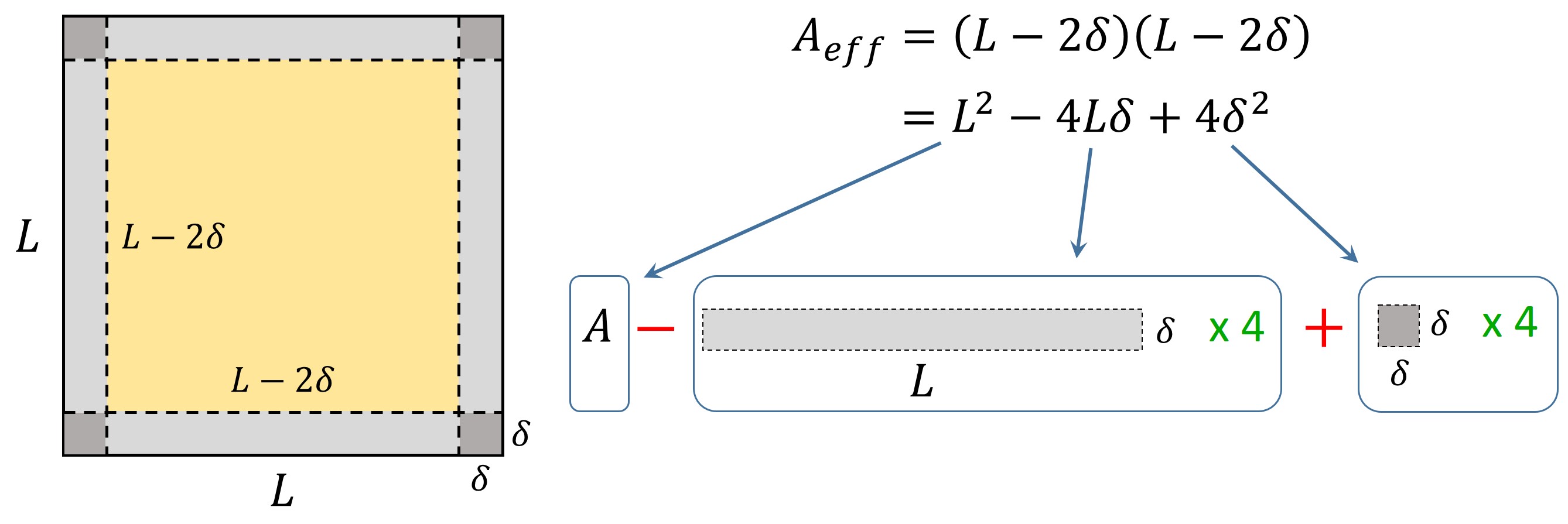}
\caption{An example showing explicitly how geometric size correction terms appear.}
\label{fig:c2f14}
\end{figure}

By the QBL procedure, effective volume in $d$-dimensions is written as $\mathcal{V}_{\mli{eff},d}=\mathcal{V}_d-\mathcal{V}_{qbl,d}$. More explicitly, effective volumes in various discrete dimensions are written as
\begin{subequations}
\begin{align}
\mathcal{V}_{\mli{eff}}&=\mathcal{V}-\delta\mathcal{A}+\delta^2\mathcal{P}-\delta^3\mathcal{N_V}, \\
\mathcal{A}_{\mli{eff}}&=\mathcal{A}-\delta\mathcal{P}+\delta^2\mathcal{N_V}, \\
\mathcal{L}_{\mli{eff}}&=\mathcal{L}-\delta\mathcal{N_V}.
\end{align}
\end{subequations}
which can be generalized by the expression $\mathcal{V}_{\mli{eff},d}=\sum_{n=0}^{d}(-\delta)^{d-n}\mathcal{V}_n$ for the $d$-dimensional effective volume.

\subsection{Obtaining quantum size effects by the quantum boundary layer}

We can now apply the QBL procedure to find the partition function, which can be approximated in $d$-dimension by the QBL method as
\begin{equation}
\begin{split}
\zeta_d=\sum_{\varepsilon}f(\varepsilon)\approx\zeta_{cl,d}\frac{\mathcal{V}_{\mli{eff},d}}{\mathcal{V}_d},
\end{split}
\end{equation}
where $\zeta_{cl,d}=\mathcal{V}_d/\lambda_{th}^{d}$ is the classical partition function expression in $d$-dimension, that can be found in textbooks. The classical partition function is found by direct conversion of summation into integral.

In terms of QBL thickness $\delta$, partition function using QBL method can be written analytically as
 \begin{equation}
\zeta_d=\sum_{n=0}^{d}\frac{(-\delta)^{d-n}}{(4\delta)^d}\mathcal{V}_n=n_{dB,d}\mathcal{V}_{\mli{eff},d}
\end{equation}
where $n_{dB,d}=(4\delta)^{-d}$ is $d$-dimensional de Broglie density and $\mathcal{V}_{\mli{eff},d}$ is $d$-dimensional effective volume. As is seen, partition function actually becomes the product of de Broglie density and the effective volume. In this sense, partition function can be interpreted also as the number of particles occupying the effective volume with the de Broglie density.

Eq. (2.17) can be written in its open form for different dimensions:
\begin{subequations}
\begin{align}
& \zeta_3=n_{dB,3}\mathcal{V}\left(1-\delta\frac{\mathcal{A}}{\mathcal{V}}+\delta^2\frac{\mathcal{P}}{\mathcal{V}}-\delta^3\frac{\mathcal{N_V}}{\mathcal{V}}\right), \\
& \zeta_2=n_{dB,2}\mathcal{A}\left(1-\delta\frac{\mathcal{P}}{\mathcal{A}}+\delta^2\frac{\mathcal{N_V}}{\mathcal{A}}\right), \\
& \zeta_1=n_{dB,1}\mathcal{P}\left(1-\delta\frac{\mathcal{N_V}}{\mathcal{P}}\right).
\end{align}
\end{subequations}
where the de Broglie density becomes $n_{dB,d}=(4\delta)^{-d}$. These are mathematically equivalent to the ones found by the WDOS method in subsection 2.2.3. Note that the expressions aren't geometry specific, i.e. they are valid for any geometry. If one can properly dress the QBL to a domain, one does not need to worry about the lower-dimensional geometric size elements. They emerge as a result of this dressing. Therefore, QBL method not only generates the same terms but also underlies the existence of these terms in the equations. Note that all quantum size effect terms depend on the thickness of QBL which is of quantum nature. When Planck constant is hypothetically set to zero, they all disappear. In this regard, they are genuinely quantum.

\subsection{Accuracy in arbitrary domains}

QBL is primarily defined by considering regular domains. Still, it is directly applicable to arbitrary domains and accurately predicts the thermodynamic properties with high precision. One can think of many arbitrary shapes. To examine, we choose three characteristic shapes which are significant for our discussion due to their unique properties. The first shape we consider is a square domain having an infinitesimally small point/dot at its center. Classically, infinitesimally small points shouldn't change any property of the system, because they don't even have any size. However, due to the quantum nature of particles, they can feel the existence of even an infinitesimal boundary where they form a quantum boundary layer. This can be seen from the density distributions given in Fig. 2.15. For comparison, we'll execute the analysis by comparing our shape (II) with a plain square (I). Temperature is taken to be $300$K for all the calculations in this subsection. Errors due to numerical calculations are ensured to be negligible by the procedures explained in Appendix A.1.

\begin{figure}
\centering
\includegraphics[width=0.55\textwidth]{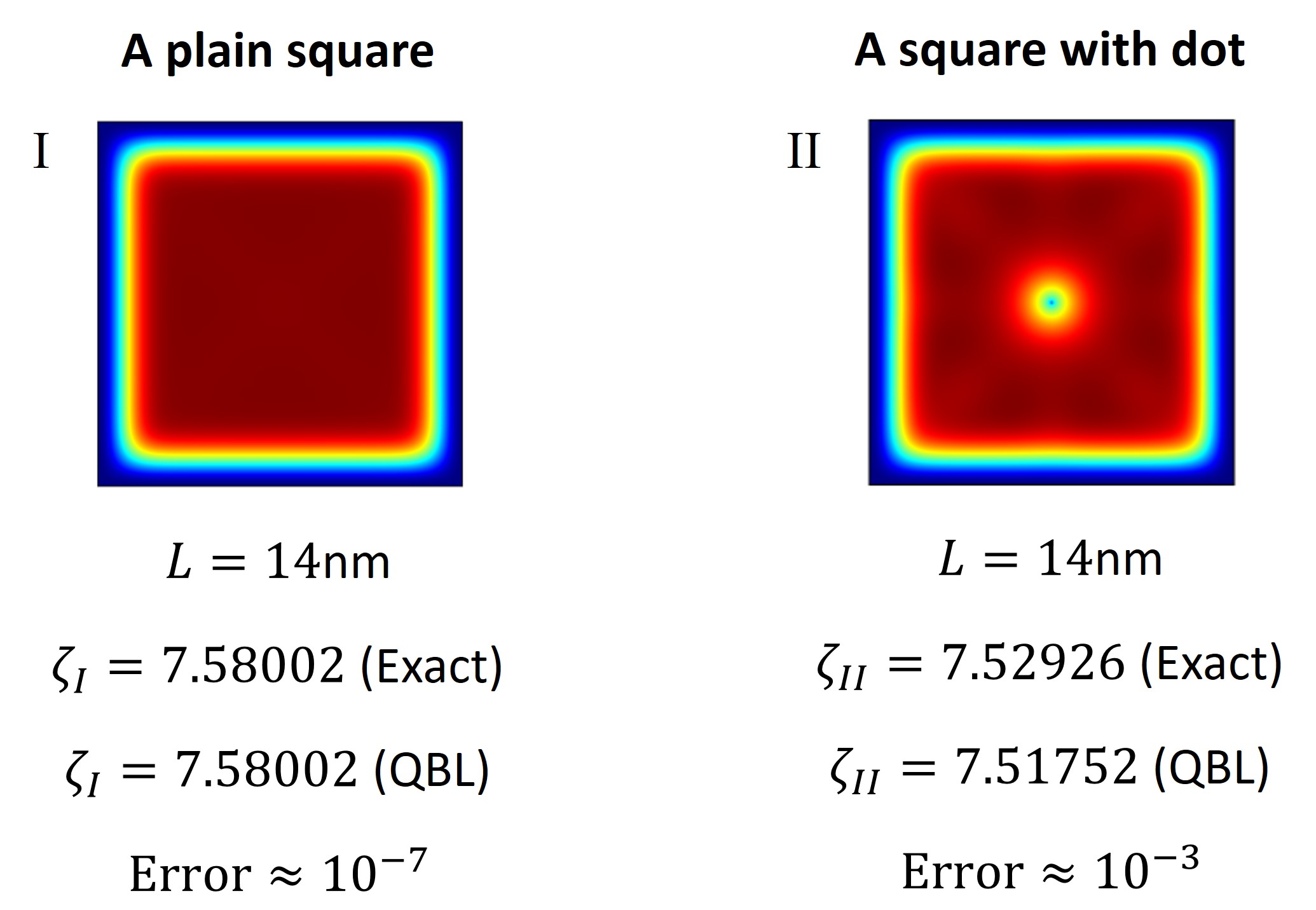}
\caption{The first arbitrary domain example. Comparison of a plain square domain and a square domain with an infinitesimally small point at the center. Even an infinitesimal point creates considerable amount of change in the partition function and QBL methodology can predict this with quite good accuracy. Blue and red colors represent the lower and higher density regions.}
\label{fig:c2f15}
\end{figure}

We calculate the partition functions for both shapes numerically (exact) first. We find more than $0.6\%$ difference between their partition functions. The difference might sound tiny at first, but when you think that this is caused just by a single point with zero thickness, that difference is not trivial at all. Then we calculate partition functions using QBL method. The calculation for plain square is easy as we have already shown before, $\zeta_{I}=\mathcal{A}_{\mli{eff}}^{I}/(4\delta^2)$ where $\mathcal{A}_{\mli{eff}}=14^2\times 10^{-18}-14\times 10^{-9} 4\delta+4\delta^2$. This expression gives less than $10^{-6}$ relative error, which is remarkably accurate (100 times smaller than the difference itself) for such a small domain size. A plain square was easy, but what about the square with a dot inside? Can QBL method provide accurate expressions for that as well? The answer is yes! The difference of this shape from a plain square is the creation of the infinitesimally small dot or a single vertex. But the tricky point is we shouldn't add this vertex point to the number of vertices! We should subtract. The reason is creation of a vertex inside the domain is quite different than creating it on the boundary (such as corners). When you create a point inside of the domain (which is considered as an internal boundary), you are reducing the effective sizes of the domain by the amount of the space evacuated by its QBL. By keeping the self-similarity of the domain, we can approximate the evacuated area near the single point as $\delta^2$. Removing one $\delta^2$ exactly means removing a single vertex point from the number of vertices. Then, the effective area in the partition function of the square with dot shape becomes $\mathcal{A}_{\mli{eff}}^{II}=14^2\times 10^{-18}-14\times 10^{-9} 4\delta+3\delta^2$. This expression, not only provides physical explanations to the changes in partition function, but also relatively accurately represents it with a relative error around $0.1\%$.

The second domain that we consider is the one shown in Fig. 2.16 where ITU (stands for the abbreviation of Istanbul Technical University) letters are formed by infinitesimally thin line boundaries inside a rectangle (constituting internal boundaries as opposed to the external ones which are the rectangle's boundaries). We again numerically solve the Schr\"{o}dinger equation for this domain and find the partition function as $\zeta_{ITU}=15.7069$. Now let's calculate what QBL method gives. As we always do, we calculate the effective area to find the partition function. So we need the surface area, periphery and number of vertices for this domain. The surface area is just the area of the rectangle, $\mathcal{A}=27.5\times 17.0$ nm$^2$, since infinitesimal lines don't contribute to the surface area. Periphery calculation in this domain needs to be handled carefully. As it is seen from the density distributions of particles in this domain, QBL of external boundaries are calculated by periphery times $\delta$, whereas QBLs of internal boundaries are twice of the periphery times $\delta$. This is because internal boundaries lead to the evacuation of particles on both sides of the boundary, as QBL is formed in both sides of internal boundaries, unlike the external ones, Fig. 2.16. Therefore, QBL's of internal boundaries have $2\delta$ thickness (except when they are closer to boundaries than $2\delta$). Periphery calculation in this domain then becomes: $\mathcal{P}=2\times(27.5+17.0)+2\times(4\times 7.0+5.0+4.5)$ nm (vertical lines have 7.0nm, horizontal lines have 5.0 and 4.5 nm lengths).

\begin{figure}[!b]
\centering
\includegraphics[width=0.85\textwidth]{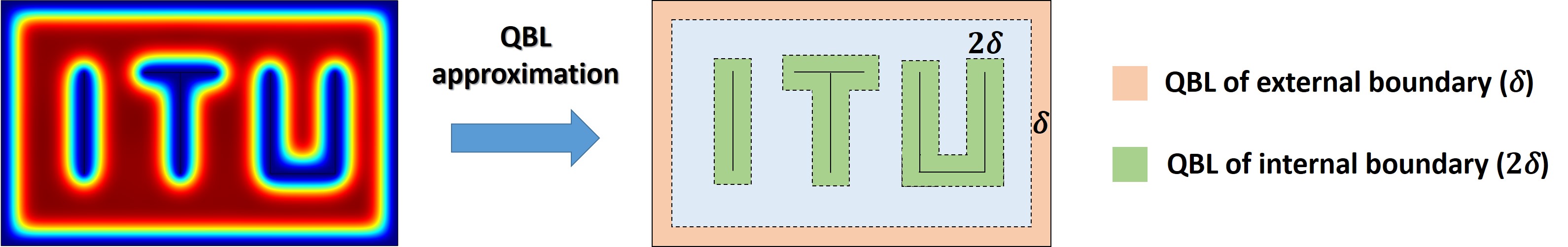}
\caption{The second arbitrary domain example. Application of the QBL methodology to an arbitrary domain. Unlike the previous domains that are considered, internal and external boundaries affect the system differently.}
\label{fig:c2f16}
\end{figure}

We are almost done with the effective area calculation. But the trickiest part comes at the last. How do we calculate number of vertices in this domain with both external and internal boundaries? Should we just sum them up? The answer is no! It is clear from the density distribution in Fig. 2.16 that QBL does not form evenly among the different vertices. For example around outer corners, QBL evacuates lots of space, whereas around inner corners (e.g. the bottom of the letter U) instead of an evacuation QBL is even invited to enter more closer into the cusps. This is the case also for the open cusps (e.g. the ends of the letter I). Is it possible to account these effects at once? Or do we have to deal with each and every case separately? Fortunately, we found an approximate but reasonable formula for the calculation of the effective number of vertices in a confined system. The effective number of vertices in its general form is written as
\begin{equation}
\mathcal{N_V}=\sum_n \cot(a_n/2),
\end{equation}
where $a$ is an interior angle of the domain and the sum is over all interior angles of the domain.

When we calculate the area of QBL covering a boundary, it is simply the subtraction of vertex contribution ($\mathcal{N_V}\delta^2$) from the periphery (that is exposed by the particles) times QBL thickness ($\mathcal{P}\delta$). Note that the vertices contribution can be positive or negative depending on the angle of the vertex. In case of an acute angle, vertices contribution is positive since it causes the overlaps of QBLs at the vertex. On the other hand, it is negative in case of a reflex angle because of the reverse effect (filling rather than evacuation). The amount of overlap area at $90^{\circ}$ angle is exactly $\delta^2$.

The number of vertices formula is approximate because it assumes that the angle $a_n$ is between infinitely long boundaries. Due to that, it diverges at $0^{\circ}$ and $360^{\circ}$ angles. Nevertheless, we can calculate the effective number of vertices with quite good precision. There are eight of $90^{\circ}$ angles (4 corners of the rectangle, 2 inner angles of the bottom of U letter and 2 faced down angles of the top of T letter), two of $270^{\circ}$ angles (2 outer angles of the bottom of U letter) and there are seven angles which are open (the ends of I and T letters and upper tips of U letter). We can treat these open cusps as $270^{\circ}$, since the formula goes to infinity for $360^{\circ}$. Then, the number of vertices becomes $\mathcal{N_V}=8\cot(90/2)+2\cot(270/2)+7\cot(270/2)=-1$. This is value and other negative values for number of vertices are perfectly fine as they should be interpreted as how the confined particles feel them. By using $\mathcal{A,P,N_V}$, we can find the effective area and then the partition function as $\zeta_{ITU}^{QBL}=15.6582$. The relative error of this result is around $0.3\%$, which is very low, considering the approximations we did.

The last domain we'd like to consider is even more complex, which is a tractricoid, see Fig. 2.17. Numerical calculations reveal that the partition function for the particles confined in this domain is $\zeta_{trac}=8.80968$. It is easy to calculate the surface area and periphery for this domain. For the number of vertices, it is clear that there are four of $90^{\circ}$ angles (corners of the rectangle). Also, the bottom tip of the tractricoid is very close to the outer boundary and evacuates all the space from particles around it. Therefore it can be thought as if it closes the bottom by creating additional two $90^{\circ}$ angles. On the top and middle of the tractricoid, there are three very small capes. These are so small that, they can be treated as $270^{\circ}$ cusps. This is justified by also the density distribution profiles. Then, number of vertices become  $\mathcal{N_V}=6\cot(90/2)+3\cot(270/2)=3$. With that the QBL approach gives $\zeta_{trac}^{QBL}=8.82385$. This value has a relative error of less than $0.2\%$. 

\begin{figure}
\centering
\includegraphics[width=0.3\textwidth]{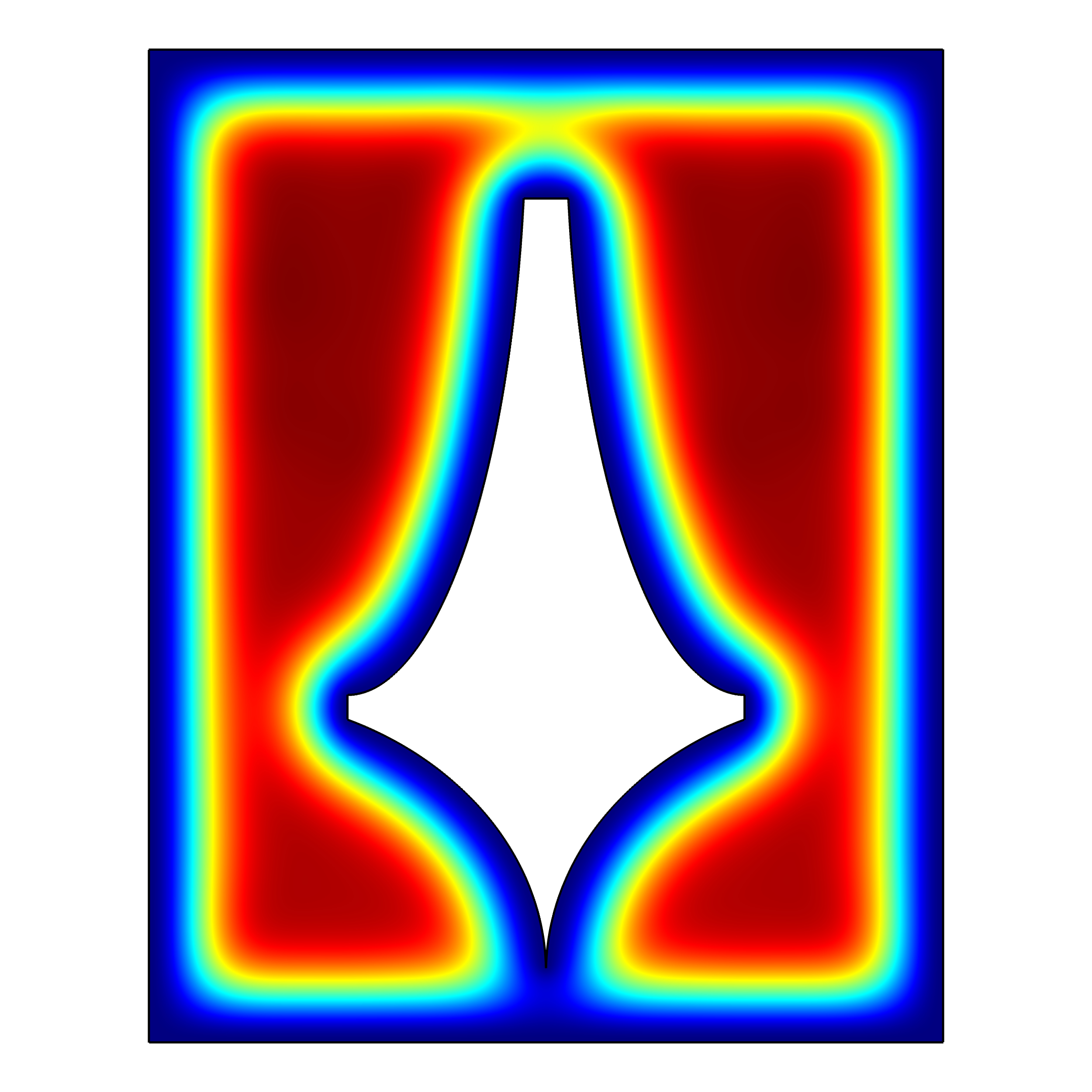}
\caption{The third arbitrary domain example. Application of the QBL methodology to an arbitrary domain. Unlike the previous domains that are considered, internal (defining the tractricoid) and external (defining the outer rectangle) boundaries affect the system differently.}
\label{fig:c2f17}
\end{figure}

We've investigated three very distinct arbitrary domains and demonstrated the accuracy of QBL method in predicting the partition function, from which all thermodynamic quantities can be derived. Note that the partition function is a basic statistical property, which suggests QBL method can be used to predict any other statistical property such as transport properties from the kinetic theory. In this sense, QBL method is a quite general and accurate tool to predict quantum size effects on the statistical properties of confined particles.

This analysis completes this chapter and our preparation for the subject. In the next chapter, we'll enter to the main topic and the results of this thesis.

%%%
%3%
%%%

%\chapter{The Nature of Quantum Shape Effects}
\chapter{The Nature of Quantum Shape Effects}

After the necessary introduction and the review, we are now in position to start the thesis subject. In this chapter, we will introduce a new effect that we've recently discovered in nanoscale confined systems \cite{aydin7}. As the name of the thesis suggests, we call it \textit{quantum shape effect}. In this chapter, we'll explore the essence of quantum shape effects by separating them from quantum size effects. We'll look at the eigenspectra of different confinement shapes and compare them with the difference in eigenspectra due to size effects. After that, the shape dependence of the single-particle partition function will be examined. We'll then advance the usual quantum boundary layer method to be able to predict the shape dependencies in an analytical way. The development and implementation of the analytical method is one of the most essential parts of this thesis, because it gives also a physical understanding to the effect that we've proposed. We conclude this introductory chapter for quantum shape effects by presenting them in various additional confinement domains and examining their occurrence conditions via playing with the sizes.

Influence of geometry manifests itself on the discrete energy spectrum of the confined system of particles, Fig. 3.1. In fact, it is somewhat natural to expect this, because the boundary conditions are defined directly by the geometry of the system. On the other hand, as we have seen before, different geometries can have the same spectrum: isospectral domains. Moreover, different energy spectra may possibly lead to the same statistical property, because of the averaging process that is inherent in the quantum statistical thermodynamics formalism. Therefore, the analysis of the influence of geometry in confined systems is by no means trivial and the physics behind this phenomenon is actually very rich and interesting. 

\begin{figure}
\centering
\includegraphics[width=0.75\textwidth]{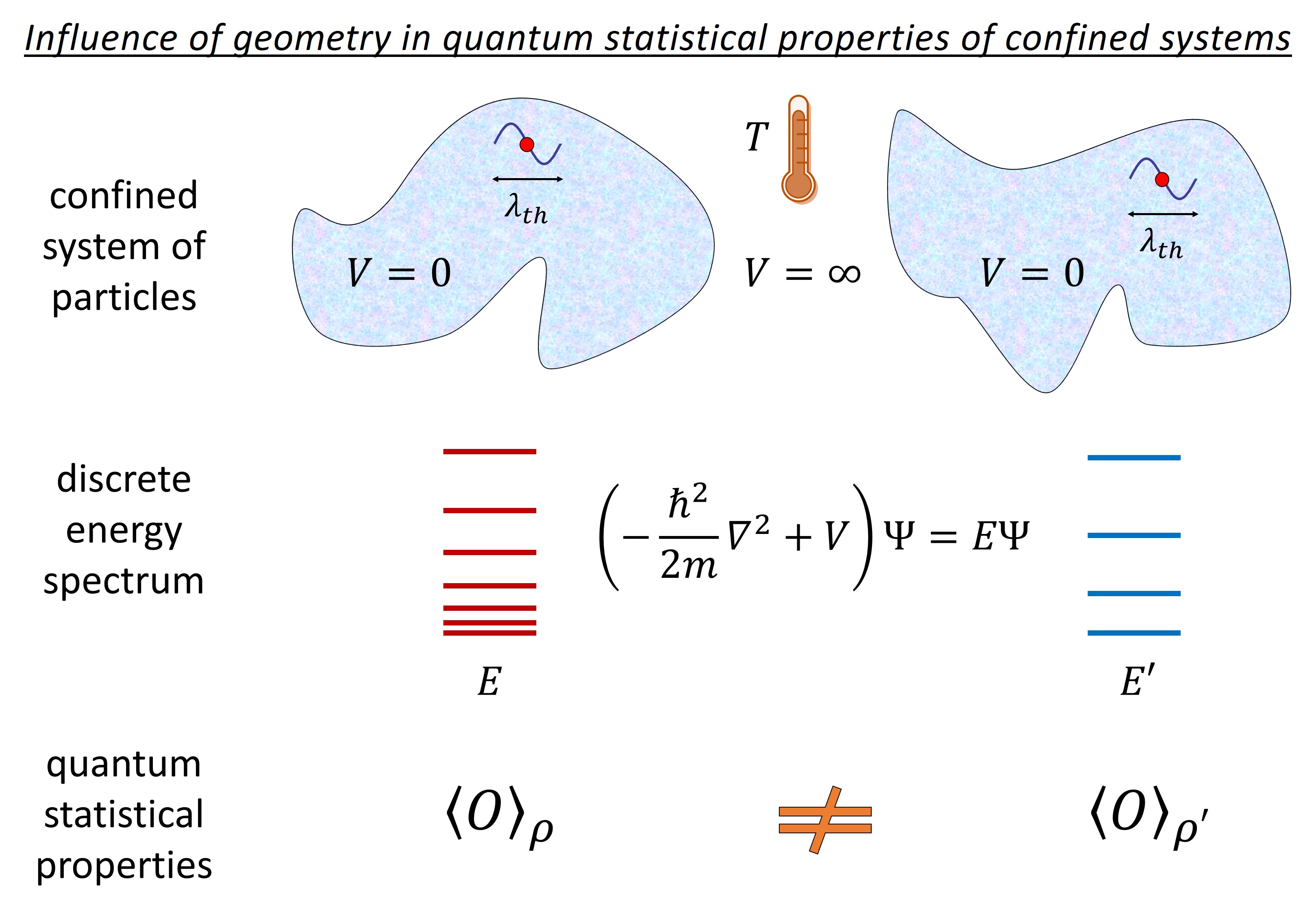}
\caption{A simplified picture of how geometry effects appear in general. When particles confined in very small domains, the differences in their discrete energy spectrum can lead to substantial variations in their quantum statistical properties.}
\label{fig:c3f1}
\end{figure}

\section{What Is Shape?}

We have explored size effects in the previous chapter. We've identified and quantified them very precisely by using geometric size variables, $\mathcal{V,A,P,N_V}$. Each size variable has a distinct role on the thermodynamic properties of a confined system, for instance their sign alternates and their magnitudes are very different in contributions to the partition function. Let's turn our attention to the shape. We'll question; How we can define the shape of an object? How does shape differ from size? Are they inherently linked to each other? Is there any way to separate them and focus on their individual effects? Do size and shape influences the thermodynamics in a similar way or not?

Although shape might be seen as a somewhat vague concept at first thought, there are rigorous mathematical definitions of it. Here we consider the shape as an object's geometric information that is invariant under Euclidean similarity transformations such as translation, rotation, reflection and uniform scaling \cite{shape1,shape2}. Left and right subfigures in Fig. 3.2 show the same and distinct shapes respectively. On the left table, a triangle undergoes Euclidean similarity transformations which preserves its shape. Note that they do not necessarily preserve the size, e.g. in uniform scaling. On the right table, all shapes are different than each other, although some or all of their geometric size variables might be the same, e.g. in isospectral domains. 

\begin{figure}
\centering
\includegraphics[width=0.7\textwidth]{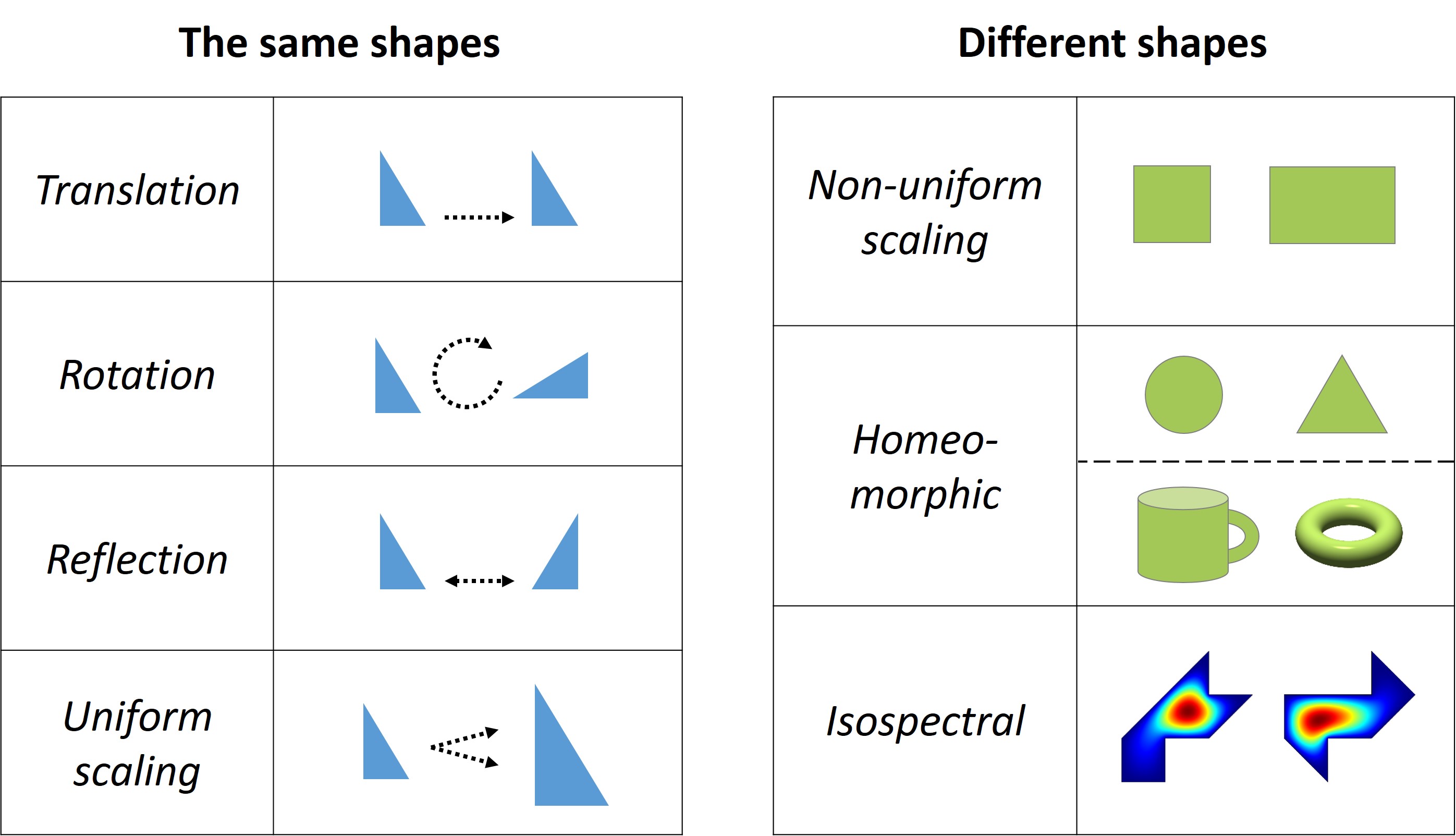}
\caption{Understanding the shape of an object. Left table shows the objects under Euclidean similarity transformations that preserve the shape. Right table shows the objects having distinct shapes, regardless of their topological or spectral properties.}
\label{fig:c3f2}
\end{figure}

Changes in geometric size variables like $\mathcal{V}$, $\mathcal{A}$, $\mathcal{P}$ and $\mathcal{N_V}$ changes the size, but also they change the shape of a particular object. Generally in literature, aspect ratios of sizes or plain geometry of structures, curvature values, etc. are considered as indicators of the shape characteristics of a domain \cite{biss,shapemat,pchenprb,nantech,nanscsh,langmu,pccp1,PhysRevB.50.17721,shapeeffnl,PhysRevLett.118.157402}. However, when aspect ratios of a domain changes, size variables also change along with shape simultaneously. Similarly, a change in plain geometry (e.g. from cube to sphere) cannot be done without changing geometric size variables of a domain, i.e. $\mathcal{V}$, $\mathcal{A}$, $\mathcal{P}$ and $\mathcal{N_V}$. In other words, in those type of processes size and shape effects are inherently linked and cannot easily be separated from each other. Even if there exist domains with distinct shapes having the same sizes (like isospectral domains), it is not quite possible to transform them from one form to another by continuous boundary deformations while still preserving the size-invariance. But what we would like to focus on is the pure shape effect. To do that we need to be able to keep all the geometric size variables constant but still being able to change the shape in a continuous fashion. We see that it is possible and rather easy.

\section{Size-invariant Shape Transformation}

It is possible to continuously change the shapes of confinement domains without changing the values of geometric size variables. The technique that we propose is called the size-invariant shape transformation and the procedure is as follows: Take a domain, for instance a 2D square domain as in Fig. 3.3. Then, remove a smaller region from inside and fill the remaining domain with particles. Performing a rotation or translation on the inner region changes the shape of the confinement domain perceived by particles. This simple process keeps all geometric size variables constant during the rotation, while still changing the shape of the confinement domain. Therefore, this is a pure shape transformation. But, what is quantum about it? As we shall see later on, in order to observe any difference on the physical properties of the particles due to this change, we need quite strong confinements and the effect vanishes in the limit of $\lambda_{th}<<L$ where $L$ is a characteristic domain size (not to mention this notion can be directly generalized into any dimension so that $\lambda_{th}^d<<\mathcal{V}_d$). Just like quantum size effects, the quantum shape effect is a direct consequence of the wave nature of particles. Things will be more concrete when we corroborate our discussion with the analytical methodology.

\begin{figure}
\centering
\includegraphics[width=0.9\textwidth]{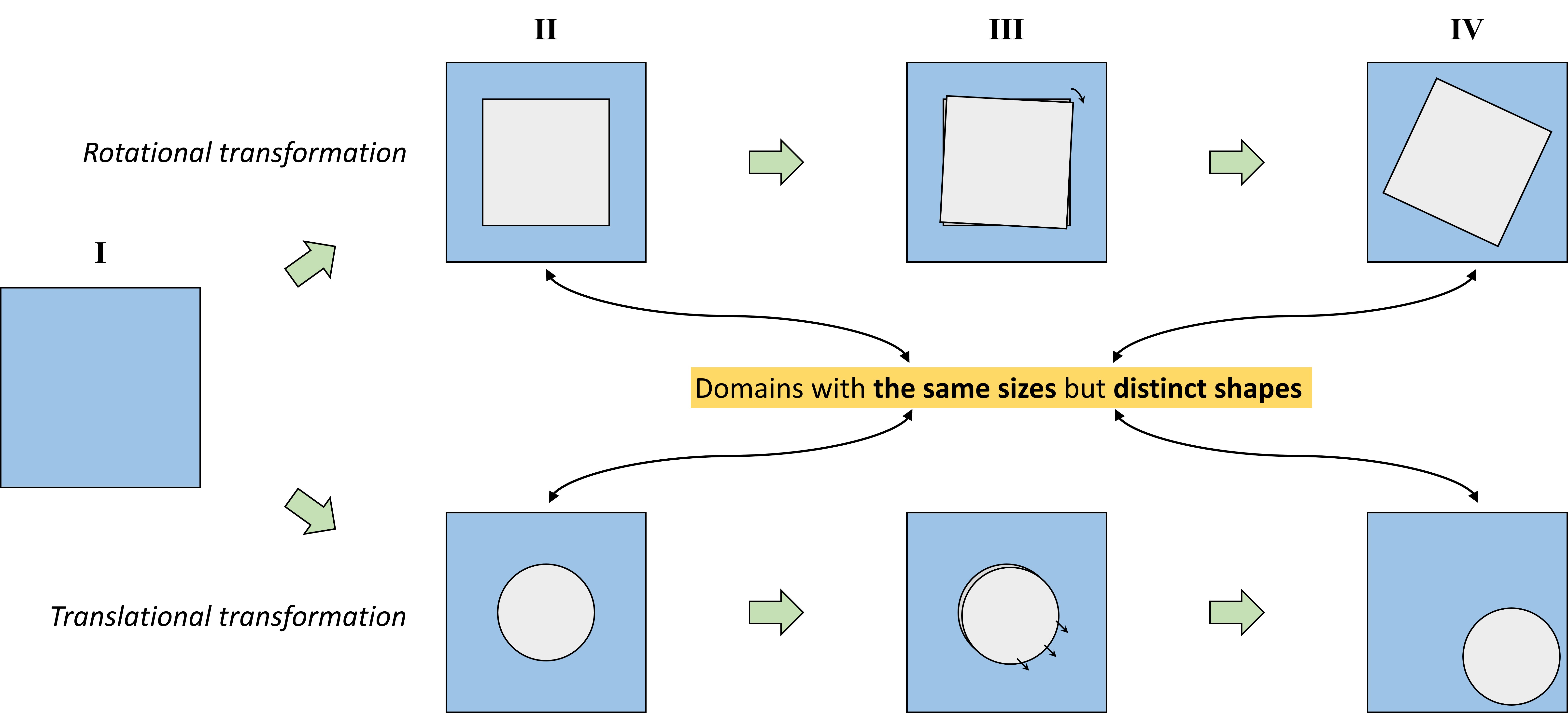}
\caption{Procedure of the size-invariant shape transformation, creating differently shaped confinement domains, without changing their sizes, out of nested objects. (I) Take a simple domain filled with particles. (II) Remove a region from the domain, where the removed part becomes a region that particles cannot penetrate. (III) Perform rotation or translation to the inner region. (IV) Resulting shape is different than the one before the transformation, while their sizes ($\mathcal{V},\mathcal{A},\mathcal{P},\mathcal{N_V}$) remain unchanged.}
\label{fig:c3f3}
\end{figure}

We choose to focus on the rotational transformation in this thesis, as it preserves symmetry with respect to the origin and thus simplifies the discussions. In rotational case, shape transformation is characterized by the rotation angle of the inner square region, which we denote by $\theta$. We choose the configuration II in Fig. 3.3 as $\theta=0^{\circ}$ and clockwise direction is chosen as the positive angle.

\section{Looking at the Eigenspectra}

Domains' shape can be changed size-invariantly, but what about their quantum statistical properties? Does their energy spectrum even change or have we just discovered some new isospectral domains? In order to explore this, we need to solve the time-independent Schr\"{o}dinger equation for different $\theta$ values corresponding to the different angular configurations shown in Fig. 3.4. After some initial tests and considerations for the determination of system sizes, side lengths of outer and inner squares of the confinement domain are chosen as $L_o=\SI{21.2}{\nano\metre}$ and $L_i=\SI{13.6}{\nano\metre}$ so that confinement parameter $\alpha$ of the domain is fixed to unity. For our 2D nested square domain, the confinement parameter is calculated as $\alpha_{*}=(\sqrt{\pi}/2)*(\lambda_{th}/L_{*})$ where $L_{*}=2\mathcal{A}/\mathcal{P}$ is the harmonic mean size of the nested domain in transverse direction. The closest distance between the outer and inner boundaries of the domain is called the apex length and its maximum and minimum values are \SI{3.8}{\nano\metre} and \SI{1.0}{\nano\metre} respectively occurring at $\theta=0^{\circ}$ and $\theta=45^{\circ}$, for the chosen parameters. The variation of apex length (see Fig. 3.8) with the rotation angle can be analytically found for the nested square domain as
\begin{equation}
a=\frac{L_o}{2}-\frac{L_i}{2}\left[\sin(\theta)+\cos(\theta)\right].
\end{equation}

For this particular shape the angular configuration ranges only from $\theta=0^{\circ}$ to $\theta=45^{\circ}$, since from $\theta=45^{\circ}$ to $\theta=90^{\circ}$ it is the mirror symmetry of the former and $\theta=0=n\pi/2$ for any integer $n$ value. Therefore, solving the 2D Schr\"{o}dinger equation for the configurations between $\theta=0^{\circ}$ and $\theta=45^{\circ}$ is sufficient. Here, $\theta=45^{\circ}$ is called the symmetric periodicity angle of this particular domain. Analytical solution for these kinds of irregular domains is not possible, so we solve the Schr\"{o}dinger equation numerically using the finite element method. The rotation discretization steps are chosen to be small. The solutions are obtained with $\Delta\theta=1^{\circ}$ and also with a higher resolution $\Delta\theta=0.25^{\circ}$. It is seen that results do not change and perfectly match with each other. The details of how we implement this numerical calculation are given in Appendix A.1.

\begin{figure}
\centering
\includegraphics[width=0.8\textwidth]{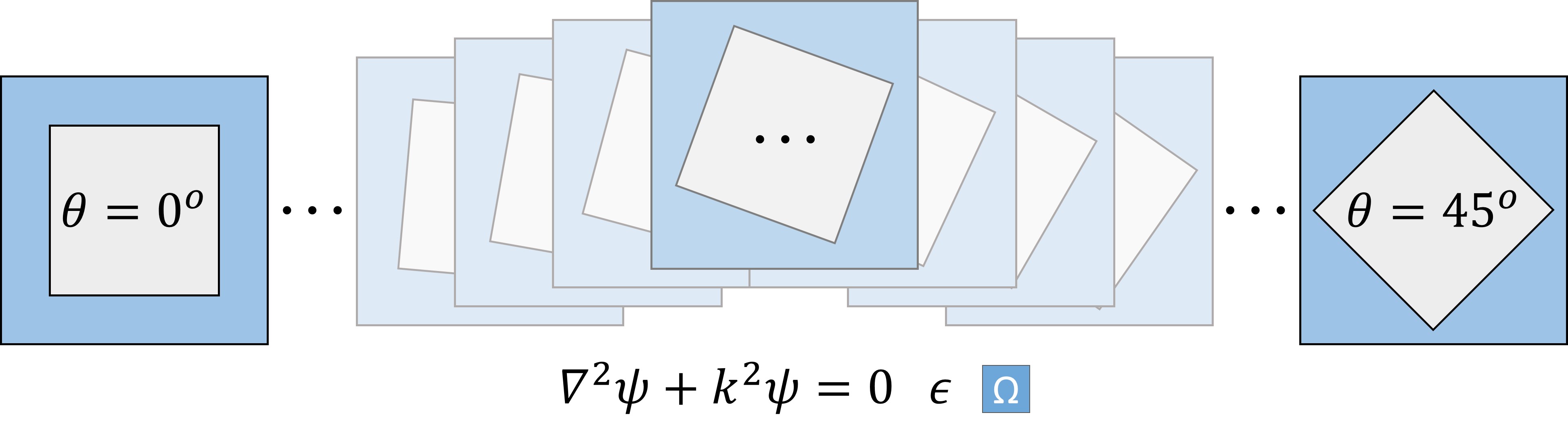}
\caption{Solving the 2D Schr\"{o}dinger equation for blue regions in the nested square confinement domain and finding the eigenvalues for each angular configuration.}
\label{fig:c3f4}
\end{figure}

\begin{figure}[!b]
\centering
\includegraphics[width=0.75\textwidth]{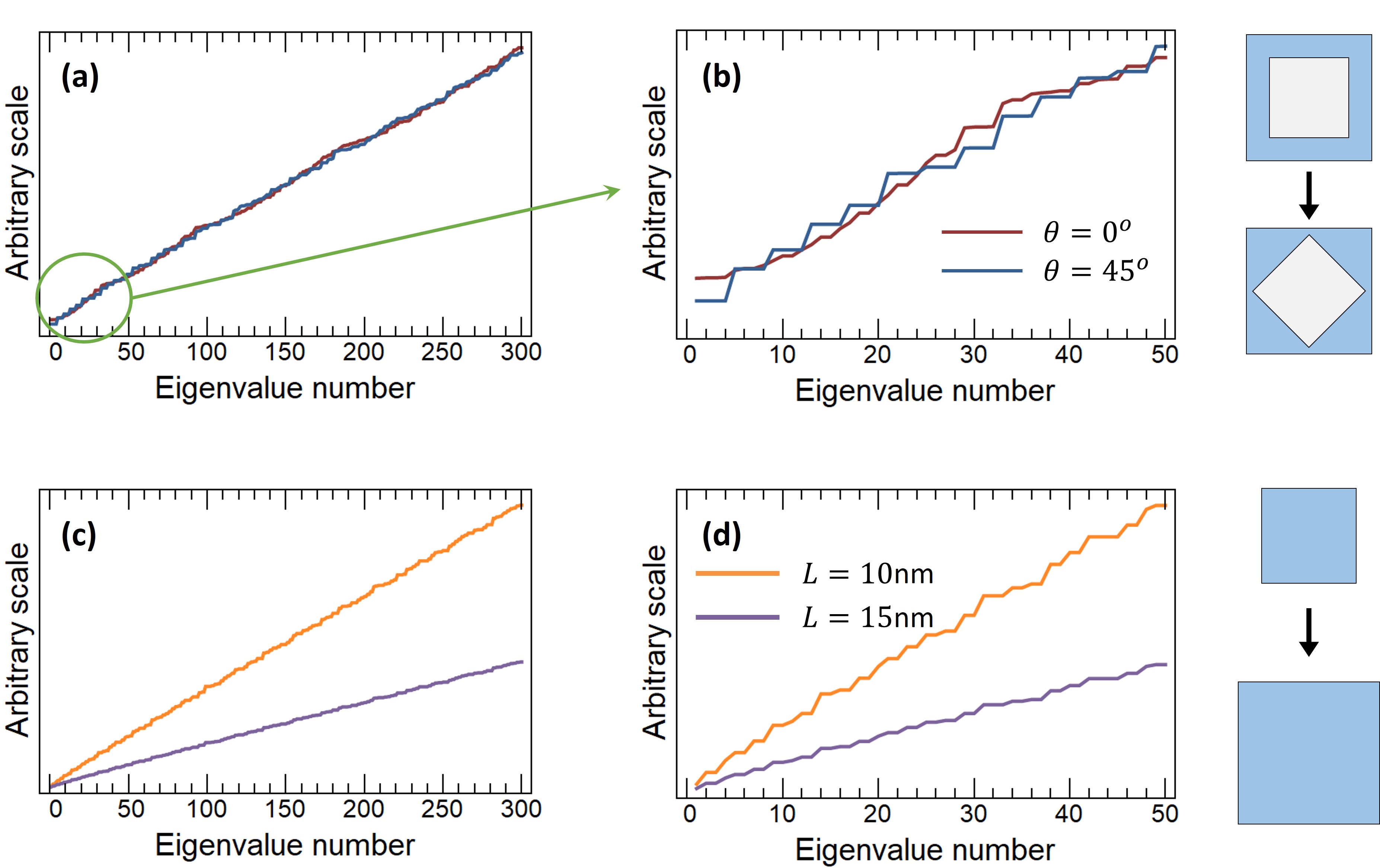}
\caption{Looking at the energy eigenspectra. (a), (b) Purely due to shape difference, different angular configurations lead to difference in eigenspectra, which is more apparent near to ground state. (c), (d) Difference in eigenspectra due to quantum size effects. The gap between spectra is much clear since energy is inversely proportional to length.}
\label{fig:c3f5}
\end{figure}

Solving the 2D Schr\"{o}dinger equation for the blue regions in Fig. 3.4 gives the energy eigenvalues of that particular angular configuration. We compare the energy spectra of $\theta=0^{\circ}$ (red curve) and $\theta=45^{\circ}$ (blue curve) configurations in Fig. 3.5a and 3.5b respectively. Although the magnitudes of their energy spectra are quite similar, the values are different than each other, which can be more explicitly seen when we zoom in the first fifty eigenvalues, Fig. 3.5b. A relatively large difference in their ground states can also be seen. During the shape transformation, eigenvalues smoothly transforms from red curve to the blue one. For comparison we plot the eigenspectra of square confinement domains with different sizes; side lengths $L=10$nm (orange curve) and $L=15$nm (purple curve), Fig. 3.5c and 3.5d respectively. This analysis allows us to compare respective influences of quantum size and shape effects in the energy eigenspectrum. Quantum size effects more apparently affect the energy spectrum because of the inversely proportional dependence of energy to the length (here in 2D shapes the area; length squared). Conversely, shape difference has a much less apparent but definitely distinct influence on the energy spectrum.

\begin{figure}
\centering
\includegraphics[width=0.95\textwidth]{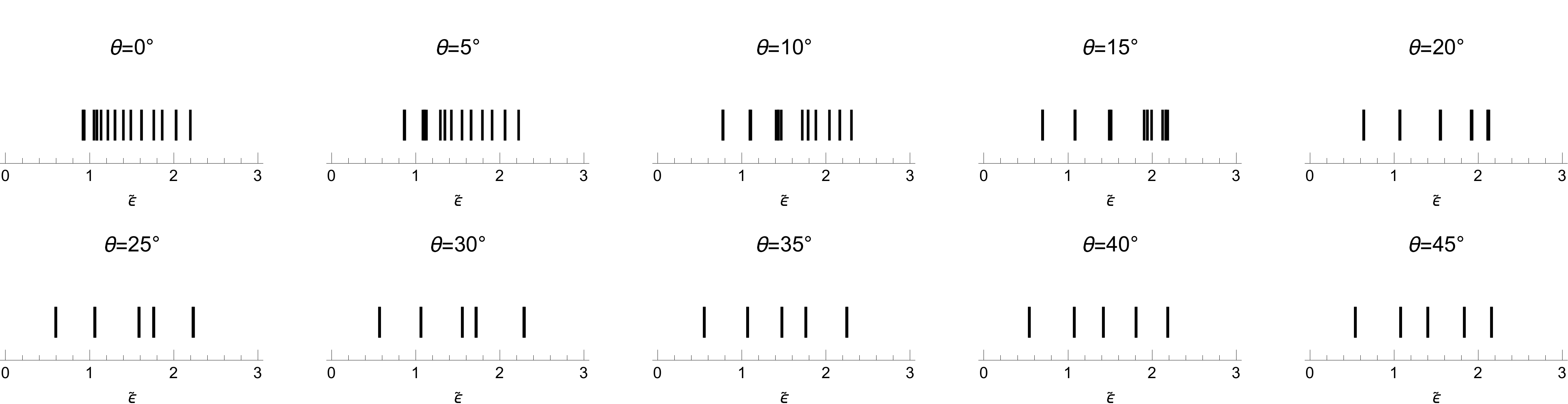}
\caption{Emergence of spectral gaps. The first twenty eigenvalues in their dimensionless form (normalized to $k_BT$) in a number line plot. The results are shown for the range between $\theta=0^{\circ}$ and $\theta=45^{\circ}$ with $\Delta\theta=5^{\circ}$ steps.}
\label{fig:c3f6}
\end{figure}

Examination of the eigenspectra in number line plot gives more information about what is going on in energy space. In Fig. 3.6, first twenty dimensionless energy eigenvalues are shown in a number line plot from $\theta=0^{\circ}$ to $\theta=45^{\circ}$ with $\Delta\theta=5^{\circ}$ steps. Note that energy level spacing of eigenvalues are much less than $k_BT$. As it can be seen, at $\theta=0^{\circ}$ the energy spectrum looks roughly uniform. At $\theta=5^{\circ}$, separation of ground state from the remaining spectrum is occurring. At $\theta=10^{\circ}$, the first and even the second excited states also separate themselves from the rest of the spectrum. Emergence of spectral gaps between eigenvalues become very much apparent after $\theta=20^{\circ}$. Note that in all subfigures of Fig. 3.6 the same number of eigenvalues (which is 20) are shown. The reason why they look like fewer is because there are degeneracies which are much more explicit especially after $\theta=15^{\circ}$. Moreover, there is a slight decrement on the magnitudes of energy eigenvalues on average, as spectral gaps emerge. We'll see the reason of this explicitly during the next section.

\begin{figure}[!b]
\centering
\includegraphics[width=0.9\textwidth]{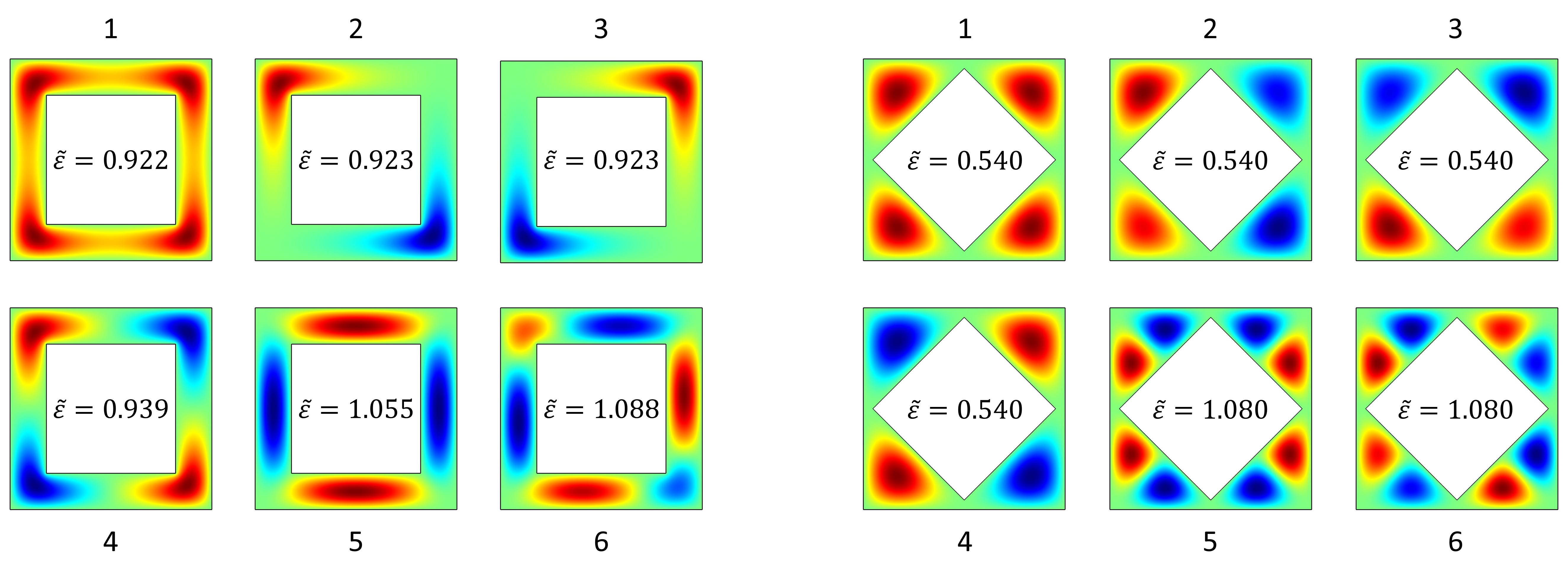}
\caption{Behavior of the wavefunctions. First six eigenvalues and eigenfunctions for $\theta=0^{\circ}$ (left) and $\theta=45^{\circ}$ (right). Red color shows the positive amplitudes, blue color denotes the negative amplitudes in the rainbow color spectrum and the green parts are zero.}
\label{fig:c3f7}
\end{figure}

To get even more intuition about how shape affects the spectrum we can plot the eigenfunctions. In Fig. 3.7, first six eigenfunctions for $\theta=0^{\circ}$ (left subfigure) and $\theta=45^{\circ}$ (right subfigure) can be seen. Higher the amplitude the redder and lower the amplitude the bluer in the rainbow color spectrum. The zero value is denoted by the green color in the middle of the spectrum. The ground state eigenfunction is always positive everywhere, while excited state eigenfunction can take positive and negative values. Energy eigenvalues in their dimensionless forms ($\tilde{\varepsilon}=\varepsilon/(k_BT)$ where $T=300$K) are given within the corresponding eigenfunction graphs. For $\theta=0^{\circ}$ configuration, energy eigenvalues are very close to each other, while there is a huge gap between the fourth and fifth eigenvalues of $\theta=45^{\circ}$ configuration compared to the $\theta=0^{\circ}$ configuration. The difference between their ground state values is nearly double. The only degeneracy in the first six eigenvalues of $\theta=0^{\circ}$ occurs in second and third eigenvalues, whereas $\theta=45^{\circ}$ has a four-fold degeneracy in its first four eigenvalues with a two-fold degeneracy following right after. These differences are really substantial ones and play significant roles in the variation of quantum statistical properties as we shall see next.

\section{Quantum Shape Dependence of the Partition Function}

We see that energy spectrum changes with the size-invariant shape transformation. The next step is to see how quantum statistical properties change. We do this analysis over the single-particle partition function having Boltzmann factor at room temperature, $T=300$K. We use bare electron mass for the particles and we can think of them as an ideal low density electron gas. This chapter constitutes more of a mathematical examination of quantum shape effects to understand their origins better.

We obtained the eigenvalues for a 2D domain because the perpendicular direction to the transverse directions does not affect quantum shape effects. Shape dependence originates from the particular 2D shape of the domain and that's why in this chapter we focus only on 2D domain. As long as one ensures the ergodicity, one can use Maxwell-Boltzmann statistics even for a single-particle system, by considering many copies of the considered system. We will open the third direction to contain large enough number of particles inside the system while keeping the density at low to satisfy the Maxwell-Boltzmann statistics, during the calculation of thermodynamic properties for the 3D structures in fourth chapter.

After solving the Schr\"{o}dinger equation repeatedly for each angular configuration and obtaining eigenvalues, we use them to calculate partition function. Indeed, partition function changes with changing rotation angle of the inner region $\theta$ in Fig. 3.8. This is the first clear demonstration of pure quantum shape effects: changing a quantum statistical quantity due to just shape difference. Variation of partition function with respect to angle has a $\tanh$ function (sigmoid) behavior. It has the lowest value at $\theta=0^{\circ}$, gradually rises from there up until $\theta=45^{\circ}$ where makes its maximum. There is $7.6\%$ difference between the minimum and maximum of the partition function, which is a considerable change. Since this is a quantum effect, its magnitude also depends sensitively on temperature. The smaller the temperature, the larger the quantum shape effect.

\begin{figure}
\centering
\includegraphics[width=0.8\textwidth]{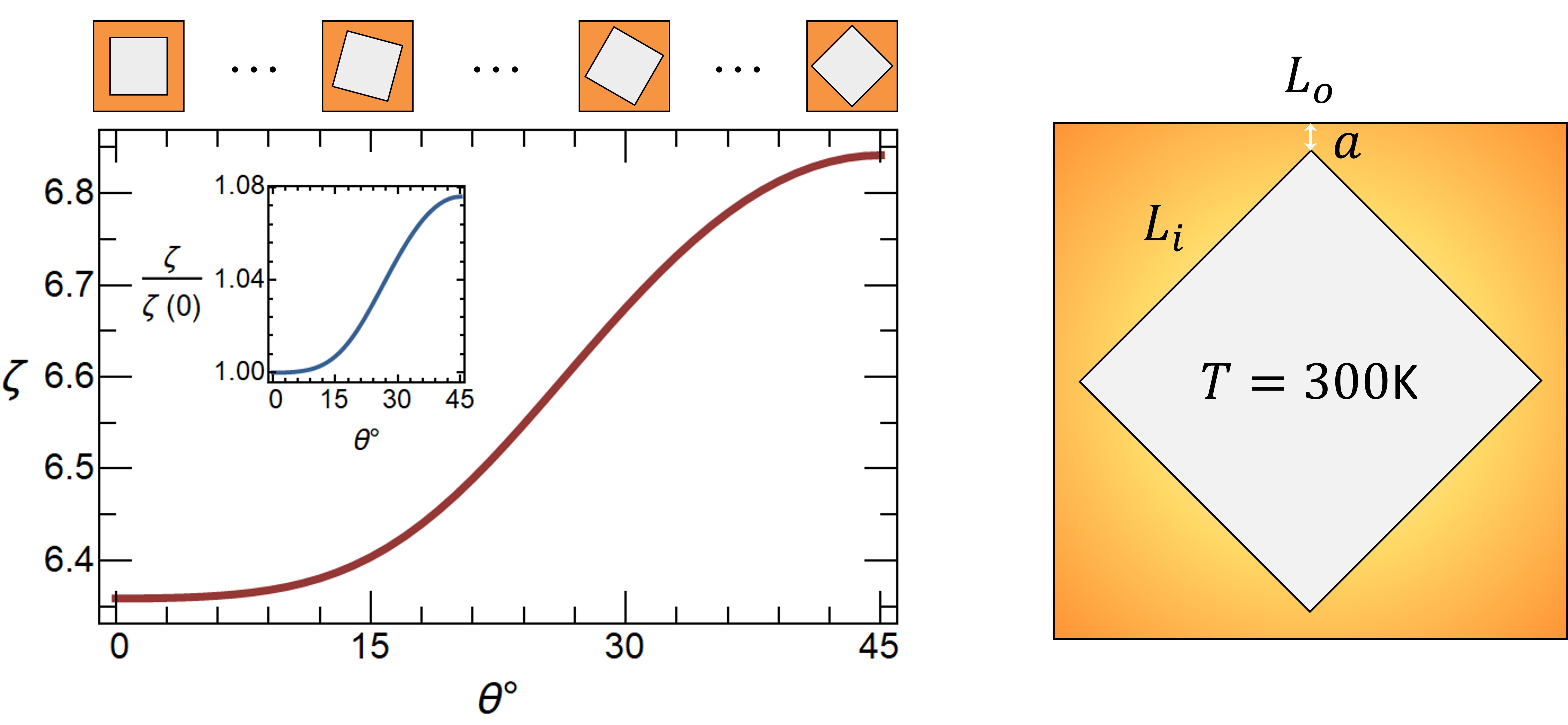}
\caption{Variation of the partition function with respect to the rotation angle of the inner square region $\theta$. Temperature is taken to be room temperature, $T=300$K. Subfigure shows the partition function normalized to its value at the initial angular configuration ($\theta=0^{\circ}$). Domain parameters are shown in the right figure. $L_o$, $L_i$ and $a$ are the outer square length, inner square length and apex length at the symmetric periodicity angle respectively.}
\label{fig:c3f8}
\end{figure}

The change in partition function suggests the changes in other thermodynamic quantities as well, but their examination will be done in the next chapter. Now we will try to understand more about quantum shape effects by exploring their underlying mechanisms as well as to attempt analytically predicting them.

\section{Overlapped Quantum Boundary Layer Method}

How quantum shape effects work? Why does the partition function change in such a way and how do we interpret its behavior? The best way to understand the mechanisms of quantum shape effects is trying to predict them analytically. Recall that quantum boundary layer method was giving very precise analytical results for quantum size effects. To see if it can help also in shape effects, let's look at first the quantum thermal density distributions of particles confined in our confinement domains. In Fig. 3.9a, quantum thermal densities of particles for four different angular configurations are plotted. It shows where the particles are most probably located and from where they stay away. Expectedly, particles stay away from the regions near to all boundaries. It is clear by looking the reddish regions that the rotation of inner object from $0^{\circ}$ to $45^{\circ}$ causes particles to confine into four local regions than their initial closed-pipe-like formation. During this transition, position of the boundaries of the inner object varies inside the domain and it comes really close to the boundaries of the outer one. Remember that quantum boundary layer was leading to an evacuation of the regions from particles near to boundaries. If, during the rotation, quantum boundary layers of outer and inner objects overlap, this would lead to a double evacuation of those regions, which is not possible of course and needs to be corrected. After this initial logic, we tested our argument by dressing the quantum boundary layer to our domain. As you can see in Fig. 3.9b, quantum boundary layers of outer and inner boundaries indeed overlap with each other for some degrees, e.g. for $\theta=25^{\circ}$ and $\theta=45^{\circ}$ in the figure. In addition, existence and the amount of overlap changes with the rotation angle $\theta$. We are looking for a parameter or a quantity to reveal the hidden information about the shape of the confinement domain analytically. We have overlap areas in our hands now. Let's see if it can give some idea about what's going on in our confinement domain during the shape variation.

\begin{figure}
\centering
\includegraphics[width=0.85\textwidth]{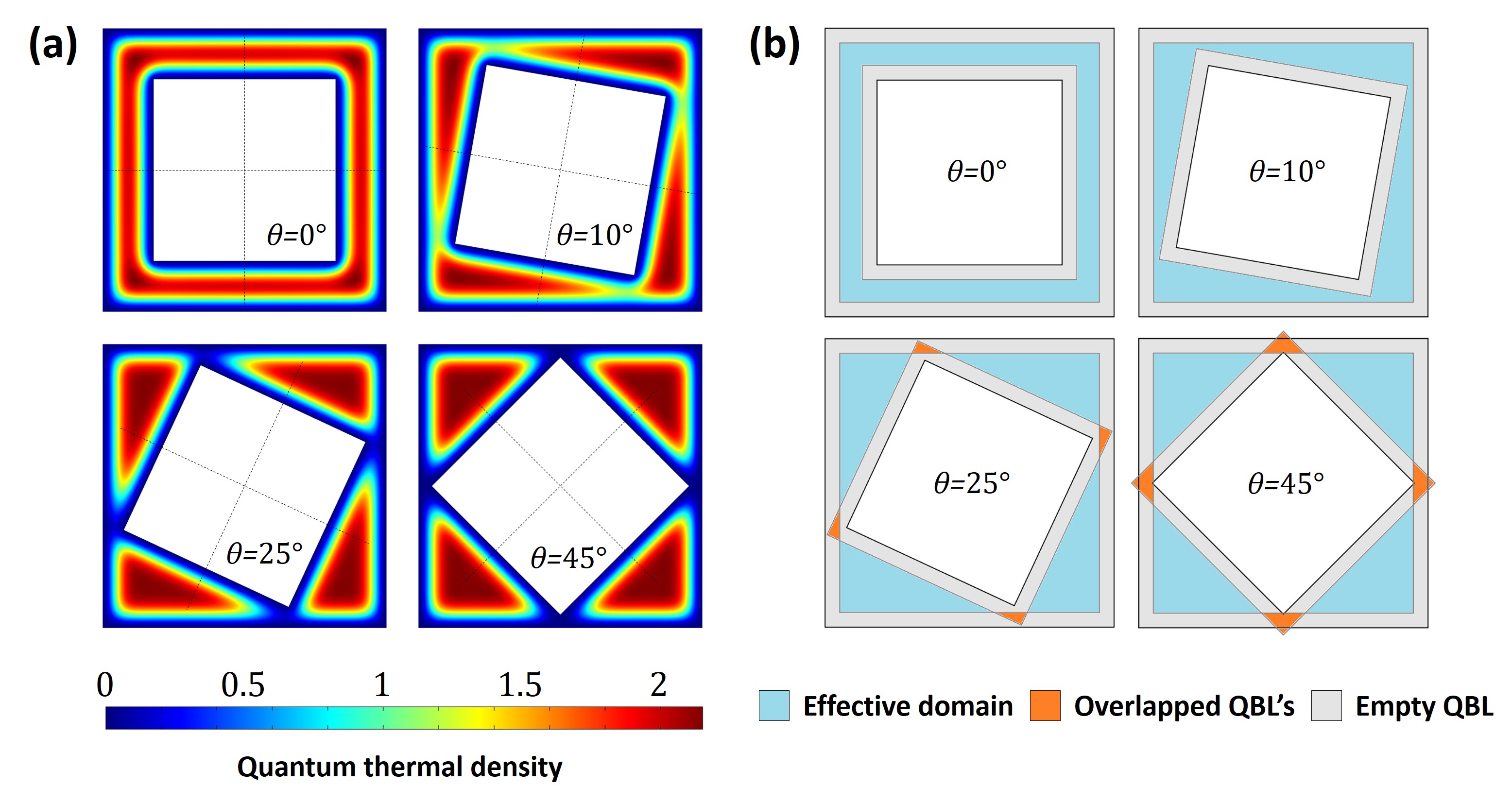}
\caption{For four different angular configurations, (a) quantum thermal density distributions of particles confined in our specially designed domain, (b) formation of the overlaps of outer and inner objects' quantum boundary layers carrying information about the intrinsic shape of the confinement domain.}
\label{fig:c3f9}
\end{figure}

Formation of overlapping areas can be seen in more detail in Fig. 3.10a. Gray regions denote the quantum boundary layers of outer and inner boundaries. After dressing the quantum boundary layer, effective boundaries arise which determine the effective area, cyan region. This is the region where the non-uniform density distribution of particles is replaced by a uniform one and particles effectively occupy that region only. As is seen, quantum boundary layers of outer and inner boundaries overlap with each other at this angular configuration, which are denoted by the orange color. Normally, as we have seen in Chapter 2, the effective area of a domain is calculated by subtracting the area evacuated by the quantum boundary layer from the actual area, e.g. in 2D case: $\mathcal{A}_{\mli{eff}}=\mathcal{A}-\mathcal{A}_{\mli{qbl}}=\mathcal{A}-\delta\mathcal{P}+\delta^2\mathcal{N_V}$. On the other hand, in this particular case, where the quantum boundary layers of outer and inner boundaries overlap, the usual QBL procedure improperly subtracts these overlap areas twice from the actual area. Therefore, we need to add these overlap regions to the usual effective domain in order to find the proper effective domain. From now on, we denote the usual non-overlap versions of effective sizes with $0$ superscripts as $\mathcal{V}_{\mli{eff}}^0$, $\mathcal{A}_{\mli{eff}}^0$, $\mathcal{P}_{\mli{eff}}^0$ or $\mathcal{N_V}_{\mli{eff}}^0$ depending on the intrinsic dimension of the domain. Then, the real effective area can be found as $\mathcal{A}_{\mli{eff}}=\mathcal{A}_{\mli{eff}}^0+\mathcal{A}_{\mli{ovr}}$ where $\mathcal{A}_{\mli{eff}}^0=\mathcal{A}-\mathcal{A}_{\mli{qbl}}$. This we call the overlapped quantum boundary layer method which may be seen as an extension of the usual quantum boundary layer methodology.

\begin{figure}
\centering
\includegraphics[width=0.95\textwidth]{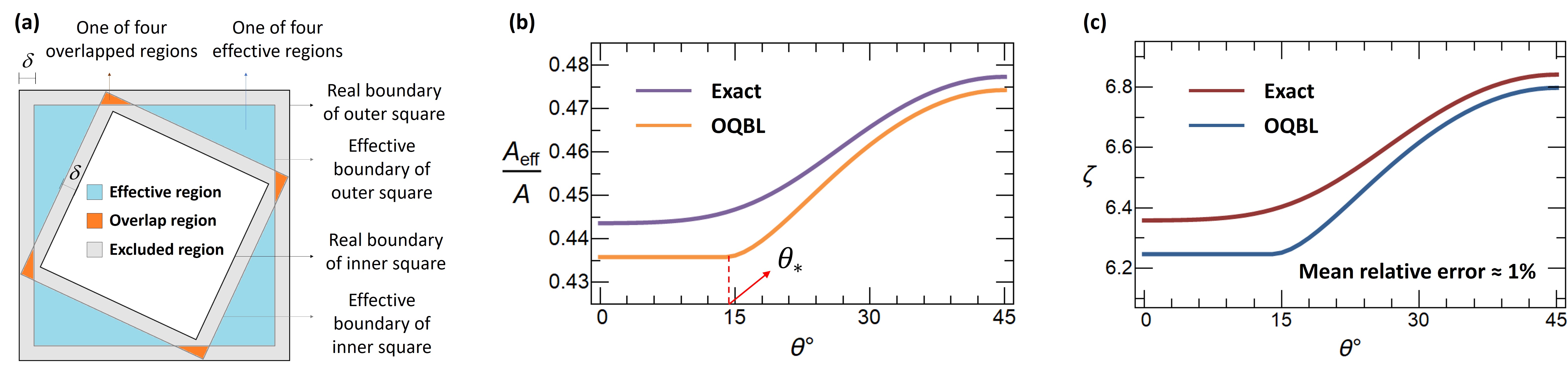}
\caption{(a) Quantum boundary layer is dressed to the confinement domain for a particular rotation angle. Cyan, orange and gray colors denote effective, overlap and excluded regions. (b) The amount of normalized effective area varies with the angular configuration. Purple and orange curves show the exact and analytical results respectively. $\theta_{*}$ denotes the angle after which overlaps start. (c) Change in partition function with respect to the rotation angle. Exact and analytical results are represented by red and blue curves respectively.}
\label{fig:c3f10}
\end{figure}

More explicitly, for the selected nested square confinement domain, the effective area is expressed in overlapped quantum boundary layer method as $\mathcal{A}_{\mli{eff}}=[(L_o-2\delta)^2-(L_i+2\delta)^2]+\mathcal{A}_{\mli{ovr}}=(L_{o,\mli{eff}}^2-L_{i,\mli{eff}}^2)+\mathcal{A}_{\mli{ovr}}=\mathcal{A}_{\mli{eff}}^0+\mathcal{A}_{\mli{ovr}}$, where the overlap area can analytically be obtained as a function of $\theta$ by using simple geometric relations as follows,
\begin{equation}
\mathcal{A}_{\mli{ovr}}=\frac{\tan\theta}{2}\left[L_{i,\mli{eff}}\left(1+\frac{1}{\tan\theta}\right)-\frac{L_{o,\mli{eff}}}{\sin\theta}\right]^2.
\end{equation}
We can calculate the accuracy of the analytically found $\mathcal{A}_{\mli{eff}}$ by comparing it with the one that can be numerically found from $\mathcal{A}_{\mli{eff}}=\zeta\lambda_{th}^2$, where $\zeta$ is the exact partition function in which the numerically solved eigenvalues are used. This equation gives the effective area if we could have had the chance of calculating it exactly. The comparison of the accuracy of our analytical method in estimating the effective area is given in Fig. 3.10b where the variation of the normalized effective area with respect to $\theta$ is plotted. Purple and orange curves represent the results of exact (numerical) and analytical (overlapped quantum boundary layer method) calculations. As is seen, the overlap areas start to be formed after the critical value of the rotation angle, $\theta_{*}$ which is found as
\begin{equation}
\theta_{*}=\text{Round}\left[\frac{180}{\pi}\arctan\left(\frac{L_{o,\mli{eff}}\sqrt{2L_{i,\mli{eff}}^2-L_{o,\mli{eff}}^2}-L_{i,\mli{eff}}^2}{L_{i,\mli{eff}}^2-L_{o,\mli{eff}}^2}\right)\right].
\end{equation}
Note that Eq. (3.2) is valid for $\theta_{*}\leq\theta\leq45^{\circ}$ and $\mathcal{A}_{\mli{ovr}}=0$ for $0^{\circ}\leq\theta<\theta_{*}$, as there is no overlap in that range of angular configurations. For the sizes ($L_o$ and $L_i$) and the temperature ($T=300$K) considered in this analysis, the critical angle corresponds to $\theta_{*}=14^{\circ}$. 

Values in Fig. 3.10b suggest that effective area actually takes up less space than half of the actual area. In other words, excluded regions due to quantum boundary layers are larger than the effective areas occupied by particles. This shows us that the confinement is extremely strong in this domain with the chosen parameters. In fact, we need such strongly confined domains to make quantum shape effects appear. Because on the contrary case, there won't be substantial overlap or any overlap at all to generate the quantum shape effects. This point will become clearer when we examine the quantum size effects on quantum shape effects at the end of this chapter.

In Fig. 3.10c, accuracy of our analytical method is tested over the partition function. The maximum and mean relative errors of analytical effective area expression are around $2\%$ and $1\%$ respectively, so it correctly predicts the functional behavior of the partition function with respect to changes in $\theta$. Our analytical method explains also why the partition function has a sigmoid variation and increases with increasing shape effects. The partition function is directly proportional with the effective area of the domain. As a result of this, it mimics the behavior of the effective area. Since the angular transformation from $0^{\circ}$ to $45^{\circ}$ creates overlaps and increases the effective area by a sigmoid-like fashion, the variation of the partition function has also the same behavior. This is a direct consequence of the chosen confinement geometry. In some other geometries, one can expect different behaviors, but partition function will always mimic the behavior of the effective areas (or volumes or whatever the effective bulk term is proportional).

Note that both Weyl density of states and the first two terms of PSF cannot predict any shape-dependent change in partition function. These methods fail to predict quantum shape effects. (Third term of PSF would include them, that term cannot be obtained analytically for arbitrary shapes.) On the other hand, our overlapped quantum boundary layer method predicts and represents the quantitative and qualitative nature of the shape dependence of partition function reasonably well. The discrepancy between the numerical and analytical methods most probably comes from the all-or-none nature of the quantum boundary layer method. A finer approximation to the non-uniform density distribution might give more accurate results, which is discussed in Appendix A.5.

\section{Various Nested Confinement Domains}

So far we've only tested our method in the nested square confinement domain. Many other types of domains can be designed using the size-invariant shape transformation technique. Let's explore for instance nested triangle and nested rectangle domains. For a fair comparison we want to choose our new confinement domains as close as possible to the nested square domain in terms of their confinement strengths. In this regard, we need to determine a parameter to compare the confinements of arbitrary domains. The usual confinement parameter defined in Chapter 2 in Eq. (2.4) cannot do the job, because it is defined only for regular rectangular domains and it is specific to a chosen direction. Conversely, in nested domains the geometry is so complex that a single parameter cannot define the confinement of even a particular direction, let alone the whole domain. Nevertheless, we won't need to be such precise in the determination of the comparison parameter, after all we just try to make a fair comparison between the thermodynamic properties of the domains. As long as we find a sound parameter and use the same parameter in the comparison of different domains, it should be fine. To this end, the characteristic length is defined as the harmonic mean size of the nested domain in transverse direction, which is found by $L_{*}=2\mathcal{A}/\mathcal{P}$. Then, the characteristic confinement parameter for nested domains becomes $\alpha_{*}=(\sqrt{\pi}/2)*(\lambda_{th}/L_{*})$. Characteristic confinement parameter of the nested square domain with the considered parameters is unity. For nested triangular domain, lengths of outer and inner triangles are chosen as  $L_o=\SI{23.0}{\nano\metre}$ and  $L_i=\SI{9.7}{\nano\metre}$ respectively. For nested rectangular domain, long and short sides of outer rectangle are $L_{o,l}=\SI{16.8}{\nano\metre}$ and $L_{o,s}=\SI{8.4}{\nano\metre}$, of inner rectangle are $L_{i,l}=\SI{6.8}{\nano\metre}$ and $L_{i,s}=\SI{2.0}{\nano\metre}$. By choosing these parameters, we managed to fix the apex length at the symmetric periodicity angles to $\SI{1}{\nano\metre}$ and confinement parameters to unity, $\alpha_{*}=1$  for all domain types.

\begin{figure}
\centering
\includegraphics[width=0.7\textwidth]{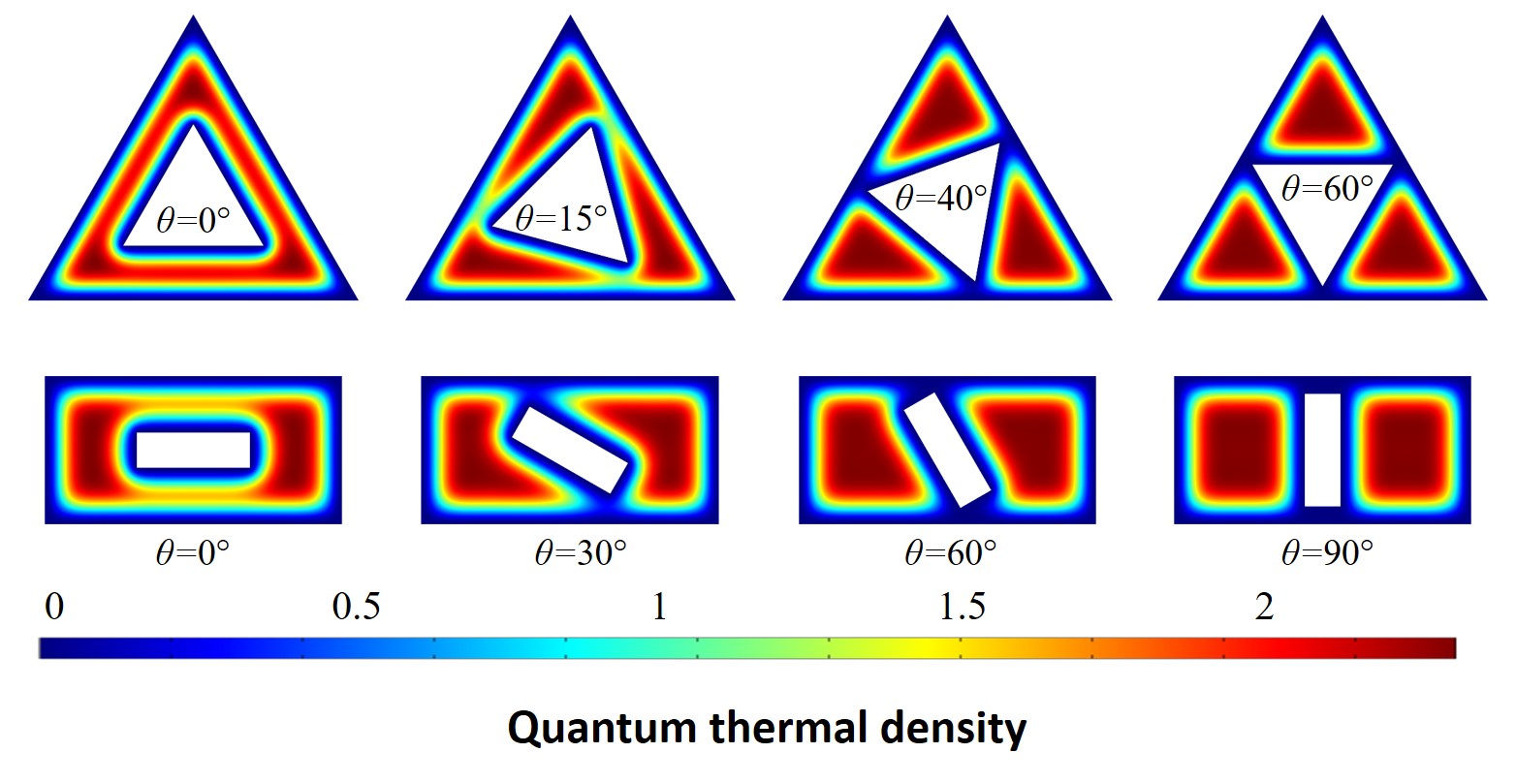}
\caption{Dimensionless quantum thermal density distributions of nested triangular and nested rectangular domains.}
\label{fig:c3f11}
\end{figure}

\begin{figure}[!b]
\centering
\includegraphics[width=0.95\textwidth]{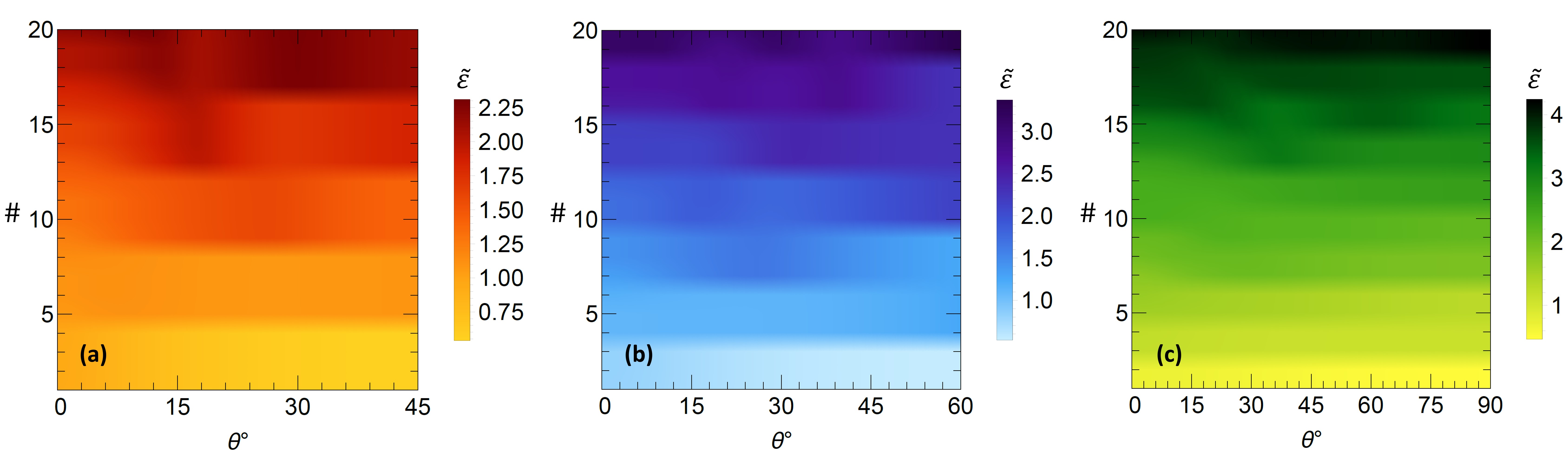}
\caption{Formation of $n$-fold degeneracies in the eigenspectra. First 20 dimensionless energy eigenvalues ($\tilde{\varepsilon}=\varepsilon/{k_BT}$) for (a) nested square ($n=4$), (b) nested triangle ($n=3$) and (c) nested rectangle ($n=2$) domains.}
\label{fig:c3f12}
\end{figure}

Quantum thermal densities of nested triangular and rectangular domains are shown in Fig. 3.11. Symmetric periodicity angles of nested triangle and rectangle domains are $\theta=60^{\circ}$ and $\theta=90^{\circ}$ respectively. During the rotation in nested square domain there was a formation of four-fold degeneracy in the ground state. In nested triangle and rectangle domains respectively three-fold and two-fold degeneracy are formed as expected, which can be inferred also from the rightmost subfigures of Fig. 3.11.

\begin{figure}
\centering
\includegraphics[width=0.75\textwidth]{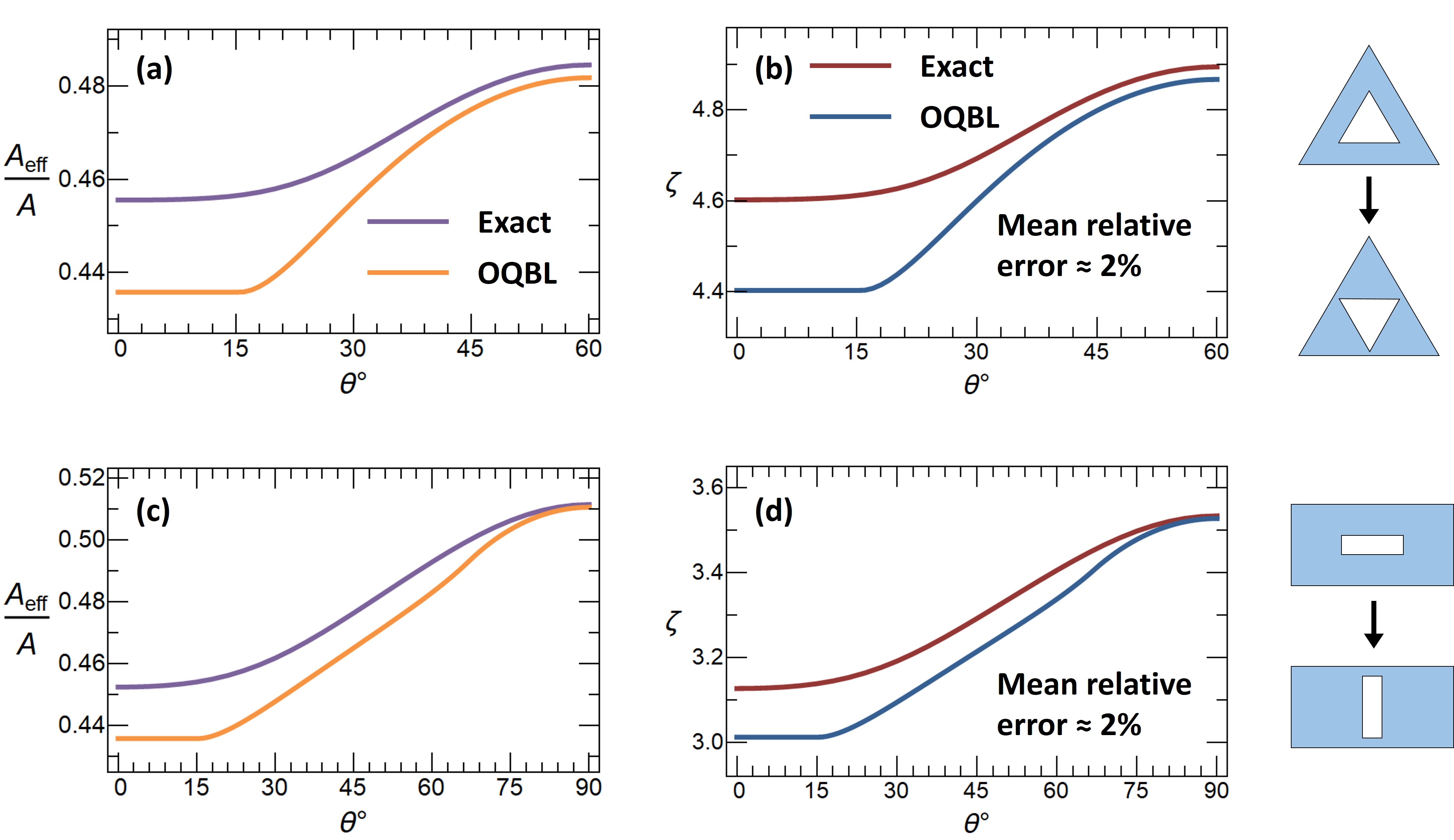}
\caption{Comparisons of numerical and analytical results for different domains. Variation of (a), (c) effective area and (b), (d) partition function due to changes in angular configuration for the nested triangle and rectangle domains respectively.}
\label{fig:c3f13}
\end{figure}

The formation of distinct degeneracies of energy levels for our considered confinement domains can be seen more explicitly in Fig. 3.12. Variations of the first twenty eigenvalues with respect to rotation angles are plotted where $y$-axis shows the number and the color bar denotes the magnitude of the dimensionless energy eigenvalues. Clear formation of lines in Fig. 3.12 occurs at the multiples of 4, 3 and 2 for nested square (Fig. 3.12a), triangle (Fig. 3.12b) and rectangle (Fig. 3.12c) respectively, providing a direct evidence of four-fold, three-fold and two-fold degeneracies from the eigenvalue spectra. While the signatures of degeneracies in the eigenspectra are nearly perfect for lower eigenvalues, deformations of the spectral lines occur at higher eigenvalues especially near the critical values of rotation angles. This analysis shows that some properties of the arbitrarily confined domains can be inferred directly from their eigenspectra, as long as the domain has distinct geometric features such as the occurrence of local confinement regions as in our domains.

In Fig. 3.13, comparisons of the overlapped quantum boundary layer method with the exact numerical results for the estimation of effective area and partition function are shown for nested triangle and rectangle domains. The accuracy of our analytical method shows similar characteristic to the one in nested square domain, though the mean relative errors are doubled in this case. Amount of effective areas compared to whole domain is similar (around half) in all three (nested square, triangle, rectangle) cases. It is seen that in all cases, estimation accuracy of the analytical method increases with the formation of degeneracies in energy levels, whereas errors are large during the transition regions from non-overlap to overlap cases. Errors decrease with the increments in overlaps. Because the increments of overlaps lead to the formation of disconnected subregions inside the domain, which makes the geometry simpler to tackle. These are noticeable from the apparent formation of the splits of high density regions for the configurations near the critical value of $\theta_{*}$, see Fig. 3.11. This suggests that finer approximation to the quantum thermal density may increase the accuracy indeed.

\section{Different Boundary Conditions}

In our considered cases, quantum confinement conditions were extremely strong in terms of two ways; the domain sizes were really small and the boundaries were perfectly impenetrable so that the wave function was zero at the boundaries, i.e. the Dirichlet boundary conditions. We can also examine the opposite and intermediate cases to understand more about the nature of quantum shape effects. The opposite case is imposing the Neumann boundary condition where the wavefunction has a finite value on boundaries but its derivative is zero.

\begin{figure}
\centering
\includegraphics[width=0.5\textwidth]{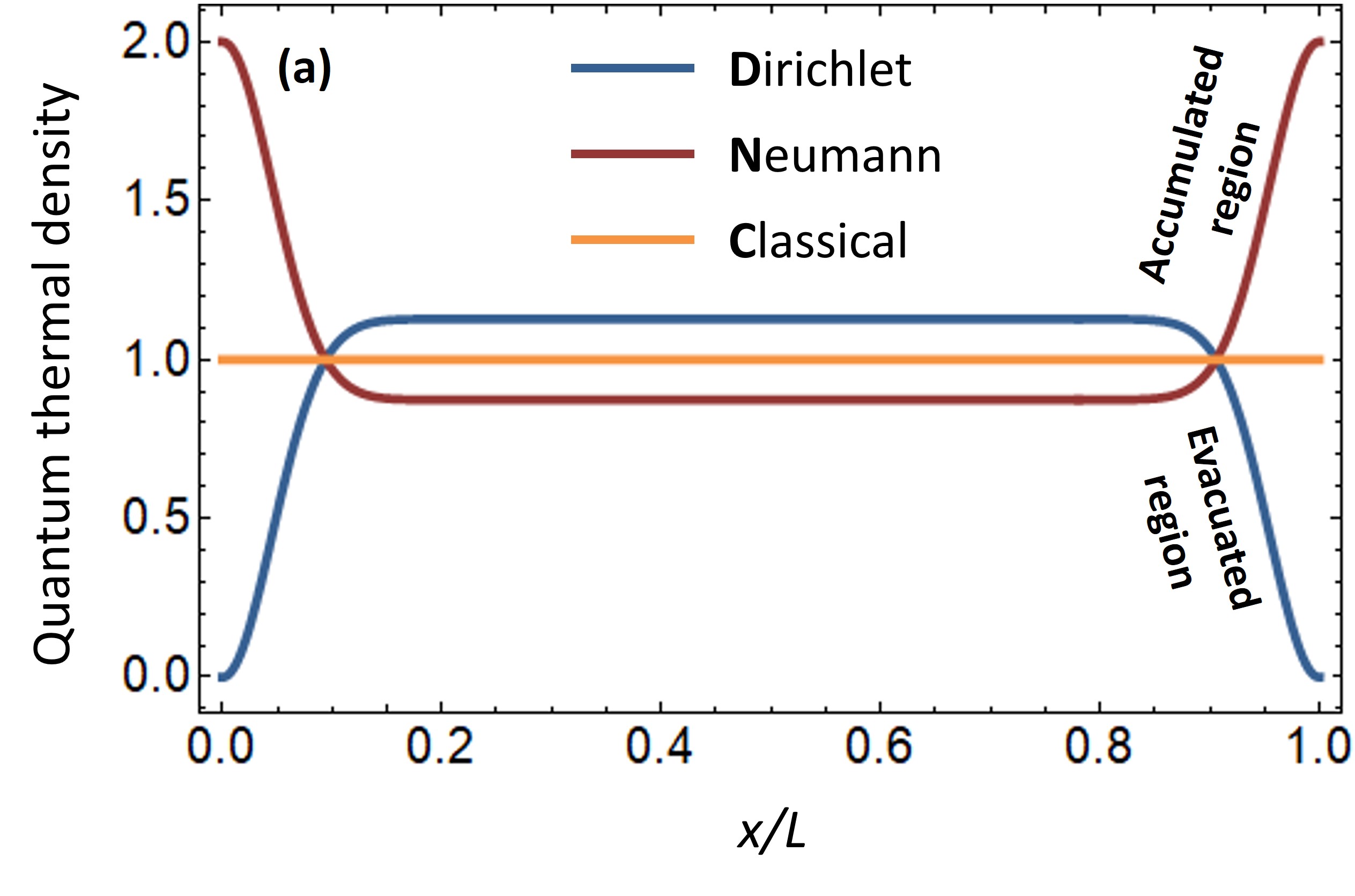}
\caption{Quantum thermal density distributions of particles confined in 1D domains with Dirichlet (blue curve) and Neumann (red curve) boundary conditions at the both ends. Uniform classical distribution is shown by the orange line. Particles evacuate (accumulate) near the (to) boundaries for Dirichlet (Neumann) boundary conditions.}
\label{fig:c3f14}
\end{figure}

\begin{figure}
\centering
\includegraphics[width=0.95\textwidth]{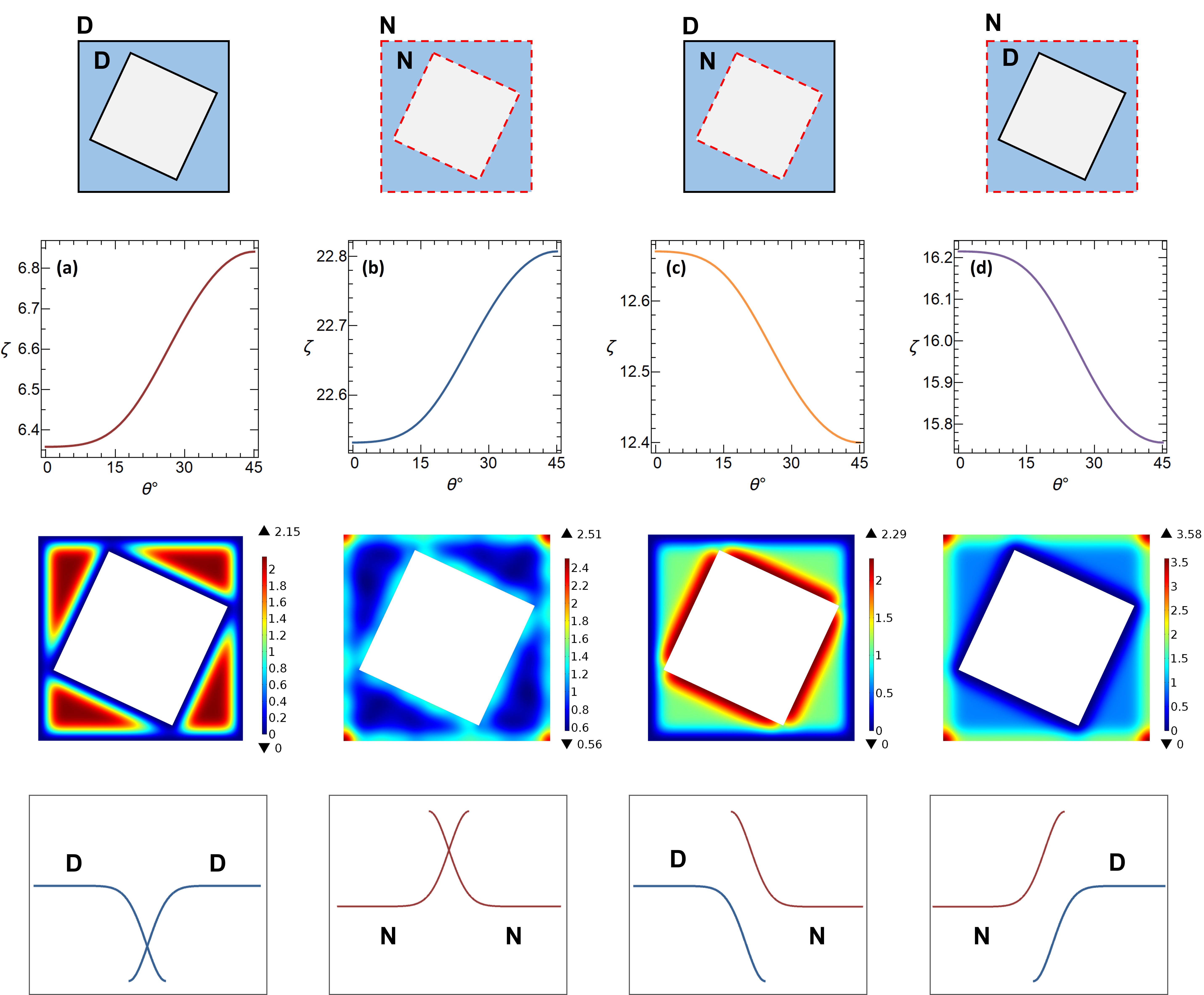}
\caption{Rows from top to bottom show the domains with different boundary conditions, variation of partition function with $\theta$, quantum thermal density distributions and different type of overlaps of quantum boundary layers. Columns from left to right show the results for the nested square domains with (a) Dirichlet, (b) Neumann, (c) outer boundary Dirichlet, inner boundary Neumann and (d) outer boundary Neumann, inner boundary Dirichlet boundary conditions. Partition function has opposite behavior in mixed boundary conditions, compared to the Dirichlet and Neumann boundary conditions.}
\label{fig:c3f15}
\end{figure}

Before examining these conditions in our confinement domain, let's compare the quantum thermal densities that are occurring in Dirichlet and Neumann boundary conditions in simple 1D domains, Fig. 3.14. This comparison allows us to see how quantum boundary layer transforms in Neumann case in comparison with the Dirichlet case. Unlike in the Dirichlet boundary conditions, particles tend to stay close to boundaries rather than staying away from them. In fact, their behavior is the exact opposite of the Dirichlet conditions.

In Fig. 3.15a and Fig. 3.15b partition function changing with the rotation angle is shown for Dirichlet and Neumann boundary conditions respectively. The functional behavior remains unchanged when the boundary conditions are changed from Dirichlet to Neumann, but the magnitude of the partition function increased because the domain is unconfined in the Neumann case where the leakage of wavefunction from boundaries is at its maximum. In Fig. 3.15c (3.15d), mixed boundary conditions, outer boundary Dirichlet (Neumann) and inner boundary Neumann (Dirichlet), are examined. Interestingly, for mixed boundary conditions the functional behavior of the partition function is reversed. The reason of this can be understood again by considering the overlaps of quantum boundary layers.

% Buraya biraz daha elabore gerekiyor. Fig. 3.15'i yorumlayabilirsin.

\section{Quantum Size Effects on Quantum Shape Effects}

Advent of quantum shape effects depends on the strength of the confinement and the strong confinement is a prerequisite for the appearance of quantum shape effects. This fact became much more clear with the investigation of quantum boundary layer overlaps. If outer and inner boundaries do not close enough to each other, overlaps won't emerge and quantum shape effects diminish. At this point, it is reasonable to ask how quantum size effects influence the magnitude and behavior of quantum shape effects.

In this section, we choose two different methods to change the characteristic confinement of the domain and then compare the variation of two different angular configurations. Remember, the characteristic confinement was defined as $\alpha_{*}=(\sqrt{\pi}/2)*(\lambda_{th}/L_{*})$ which was giving a rough estimate (the roughness comes from the fact that a 2D information is reduced into a single parameter) of the information about the size-confinement of the domain. The first method of changing the characteristic confinement is keeping the outer square length fixed and changing only the inner square length, see top-left subfigure in Fig. 3.16. It can be inferred even from the image that size-invariant shape transformation's influence on the shape of the domain decreases when the confinement is reduced (from right to left).

Variation of partition function by increasing the characteristic confinement this way can be seen in Fig. 3.16a. Red and blue curves represent the $\theta=0^{\circ}$ and $\theta=45^{\circ}$ configurations respectively. When the confinement increases two configurations start to differ. Note that the previous examination of partition function in this chapter is done considering $\alpha_{*}=1$ value, which corresponds to the end points of the curves in Fig. 3.16a. When the confinement is decreased, the difference between two most distinct configurations vanishes, see the subfigure showing the percentage difference in Fig. 3.16a. This is because the domain becomes less and less sensitive to the changes in inner object, when inner object size is so small. Imagine yourself as a particle traveling inside the domains of $\theta=0^{\circ}$ and $\theta=45^{\circ}$ configurations with confinement $\alpha_{*}=0.6$ and then compare your experience with the $\alpha_{*}=1$ one. In which one you will feel more difference between the $\theta=0^{\circ}$ and $\theta=45^{\circ}$ configurations? For instance, imagine you started from the bottom and you want to reach the top, there is a very narrow bottleneck at $\theta=45^{\circ}$ for $\alpha_{*}=1$, whereas you will experience a more or less similar path in between $\theta=0^{\circ}$ and $\theta=45^{\circ}$ for $\alpha_{*}=0.6$. 

\begin{figure}
\centering
\includegraphics[width=0.95\textwidth]{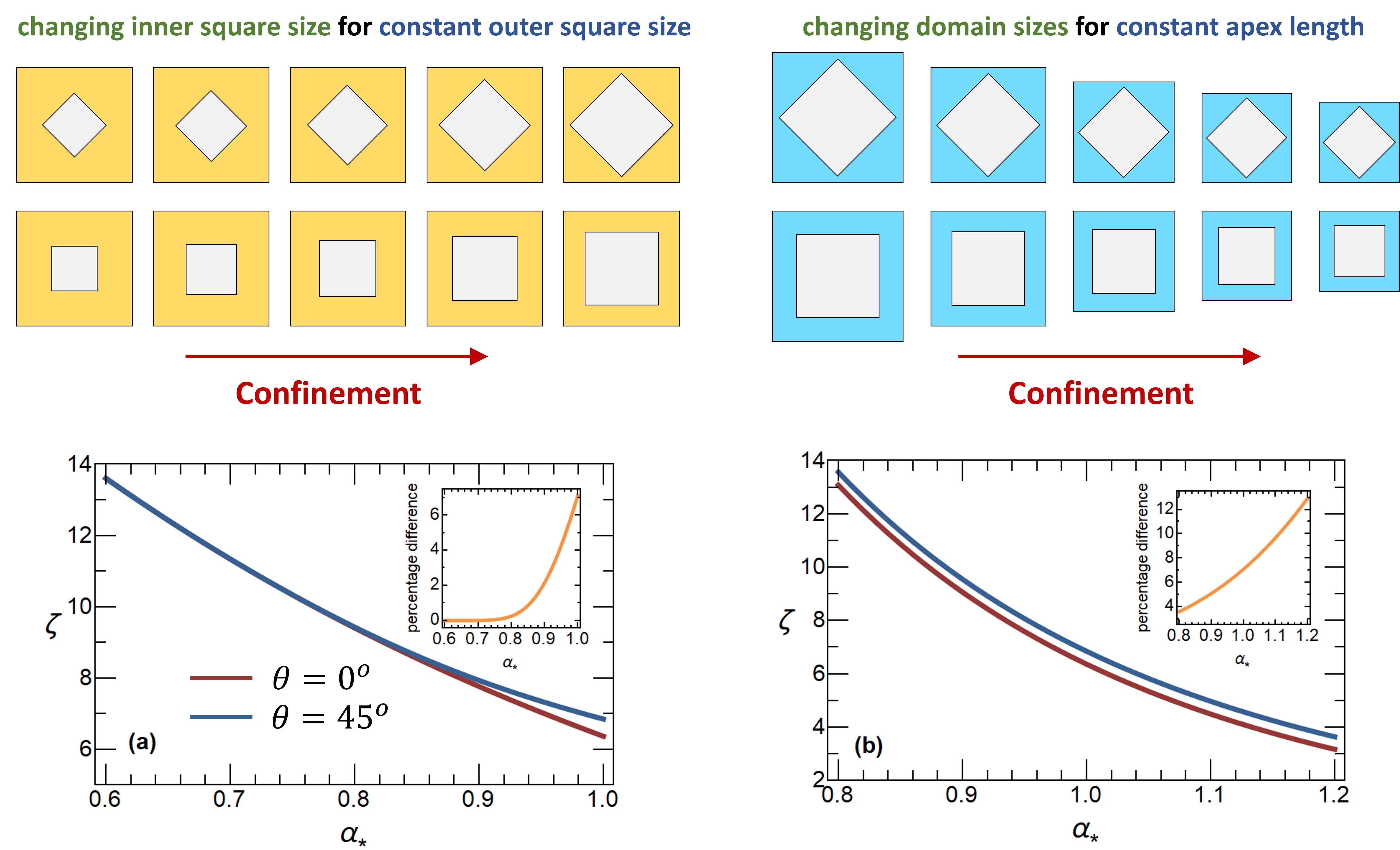}
\caption{Influence of quantum size effect on quantum shape effect. Changing confinement via changing inner square size by keeping the outer square size fixed (top-left) and changing outer and inner square sizes by keeping the apex length constant (top-right). Partition function vs characteristic confinement for (a) top-left and (b) top-right confinement methods. The legend applies to both figures.}
\label{fig:c3f16}
\end{figure}

The other type of confinement variation is changing both the outer and inner square lengths together so that the apex length is kept constant, which is seen in the top-right subfigure of Fig. 3.16. In this type of change, the self-similarity of the domain is preserved much better than the previous one. As is seen in Fig. 3.16b, even for $\alpha_{*}=0.8$, the difference between $0^{\circ}$ and $45^{\circ}$ configurations is around $3\%$ (see the percentage difference subfigure), which was near to zero in the previous confinement type for the same alpha value of 0.8. The percentage difference between the partition function values of $0^{\circ}$ and $45^{\circ}$ configurations gradually increases with the confinement, but in a slow pace. By the way, it is also seen that the characteristic confinement parameter that we choose did a good job in comparing different type of confinements. Because the values of partition function are very close to each other despite the confinement type is different in both pictures. This shows that the harmonic mean for calculating the characteristic sizes is a reasonable choice in these type of comparisons. We conclude this chapter here with this analysis, and we'll go into the thermodynamics of quantum shape effects in the next chapter.

%%%
%4%
%%%

%\chapter{Quantum Shape Effects in Nanoscale Thermodynamics}
\chapter{Quantum Shape Effects in Nanoscale Thermodynamics}

Last chapter we have seen that the partition function has distinct values per rotation angle of the inner object in nested confinement domains. In this chapter we'll go further to examine the thermodynamic properties of the particles confined in domains that undergo size-invariant shape transformation. We assume quasi-static processes all the time so that any change is happening slowly enough to keep the system always in equilibrium with the environment. This allows us to focus on the thermodynamic properties and their shape dependence in strongly confined systems. Throughout our analysis in this thesis, we use this assumption. Non-equilibrium and finite-time processes are out of scope of this thesis. Even for equilibrium processes, as we shall see, many undiscovered and fascinating novel thermodynamic behaviors are hidden to be come out by quantum shape effects.

\section{Shape as a New Thermodynamic Control Variable}

Thermodynamic state space of simple systems (consisting of unconfined particles in the absence of external force fields) has two macroscopic degrees of freedom or two dimensions because classically thermodynamic state functions have only temperature and volume dependencies (or any other set of two state variables). Quantum size effects add three new variables to this state space. In addition to the volume, we have surface area, periphery and vertices dependencies, making the thermodynamic state space five dimensional. One can go beyond these dimensions by keeping them constant and playing with another variable, and so dimension. Remember we have changed the partition function of a system by keeping all these five degrees of freedom constant, but changing only the shape which is characterized by the rotation angle $\theta$. Thus, shape of a system becomes the new thermodynamic control variable. It can characterize thermodynamic properties of a system independently of any other thermodynamic state variables. An example of this can be seen in Fig. 4.1 where $\theta$ plays the role of opening a new dimension in the conventional $T$-$S$ diagram, thereby allowing the access of taking advantage from a three-dimensional space of thermodynamic states even if all size parameters are kept constant. The resulting diagram is a sketch of a novel three-dimensional thermodynamic temperature-entropy-shape ($T$-$S$-$\theta$) diagram. Similar diagrams can also be drawn for pressure-volume relationship, allowing to exploit the opening of a new dimension for the amount of possible extractable work from a system with the help of quantum shape effects.

\begin{figure}
\centering
\includegraphics[width=0.45\textwidth]{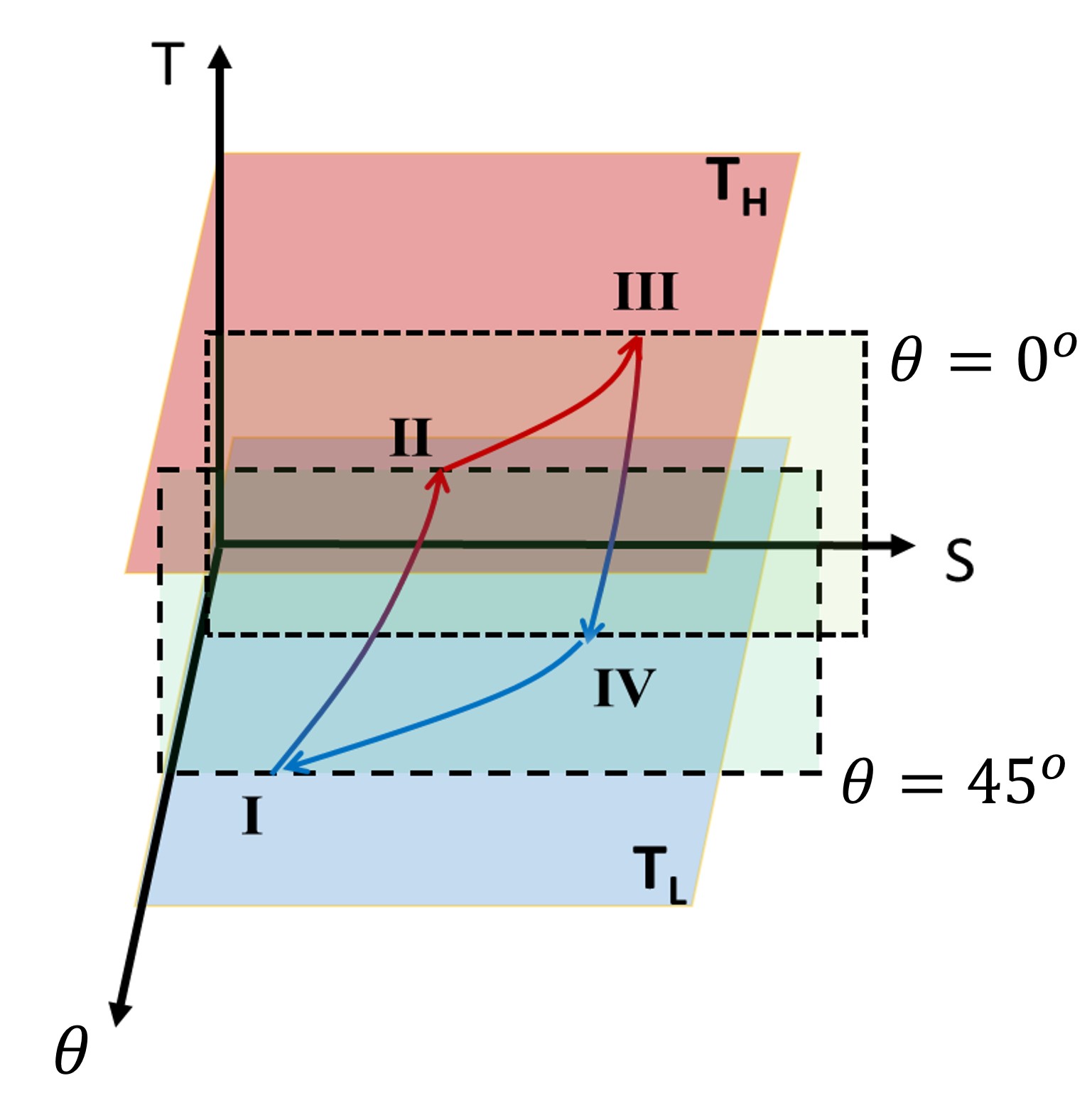}
\caption{An example of a higher dimensional thermodynamic diagram. Shape variation adds a new dimension to the conventional temperature-entropy ($T$-$S$) diagram, making a three-dimensional thermodynamic temperature-entropy-shape ($T$-$S$-$\theta$) diagram.}
\label{fig:c4f1}
\end{figure}

For a given number of particles in a system, thermodynamic state space now becomes six-dimensional,
\begin{equation}
X\left(T,\mathcal{V}\right)\xrightarrow{\mli{QSE}}X\left(\ldots,\mathcal{A},\mathcal{P},\mathcal{N_V}\right)\xrightarrow{\mli{QShE}}X\left(\ldots,\theta\right),
\end{equation}
where $X$ is any thermodynamic state function with particle number $N$. Note that shape parameter does not necessarily have to be $\theta$. As long as a shape transformation is size invariant, the relevant parameter can be considered as the shape variable, e.g. coordinate positions in the translational change of inner object instead of rotational. In translational changes it can be even more complicated by adding many additional control variables to the thermodynamic state space. Extension of thermodynamic state space paves the way for designing new thermodynamic processes and cycles with interesting and useful properties, as we shall see in next chapter where we discuss the applications. 

\section{Satisfying the Maxwell-Boltzmann Statistics}

Thermodynamic analysis of confined systems relies on the methods of statistical physics. Therefore, we would like to construct a reliable model that we can characterize by the statistical methods. Our first and foremost condition is to satisfy the basic statistical conditions. To be able to perform a statistical analysis, we need to have a large number of particles inside the confinement domain, so that $N>>1$. In fact, even for a single-particle system, one can use the statistical methods because the ensemble averages would be equal to the time averages, as long as the system behaves ergodic. However, we will ensure the sufficiently large number of particles to safely use the methods of statistical mechanics and to cope with the fluctuations. Throughout our analysis, we choose the particle density as $n=5\times 10^{24} \text{m}^{-3}$ which gives the particle number in our systems as $N=1012$, unless it is stated otherwise. Beside physical constraints, the reason why we don't want to choose a much larger particle number is that for our initial analysis, we would like to use the Maxwell-Boltzmann particle statistics, due to the reasons of its simplicity in expressions and possibility of getting analytical results easier. Besides, understanding of quantum shape effects in Maxwell-Boltzmann statistics will shed light into its understanding under other statistics.

Maxwell-Boltzmann statistics is used to describe the average distribution of particles on energy states in thermal equilibrium for high temperature and low density conditions. These conditions can be quantified by comparing the classical density of the particles inside the domain with their de Broglie densities. Classical density ($n_{cl}=N/\mathcal{V}$) should be much smaller than the de Broglie density, $n_{cl}<<n_{dB}$, for Maxwell-Boltzmann conditions to be satisfied. Increasing the temperature decreases the de Broglie wavelengths of particles which causes a higher de Broglie density. Similarly, decreasing the density basically means smaller $n_{cl}$, which again enlarges the gap between the classical and de Broglie densities, therefore contributing to the $n_{cl}<<n_{dB}$ condition. Since volume is constant throughout our analyses, decreasing the density corresponds to decreasing the number of particles. If $n_{cl}<<n_{dB}$ condition is not satisfied, indistinguishability property of particles becomes prominent, so Bose-Einstein or Fermi-Dirac statistics needs to be used depending on the symmetric or antisymmetric nature of the particles' wavefunctions respectively under a particle exchange. In this regard, Bose-Einstein or Fermi-Dirac statistics are inherently quantum-mechanical and called together as quantum statistics. This terminology sometimes is extended to call the Maxwell-Boltzmann statistics as classical statistics. However, we think this is not quite appropriate terminology as it may cause reader not to expect any quantum behaviors in the systems obeying Maxwell-Boltzmann statistics, which is definitely not true. Although the statistics itself is derived using the classical mechanics, it's just an asymptotic case of the quantum statistics and quantum nature of particles can still be incorporated to the system by their discrete eigenspectrum that is determined by the Schr\"{o}dinger equation.

We can only satisfy $n_{cl}<<n_{dB}$ condition, at room temperature with bare electron mass, by increasing the sizes of at least one direction. We extend the domain in the third direction longitudinally (see Fig. 4.2) without making any changes in the transverse directions which will be responsible from quantum shape effects. By this way, we fill the domain with many particles. Adding the third dimension not only gives us possibility to satisfy the conditions of Maxwell-Boltzmann statistics, but also makes our discussions more realistic, since all materials are intrinsically 3D. Moreover, there are some nanoarchitectures, e.g. core-shell nanostructures, which are suitable candidates for the demonstration of quantum shape effects and very close to our models\cite{coreshell1,coreshell2,coreshell3,PhysRevB.98.085419}.

\begin{figure}[!b]
\centering
\includegraphics[width=0.7\textwidth]{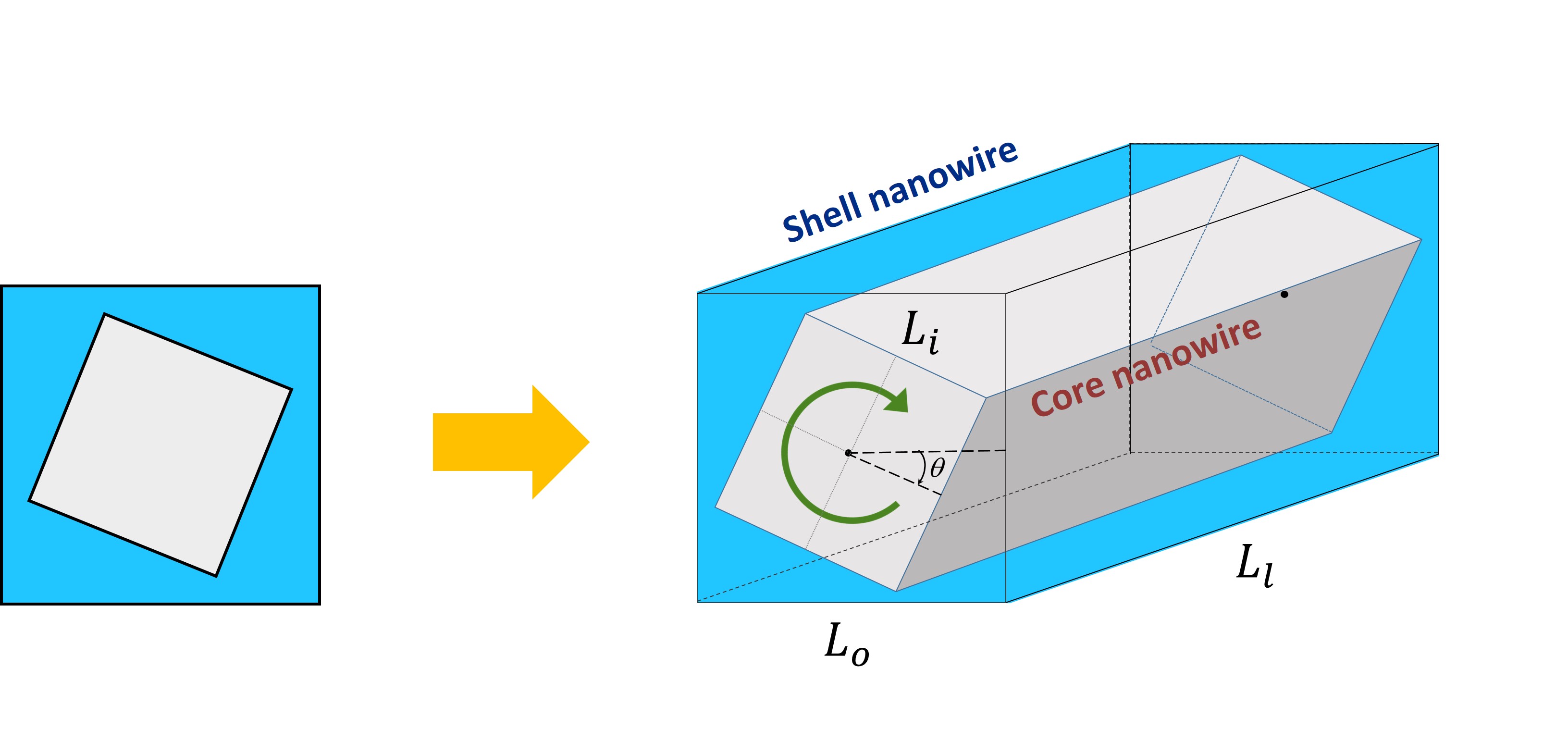}
\caption{Addition of the third dimension and turning the domain into a core-shell nanowire. The outer (shell) structure is fixed and the inner (core) structure is rotatable.}
\label{fig:c4f2}
\end{figure}

The length of the longitudinal direction is chosen as $L_l=\SI{762.8}{\nano\metre}$. If we calculate the confinement parameter of the longitudinal direction at room temperature, it gives $\alpha_l=0.005$, which means the third direction is nearly free (unconfined). Having at least one direction unconfined is in fact necessary to be able to satisfy the conditions of Maxwell-Boltzmann statistics as it is impossible to do it with all directions confined. Nevertheless, the shape of the domain is preserved in transverse direction, so quantum shape effects imposed by the transverse direction eigenvalues are still there.

Remember we have solved the Schr\"{o}dinger equation for 2D domain in the previous chapter. We changed our domain into a 3D one now, so we need to solve the Schr\"{o}dinger equation for our new 3D confinement domain. However, this is actually an unnecessarily hard task in terms of computational point of view. The first difficulty is we need a lot more mesh points in a 3D domain, which will increase the numerical solution time substantially compared to 2D case. The second difficulty is the truncation point of the summation over longitudinal momentum values would be much higher than the 2D transverse case, which means we have to calculate much more eigenvalues.

Fortunately, we don't need to worry about any of these numerical difficulties, because choosing a relatively long length in longitudinal direction gives us possibility to obtain analytical expressions without solving the Schr\"{o}dinger equation explicitly for that direction. Using the orthogonality of eigenstates and the quadratic energy-momentum dispersion relation as well as taking the advantage from the product rule of exponents in Maxwell-Boltzmann distribution function give us opportunity to separate the partition function into two parts as numerical and analytical and calculate each part individually. By this way, the partition function for our 3D domain can be captured as
\begin{equation}
\begin{split}
\zeta=\sum_k \exp\left(-\tilde{\varepsilon}_k\right) &\approx \left(\int\exp\left(-\tilde{\varepsilon}_l\right)di_l -\frac{f_l(0)}{2}\right)\sum_t \exp\left(-\tilde{\varepsilon}_t\right) \\
&=\underbrace{\left(\frac{\sqrt{\pi}}{2\alpha_l}-\frac{1}{2}\right)}_{\text{analytical}} \underbrace{\sum_t \exp\left(-\tilde{\varepsilon}_t\right)}_{\text{numerical}}
\end{split}
\end{equation}
where energy eigenvalues are represented in their dimensionless form as $\tilde{\varepsilon}_k=\varepsilon_k\beta=\varepsilon_k/(k_BT)$ with $k$ representing the 3D domain eigenvalues (which are hard to obtain with acceptable precision in a reasonable time, but as we shall see there is no need for that). $\tilde{\varepsilon}_l$ and $\tilde{\varepsilon}_t$ are the dimensionless energy eigenvalues for 1D longitudinal direction and 2D transverse directions respectively. Since the structure in longitudinal direction is generated by a straightforward extension of the structure in transverse direction, the translational energy eigenvalues are analytically known for the longitudinal direction, which are in their dimensionless form $\varepsilon_l/(k_BT)=\tilde{\varepsilon}_l=(\alpha_l i_l)^2$ where $i_l$ is the quantum state variable for momentum component in longitudinal direction. Then, the summation of longitudinal direction can be analytically approximated by the first two terms of PSF, which results a very simple expression shown within parentheses in Eq. (4.2). Although the usual continuum approximation (the first term in the parenthesis) would also work for the longitudinal direction, we will use the safer one and consider also the quantum size effect corrections coming from the nearly free direction. The numerical part has already been solved in the previous chapter, so the partition function is successfully constructed.

Note that our approach to finite-size thermodynamics is different than Hill's approach. In Hill's nanothermodynamics, the thermodynamic limit is taken and the system is partitioned into an ensemble of identical subsystems \cite{hillbook,bigworld, trondheim}. In that case, extensivity is broken due to the interactions of subsystems via the subdivision potential \cite{rubifinite}. In non-interacting case, the subdivision potential is associated with the addition or removal energy cost of the subsystems \cite{kitaevc}. It is not possible to predict the existence of quantum shape effects by using Hill's nanothermodynamic approaches, because it is a limited approach where the thermodynamic limit is used and finite-size corrections are manually added into the expressions of continuum approximation. On the other hand, in our approach we are not using thermodynamic limit at all, since the volume and number of particles are both finite in our system. We use the fundamental thermodynamic expressions in their "summation-over-all-states" form, which are coming directly from the statistical mechanics framework via the entropy maximization under constant number of particles and energy. The information about the finiteness of the system is embedded within the eigenstates that we considered in our distribution functions. The thermodynamic expressions that we used are true independently from the fact that whether the system is infinite or finite. Extensivity in our case is naturally (i.e. without manually adding any terms) broken due to the quantum-mechanical corrections of non-existing zero-energy states.

We won't examine the partition function again in this chapter, because adding the longitudinal direction does not change the analysis of the shape dependence of partition function. Now it's time to explore the thermodynamic properties of the particles confined in our nested domains.

\section{Internal Energy}

When a system is in thermodynamic equilibrium, we can define the thermodynamic state functions such as internal energy, free energy and entropy. Thermodynamics is concerned with the changes in state functions, rather than their absolute values. Nevertheless, we give the normalized values of each thermodynamic state function rather than their relative values, to provide an insight about their magnitudes in different systems. Their relative values can also be easily seen from the figures that are given. Each thermodynamic state function is normalized by either $Nk_BT$ or $Nk_B$ depending on the appropriate normalization factor. We used per particle normalization to prevent unnecessary amplification of magnitudes by the particle number which is chosen to be not so small.

The primary thermodynamic quantity we would like to investigate is the internal energy, which is the energy of a system associated with the average random motion of its microscopic constituents. For an ideal monoparticular gas, the internal energy consists of only the translational kinetic energies of particles. The internal energy of a system with $N$ identical particles is defined in statistical mechanics as
\begin{equation}
U=N\sum_k p_k \varepsilon_k
\end{equation}
where $p_k$ is the probability of the system occupying the momentum state $k$, which is given by the probability density function
\begin{equation}
p_k=\frac{\exp\left(-\tilde{\varepsilon}_k\right)}{\sum_k \exp\left(-\tilde{\varepsilon}_k\right)}=\frac{f_k}{\sum_k f_k}=\frac{f_k}{\zeta}
\end{equation}
where $f_k=\exp\left(-\tilde{\varepsilon}_k\right)$ is the Maxwell-Boltzmann distribution function. Again, this becomes separable because the product rule of exponents can be exploited in Maxwell-Boltzmann statistics. The partition function here serves as a normalization constant of the probabilities, ensuring that they add up to one. Putting Eq. 4.4 into Eq. 4.3 shows that the internal energy is probabilistically the ensemble averaged expectation value of single particle energy eigenvalues, $U=N\left\langle\varepsilon_k \right\rangle_{\mli{ens}}$. Thermodynamically, there are two types of energy exchanges between a system and its environment: the work exchange (organized form of energy) and the heat exchange (unorganized form of energy). The change in internal energy corresponds to the sum of work and heat exchanges by the first law of thermodynamics ($dU=\dbar W+\dbar Q$).

Internal energy is related also with the confinement of the domain. When the confinement increases, energy eigenvalues shift up on the spectrum (due to their inverse proportionality with size) and since the internal energy is basically their ensemble average, the internal energy increases as well. Increase in internal energy under an isothermal process is also called confinement energy, since it is solely due to the confinement.

Similar to the partition function, internal energy can also be decomposed (thanks to MB statistics) into the numerical and analytical parts as follows,
\begin{subequations}
\begin{align}
\frac{U}{Nk_BT}&=\sum_{k}p_k\tilde{\varepsilon}_k=\frac{\sum_{k}f_k\tilde{\varepsilon}_k}{\sum_{k}f_k}=\frac{\sum_t\sum_l f_t f_l (\tilde{\varepsilon}_t+\tilde{\varepsilon}_l)}{\sum_t\sum_l f_t f_l} \\
& \approx\frac{\sum_t f_t\tilde{\varepsilon}_t\left(\int f_l di_l-f_l(0)/2\right)+\sum_t f_t\left(\int f_l\tilde{\varepsilon}_l di_l\right)}{\sum_t f_t\left(\int f_l di_l-f_l(0)/2\right)} \\
& =\underbrace{\frac{\sum_{t}f_t\tilde{\varepsilon}_t}{\sum_{t}f_t}}_{\text{numerical}}+\underbrace{\frac{1}{2-2\alpha_l/\sqrt{\pi}}}_{\text{analytical}}
\end{align}
\end{subequations}
where $f_l=\exp\left(-\tilde{\varepsilon}_l\right)$ and $f_t=\exp\left(-\tilde{\varepsilon}_t\right)$ are the distribution functions for longitudinal and transverse directions respectively. For a better representation than the continuum approximation, we always use bounded continuum approximation (using the first two terms of PSF) for the analytical parts. Note that sometimes the second term can give exactly zero (e.g. in the energy integrals in Eq. 4.5b), in those cases the bounded continuum approximation does not lead to any improvement.

Change in internal energy with rotation angle $\theta$ is shown in Fig. 4.3 by green curve for three different nested confinement domains, calculated from Eq. (4.5c). Dashed-black curves show the results of analytical calculation by the overlapped quantum boundary layer method. Internal energy of the system smoothly decreases from $\theta=0^{\circ}$ to $\theta=45^{\circ}$ in a sigmoid-like fashion. During this variation, all physical parameters like number of particles, temperature of the system as well as all geometric sizes are constant, except the rotation angle of the core structure.

\begin{figure}
\centering
\includegraphics[width=0.9\textwidth]{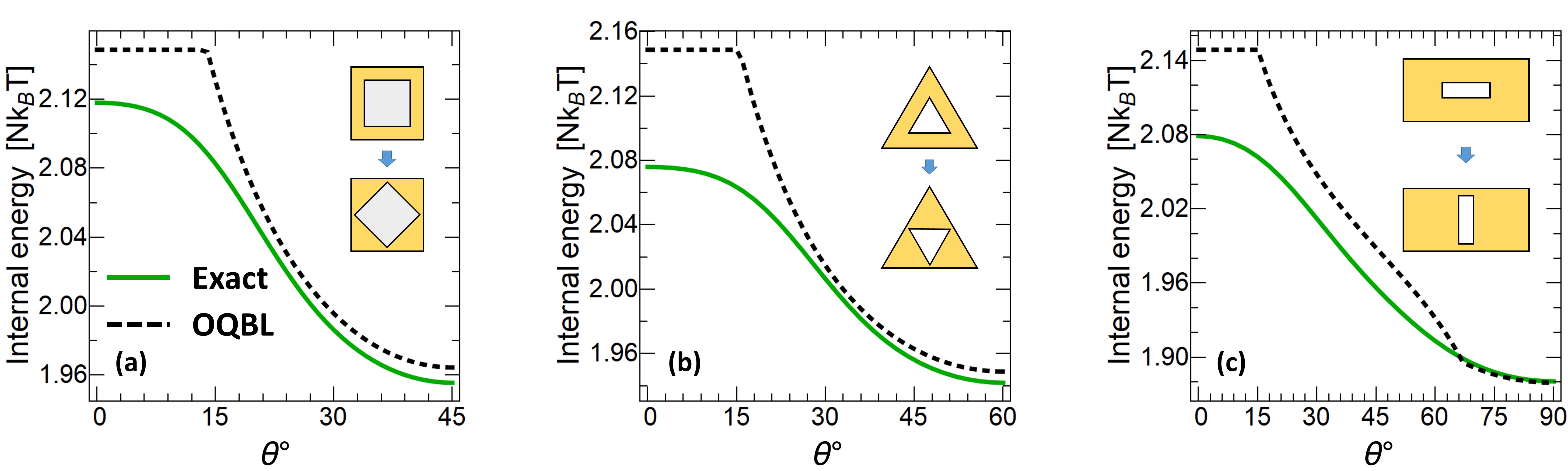}
\caption{Variation of internal energy (normalized to $Nk_BT$) with respect to the shape change via $\theta$, for (a) nested square, (b) nested triangle and (c) nested rectangle domains. Green and dashed-black curves represent the exact (Eq. 4.5) and analytical (Eq. 4.6) calculations respectively.}
\label{fig:c4f3}
\end{figure}

The shape-dependent behavior of internal energy is very similar in all three examined nested confinement domain cases. As an indicator of the reasonable selection of characteristic confinement parameter, the internal energy values for all cases are quite close to each other. Internal energy decreases with the rotation angle of the core structure. Higher internal energy means that the system is effectively more confined in higher internal energy states. The reason of this behavior can be very well understood under the overlapped quantum boundary layer framework. The amount of overlap increases with the rotation angle and increase in overlap volume (since the domain is 3D, now we have overlap volumes all along the longitudinal direction instead of overlap areas which was in the 2D case). More overlap volume means there are more available effective volume inside the domain which can be occupied by particles and so the domain is effectively less confined. Note that now we are talking about an effective confinement instead of an apparent confinement, because the confinement is initially defined to be depend only on the sizes of the domain but not its intrinsic shape. Keeping the sizes constant means the confinement does not change. On the other hand, effectively the space occupied by particles increases due to the overlap regions and so contributing to the decrease in confinement and internal energy. This decrease of internal energy explicitly shows deconfinement just by change of shape.

We realize that the dependency of quantum size and shape effects on the confinement of the system is quite opposite of each other. Both effects appear under confinements of particles. However, unlike quantum size effects, appearance of quantum shape effects leads to a decrement in the effective confinement of the domain. This is because quantum shape effects are caused by the overlap volumes which are contributing to the relaxation of the effective confinement by increasing the effective volume inside the domain. This behavior is evident also from the decrease in internal energy due to quantum shape effects. Overlaps of quantum boundary layers provide a nice physical explanation for this relaxation of confinement due to quantum shape effects.

The analytical expression that is obtained for internal energy to plot dashed-black curves is given by the overlapped quantum boundary layer method as
\begin{equation}
\begin{split}
\frac{U_A}{Nk_BT}=\left(\frac{3}{2}\right)_{\mkern-7mu\mli{bl}}+\left(-\frac{T}{\mathcal{V}_{\mli{eff}}^{\mli{0}}}\frac{\partial \mathcal{V}_{\mli{qbl}}}{\partial T}\right)_{\mkern-7mu\mli{QSE}}+\left(\frac{\mathcal{V}_{\mli{eff}}^{\mli{0}} T}{\mathcal{V}_{\mli{eff}}}\frac{\partial}{\partial T}\frac{\mathcal{V}_{\mli{ovr}}}{\mathcal{V}_{\mli{eff}}^{\mli{0}}}\right)_{\mkern-7mu\mli{QShE}}.
\end{split}
\end{equation}
where $A$ subscript denotes that the expression is completely analytical. The first term on the right-hand side represents the bulk term, as denoted by the subscript $bl$, which is a result from continuum approximation. The second term is the quantum size effect term that is characterized by the excluded quantum boundary layer volume ($\mathcal{V}_{\mli{qbl}}$) and the conventional effective volume $\mathcal{V}_{\mli{eff}}^{\mli{0}}$. It gives the usual quantum size effect corrections to the internal energy of Maxwell-Boltzmann gases. The third term is the quantum shape effect term, the only term that depends on the overlap volume and the shape characterization parameter $\theta$. Thanks to the mathematical structure of the Maxwell-Boltzmann distribution, it is possible to express classical, quantum size effect and quantum shape effect terms explicitly and separately from each other. Note that this additivity of quantum size and shape effect terms are not an assumption but directly comes from the nature of MB statistics. Bulk term coming from the continuum approximation incorrectly calculates surface modes as well. The quantum size effect corrections actually fix this miscalculation by excluding the zero energy modes. Surface modes are quantum mechanically forbidden because they correspond to $i=0$ states. On the other hand, in bounded systems states start from a non-zero value, the ground state value, which is $i=1$. Additive quantum size effect term naturally corrects this by use of PSF or Weyl formulas. This is very convenient for the examination of distinct features of quantum size and shape effects.

In addition, since we consider a single-particle picture (not the many-body picture), we assume no quantum correlations between the wave functions of the particles. All states are completely independent from each other. Hence, all the sums are diagonal. This also justifies the additive nature of the quantum size and shape effect corrections which are semi-classical, not genuinely quantum thermodynamic.

It is also worth to mention that averaged internal energy of particles in our confinement domains is much higher than its classical value based on continuum approximation which is 3/2. Quantum size effects are responsible from this confinement energy, quantum shape effects just cause a reduction of it. In fact, if we wouldn't have considered the overlap volumes, the dashed black curves in Fig. 4.3 would have stayed constant on their $0^{\circ}$ values (with no overlap) which corresponding to the first two terms of Eq. (4.6). 

Lastly, the maximum and mean relative errors of the analytical internal energy expression are respectively $2.7\%$ and $1.1\%$ for nested square, $3.9\%$ and $1.5\%$ for triangle and $4.0\%$ and $1.6\%$ for rectangle domains. We give both the maximum and the mean relative errors in order to provide a more detailed information about the nature of errors which speak for the accuracy of the analytical expressions that we derived.

\section{Free Energy}

Investigation of internal energies of particles inside our nested confinement domains allows us to learn interesting things about the nature of quantum shape effects. Now we move on to another important thermodynamic potential which is the Helmholtz free energy (we simply call free energy). Under constant temperature condition, change in free energy equals to the total reversible work exchange (means the maximum possible total work exchange). Under constant temperature as well as pressure conditions, on the other hand, the related quantity is the Gibbs free energy in which the changes represent the maximum non-mechanical work (the work exchange except expansion/contraction process). 

For $N$ number of identical particles obeying the Maxwell-Boltzmann statistics, free energy is written as
\begin{equation}
F=-k_BT\ln Z
\end{equation}
where $Z=\zeta^N/N!$ is the $N$-particle partition function. Here it is assumed that single-particle partitions are independent and the indistinguishability of particles are ensured by $N!$. Although this is an approximation, as we shall see later during the comparison of results with the ones using quantum statistics (FD in particular), it is a good approximation for the systems that are considered here. Eq. (4.7) then can be rewritten as $F=-Nk_BT\ln\zeta+k_BT\ln N!$. Usually the term $\ln N!$ is approximated by Stirling's approximation as $\ln N!\approx N\ln N-N$, but we won't use this approximation since the calculation is already easy in its exact form. Like we have done in internal energy expression, we can write the free energy as the sum of numerical and analytical parts,
\begin{equation}
\begin{split}
\frac{F}{Nk_BT}&=-\ln\left[\sum_t f_t\left(f_l di_l-f_l(0)/2\right)\right]+\frac{\ln N!}{N} \\
& =\underbrace{-\ln\left(\sum_t f_t\right)}_{\text{numerical}}-\underbrace{\ln\left(\frac{\sqrt{\pi}}{2\alpha_l}-\frac{1}{2}\right)+\frac{\ln N!}{N}}_{\text{analytical}}
\end{split}
\end{equation}
where the numerical part consists of the logarithm of transverse direction partition function with negative sign and the analytical part contains both the longitudinal counterpart and the contribution due to the indistinguishability of particles.

\begin{figure}[!b]
\centering
\includegraphics[width=0.9\textwidth]{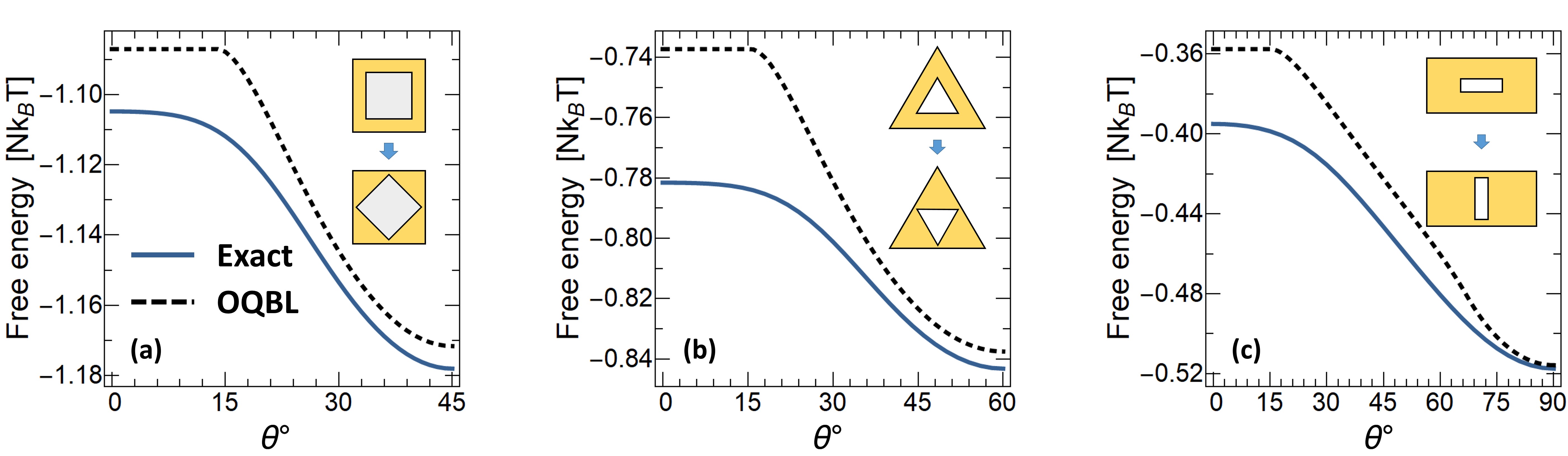}
\caption{Variation of free energy (normalized to $Nk_BT$) with respect to the shape change via $\theta$, for (a) nested square, (b) nested triangle and (c) nested rectangle domains. Blue and dashed-black curves represent the exact (Eq. 4.8) and analytical (Eq. 4.9) calculations respectively.}
\label{fig:c4f4}
\end{figure}

Nature tends to minimize the free energy. Therefore, free energy specifies also the direction of a spontaneous thermodynamic transition from one state to the other. Examining the free energy of our system under the shape variation allows us to determine which angular configuration minimizes the free energy of the system and so to which configuration the system "wants" to transform to become stable. The variation of free energy with respect to $\theta$ is shown in Fig. 4.4. Free energy of all systems decreases with increasing rotation angle or in other words increasing overlap volumes (minimizing the effective confinement). This means nature favors the existence of quantum shape effects. The reason is quite clear from the perspective of the overlapped quantum boundary layer theory. Minimization of free energy is equivalent to the maximization of the effective volume (minimization of the effective confinement) under constant temperature. Increment in overlap volumes means there are effectively more available space to be occupied by particles. Nature prefers to increase the overlap volumes to minimize the free energy.

The minimum free energy configurations correspond to the symmetric periodicity angles of the domains ($\theta=45^{\circ}$, $\theta=60^{\circ}$ and $\theta=90^{\circ}$ for nested square, triangle and rectangle domains respectively). Consider the outer shell is fixed and the inner core is free to rotate. Assuming zero friction and neglecting the interactions, if the initial state of a system is prepared in a configuration other than the minimum free energy one, the inner structure would spontaneously start to rotate until it reaches to the symmetric periodicity angle. Even an external torque needs to be applied to keep the inner structure stable under unstable angular configurations. In short, the rotating thing will rotate! Though it seems counter-intuitive at first, this spontaneous movement of course does not violate any physical law. Configurations except the minimum free energy one are actually unstable, despite the existence of thermodynamic equilibrium. A classical analog of this situation can be seen in the example of squeezing a spring. When you squeeze a spring, you store potential energy which can turn into kinetic energy when you release it. Similarly, preparing the confinement domain in a configuration other than the symmetric periodicity angle puts system into an unstable state, provided that the inner structure is free to rotate without any friction. 

By using the overlapped quantum boundary layer method, fully analytical expression for free energy can be obtained as
\begin{equation}
\begin{split}
\frac{F_A}{Nk_BT}=\left(-\ln\frac{\mathcal{V}}{N(4\delta)^3}-1\right)_{\mkern-7mu\mli{bl}}+\left[-\ln\left(1-\frac{\mathcal{V}_{\mli{qbl}}}{\mathcal{V}}\right)\right]_{\mkern-2mu\mli{QSE}}+\left[-\ln\left(1+\frac{\mathcal{V}_{\mli{ovr}}}{\mathcal{V}_{\mli{eff}}^{\mli{0}}}\right)\right]_{\mkern-2mu\mli{QShE}},
\end{split}
\end{equation}
where the first term is the classical free energy expression, while second and third terms correspond to the quantum size and shape effects corrections respectively. Just like size effect terms, shape effect terms also cause a breakdown of extensive nature of the thermodynamic state functions. Dashed black curves in Fig. 4.4 are the results of this expression in Eq. (4.9). It captures the true behavior of free energy with respect to a shape variation very well. The maximum and mean relative errors of the analytical free energy expression are respectively $2.2\%$ and $1.3\%$ for nested square, $6.3\%$ and $3.3\%$ for triangle and $11.5\%$ and $6.0\%$ for rectangle domains.

Existence of free energy variation suggests that there should be some mechanical forces acting and a torque that must be generated inside the confinement domain. We will explore this torque along with its quantum-mechanical origins in Sections 4.7 and 4.8. Before going into that, we will examine another very important thermodynamic property of our systems, called entropy. 

\section{Entropy}

Entropy is one of the primary concepts of thermodynamics and in fact of all physics, and it may be difficult to grasp as there are several definitions, explanations and even derivations of it. Entropy is an extensive measure of the number of microscopic configurations corresponding to the macroscopic state of a thermodynamic system.

In Maxwell-Boltzmann statistics, entropy reads
\begin{equation}
S=-Nk_B\sum_k p_k\ln p_k
\end{equation}
Putting Eq. (4.4) into the above equation generates the terms in internal energy and free energy in a way: $S=U/T-F/T$, indeed it is true that entropy is written by taking the difference of Eq. (4.5) and Eq. (4.8). We avoid rewriting all these terms again for brevity. In Fig. 4.5, entropy of the confined system with respect to the rotation angle is plotted for nested square, triangle and rectangle domains. In all cases, entropy has a decreasing behavior with respect to the starting position of the inner structures, which is $\theta=0^{\circ}$ for all cases. We have seen the decreasing behavior of free energy as well during the previous free energy section. The spontaneous decrease of both free energy and entropy is very interesting and a unique behavior in the thermodynamics of ideal gases. It even almost somewhat counter-intuitive at first thought. The direction of thermodynamic processes, which is determined by the minimum of free energy, is usually associated with the higher entropy state. Moreover, quantum shape effects cause an effective expansion of the domain via overlap volumes and entropy should, in general, increase in the direction of expansion. The larger the occupiable space, the larger the number of microscopic possibilities and larger the entropy. On the other hand, this logic seems to be not working in systems exhibiting quantum shape effects. 

\begin{figure}[!b]
\centering
\includegraphics[width=0.9\textwidth]{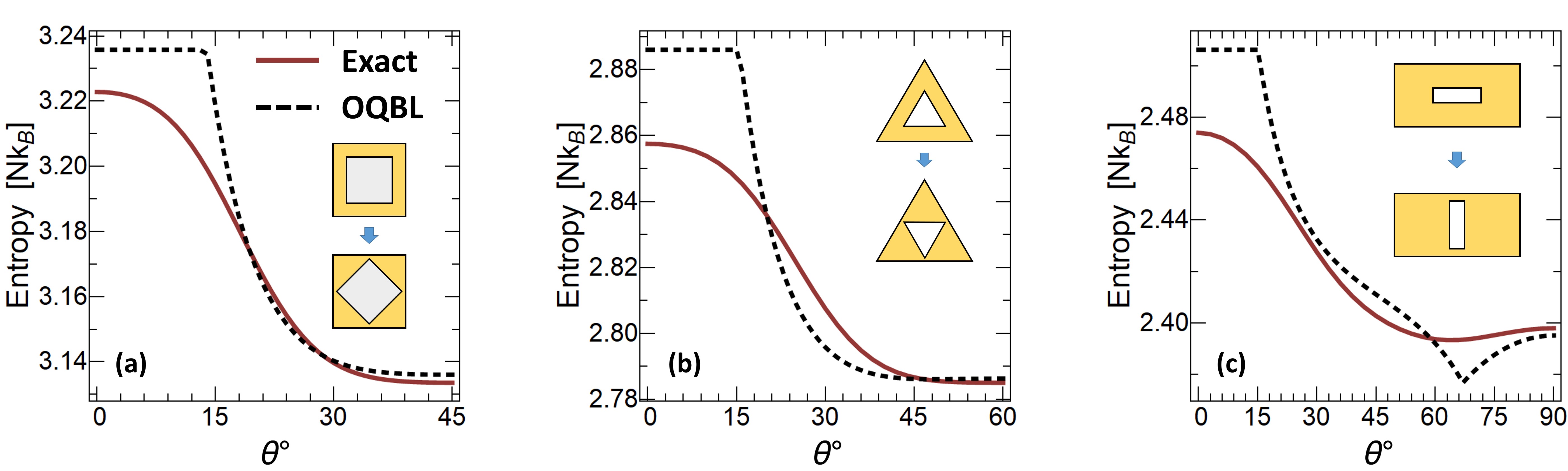}
\caption{Variation of entropy (normalized to $Nk_B$) with respect to the shape change via $\theta$, for (a) nested square, (b) nested triangle and (c) nested rectangle domains. Red and dashed-black curves represent the exact (Eq. 4.10) and analytical (Eq. 4.11) calculations respectively.}
\label{fig:c4f5}
\end{figure}

In order to understand what's going on, we need to look closer to the thermodynamic state functions of the system. From the initial configuration angle to the symmetric periodicity angles, not only free energy (Fig. 4.4) and entropy (Fig. 4.5) of the system decreases, but also the internal energy (Fig. 4.3) decreases. In fact, the decrease in internal energy is larger than the decrease in entropy and this is why free energy also decreases. Since the decrement in internal energy of the system is larger than the decrement in its entropy, free energy must decrease during a shape transformation.

Recall that we are dealing with a system that is in contact with a heat bath so that the processes are isothermal. This means, in order to keep the temperature constant during the prescribed shape transformation the system has to exchange heat with the environment. Indeed, the decrease in internal energy is due to the heat loss of the system to the environment. The system decreases its entropy, by transferring it to the environment in a heat transfer process. Total entropy of system plus environment is unchanged as we assume the processes are quasi-static and so reversible. Therefore, everything is perfectly consistent also with the second law of thermodynamics.

In fact, there are examples of this phenomenon, spontaneous decrease of entropy, in nature. A perfect example is the formation of a snowflake. A snowflake has much lower entropy than a liquid drop of water, yet it forms spontaneously. At temperatures below the freezing point of water, the rate of energy reduction overcomes that of entropy, resulting a lower free energy state for a snowflake than the liquid water. Thus, nature favors the formation of a snowflake. Conversely at higher temperatures, entropy term overcomes again and frozen water turns into the liquid water \cite{seclaw}. The crucial point here is that nature does not follow the direction of increasing entropy but rather decreasing free energy. 

The reason why this is happening in our systems under quantum shape effects can be understood through the analytical analysis of entropy. Analytical expression of entropy is obtained by overlapped quantum boundary layer method as
\begin{equation}
\begin{split}
\frac{S_A}{Nk_B}= &\left(\ln\frac{\mathcal{V}}{N(4\delta)^3}+\frac{5}{2}\right)_{\mkern-7mu\mli{bl}}+\Bigg[\underbrace{\ln\left(1-\frac{\mathcal{V}_{\mli{qbl}}}{\mathcal{V}}\right)}_{S_{\mli{QSE}}^{\mli{I}}}\underbrace{-\frac{T}{\mathcal{V}_{\mli{eff}}^{\mli{0}}}\frac{\partial \mathcal{V}_{\mli{qbl}}}{\partial T}}_{S_{\mli{QSE}}^{\mli{II}}}\Bigg]_{\mkern-2mu\mli{QSE}} \\
&+\Bigg[\underbrace{\ln\left(1+\frac{\mathcal{V}_{\mli{ovr}}}{\mathcal{V}_{\mli{eff}}^{\mli{0}}}\right)}_{S_{\mli{QShE}}^{\mli{I}}}+\underbrace{\frac{\mathcal{V}_{\mli{eff}}^{\mli{0}} T}{\mathcal{V}_{\mli{eff}}}\frac{\partial}{\partial T}\frac{\mathcal{V}_{\mli{ovr}}}{\mathcal{V}_{\mli{eff}}^{\mli{0}}}}_{S_{\mli{QShE}}^{\mli{II}}}\Bigg]_{\mkern-2mu\mli{QShE}}
\end{split}
\end{equation}
In Fig. 4.5, comparison of the numerical and analytical results, which are in very good agreement, can be seen. The maximum and mean relative errors of the analytical entropy expression are respectively $1.1\%$ and $0.3\%$ for nested square, $1.3\%$ and $0.5\%$ for triangle and $1.8\%$ and $0.5\%$ for rectangle domains.

Unlike in free energy and internal energy, both quantum size and shape effect corrections consist of two distinct terms in entropy expression. Let's examine these terms individually to understand their contributions to entropy. This will allow us also to understand the behaviors of internal and free energies better, as they contain the same terms, compare Eqs. (4.6), (4.9) and (4.11). The first term in the analytical entropy expression in Eq. (4.11) is the classical term which is also known as Sackur–Tetrode equation. The second term (denoted by QSE subscript and within the square brackets) is the QSE correction to entropy and it has two distinct terms. The first term of QSE correction to entropy is exactly the same term that free energy has only with the opposite sign as expected. It has a negative contribution to entropy because $\mathcal{V}_{\mli{qbl}}/\mathcal{V}$ is always smaller than one. The other QSE term in entropy expression is exactly the same term that internal energy has as a QSE correction. This term has a positive contribution to the entropy because the derivative of $\mathcal{V}_{\mli{qbl}}$ with respect to temperature is negative. All in all, both terms contribute and the end result is the decrement of entropy due to QSE terms for the given temperature and sizes.

There are also two shape-dependent terms in entropy expression, which are competing with each other on the determination of the overall behavior. Again, the first shape-dependent term is the free energy correction with the opposite sign. Its contribution to entropy is always positive because of $\mathcal{V}_{\mli{eff}}^{\mli{0}}>\mathcal{V}_{\mli{ovr}}$ by definition. Hence, this term does not cause any unconventional behaviors in free energy and entropy. In other words, when effective volume increases, entropy also increases and free energy decreases. The second shape-dependent term, on the contrary, contribute to entropy negatively. Since increasing temperature reduces the thickness of QBLs and cause overlap volumes to diminish, $\mathcal{V}_{\mli{ovr}}$ is inversely proportional with temperature and so the contribution of $S_{\mli{QShE}}^{\mli{II}}$ term is negative. At this point it should be stressed that these terms are not entropy but size/shape corrections inside the genuine entropy expression which is always positive. The $S$ notation is for convenience.

The unusual behavior of entropy directly comes from this second QShE term. Of course in overall, $S_{\mli{QShE}}^{\mli{I}}$ and $S_{\mli{QShE}}^{\mli{II}}$ terms compete with each other (as one has positive, the other has negative sign). However, the course of competition is determined by the second QShE term. $S_{\mli{QShE}}^{\mli{II}}$ is always larger than $S_{\mli{QShE}}^{\mli{I}}$ within the spanned range of $\theta$, so its functional behavior has a larger effect on the sum of two. For the nested rectangle domain, however, $S_{\mli{QShE}}^{\mli{II}}$ decreases its decreasing trend (so the derivative) with $\theta$ after around $\theta\approx 60^{\circ}$. $S_{\mli{QShE}}^{\mli{II}}$ varies much less with shape after that point and the functional behavior of $S_{\mli{QShE}}^{\mli{I}}$ starts to dominate which is causing the entropy to increase rather than decrease with $\theta$, see Fig. 4.5c. But this functional change is solely due to the behavior of $S_{\mli{QShE}}^{\mli{II}}$ term that depends on the temperature sensitivity of overlap volumes. Let's recall how we define the overlap volumes. We shall give the analytical expressions of overlap volumes and effective volumes only for the nested square domain, because they are much simpler and easier to grasp. The effective volume of the nested square domain is found as $\mathcal{V}_{\mli{eff}}=[(L_o-2\delta)^2-(L_i+2\delta)^2](L_l-2\delta)+\mathcal{V}_{\mli{ovr}}=(L_{o,\mli{eff}}^2-L_{i,\mli{eff}}^2)L_{l,\mli{eff}}+\mathcal{V}_{\mli{ovr}}$. The overlap volume can be obtained as,
\begin{equation}
\mathcal{V}_{ovr}=L_{l,\mli{eff}}\frac{\tan\theta}{2}\left[L_{i,\mli{eff}}\left(1+\frac{1}{\tan\theta}\right)-\frac{L_{o,\mli{eff}}}{\sin\theta}\right]^2,
\end{equation}
for $\theta_{*}\leq\theta\leq45^{\circ}$, where $\theta_{*}$ is given by Eq. (3.3). For $0^{\circ}\leq\theta<\theta_{*}$, $\mathcal{V}_{ovr}=0$ and there is no overlap. After this recall, the temperature sensitivity of the overlap volume (see in Eq. (4.11)) can then easily be found by its temperature derivative,
\begin{equation}
\begin{split}
\frac{\partial \mathcal{V}_{\mli{ovr}}}{\partial T}=-\sqrt{2\tan\theta}\left(1+\cot\frac{\theta}{2}\right)L_{l,\mli{eff}}\frac{\delta}{T}\sqrt{\frac{\mathcal{V}_{\mli{ovr}}}{L_{l,\mli{eff}}}}+\frac{\delta}{T}\frac{\mathcal{V}_{\mli{ovr}}}{L_{l,\mli{eff}}}.
\end{split}
\end{equation}
The first term dominates as long as $L_{*}$ (the characteristic length of transverse sizes, defined in Section 3.6) is much smaller than the longitudinal size of the domain, which is the case here. The dominant first term in Eq. (4.13) is negatively proportional with the square root of the overlap volumes. That's why entropy decreases while overlap volume increases. To put it differently, increase in overlap volume affects entropy in two different competing ways: one is the increase in entropy due to increment of effective volume, the other is decrease in entropy due to negative temperature sensitivity of effective volume. The overall behavior depends on the result of the competition of these two terms. This is what happens in the nested rectangle case. Depending on the shape of the domain and the angular configuration, different QShE terms determine the behavior of entropy.

Above, we explained the peculiar behavior of entropy by examining the analytical expressions that we derived. The physical intuition for such a behavior can be conceived by a thermodynamic temperature analysis. Temperature is fundamentally associated with the distribution of the population of particles on the energy states of the system. When a system's temperature drops, the particles accumulate into the ground state and lower excited states, whereas if temperature rises, they accumulate into the higher excited states. In size variations (such as a volumetric expansion), energy levels mainly shift (see Fig. 3.5). On the other hand, when there is an effective shape expansion (the change of effective volume by the formation of overlap volumes), energy levels behave in a much more complicated way. When we rotate the system from $\theta=45^{\circ}$ to $\theta=0^{\circ}$ configuration, we cause ground state energy to rise and lower excited states become near ground state in $\theta=0^{\circ}$ configuration. Higher excited states do not change much during shape transformation. If we look at the picture from the distribution point of view, now there are more particles occupying the ground state and lower excited states than before, so the temperature drops. Therefore, distribution of occupied states changes in favor of a temperature drop from $\theta=45^{\circ}$ to $\theta=0^{\circ}$ configuration. As an analogy for the process, one can consider the population inversion in a laser in which case the most of the constituents in the system occupies higher excited states than the lower ones. In our case, this is not the case, but the physical reason of this behavior is again the change in population distribution over the states. With this analogy, we may call the process happening in our system as "population mixing".

At this point, we would like to explain a small misprediction of our proposed overlapped QBL approach. Depending on the type of nested confinement geometry, there might be some sharp changes (e.g. the cusp in Fig. 4.5c black dashed curve), in the analytical calculations. There are no sharp changes in actual behaviors of the functions (Fig. 4.5c red curve). This type of cusps may be apparent only for certain geometry types and there may be additional ones in some other more exotic ones. The misprediction of the analytical method is also very clear for the regions where the overlaps does not start ($\theta\leq\theta_{*}$). This is of course due to the simplistic nature of the approximation to define the QBL. In Appendix A.5, we will mention the possible improvements that can be done on the developed methodology, which can resolve these mispredictions, though the geometrical calculations become extremely complicated. Nevertheless, even the simplest QBL approach correctly captures the functional behaviors of the thermodynamic state functions.

\section{Specific Heat Capacity at Constant Weyl Parameters}

Examination of heat capacity is essential for a better understanding of the thermal properties of a particular system. Heat capacity is defined as the amount of heat that is required to change the temperature of a system by a certain amount. Under constant volume, it is described mathematically as the derivative of internal energy with respect to temperature. Specific heat capacity gives the heat capacity per particle, providing a quantity that is independent of how large the system is. 

An important distinction we would like to mention here is we actually calculate the specific heat capacity at constant Weyl parameters. Note that in quantum shape effect calculations, not only volume, but also surface area, peripheral lengths and number of vertices are constant. These geometric size parameters are called Weyl parameters and thus we calculate the heat capacity at constant Weyl parameters. 

In Fig. 4.6, we plot the specific heat changing with the rotation angle for our three different domain configurations. Specific heat does not monotonically increase in all nested domain configurations, rather it makes a small peak after its initial increase and then slightly decreases forming a plateau at the higher angles. Despite the slight decrease at the higher angles, the figures show that the specific heat grows when quantum shape effects increase. This means a confined system's capacity to store heat can be increased by help of quantum shape effects. Specific heat is also a convenient property for the experimental demonstrations of quantum shape effects.

\begin{figure}
\centering
\includegraphics[width=0.9\textwidth]{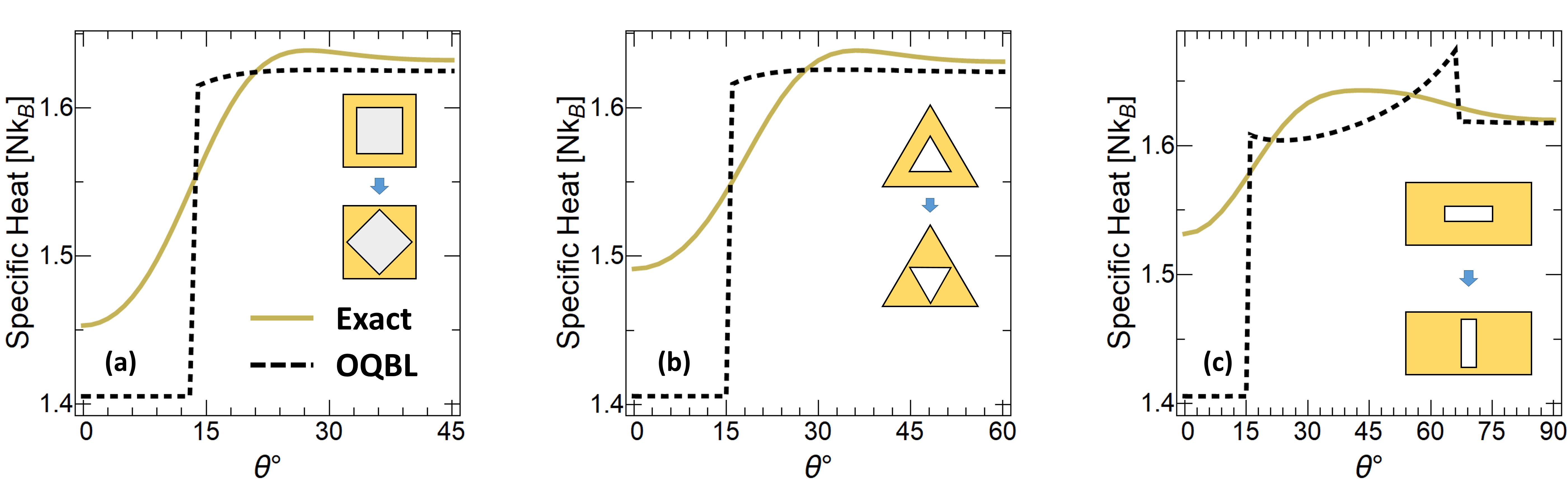}
\caption{Variation of specific heat capacity at constant Weyl parameters (normalized to $Nk_B$) with respect to the shape change via $\theta$, for (a) nested square, (b) nested triangle and (c) nested rectangle domains. Solid-gold and dashed-black curves represent the exact and analytical calculations respectively.}
\label{fig:c4f6}
\end{figure}

Dashed-black curves in Fig. 4.6 are the results of the analytical overlapped QBL approach. Our analytical model correctly predicts the saturation values of the specific heat at the higher angles. It deviates from the exact calculations for the lower angles, a behavior that is similar to the previously examined thermodynamic state functions. The maximum and mean relative errors of the analytical specific heat capacity result are respectively $9.9\%$ and $2.3\%$ for nested square, $9.3\%$ and $2.6\%$ for triangle and $12.0\%$ and $2.7\%$ for rectangle domains. Since internal energy is written in a piecewise form under the overlapped QBL approach, a discrete jump occurs at the critical angle in heat capacity, which is the main reason of larger errors. The second jump in Fig. 4.6c is again due to the temperature sensitivity of the overlap volume. Improvements on overlapped QBL approach may resolve these mispredictions, please refer to Appendix A.5.

%%%%%%%%%%%%%%%%%%%%%%%%%%%%%%%%

\section{Quantum Torque}

Let's recall the change in free energy in our systems. We argue that under zero friction, inner structure should rotate spontaneously to the equilibrium angle which minimizes the free energy, if the system is prepared at any other angular configuration in the first place. But in order for the inner structure to be able to rotate, there needs to be a torque exerted on the inner structure by the confined particles. Mathematically, negative derivative of free energy with respect to the rotation angle should give us the amount of torque that is exerted by particles to the inner structure. Consequently, the torque at a given configuration angle $\theta$ reads
\begin{equation}
\tau=-\frac{\partial F}{\partial\theta}\approx-\frac{F_{\theta+\Delta\theta}-F_{\theta}}{\Delta\theta}=-\frac{Nk_BT}{\Delta\theta}\ln\left(\frac{\zeta_\theta}{\zeta_{\theta+\Delta\theta}}\right),
\end{equation}
where $\theta$ and $\theta+\Delta\theta$ subscripts denote initial and perturbed rotational states respectively. For the numerical solution continuous derivative is not possible. To tackle with this, we perturb the system from its initial configuration angle a tiny amount and then solve the Schr\"{o}dinger equation again for the resulting perturbed system. $\Delta\theta$ denotes the amount of the rotational perturbation, which has to be very small for a good approximation of the derivative. Here it is chosen as $\Delta\theta=0.1$. We have tested smaller and larger $\Delta\theta$ values and up to $\Delta\theta=1$, the error due to the discrete differentiation is negligible.

The analytical expression of torque can be found by taking the derivative of Eq. (4.9) with respect to $\theta$, which quite simply gives $\tau_A=Nk_BT(\partial \mathcal{V}_{\mli{ovr}}/\partial\theta)/\mathcal{V}_{\mli{eff}}$. For nested square domain, it is derived in its open form as
\begin{equation}
\frac{\tau_A}{Nk_BT}=\frac{\sec^2\theta\frac{\mathcal{V}_{\mli{ovr}}}{L_{l,\mli{eff}}\tan\theta}+\tan\theta\csc^2\theta(L_{o,\mli{eff}}\cos\theta-L_{i,\mli{eff}})\sqrt{\frac{2\mathcal{V}_{\mli{ovr}}}{L_{l,\mli{eff}}\tan\theta}}}{\tan\theta\frac{\mathcal{V}_{\mli{ovr}}}{L_{l,\mli{eff}}\tan\theta}+L_{o,\mli{eff}}^2-L_{i,\mli{eff}}^2}.
\end{equation}

In Fig. 4.7, we plot the torque versus rotation angle graphs by comparing the numerical and analytical results for nested square, triangle and rectangle domains. The torque is normalized by considering maximum value of the exact solutions in each case. The maximum values predicted by Eq. (4.15) are not the same with exact ones, but the functional behaviors are captured quite well. The amount of maximum torque exerted by the particles in nested square domain is around 1 nNnm (nano-Newton nanometer). We explore the reasons why the maximum torque occurs at those particular $\theta$ values will be clear during the next section where we examine the non-uniformity of the pressure distributions along the walls of inner structure.

The torque equals to zero for $0^{\circ}$ and $45^{\circ}$ configurations. This means in order for inner structure to rotate from $0^{\circ}$ to $45^{\circ}$, one needs to generate an infinitesimally small perturbation from the unstable $0^{\circ}$ configuration to initiate the symmetry breaking. With this, one can also choose the direction of the rotation which is significant for engine/rotor systems.

\begin{figure}
\centering
\includegraphics[width=0.9\textwidth]{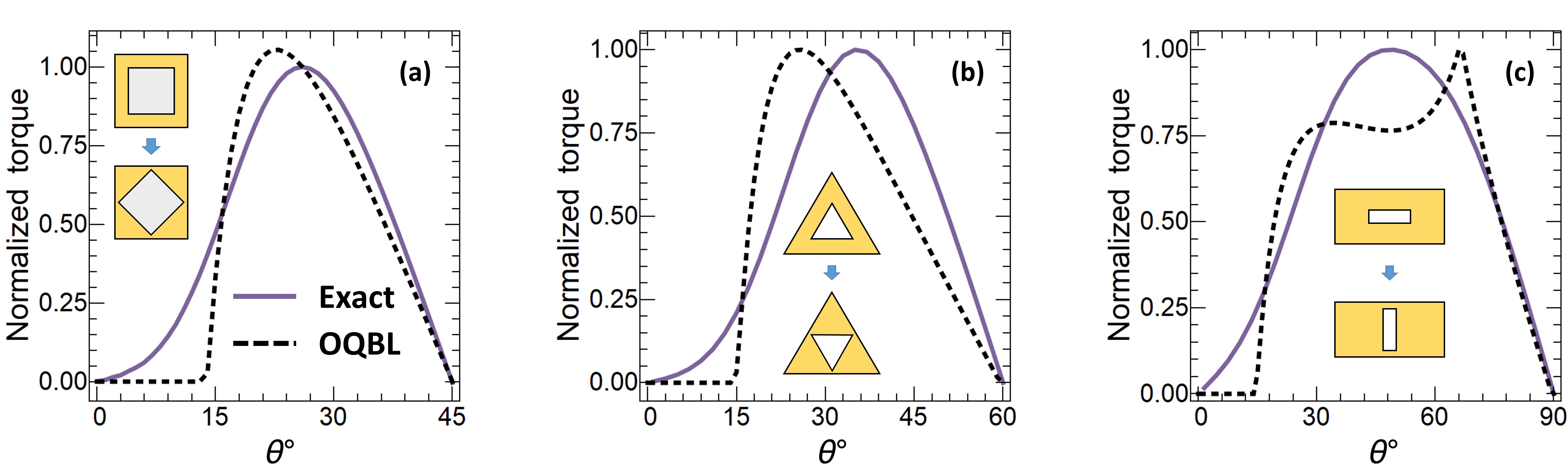}
\caption{Variation of torque (normalized to maximums of the exact ones in each case) with $\theta$, for (a) nested square, (b) nested triangle and (c) nested rectangle domains. Purple and dashed-black curves represent the exact (Eq. 4.14) and analytical (Eq. 4.15) calculations respectively.}
\label{fig:c4f7}
\end{figure}

In Fig. 4.8, we investigate the quantum torque generated by the system for various confinement strengths in nested square domain. Since this torque is a result of quantum shape effects, its value should strictly depend on the magnitude of the confinement. The confinement parameter of the transverse direction for all of our cases we examined so far was unity $\alpha_t=1.0$. We explore two additional confinements of $\alpha_t=0.9$ and $\alpha_t=1.2$ by scaling up and down the domain sizes. Variation of torque with the rotation angle for these various confinement cases is plotted in Fig. 4.8. As we expect, higher the confinement strength, larger the torque. Another interesting thing in this analysis is the maximum torque angle is not fixed for a particular domain geometry but it also changes with the amount of confinement. 

\begin{figure}
\centering
\includegraphics[width=0.9\textwidth]{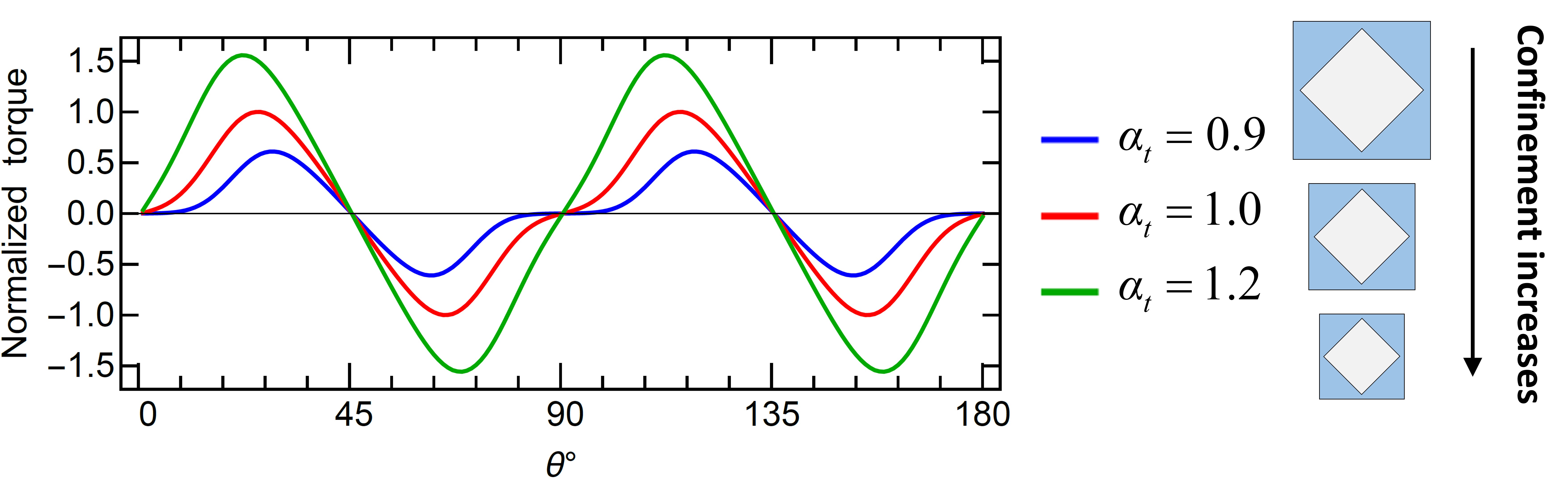}
\caption{Dependence of the torque on the magnitude of confinement. Torque varies with rotation angle for various confinement strengths for the nested square domain. The values are normalized to the maximum value of the torque for $\alpha_t=1$.}
\label{fig:c4f8}
\end{figure}

It was already clear from the variation of free energy with $\theta$ that the torque can be calculated. However, this doesn't explain how such a torque can appear. This torque is of quantum origin. There is no mechanical or any other classical kind of force take place here. Confined particles should exert pressure to the walls of the inner square structure at equilibrium. The only possible explanation for the existence of a finite amount of torque is that the pressure exerted to the walls of the inner square structure should be non-uniform so that there would be a moment imbalance which would generate a torque. Next step is to check how is the pressure profile on the walls of inner square domain and whether it is non-uniform or not.

\section{Pressure}

How do we find the pressure distribution along a surface? Classically, a gas pressure at equilibrium is uniform and the pressure is equal at any point on the boundaries. On the other hand, in our case the pressure is caused by the confined particles at nanoscale which are behaving according to the laws of quantum mechanics. We have already seen that the density distributions of these particles are non-uniform as opposed to the classical picture. We could not directly use the density distribution to obtain the pressure distribution without incorporating some additional tools like QBL method perhaps. Classical thermodynamic equation of state is not valid at nanoscale where the pressure becomes a tensorial quantity. Moreover, the equation of state becomes strongly depend on the sizes and the types of boundary deformation in strongly confined systems.

Thermodynamically, pressure is the free energy response of a system against a boundary deformation. Therefore, the most direct and fundamental way to investigate a pressure distribution is making small deformations (perturbations) on the boundaries along which the pressure distribution we want and looking to the system's free energy response to these deformations. To this end, we can create either local or global perturbations as it's shown in Fig. 4.9. First, we create tiny local perturbations on the walls of the inner square region. For convenience, we conduct the analysis on pure 2D system, as the third direction does not contribute to the creation of the non-uniform pressure profile. We already have the solutions for the unperturbed case. To probe the free energy response of the system, we need to separately solve the Schr\"{o}dinger equation again and again for each and every single perturbation we apply to the system. We have done a quite extensive analysis of the proper nature of the perturbations for the calculation of the accurate pressure distribution. The details of these analyses are given in Appendix sections A.1, A.2 and A.3. We repeat this procedure of creating local perturbations along the length of the wall. To generate its profile, we calculate the pressure exerted by particles onto a single perturbation by
\begin{equation}
P_L=\frac{1}{\Delta w}\left(-\frac{\Delta F_{0\rightarrow p}}{\Delta h}\right)=-\frac{F_p-F_0}{\Delta w\Delta h}
\end{equation}
where $P_L$ is the local pressure exerted on a single perturbation, $\Delta w$ is the width and $\Delta h$ is the height of the perturbation. $0$ and $p$ subscripts denote the unperturbed and perturbed versions of the free energy respectively. 

In Fig. 4.9, we plot the pressure profile along the inner domain wall of nested square case for $26^{\circ}$ (maximum torque angle). $\tilde{L}_i(r)$ is the normalized position along the inner square wall with respect to the middle point. As we expected, the pressure distribution along the boundary is indeed non-uniform. In fact, it resembles to the density distribution near to the boundary. (At this point, one might ask how can particles transfer momentum and exert pressure without being able to touch to the walls. It is clear that particle density is zero at the boundaries and very close to zero within $\delta$ (QBL thickness) away from the boundaries. Yet, particles can transfer momentum because they are actually not point particles but matter waves. They are remotely interacting with the boundaries through their wavefunctions. The same pressure has been examined in literature also under the name of matter-wave pressure. We use the notion of confined particles, it is a matter of choice to call them matter-waves or confined particles as they are equivalent by the de Broglie relation.) The non-uniform pressure distribution that is shown in Fig. 4.10 is consistent with both the density distributions and also the torque values that we calculated earlier. Moreover, this non-uniform pressure distribution explains the very existence of the torque. From the Fig. 4.10, we expect the inner square to rotate clockwise and this is what the torque values say in Fig. 4.7. 

\begin{figure}[t]
\centering
\includegraphics[width=0.6\textwidth]{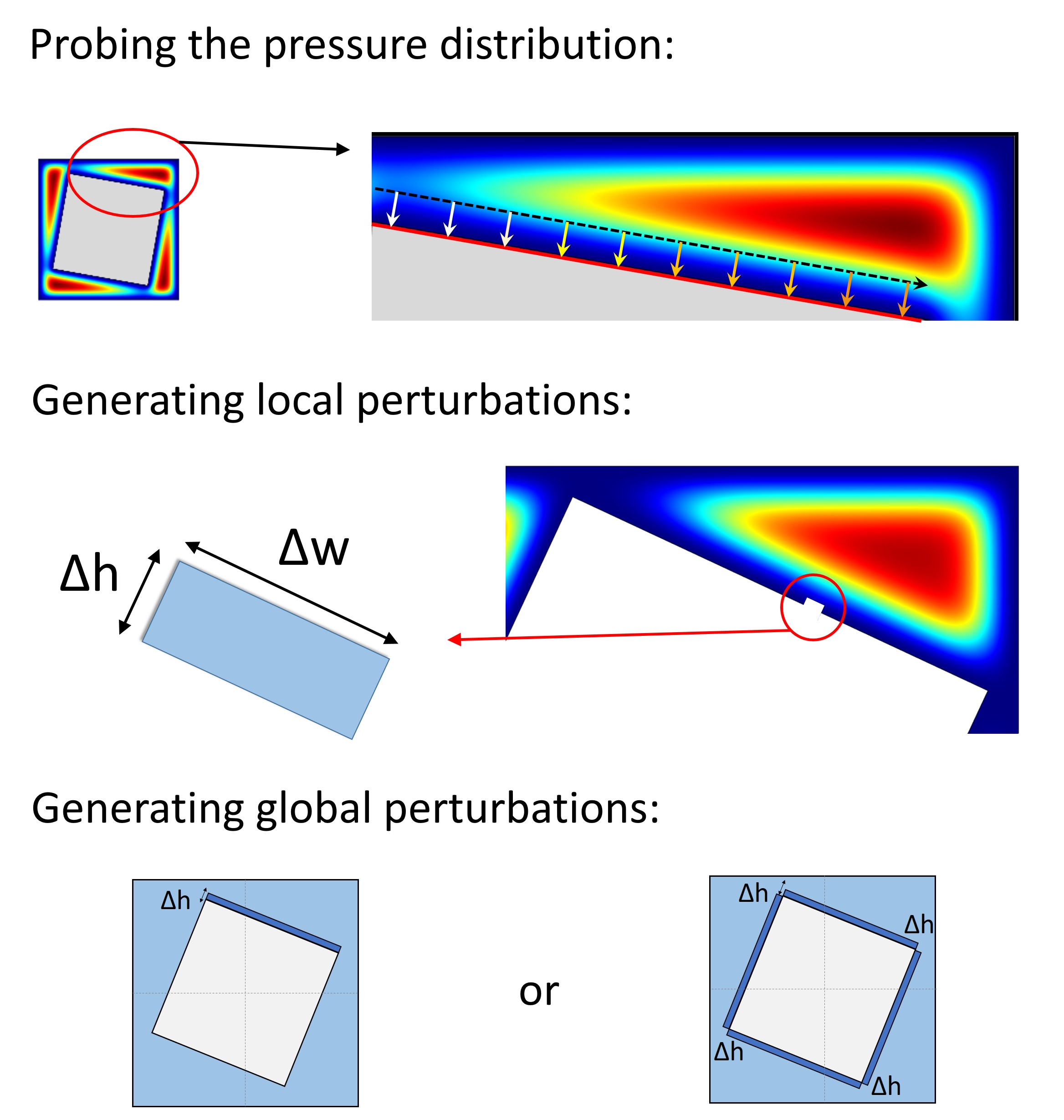}
\caption{Probing the pressure profile along the inner domain walls by generating tiny local perturbations and looking to the free energy response of the system. By a similar procedure one can create global perturbations as well to verify the integral of local pressure distribution.}
\label{fig:c4f9}
\end{figure}

\begin{figure}[!b]
\centering
\includegraphics[width=0.8\textwidth]{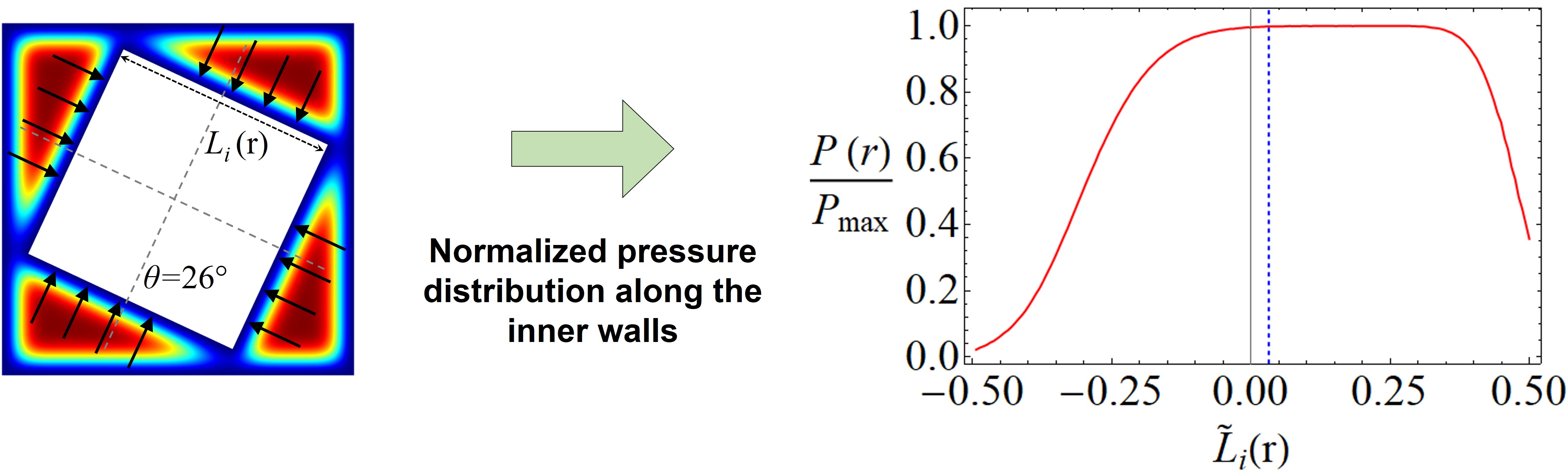}
\caption{Normalized pressure distribution along the inner boundary of nested square domain prepared at $26^{\circ}$ angular configuration. Blue dashed line shows where the point of application is.}
\label{fig:c4f10}
\end{figure}

\begin{figure}
\centering
\includegraphics[width=0.95\textwidth]{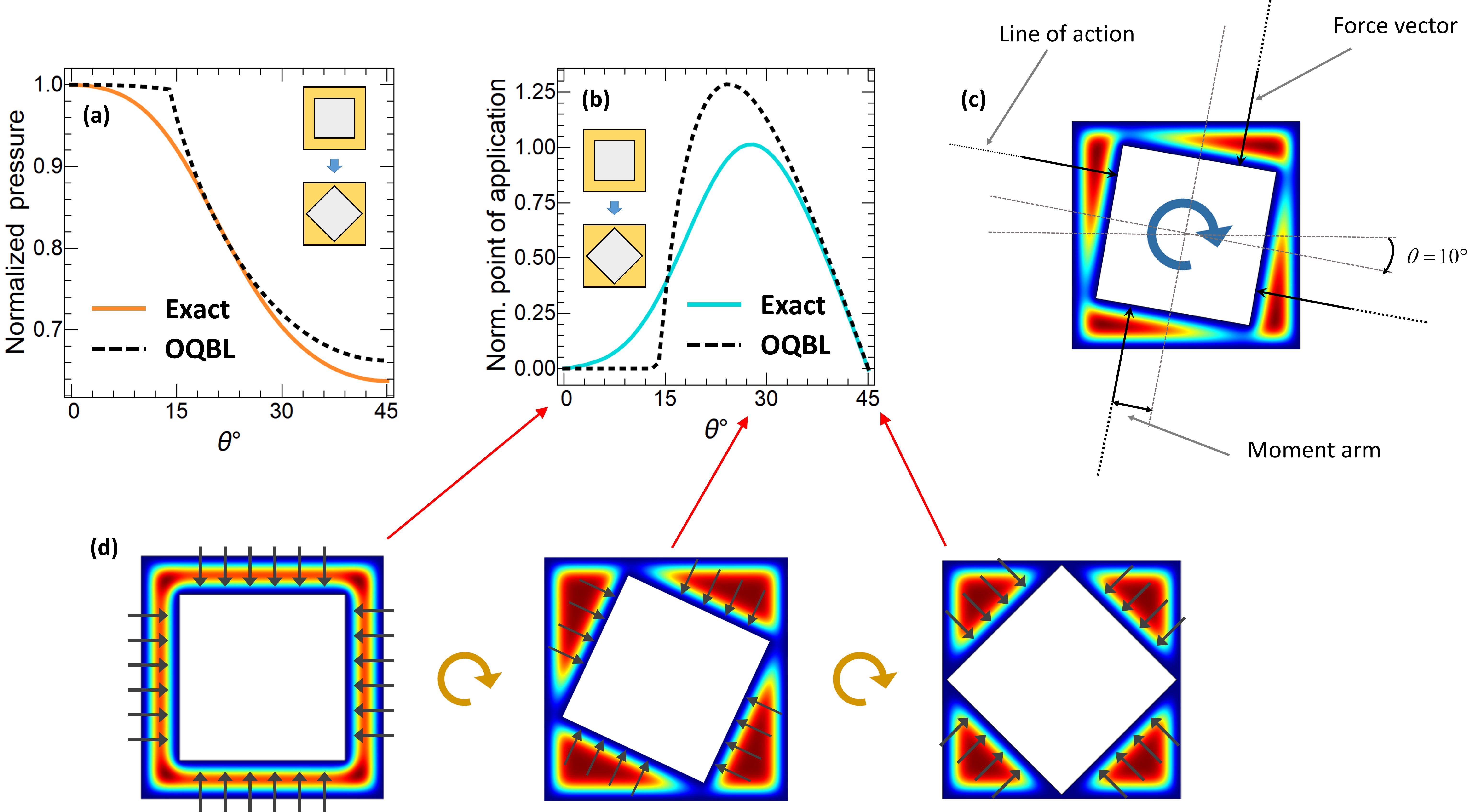}
\caption{(a) Normalized pressure, (b) normalized application points changing with $\theta$. Orange and cyan colors represent exact and black dashed curves represent the analytical results. (c) Analysis of the torque. (d) Pressure forces on the inner boundaries for various angular configurations.}
\label{fig:c4f11}
\end{figure}

In addition to this approach, we analytically showed the equivalence of pressure and momentum flux for all types of particle statistics, see the derivation in Appendix A.6. This shows the consistency of our methodology from two separate physical approaches; thermodynamics and transport theory. Therefore, the non-uniform pressure distribution is exactly equivalent to the momentum flux distribution across the boundary.

We can analyze further and look for the total amount of pressure exerted on the inner boundaries. To do that, we can either sum the local pressures generated by the local perturbations, or quite simply we can generate a perturbation with the width of the domain wall itself and look for the free energy response of the system. Inner square has 4 sides. We can either choose a side to apply this perturbation and multiply by 4 later, or we can even completely increase the outer length size at once, see the last row of Fig. 4.9. All the methods that we mention here give the same result under a negligible error margin (summing local perturbations give a larger error since the errors of individual calculations add up). Global pressure then can be calculated by
\begin{equation}
P_G=-\frac{F_p-F_0}{4L_i\Delta h},
\end{equation}
where here $F_p$ is the free energy under global perturbation, different than the one in Eq. (4.16) which was local. Since we are interested in total pressure exerted on the inner square, we divide the expression with $4L_i$. By using torque and pressure, we can also find the application points of the forces. Dividing Eq. (4.14) by Eq. (4.17) gives the point of application of the pressure. Variation of pressure and application points with respect to the rotation angle $\theta$ is given for nested square domain in Figs. 4.11a and 4.11b along with the comparison of numerical and analytical results. The analytical expression for pressure is simply given by
\begin{equation}
P_G^A=\frac{Nk_BT}{\mathcal{V}_{\mli{eff}}}.
\end{equation}
It should be noted that Eq. (4.18) contains the effective volume instead of the apparent one. Ratio of Eq. (4.15) to Eq. (4.18) gives the application points of forces analytically. Analytical pressure represents the true behavior extremely well and the accuracy of the analytical expression for application points is similar to the accuracy of the torque expression.

\begin{figure}
\centering
\includegraphics[width=0.85\textwidth]{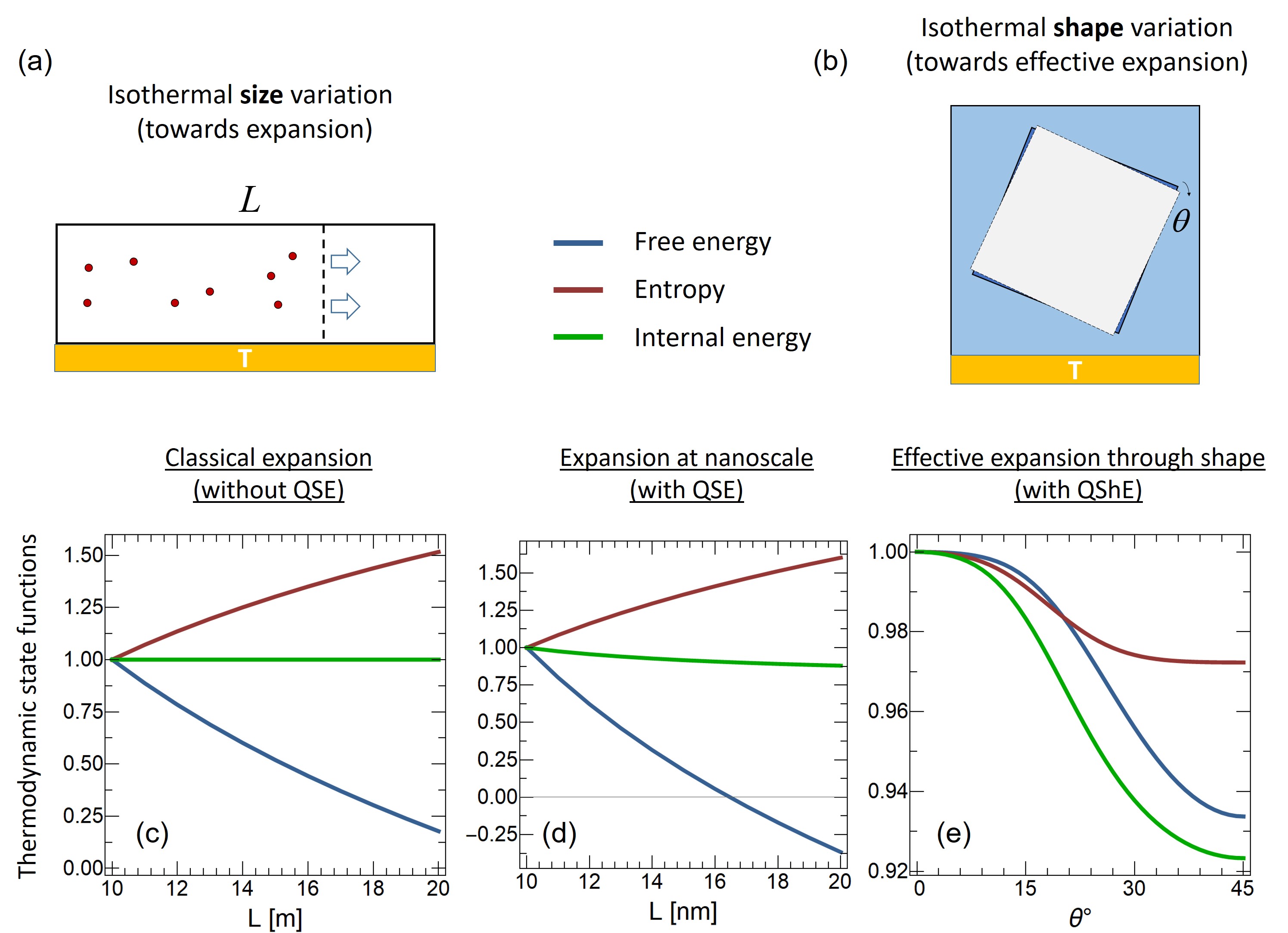}
\caption{(a) Isothermal expansion (size variation) (b) Isothermal effective expansion (shape variation at constant sizes) Changes in thermodynamic state functions with respect to (c) size (macroscopic), (d) size (nanoscale), (e) rotation angle. Blue, red and green curves represent free energy, entropy and internal energy respectively.}
\label{fig:c4f12}
\end{figure}

The point of application for nested square domain has already shown as an example in Fig. 4.10. Application points of forces are shifted from center due to non-uniform pressure and so create a moment on the body causing the rotation of the inner structure, Fig. 4.11c. As expected, the variation of application point with respect to rotation angle has the same character of the variation of the torque with respect to angle, Fig. 4.11b. The reason why the torque is zero for $0^{\circ}$ and $45^{\circ}$ degree configurations can be clearly inferred from Fig. 4.11d, where it is shown that pressures distributions are symmetric in those configurations, unlike in between angles.

It should also be noted that even though pressure forces decrease with increasing angle, the increase in moment arm length is much larger. Therefore, the main mechanism determining the torque is the change of moment arm length or the application points. The farther the moment arm, the higher the torque, vice versa.

Before passing on to the next section, we would like to mention another novelty of the quantum shape effect which is related with the spontaneous decrease of entropy and free energy. In the classical isothermal expansion of the gases, internal energy stays constant, free energy decreases since entropy of the system increases, Figs. 4.12a and 4.12c. We know that due to quantum size effects, internal energy of a system does not stay constant during an isothermal expansion, Fig. 4.12d. Conversely, in an isothermal shape expansion (which corresponds to an effective expansion due to decrement of overlap volumes), internal energy decreases so much that not only free energy, but also entropy decreases, Figs. 4.12b and 4.12e. Fig. 4.12 summarizes how quantum shape effects cause novel behaviors at nanoscale which haven't seen before in classical systems or confined systems with only size effects.

We complete our analysis here for the thermodynamic properties of Maxwell-Boltzmann gases confined in nested nanodomains. In the next section, we explore the thermodynamic properties of systems obeying Fermi-Dirac statistics. Particularly, we focus on the free electron gas. Most of the candidate systems for quantum shape effects such as semiconductors and metals can be modelled by free electron approximation and it is easier to focus on how QShE affects these systems as the model itself does not complicated with the details which are unnecessary for our investigation here.

\section{Quantum Shape Dependent Confined Electron Gas}

Statistical behaviors of particles in nature are captured by two very distinct types of statistics, namely Fermi-Dirac and Bose-Einstein. Charge carriers in materials behave according to the Fermi-Dirac statistics. Electrons in various semiconductor and metallic systems have been studied extensively due to their paramount importance on nanoscale electronics and nano-engineered devices in general. Based on the research in the literature, we think that our nested domain architectures could possibly be realized in such systems. Besides, thermodynamic and transport behaviors of charge carriers in semiconductors and metals carry valuable information about the material's mechanical, electronic and energetic properties. Due to these reasons it is crucial to investigate quantum shape effects on specific and more realistic systems such as electrons in low-dimensional semiconductors/metals. To accomplish that, we cannot rely on Maxwell-Boltzmann statistics anymore of course and we will use the Fermi-Dirac statistics. Even though Maxwell-Boltzmann statistics is still valid on those systems, we would like to extend our investigation and be not restricted by low density and/or high temperature conditions. As a concrete example, we consider the electrons in a Gallium Arsenide (GaAs) semiconductor nanowires in this section. 

Fermi-Dirac distribution function reads $f=1/[\exp(\tilde{\varepsilon}-\Lambda)+1]$ where $\Lambda=\mu/(k_BT)$ and $\mu$ is chemical potential. Spin degree of freedom is just a factor of two and ignored in our calculations and in the expressions for brevity. Note that we don't have the restriction of longitudinal size to be relatively long anymore. Nevertheless, we keep our domain as it is in Maxwell-Boltzmann case, to provide a better comparison with the previous cases. We analyze only the nested square domain in this section.

Unlike in Maxwell-Boltzmann case, we cannot separate the distribution function and partition function into the products of numerical and analytical parts unfortunately. This is because in Fermi-Dirac (also in Bose-Einstein) statistics, one cannot separate the chemical potential from the energy term on the exponential due the structure of their distribution functions. Either we have to solve the Schr\"{o}dinger equation for the whole 3D confined system, which is computationally very demanding, or we have to find a way around.

Choosing one direction relatively long again favors on the simplicity of this problem. Energy eigenvalues of this kind of system can be separated as $\tilde{\varepsilon}=\tilde{\varepsilon}_t+\tilde{\varepsilon}_l$. Although we cannot completely separate the analytical part from the numerical, since we know the eigenvalues of the longitudinal part analytically, we don't have to sum, instead we can integrate them. The longitudinal part is relatively long so that the first two terms of PSF work very well on its representation. Therefore, we use the first two terms of PSF to the longitudinal direction, without separating them from the other parts. The expression for the Fermi-Dirac partition function becomes
\begin{subequations}
\begin{align}
Z& =\sum_{\varepsilon}\ln\left[1+\exp(\Lambda-\tilde{\varepsilon})\right] \\
& =\sum_{\varepsilon_t}\int{d \tilde{\varepsilon}_l}\ln\left[1+\exp(\Lambda-\tilde{\varepsilon}_t-\tilde{\varepsilon}_l)\right] \\
& =\sum_{\varepsilon_t}{-\frac{L_l}{\lambda_{th}}Li_{\frac{3}{2}}\left(-e^{\Lambda-\tilde{\varepsilon}_t}\right)}+\frac{1}{2}Li_{1}\left(-e^{\Lambda-\tilde{\varepsilon}_t}\right),
\end{align}
\end{subequations}
where $\lambda_{th}=h/\sqrt{2\pi m_{\mli{eff}}k_BT}$ and $m_{\mli{eff}}=0.067m_e$ is the effective mass of conduction band electrons of GaAs, where $m_e$ being the bare electron mass. The summation is over the eigenvalues of the transverse part whereas the kernel terms are the results of the PSF. In this sense, expressions that are derived this way can be considered as semi-analytical.

Until this section, we always used the bare electron mass in our calculations. Having the specified effective mass in GaAs means thermal de Broglie wavelength of electrons in these structures are roughly four times larger than the case we considered before. That is to say, we can enlarge our domain sizes a bit more than the previous cases so that the manufacturing these kinds of materials in labs might be more realizable. With ease we can scale the transverse domain sizes in all directions by a scale parameter $s$. So, instead of $L_o$ ($L_i$), we have $L_os$ ($L_is$), $s$ being the scale parameter we would like to choose. For $s=1$, the previously examined domain sizes are recovered. We choose $s=3.02$ for the calculations in this section. This value gives outer and inner square lengths as 64nm and 41nm respectively. Since we apply this to both directions in transverse plane, we don't even have to calculate new eigenvalues for our new domains, we can scale the numerically solved transverse eigenvalues easily by $\tilde{\varepsilon}/s$. Naturally, we tested these scaled eigenvalues for selected numerical results and no errors have been found. Temperature is again 300K in all calculations.

\begin{figure}[!b]
\centering
\includegraphics[width=0.9\textwidth]{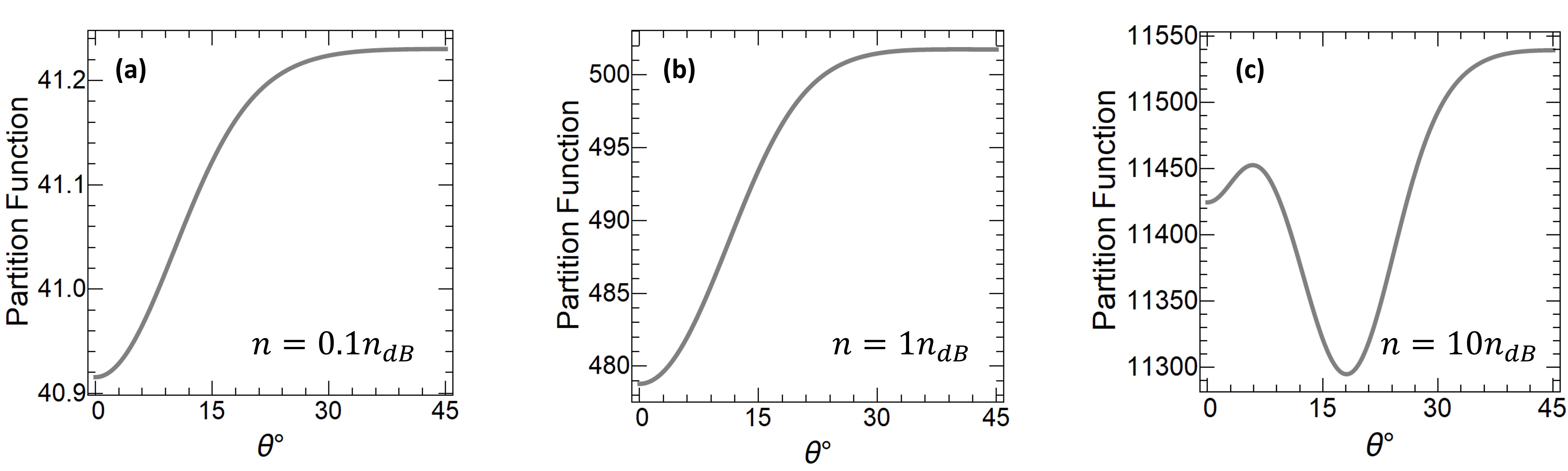}
\caption{Change in partition function with rotation angle for various electron densities (a) $0.1n_{dB}\approx 2.2\times 10^{22}$ (non-degenerate regime), (b) $n_{dB}\approx 2.2\times 10^{23}$ (weakly degenerate regime) and (c) $10n_{dB}\approx 2.2\times 10^{22}$ (moderately degenerate regime).}
\label{fig:c4f13}
\end{figure}

The number of particles ($N$) is defined by the summation of the distribution function over all eigenvalues, $N=\sum_{\varepsilon}f$. The semi-analytical expression for number of particles is
\begin{equation}
N=\sum_{\varepsilon_t}{-\frac{L_l}{\lambda_{th}}Li_{\frac{1}{2}}\left(-e^{\Lambda^{'}}\right)}+\frac{1}{2}Li_{0}\left(-e^{\Lambda^{'}}\right),
\end{equation}
where $\Lambda^{'}=\Lambda-\tilde{\varepsilon}_t$ is shortened for brevity. From the expression above (Eq. 4.20), chemical potential can be numerically calculated as an inverse problem. To ensure large number of particles inside our confinement domain as well as to comply with doped GaAs semiconductor conduction band electron concentrations, we choose three different electron densities to examine, ranging from $0.1n_{dB}$ to $10n_{dB}$ where $n_{dB}$ is the de Broglie density which is the reciprocal of the thermal de Broglie wavelength $\lambda_{th}$. Note that these $n_{dB}$ and $\lambda_{th}$ contain now the effective mass for GaAs electrons, instead of the bare electron mass which was used until this subsection.

We've examined partition function in Maxwell-Boltzmann case. For comparison, in Fig. 4.13 we plot the variation in partition function with $\theta$. The functional behavior is relatively similar to the Maxwell-Boltzmann case for low densities. In degenerate regime however, we see an oscillatory functional behavior, which is not expected at first sight. We explore more on this oscillatory behavior during the following subsections.

To compare the relative changes in the partition function with respect to shape between cases, we calculate the relative differences of the minimum and maximum values of the function in each case. From lowest to highest density cases, the relative differences are $0.8\%$, $4.6\%$ and $2.1\%$ respectively. The trend of the relative differences with density reflects the oscillatory dependence of the quantities in degenerate cases.

In order to prevent any confusion, it should be stressed here that the degeneracy concept used in this section of 4.9 is different than the degeneracy of energy levels which we've discussed in the previous chapter. Degeneracy in the energy level context means the energy levels corresponding to the same energy. On the other hand, degenerate regime in the context of electrons in solids (e.g. in semiconductors or metals) means the density of the electrons is very large so that the effect of Pauli exclusion principle and the exchange interaction is prominent. In statistical mechanics, the Pauli exclusion principle is taken into account by the Fermi-Dirac distribution function. 

\subsection{Chemical potential}

Chemical potential is defined as the rate of change in free energy with respect to the change in particle number. Under constant temperature and volume, it is the derivative of Helmholtz free energy with respect to particle number. The physical meaning of chemical potential is subject to various definitions. To grasp it as a thermodynamical concept, it may be better if we establish some analogies with the other concepts like temperature and pressure. The heat transfers from higher temperature to lower temperature to equalize the temperatures. By the same token, volume expands from higher pressure to lower pressure to equalize the pressures. Similarly, particles move from higher chemical potential to lower chemical potential to equalize the chemical potentials. Therefore, the chemical equilibrium is represented by the equality of chemical potentials. It has a major role in solid state physics, especially in the physics of semiconductors, among all other fields.

From Eq. (4.20), we can calculate the chemical potential $\mu$ as an inverse solution for a fixed number of particles. Since there are no analytical expressions for inverses of the polylogarithm functions, we solve this inverse problem numerically. As a result, the shape-dependence of the chemical potential for three different electron densities is given in Fig. 4.14. At low densities, Fig. 4.14a, chemical potential does not show any oscillation. The negativity of chemical potential suggests that electrons' Fermionic behavior is not present and they almost obey Maxwell-Boltzmann statistics. In Fig. 4.14b, the density is increased by a factor 10 and chemical potential becomes positive. This means the electrons are in weakly degenerate conditions which loosely corresponds to a transition regime from Maxwell-Boltzmann to Fermi-Dirac statistics. The peak that is shown in Fig. 4.14b is the first and the single peak and it monotonically decreases after the peak. When we increase the density further in Fig. 4.14c, chemical potential starts to oscillate with the shape variation. This is a quite interesting result, because normally chemical potential is not an intrinsically oscillatory quantity. What we mean by intrinsically is, some physical quantities exhibit quantum oscillations with respect to changes in sizes and the reason of these oscillations are due to the proximities and weights of the states near to ideal Fermi surface (or Fermi line in 2D, Fermi point in 1D nanostructures). The contributions of states near to Fermi surface are weighted by the occupancy variance function. Therefore, the physical quantities containing this occupancy variance function as a factor exhibit size- or density-dependent quantum oscillations. During my PhD, we have introduced a semi-analytical model called half-vicinity model to understand and predict the oscillatory behaviors of quantum confined systems \cite{aydin4,aydin5,aydin6}. Differently from size-dependent oscillations, this chemical potential oscillations shown in Fig. 4.14c is not coming from the behavior of occupancy variance, but they are a result of the influence of a complex shape variation on the system's thermodynamic behavior. In other words, when we decrease the size of the system while keeping the number of particles constant, chemical potential increases just monotonically with confinement and no oscillatory behavior is observed. On the other hand, in order to fix the number of particles to a certain value, while changing the domain shape smoothly with $\theta$, requires chemical potential to behave oscillatory. This is another novel type of behavior that is seen in the thermodynamics of confined nanostructures, due to quantum shape effects.

\begin{figure}
\centering
\includegraphics[width=0.9\textwidth]{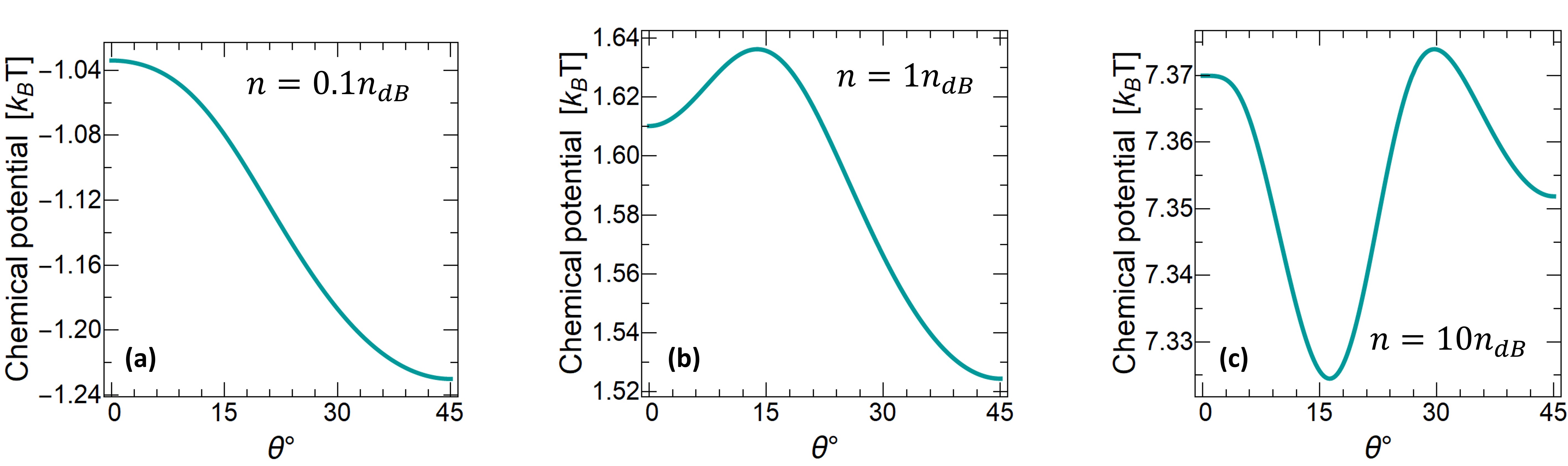}
\caption{Variation of dimensionless chemical potential ($\Lambda$) with rotation angle for electron densities (a) $0.1n_{dB}\approx 2.2\times 10^{22}$ (non-degenerate regime), (b) $n_{dB}\approx 2.2\times 10^{23}$ (weakly degenerate regime) and (c) $10n_{dB}\approx 2.2\times 10^{22}$ (moderately degenerate regime). Figure shows the change in behavior of chemical potential as the density goes from Maxwell-Boltzmann statistics conditions to Fermi-Dirac statistics ones.}
\label{fig:c4f14}
\end{figure}

Relative differences of min-max values in each case are found as $16.0\%$, $7.0\%$ and $0.7\%$ respectively for Fig. 4.14a, 4.14b and 4.14c. This suggests quantum shape effects have much more dominant behavior for low density conditions where Maxwell-Boltzmann statistics is obeyed. When the degeneracy increases, the influence of shape reduces. This is understandable, because the degeneracy is inversely proportional to the Fermi wavelengths of particles. Like thermal de Broglie wavelength, there is another concept called Fermi wavelength which includes the influence of degeneracy in terms of chemical potential into the de Broglie wavelength. Although de Broglie wavelength is still an important quantity for Fermionic particles, Fermi wavelength provides more detailed and more accurate information about the quantum statistical nature of particles. Fermi wavelength can be given in terms of the de thermal de Broglie wavelength as $\lambda_{F}=\lambda_{th}\sqrt{\pi/\Lambda}$ for 1D Fermi systems. The larger the chemical potential, the smaller the Fermi wavelength and lesser the quantum wave behaviors of particles. In fact, it is possible to define another confinement parameter for strongly degenerate ($\Lambda>>1$) Fermionic systems, $\alpha_F=\alpha/\sqrt{\Lambda_F}$ for 1D systems for example, where $\Lambda_F$ is the dimensionless chemical potential at 0K\cite{aydin6}. However, we would like to continue our discussion of the conditions of MB and FD statistics based on the comparison of the de Broglie densities. $n<<n_{dB}$ corresponds to MB, $n\approx n_{dB}$ and $n>>n_{dB}$ corresponds to FD statistics. These conditions correspond to a comparison between the mean distance between particles and their de Broglie wavelengths. This kind of comparison gives a measure of the degeneracy in solid state systems. Our analysis shows that low density conditions are favorable for the maximization of quantum shape effects. When Fermi level rises (i.e. degeneracy increases), average de Broglie wavelength of particles decreases and quantum shape effect decreases. In degenerate conditions, however, appearance of a new type of oscillatory behavior is also worthy of note. 

\subsection{Internal energy, free energy, entropy and specific heat of electrons}

We've seen that examination of quantum shape effects in Fermi-Dirac statistics has revealed new behaviors. We wonder how thermodynamic state functions of electrons behave under quantum shape effects. Semi-analytical expressions of internal energy, free energy and entropy are derived as follows respectively

\begin{equation}
\begin{split}
U=k_BT\sum_{\varepsilon_t}-\frac{L_l}{2\lambda_{th}}Li_{\frac{3}{2}}\left(-e^{\Lambda^{'}}\right)-\tilde{\varepsilon}_t\frac{L_l}{\lambda_{th}}Li_{\frac{1}{2}}\left(-e^{\Lambda^{'}}\right)+\tilde{\varepsilon}_t\frac{1}{2}Li_{0}\left(-e^{\Lambda^{'}}\right),
\end{split}
\end{equation}

\begin{equation}
F=k_BT(N\Lambda-Z),
\end{equation}

\begin{equation}
\begin{split}
S=k_B\sum_{\varepsilon_t}-\frac{3L_l}{2\lambda_{th}}Li_{\frac{3}{2}}\left(-e^{\Lambda^{'}}\right)+\frac{1}{2}Li_1\left(-e^{\Lambda^{'}}\right)+\Lambda^{'}\frac{L_l}{\lambda_{th}}Li_{\frac{1}{2}}\left(-e^{\Lambda^{'}}\right)-\Lambda^{'}\frac{1}{2}Li_0\left(-e^{\Lambda^{'}}\right).
\end{split}
\end{equation}

In Figs. 4.15, 4.16 and 4.17, variations of internal energy, free energy and entropy with the shape parameter $\theta$ is presented respectively. The analysis shows that for non-degenerate or very weakly degenerate conditions (subfigures (a) and (b)), the functional behaviors of thermodynamic state functions are similar to the ones in Maxwell-Boltzmann statistics, as expected. On the other hand, all thermodynamic state functions exhibit oscillatory behaviors in the degenerate regime. Internal energy and free energy are the quantities that do not contain occupancy variance function as a factor in their expressions. Hence, these oscillations are not the already known density or size oscillations. Origin of these oscillations is the chemical potential oscillations due to quantum shape effects. The fact that chemical potential oscillates with shape for fixed number of particles, causes oscillations in all other quantities that depend on chemical potential.

\begin{figure}
\centering
\includegraphics[width=0.9\textwidth]{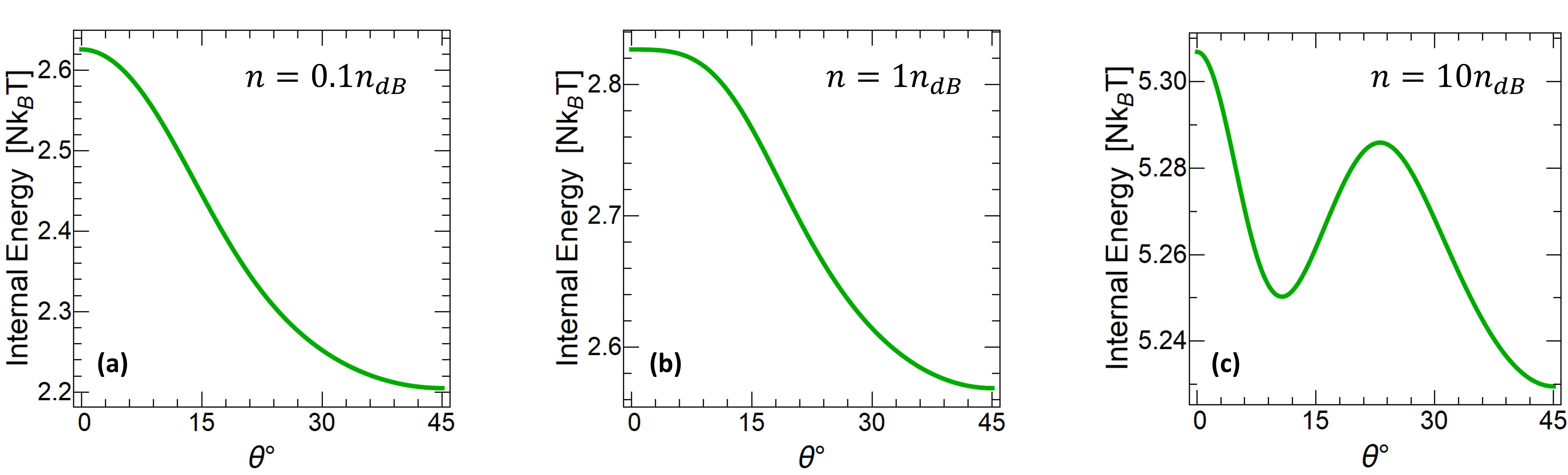}
\caption{Change in internal energy with shape for densities (a) $n=0.1n_{dB}$, (b) $n=n_{dB}$ and (c) $n=10n_{dB}$.}
\label{fig:c4f15}
\end{figure}

It is seen in Figs. 4.15c and 4.16c that the functional behaviors of internal energy and free energy are very similar to each other. On the other hand, oscillatory behavior of entropy, Fig. 4.16c, is different than those of internal energy and free energy. Combined effect of chemical potential oscillations and shape-dependent occupancy variance oscillations might be the result of this different oscillatory behavior.

\begin{figure}
\centering
\includegraphics[width=0.9\textwidth]{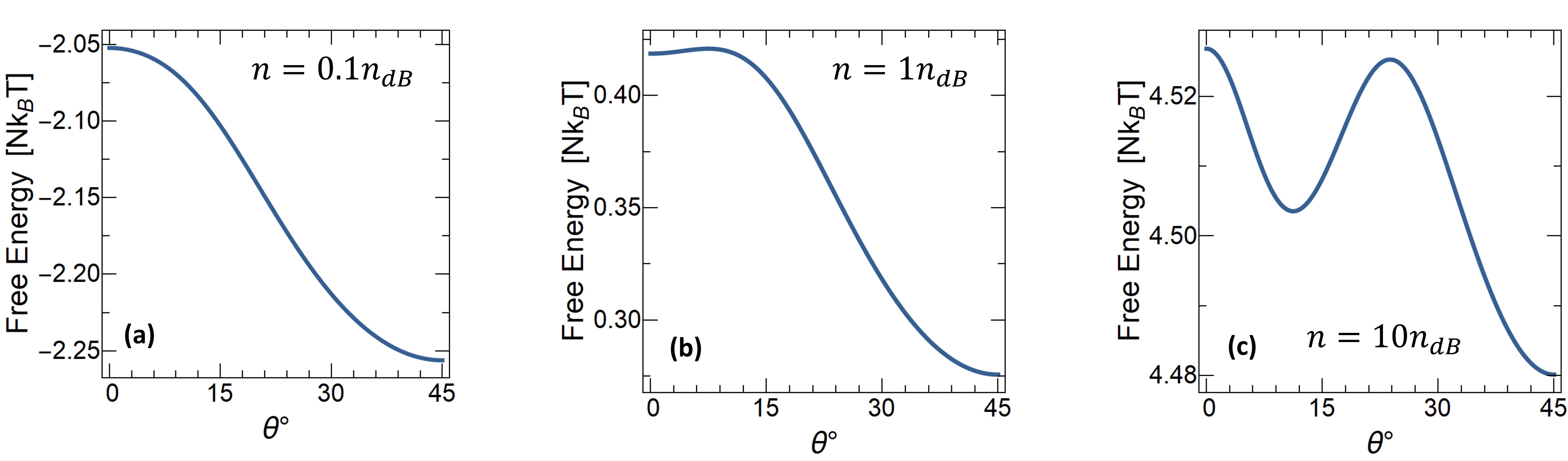}
\caption{Change in free energy with shape for densities (a) $n=0.1n_{dB}$, (b) $n=n_{dB}$ and (c) $n=10n_{dB}$.}
\label{fig:c4f16}
\end{figure}

\begin{figure}
\centering
\includegraphics[width=0.9\textwidth]{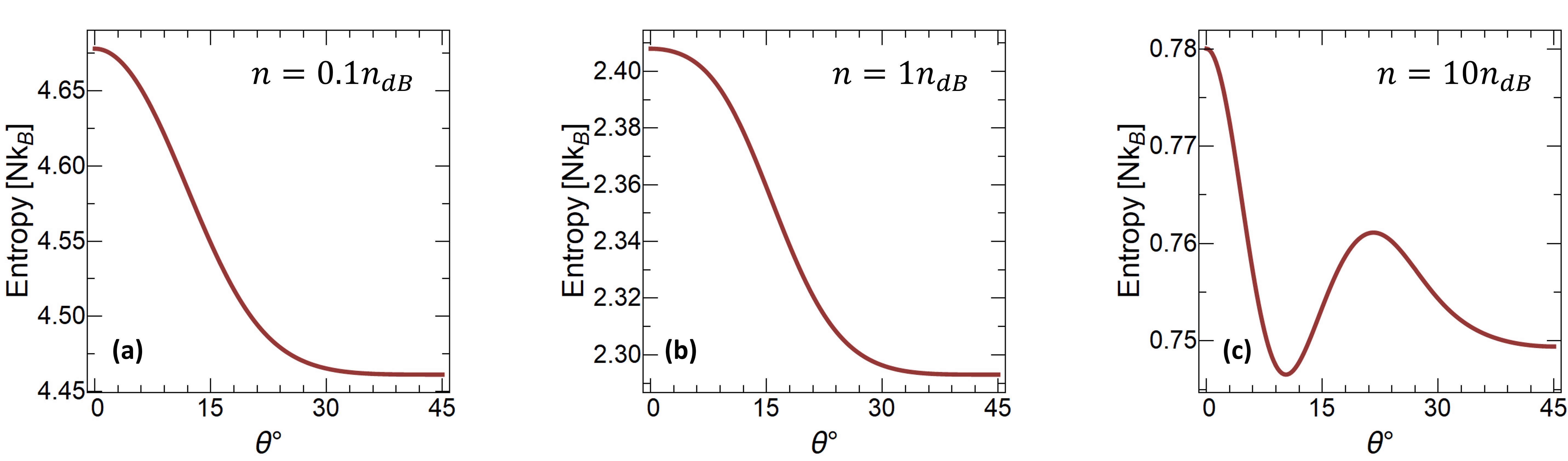}
\caption{Change in entropy with shape for densities (a) $n=0.1n_{dB}$, (b) $n=n_{dB}$ and (c) $n=10n_{dB}$.}
\label{fig:c4f17}
\end{figure}

Differences of extremum points of internal energy function changing with shape are given for $n=0.1n_{dB}$, $n=n_{dB}$ and $n=10n_{dB}$ respectively as $16.0\%$, $9.0\%$ and $1.5\%$. Similarly, differences of extremum values of free energy are $9\%$, $34\%$ and $1\%$. Finally, the differences in extremum values of entropy are $4.6\%$, $4.8\%$ and $4.3\%$. Influence of quantum shape effects are larger to internal energy for low density conditions consistently, whereas changes in entropy due to quantum shape effects are more or less the same in all cases. Interestingly, variation in free energy due to quantum shape effects has somewhat unusual behavior. There is a huge influence of quantum shape effects for weakly degenerate conditions. The reason of this is because free energy changes sign when chemical potential passes from zero. That's why, it has a higher sensitivity to the changes in shape around those values of chemical potential. Almost negligible response of entropy to the changes in electron density might be the result of the fact that the configurational entropy doesn't change much during density variations. The sizes and the shape of the domain that is considered are the same in all cases, which doesn't change the configurational entropy much.

\begin{figure}
\centering
\includegraphics[width=0.9\textwidth]{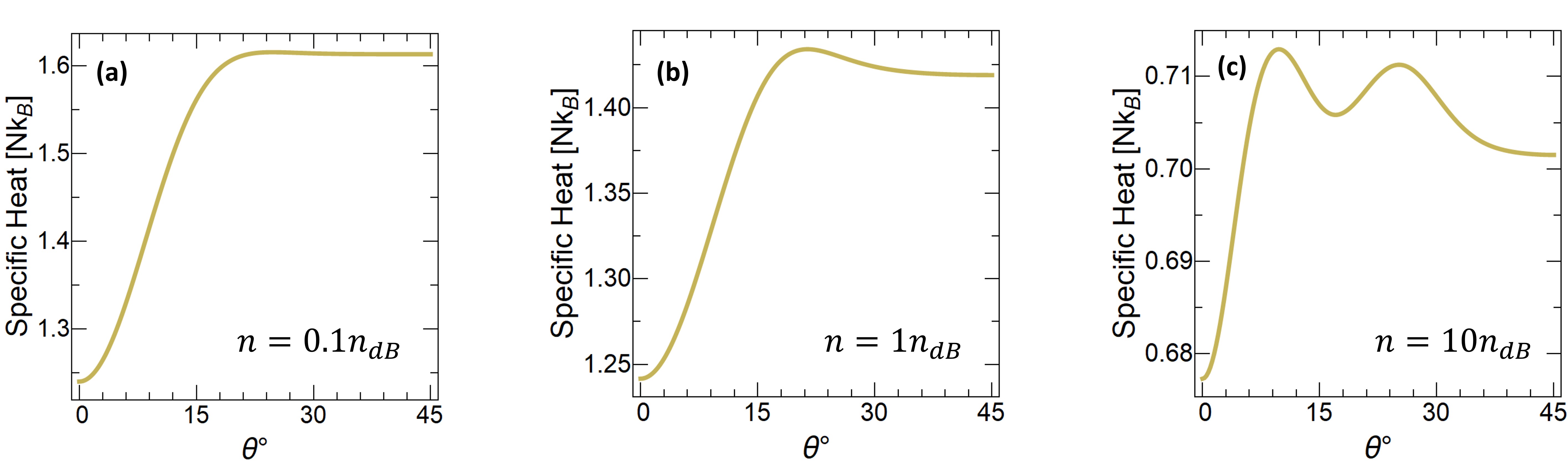}
\caption{Change in specific heat capacity at constant Weyl parameters with shape for densities (a) $n=0.1n_{dB}$, (b) $n=n_{dB}$ and (c) $n=10n_{dB}$.}
\label{fig:c4f18}
\end{figure}

Another important property to examine is the electronic heat capacity at constant Weyl parameters (includes the constant volume condition). A semi-analytical expression for electronic heat capacity at constant Weyl parameters is obtained as follows,
\begin{equation}
\begin{split}
C_W=k_B&\sum_{\varepsilon_t}{-\frac{3L_l}{4\lambda_{th}}Li_{\frac{3}{2}}\left(-e^{\Lambda^{'}}\right)}-2\tilde{\varepsilon}_t\frac{L_l}{\lambda_{th}}Li_{\frac{1}{2}}\left(-e^{\Lambda^{'}}\right) \\
&-\tilde{\varepsilon}_t^2\frac{L_l}{\lambda_{th}}Li_{-\frac{1}{2}}\left(-e^{\Lambda^{'}}\right)+\frac{1}{2}Li_{-1}\left(-e^{\Lambda^{'}}\right) \\
&-\frac{\left[\sum_{\varepsilon_t}{-\frac{L_l}{2\lambda_{th}}Li_{\frac{1}{2}}\left(-e^{\Lambda^{'}}\right)}-\tilde{\varepsilon}_t\frac{L_l}{\lambda_{th}}Li_{-\frac{1}{2}}\left(-e^{\Lambda^{'}}\right)+\frac{1}{2}Li_{-1}\left(-e^{\Lambda^{'}}\right) \right]^2}{\sum_{\varepsilon_t}{-\frac{L_l}{\lambda_{th}}Li_{-\frac{1}{2}}\left(-e^{\Lambda^{'}}\right)}+\frac{1}{2}Li_{-1}\left(-e^{\Lambda^{'}}\right)}.
\end{split}
\end{equation}

In Fig. 4.18, specific heat capacity varies with shape for various densities. The increase in heat capacity in low degrees of angle is preserved in all cases, whereas for high degrees of angle oscillations start to appear in degenerate case.

Differences between the extremum values of specific heat function are given respectively as $23\%$, $13\%$ and $5\%$ for $n=0.1n_{dB}$, $n=n_{dB}$ and $n=10n_{dB}$.

%%%
%5%
%%%

%\chapter{Applications For Nano Energy Science and Technology}
\chapter{Applications For Nano Energy Science and Technology}

Up to now, we've introduced a novel effect, which we call the quantum shape effect, appearing in the thermodynamics of nanoscale systems that are confined in a particular way. During Chapter 3 we explore the fundamentals of the quantum shape effects and constructed its theory along with an analytical approach. During the fourth chapter we investigate quantum shape effects in the thermodynamics of nanoscale confined systems. In this chapter, we examine the possible applications of quantum shape effects, specifically focusing on their energy applications.

Nano energy science and technology is developing as a research field on its own and it deals with understanding the workings of energy devices and thermodynamic machines at nanoscale. Exploration of quantum heat engines and nanoscale energy conversion devices constitutes another part of the thesis. Both theoretically and from the application point of view, investigation of the role of quantum shape effects in such systems is crucial for the development of new nano energy technologies. In this chapter we focus on how the newly introduced quantum shape effects can play a role on the nano energy applications.

\section{Isoformal Process and Novel Thermodynamic Cycles}

With the addition of shape as a control parameter in thermodynamics, it is possible to define new thermodynamic processes. Recall Fig. 4.1 where new parameters open up new dimensions in thermodynamic state space. Each variable gives rise to distinct thermodynamic processes. To give an example, keeping volume constant in a thermodynamic cycle, leads to isochoric process. Similarly, the constancy of pressure, temperature or entropy as well as zero heat transfer give rise to isobaric, isothermal, isentropic and adiabatic processes respectively. In a similar manner, we shall call the isoformal process when the shape of the working substance is constant in a thermodynamic process. In this section, we construct quantum heat engines driven by quantum shape effects. The working substance of our engine is the particles confined in our nested square domain shape. In our analyses, all processes are assumed to be infinitely slow (quasistatic), hence reversible, and no coherence exists among energy levels. For convenience, we execute our analyses in this chapter for the conditions where Maxwell-Boltzmann and Fermi-Dirac statistics give the same results, hence the density, temperature and the sizes are $n=5\times 10^{24} \text{m}^{-3}$, $T=300$K and $s=1$ (the scale parameter defined in Sec. 4.9). 

\subsection{Stirling-like heat engine}

The first cycle we would like to introduce is a Stirling-like heat engine cycle. In the usual Stirling cycle, there are two isothermal (constant temperature) and two isochoric (constant volume) processes. In our engine, rather than two isochoric processes, we have two isoformal processes, where all geometric size variables and most importantly the shape of the confined system are kept constant.

The cycle consists of four consecutive steps as it's seen in Fig. 5.1: (\textbf{1}) \textit{Isoformal heat addition} $\textbf{1}\rightarrow \textbf{2}$: The temperature of the system increases from $200$K to $300$K. (\textbf{2}) \textit{Isothermal shape-confinement (isothermal effective compression)} $\textbf{2}\rightarrow \textbf{3}$: Inner structure is rotated from $45^{\circ}$ to $0^{\circ}$ position by performing work. During this process heat is also given to the system to keep the temperature constant at $300$K. This is contrary to the classical Stirling cycle, in which the corresponding process is the isothermal compression of the gas that causes heat to be released to the environment to keep the temperature constant. Contrarily, here system absorbs heat from the environment, even though it is an isothermal effective compression process. (\textbf{3}) \textit{Isoformal heat rejection} $\textbf{3}\rightarrow \textbf{4}$: The temperature of the system decreases from $300$K back to $200$K, while keeping the inner structure at the $0^{\circ}$ position. (\textbf{4}) \textit{Isothermal shape-deconfinement (isothermal effective decompression)} $\textbf{4}\rightarrow \textbf{1}$: Inner structure spontaneously rotates from $0^{\circ}$ to $45^{\circ}$ degree position by doing work on its surroundings and rejects heat by keeping the temperature constant. This is again a quite uncommon process since both work generation and heat rejection take place during the same isothermal process.

\begin{figure}
\centering
\includegraphics[width=0.65\textwidth]{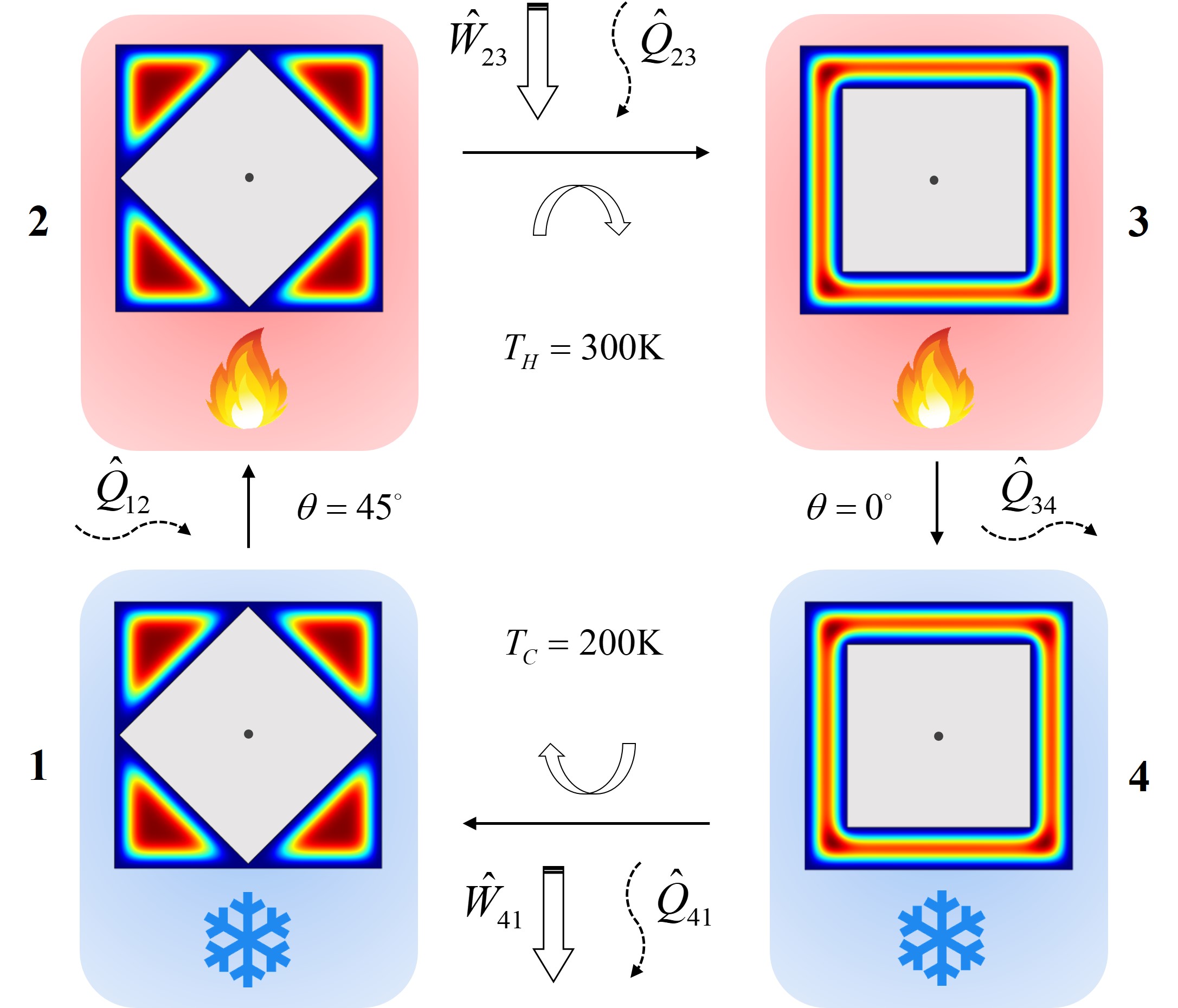}
\caption{Stirling-like thermodynamic cycle based on quantum shape effects. The cycle contains four processes: (\textbf{1}) isoformal (shape preserving) heat addition, (\textbf{2}) isothermal shape-confinement, (\textbf{3}) isoformal heat rejection, (\textbf{4}) isothermal shape-deconfinement. Work generation occurs at cold side, during the isothermal shape-deconfinement process.}
\label{fig:c5f1}
\end{figure}

In Fig. 5.2, $T$-$S$ and $\tau$-$\theta$ diagrams of the cycle are shown. Unlike in classical thermodynamic cycles, the work exchange during high temperature shape transformation (from steps \textbf{2} to \textbf{3}) is less than that of low temperature one (from steps \textbf{4} to \textbf{1}), Fig. 5.2b. Additionally, the directions of work and heat exchanges under isothermal shape transformation processes are the same, unlike the ones in isothermal volume variation, Fig. 5.2a and 5.2b. In other words, the directions of work are exactly opposite of the Stirling cycle. In the classical Stirling cycle, the isothermal expansion is carried out by means of volume increase. Work output occurs due to expansion, at the higher temperature (higher pressure) side whereas work input is given during the contraction, at the lower temperature (lower pressure) side. Since the work output is larger than the work input, the net work is positive. On the other hand, in this modified version of Stirling cycle, there is no expansion but an increment of effective confinement (or decrement of effective volume) as a result of quantum shape effects. That's why, instead of work extraction, work has to be done on the system during $\textbf{2}\rightarrow \textbf{3}$. Since quantum shape effects are larger at lower temperature, work extraction occurs on the lower temperature side of the cycle.

It should also be noted that, one does not necessarily should mechanically rotate the inner structure in a closed system to be able to realize $\textbf{2}\rightarrow \textbf{3}$ and $\textbf{4}\rightarrow \textbf{1}$ processes. Instead, a particle flow can also be considered in a nano channel made by nested structures where the inner one is twisted from $45^{\circ}$ to $0^{\circ}$ and from $0^{\circ}$ to $45^{\circ}$ along the channel. So, the particles would be exposed to varying confinement shapes during their flow through the channel. 

\begin{figure}
\centering
\includegraphics[width=0.7\textwidth]{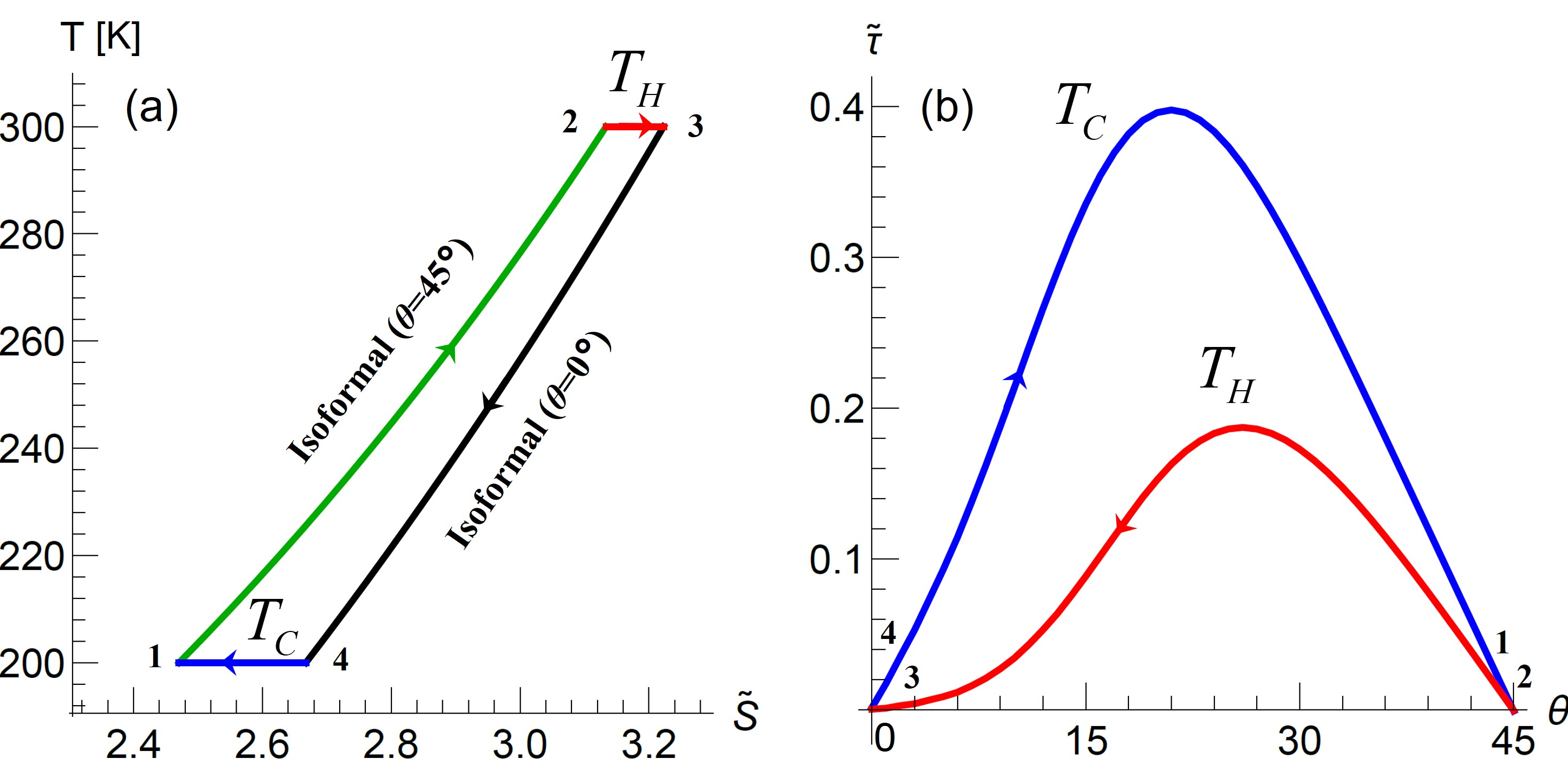}
\caption{Temperature-entropy ($T$-$S$) and torque-shape ($\tau$-$\theta$) diagrams. (a) $T$-$\tilde{S}$ diagram, where $\tilde{S}=S/(Nk_B)$. (b) $\tilde{\tau}$-$\theta$ diagram, where $\tilde{\tau}=\tau/(Nk_BT)$. From state \textbf{1} to \textbf{2} and \textbf{3} to \textbf{4}, no change occurs in $\tau$-$\theta$ diagram since $\theta$ is kept constant during those changes. In addition, the torque is zero at $\theta=0^{\circ}$ and $\theta=45^{\circ}$ configurations, as we've seen before. When the amount of overlap volumes increases with decreasing temperature, higher amount of torque occurs and this happens at lower temperatures (low temperature side of the cycle in Fig. 5.1) on the contrary to classical expectations.}
\label{fig:c5f2}
\end{figure}

Note that heat and work exchanges are possible without changing the volume or other size parameters. All of them are kept as constant during the cycle, so the only work exchange mechanism is the shape transformation through the rotation process. Hence, by using the first and second laws of thermodynamics, we can write $dU=\delta Q+\delta W=TdS-\tau d\theta$ for the derivations of heat and work exchanges during the processes. For this Stirling-like cycle, heat and work exchanges at each process are then determined as follows
\begin{equation}
\begin{split}
Q_{12}=& U\left(T_H,45^{\circ}\right)-U\left(T_C,45^{\circ}\right) \\
Q_{23}=& T_H\left[S\left(T_H,0^{\circ}\right)-S\left(T_H,45^{\circ}\right)\right] \\
W_{23}=& -\int_{45}^{0}{\tau\left(T_H,\theta\right)d\theta}=U\left(T_H,0^{\circ}\right)-U\left(T_H,45^{\circ}\right)-Q_{23} \\
Q_{34}=& U\left(T_C,0^{\circ}\right)-U\left(T_H,0^{\circ}\right) \\
Q_{41}=& T_C\left[S\left(T_C,45^{\circ}\right)-S\left(T_C,0^{\circ}\right)\right] \\
W_{41}=& -\int_0^{45}{\tau\left(T_C,\theta\right)d\theta}=U\left(T_C,45^{\circ}\right)-U\left(T_C,0^{\circ}\right)-Q_{41}.
\end{split}
\end{equation}

Using Eq. (5.1), heat input and net work output can be determined respectively as $Q_{in}=Q_{23}+(Q_{12}-Q_{34})$ and $W_{\mli{net}}=W_{23}+W_{41}$. We obtain the cycle efficiency ($W_{\mli{net}}/Q_{in}$) as $25\%$ for $T_H=300$K and $T_C=200$K. A refrigeration cycle can readily be obtained just by reversing the power cycle presented in Fig. 5.1.

\subsection{Otto-like heat engine}

The second cycle we would like to introduce is an Otto-like heat engine cycle. Otto cycle constitutes the working principle of Otto engines which are used in automobiles with gasoline. In the usual Otto cycle, there are two isentropic (adiabatic reversible) and two isochoric processes. But now like we did in the previous subsection, rather than isochoric processes we have isoformal processes (constant shape).

The cycle consists of four consecutive steps as it's seen in Fig. 5.3: (\textbf{1}) \textit{Adiabatic shape-confinement} $\textbf{1}\rightarrow \textbf{2}$: Inner structure is rotated from $45^{\circ}$ to $0^{\circ}$ position by performing work on the system without any heat addition. On the contrary to our classical expectations, the temperature of the system drops from 300K to 282K, in spite of the fact that the system undergoes an effective compression. The peculiarity of the process should be expounded more here. As we saw in Section 4.5 the entropy of a Maxwell-Boltzmann gas increases when we change the configuration from 45 to 0 degree isothermally. Accordingly, under an isentropic process, to keep the entropy constant, the system has to decrease its temperature. However, since the process is isentropic, there is no heat exchange possibility. In that case, without any heat rejection, decrement of temperature during work input is completely an unexpected behavior. What happens is, by changing the configuration from 45 to 0 degree, we are actually forcing the particles to redistribute themselves over energy states and to fill more heavily the states near to ground state. This is why temperature decreases during this process. (\textbf{2}) \textit{Isoformal heat removal} $\textbf{2}\rightarrow \textbf{3}$: On the contrary to the usual Otto cycle instead of heat addition, we have heat removal after the first adiabatic process in this cycle. Hence, the temperature of the system drops further from 282K to 250K, for constant shape. (\textbf{3}) \textit{Adiabatic shape-deconfinement} $\textbf{3}\rightarrow \textbf{4}$: Inner structure spontaneously rotates from $0^{\circ}$ to $45^{\circ}$ degree position and does work on its surroundings. Since no heat added during the process, the temperature of the system raises from 250K and 270K. (\textbf{4}) \textit{Isoformal heat addition} $\textbf{4}\rightarrow \textbf{1}$: In order to complete the cycle we raise the temperature from 270K to 300K by adding heat to the system at a constant size and shape. The sign of the heat exchange is the opposite of the usual Otto cycle.

\begin{figure}
\centering
\includegraphics[width=0.65\textwidth]{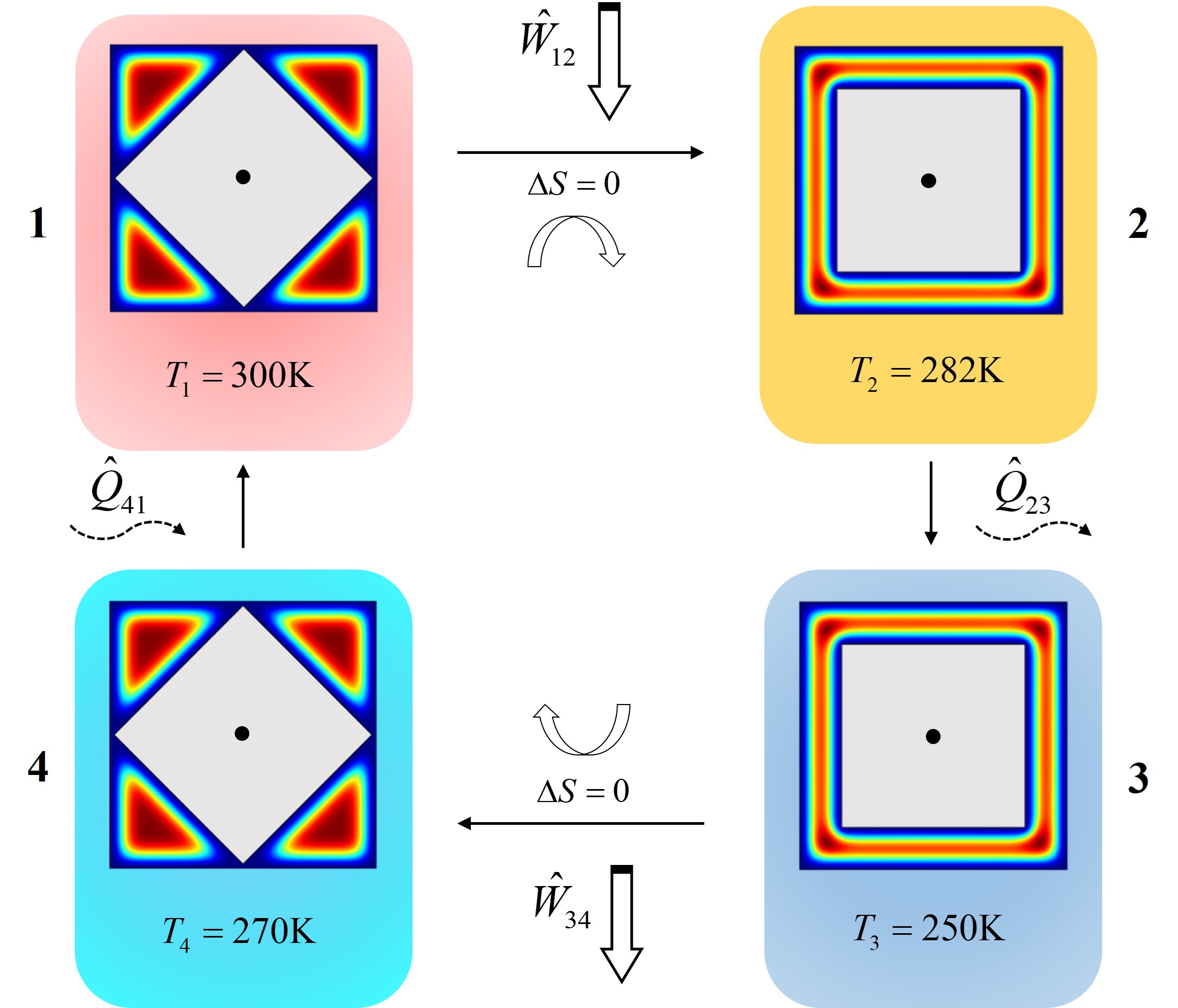}
\caption{Otto-like thermodynamic cycle based on quantum shape effects. The cycle consists of four processes and four temperatures denoted by different colors: (\textbf{1}) adiabatic shape-confinement, (\textbf{2}) isoformal (shape preserving) heat removal, (\textbf{3}) adiabatic shape-deconfinement, (\textbf{4}) isoformal heat addition. The cycle generates work during adiabatic shape-deconfinement process.}
\label{fig:c5f3}
\end{figure}

Transition from state 1 to 2 corresponds to the compression stage in the usual Otto cycle, since there is an effective shape confinement when the structure rotates from $45^{\circ}$ to $0^{\circ}$. In classical heat engines, when you compress the system, its temperature rises by the thermodynamic equation of state. In cycles featuring isoformal process, however, the temperature of the system actually drops during an effective "compression". The origin of this peculiar behavior along with the sign changes in work and heat exchanges of Stirling-like and Otto-like heat cycles respectively lies in the forenamed behavior of entropy.

\begin{figure}
\centering
\includegraphics[width=0.7\textwidth]{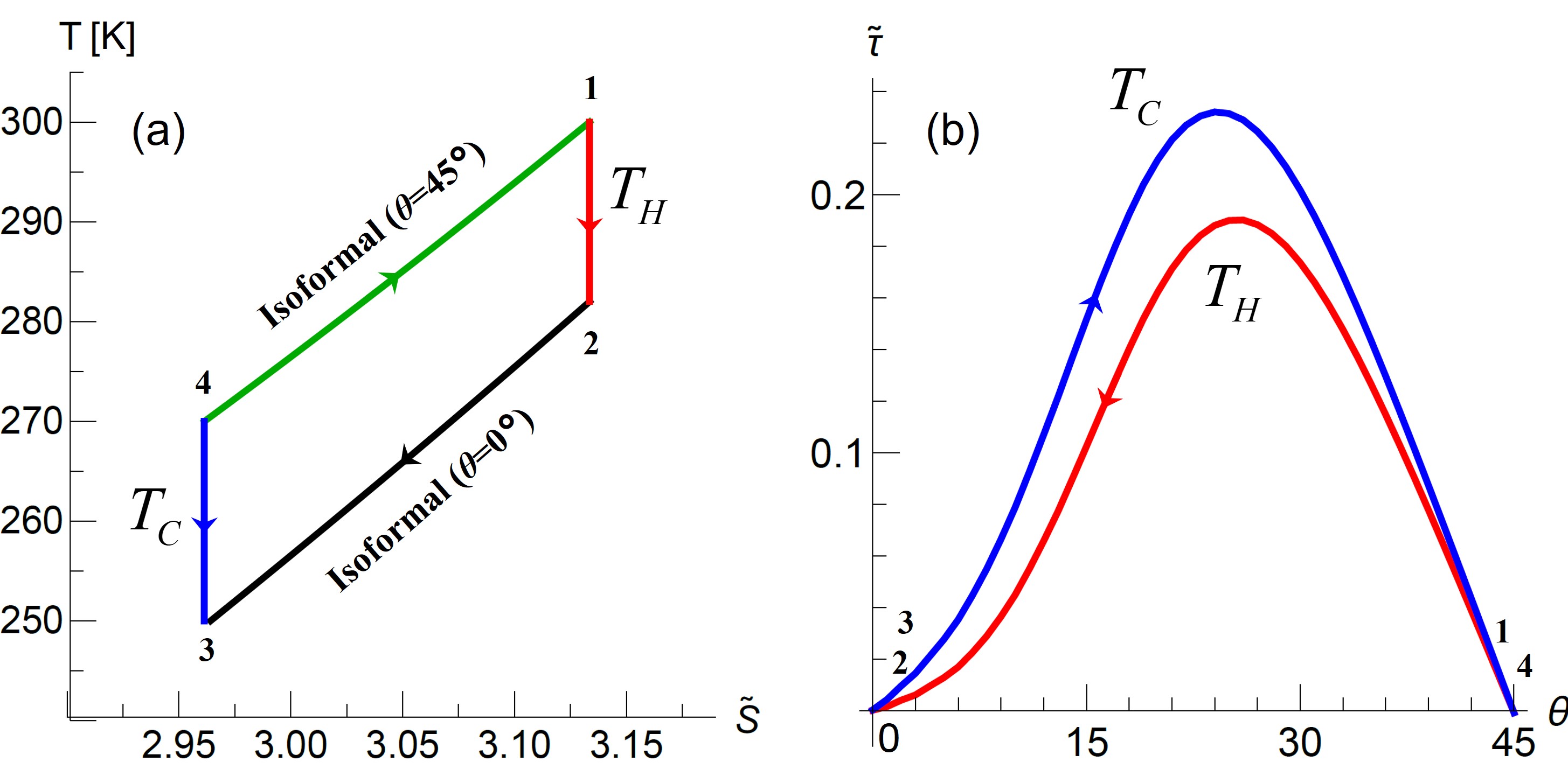}
\caption{Temperature-entropy ($T$-$S$) and torque-shape ($\tau$-$\theta$) diagrams. (a) $T$-$\tilde{S}$ diagram, where $\tilde{S}=S/(Nk_B)$. (b) $\tilde{\tau}$-$\theta$ diagram, where $\tilde{\tau}=\tau/(Nk_BT)$.}
\label{fig:c5f4}
\end{figure}

For this Otto-like cycle, heat and work exchanges at each process are then determined by differences of internal energies of each state
\begin{equation}
\begin{split}
W_{12}=& U\left(T_2,0^{\circ}\right)-U\left(T_1,45^{\circ}\right) \\
Q_{23}=& U\left(T_3,0^{\circ}\right)-U\left(T_2,0^{\circ}\right) \\
W_{34}=& U\left(T_4,45^{\circ}\right)-U\left(T_3,0^{\circ}\right) \\
Q_{41}=& U\left(T_1,45^{\circ}\right)-U\left(T_4,45^{\circ}\right).
\end{split}
\end{equation}

Using Eq. (5.2), heat input and net work output can be determined respectively as $Q_{in}=Q_{41}$ and $W_{\mli{net}}=W_{12}+W_{34}$. We obtain the cycle efficiency ($W_{\mli{net}}/Q_{in}$) as $7\%$. Again, the cooling cycle can easily be obtained just by reversing the heat pump cycle presented in Fig. 5.2.

\section{A Quantum Szilard Heat Engine Variant}

In this section, we would like to mention another heat engine called the quantum Szilard engine which constitutes a backbone for the reconciliation of thermodynamics, information theory and quantum mechanics. It has become particularly popular topic during the recent decade, especially because of its interdisciplinary nature and fundamental importance to the many open questions in physics today. Here we construct a quantum Szilard engine without featuring an explicit Maxwell's demon and we explore the role of quantum size and shape effects in this quantum Szilard heat engine variant. Our purpose in this section is to mention the relevance of this important problem to confinement effects. Introduction and examination of the problem in full details can be found in our published article in Ref. \cite{aydin10}.

First, let's start with a brief background of the problem, the Szilard's paradox. Consider a container with a single molecule confined inside. Then divide the container into two equal parts by a partition having zero thickness. Classically, insertion of the partition can be done without any work consumption due to its zero thickness. Depending on where particle is (which is determined by a measurement), the partition moves left or right like a piston, and by the isothermal expansion one can extract work from the engine. In order to complete the cycle, one needs to remove the partition from the system, which also can be done without any work consumption. One can repeat this cycle in a cyclic manner. As is seen, one can extract work by only taking heat from the heat reservoir. This clearly violates Kelvin–Planck statement of the second law of thermodynamics, as there is no heat rejected to the environment and all the heat absorbed is converted to work with $100\%$ efficiency. For the resolution of the paradox, many attempts have been done over the course of the years. The widely accepted resolution of the problem is this: in order to operate the cycle, there has to be a "demon" (an information processing being historically called as demon) who records the which-side information of the particle and in order to complete the cycle, the demon needs to erase it. According to Landauer's principle, it is this erasure process that resulting a heat dissipation and the corresponding work exchange in an isothermal process.

In the quantum versions of the Szilard engine, the same logic (erasure resolution) has been followed in the literature. However, we see that there are other additional mechanisms in a quantum Szilard engine cycle that saves the second law of thermodynamics. By placing solenoids and attaching magnetic rods to the piston at both sides, regardless of which side the piston moves, one can convert the motion of the piston into electrical power by the solenoids, and even into direct current power by a passive diode bridge. This technique allows one to extract work from a Szilard engine, without knowing the which side the particle is and therefore without using any information processing device like Maxwell's demon.

Let's consider our quantum Szilard engine variant which has this type of setup with solenoids. There are four stages in the cycle: insertion, measurement, expansion and removal of the partition, Fig. 5.5. Because of quantum size and shape effects, there has to be work done to insert the partition into the system even if the partition has zero thickness. This is because even a zero-thickness partition has a finite effective size due to quantum boundary layer in confined systems\cite{qbl,qforce}. 

\begin{figure}
\centering
\includegraphics[width=0.85\textwidth]{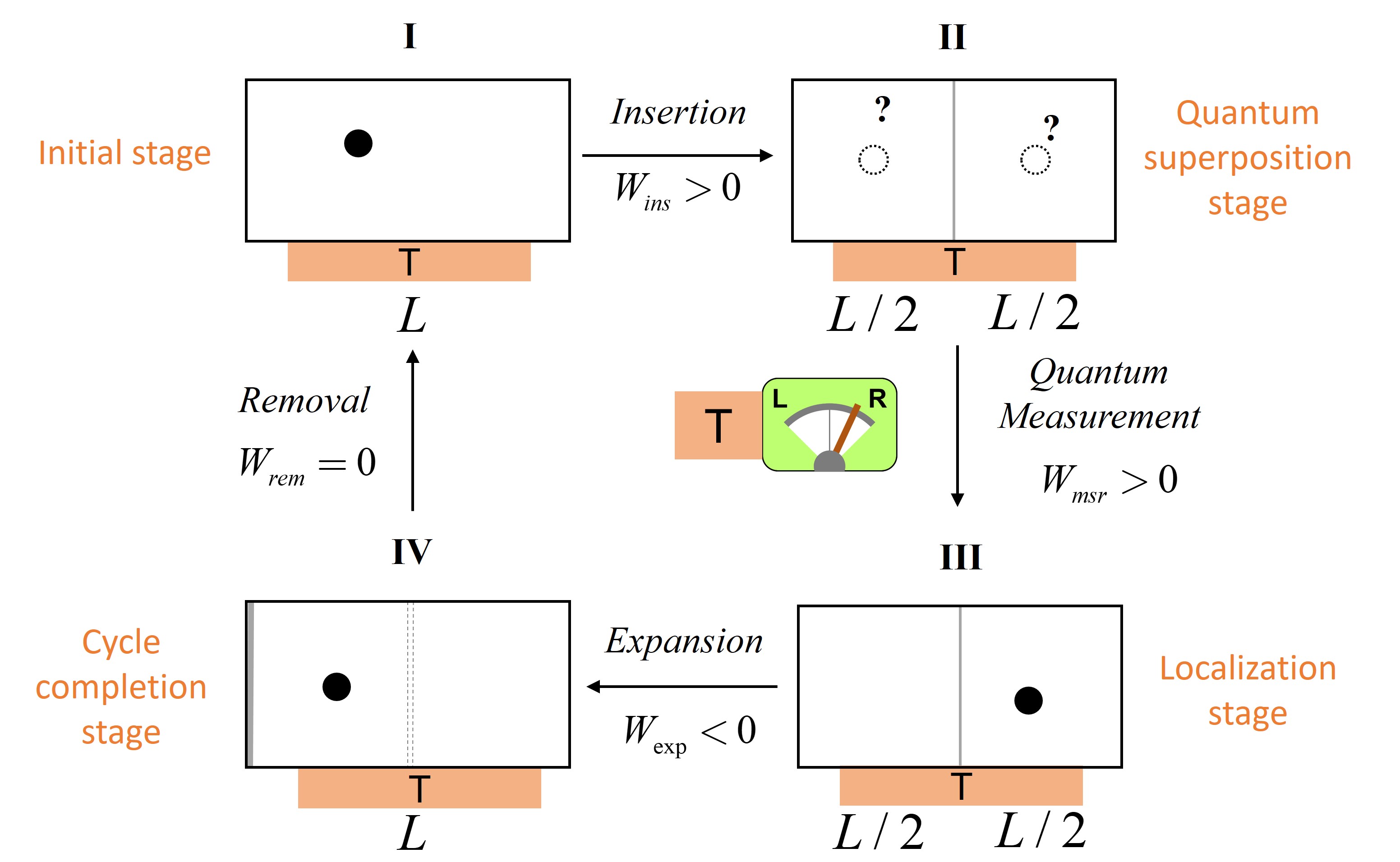}
\caption{A schematic of the quantum Szilard engine from our study in Ref. [248]. A quantum Szilard engine setup composed of three components, system (the box with a single particle inside), measuring device (whose sole purpose is to localize the particle via quantum measurement) and heat bath (keeps all processes isothermal). (I$\rightarrow$II) Partition is inserted into the container. Insertion divides the container into two equal parts and generates an entangled state of the particle's position. (II$\rightarrow$III) Quantum measurement is performed to localize the particle into one side of the box. (III$\rightarrow $IV) Particle expands the partition and work can be extracted from the system. (IV$\rightarrow $I) The partition is removed from the container at the boundary, which completes the cycle.}
\label{fig:c5f5}
\end{figure}

After the insertion of the partition, particle finds itself in superposition of being two sides at the same time. In that stage, the cycle won't operate, because the pressure at both sides will be exactly equal to each other. In order to break the quantum superposition and localize the particle, one needs to perform a quantum measurement on the particle (which destroys the entanglement that is formed with the separation), only then particle is localized and one can extract work by expansion. This fact is independent of whether we gather, process or even use the which-side information or not. After the expansion, removal of the partition is a trivial process in a single-particle Szilard engine. Since the partition will be exactly at boundary of the container where the wavefunction goes to zero, removal of the partition does not change any thermodynamic property of the system. 

In our article\cite{aydin10}, we quantify work, heat and energy exchanges during the thermodynamic cycle of quantum Szilard engine and obtained them analytically using the quantum boundary layer method. In this section, we won't examine the cycle fully, but we will give a brief example of how quantum size and shape effects play a role in the quantum version of the Szilard engine, as well as how quantum boundary layer helps us to understand the physical mechanisms of the processes. What we will focus on is the insertion process, where size and shape effects make their explicit difference. Numerical simulation of the insertion process is given in Fig. 5.6 along with the changes in free energy, entropy and internal energy during the insertion.

\begin{figure}[h]
\centering
\includegraphics[width=0.8\textwidth]{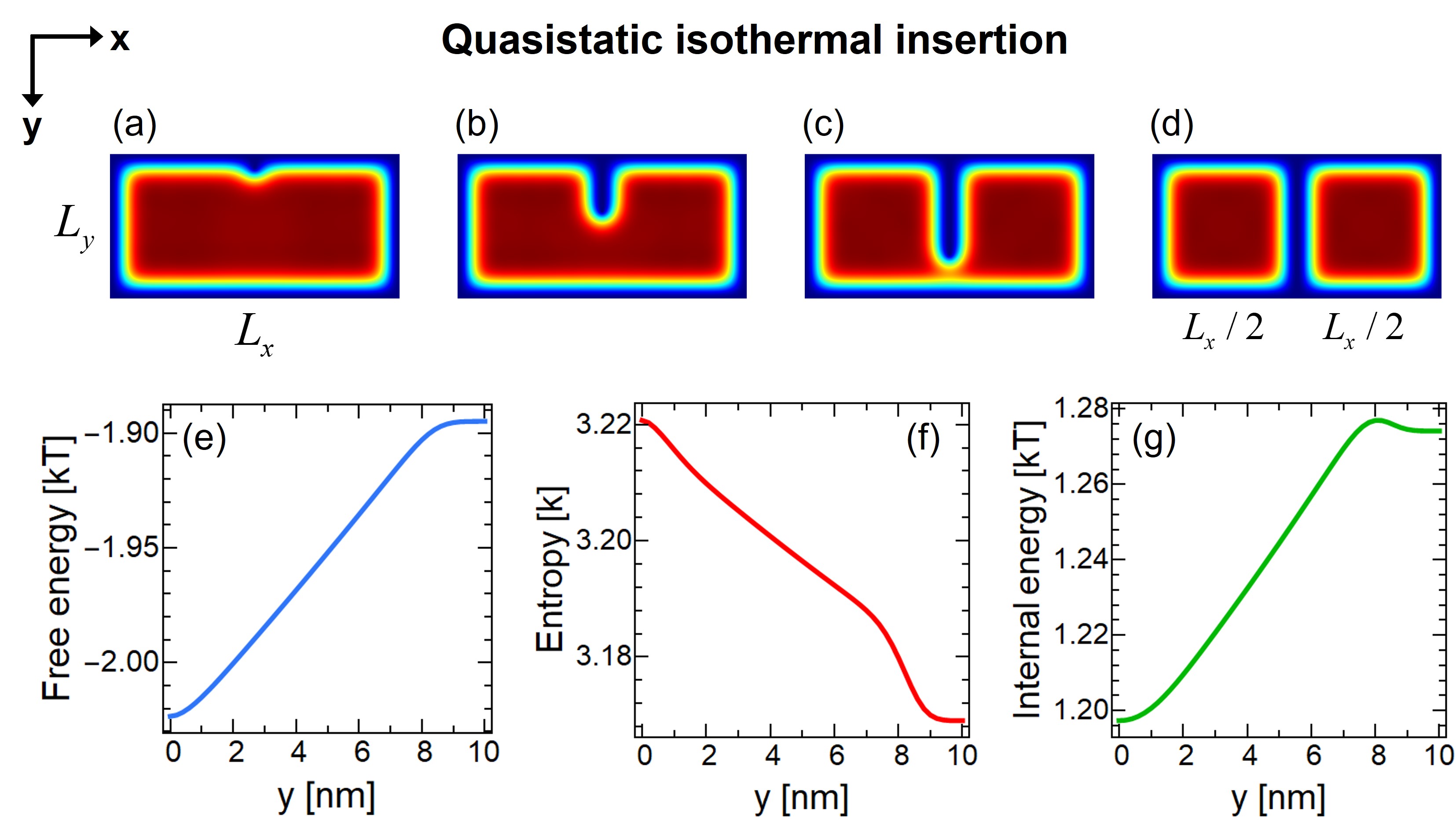}
\caption{Numerical simulation of the quasistatic isothermal insertion process for a Szilard box with sizes $L_x=20$nm, $L_y=10$nm and at temperature $T=300$K. $y$ is the depth of the piston inserted into the domain. (a) Quantum-mechanical thermal probability density distribution of the particle is non-uniform inside the container due to confinement effects. Magnitudes of the probability density distributions are represented by the rainbow color scale, where red and blue colors denote higher and lower density regions respectively. Partition with zero thickness enters the container in $y$-direction, $y=1$nm. (b) Partition enters almost halfway of the container, $y=4$nm. Although it has no thickness, confined particle perceives a finite effective thickness of $2\delta$. (c) Partition is at $y=7$nm depth. (d) Partition separates the container into two equally sized parts, $y=10$nm. Particle has equal probability to be at both sides. Variation of (e) free energy, (f) entropy and (g) internal energy with respect to partition's penetration depth $y$ in nm's.}
\label{fig:c5f6}
\end{figure}

As is seen from the thermal probability density distribution of the particle inside the domain, in Fig. 5.6(a,b,c and d), although the partition has zero thickness, the particle perceives an effective thickness of $2\delta$ as quantum boundary layer methodology suggests. Insertion of the zero-thickness wall reduces the effective domain inside the box, even though the apparent volume (here it is area) stays the same. This is a size effect. Thermodynamic state functions act accordingly by the influence of quantum size effects. However, when the partition comes very close to the bottom boundary, we see overlaps of quantum boundary layers of the partition and the bottom boundary. These overlaps suggest there is a quantum shape effect as well in this domain. Accordingly, after around $y=8$nm depth, we start to see deviations from the usual behaviors of thermodynamic state functions, Fig. 5.6(e,f and g). Work and heat exchanges during insertion is respectively calculated by the differences of free energy and entropy in the initial and final positions of the partition. Therefore, both quantum size and shape effects play role in the determination of the insertion work which is finite, contrary to the classical case where insertion work is said to be zero. In the future, we are planning to investigate more on the quantum shape effects in 1D confined systems to understand their behaviors more fundamentally. We won't go into further detail in this subject as we said before and will continue to this chapter with a novel energy conversion approaches that exploit quantum shape effects.

\section{Single-Material Unipolar Thermoelectrics}

Thermoelectricity is the direct conversion of thermal energy to electric energy and vice versa. When two materials with different charge transport properties like different density of charge carriers combined in a junction under a temperature difference, an electrical potential difference is induced in response to the temperature gradient. Building up of the electrical potential from a temperature gradient is called the Seebeck effect, whereas generation of cooling and heating from an electrical current is called the Peltier effect. Usually thermoelectric junctions consist of n-type and p-type thermocouples where electrons and holes are responsible for charge transport respectively. Provided that one end is at fixed chemical potential, difference in the electronic transport properties of electrons and holes causes a difference in the electrochemical potential at the other end of the junction. This potential difference can be utilized as an electric voltage at the separate end of the thermocouple and it is called the thermoelectric voltage. In a similar fashion, when operated reversely, the electric voltage can be converted into a temperature difference at both ends. Due to their durability, silent operation, scalability, precision and eco-friendliness, thermoelectric generators and coolers are of great interest both academically and commercially. Downsides of thermoelectrics are their high costs and relatively low energy conversion efficiency rate, although the expectation for their theoretical efficiency was much higher due to the nature of direct conversion process (being free of energy losses that happen in other multi-stage conversions of energy from one form to another).

During last two decades, numerous studies have been done in thermoelectric research. It has been shown that performance of thermoelectric devices can be improved by taking advantage of materials at nanoscale. Rather than focusing the conventional route, some studies have been focused on designing different kind of junctions both to understand the thermoelectric effect better at small scales and to contribute to the development of novel energy devices. One approach is utilizing quantum size effects not only for improving the thermoelectric performance of the materials individually, but also to directly design thermoelectric devices based on quantum size effects. Instead of creating p-n junctions, it has been proposed to design junctions with the same material but having different sizes. When the temperature gradient is applied, an electrochemical potential difference is induced in those materials even if all their material properties are the same except their sizes. The electrochemical potential difference occurs due to difference in quantum size effects on Seebeck coefficients of the pillars of the junction. This is called thermosize effect and it has been proposed by Sisman and M\"{u}ller. Thermosize effect has been studied during last decades\cite{sismanmuller,tsef2008a,tsef2008b,tsef2010a,tse2,tse3,tsef2011a,tsef2012a,tsef2012b,tse4,tsef2013a,tsef2013b,tsef2014a,tse5,tsef2018a,tsef2018b,aydin9}. We can aggregate these type of thermoelectric mechanisms under the umbrella term that we call single-material unipolar thermoelectrics. An example of such setup is shown in Fig. 5.7 where we examined the thermosize effect in graphene nanoribbon junctions.

\begin{figure}
\centering
\includegraphics[width=0.7\textwidth]{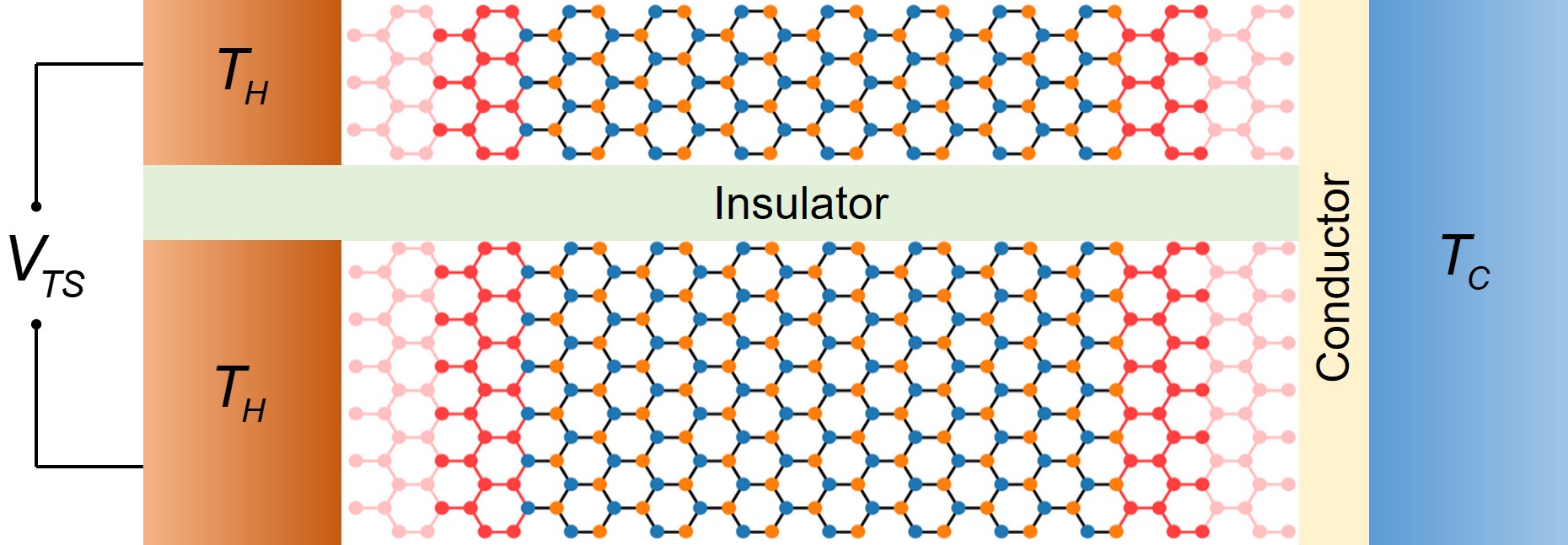}
\caption{A schematic of a thermosize junction with narrow and wide graphene nanoribbons. Narrow and wide ribbons are separated by insulator on the middle, thermally and electrically connected on the cold side and only thermally connected on the hot side. Although the junction is made of the same material, their size difference leads to an electric voltage on the hot side.}
\label{fig:c5f7}
\end{figure}

\begin{figure}[!b]
\centering
\includegraphics[width=0.7\textwidth]{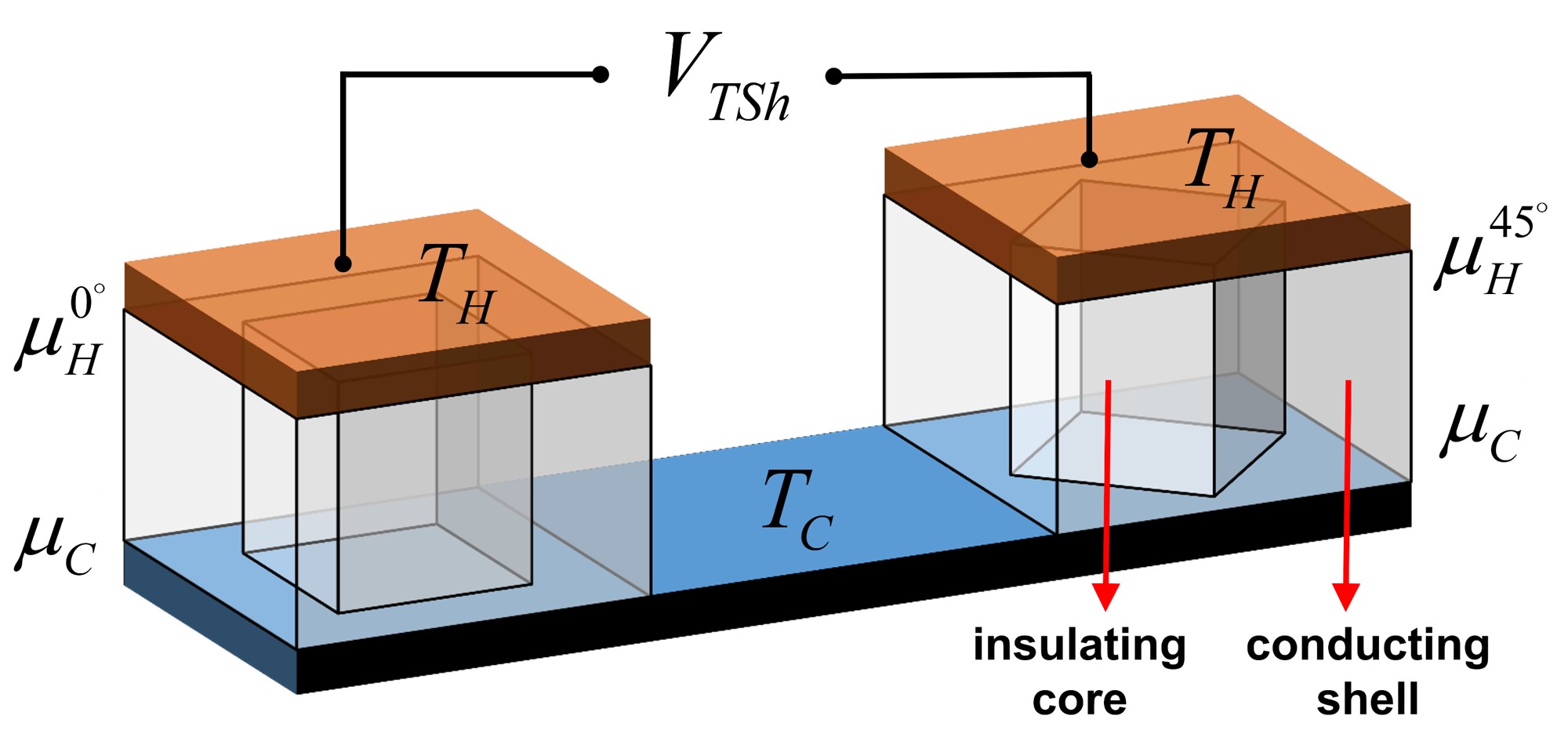}
\caption{A schematic of a thermoshape junction with core-shell nanostructures having $0^{\circ}$ and $45^{\circ}$ core structure angular configurations on the left and right pillar respectively. Even though each pillar of the junction is made of the same material having exactly the same sizes, only the shape difference leads to an electric voltage on the hot side.}
\label{fig:c5f8}
\end{figure}

Apart from the topics about quantum shape effects on thermodynamics considered in this thesis, recently we investigated whether quantum shape effects can also affect transport properties of charge carriers and result a new type of single-material unipolar thermoelectric device \cite{aydin11}. We designed a thermoelectric-like junction shown in Fig. 5.8. We used Landauer formalism along with Datta's number of modes approach and tight-binding model to calculate transport properties of core-shell nanostructures (basically the same nested-square domain we considered in this thesis) \cite{dattabook1995,dattabook2005,tbmodel,datta1,datta2,aydin11}. To maximize the influence of shape difference on the system, we considered pure ballistic weakly degenerate transport regime with positive but very low chemical potential. We used GaAs as the shell structure which has a low effective mass. We conducted our analysis at low temperatures relative to room temperature.

We showed that due to quantum shape effects on Seebeck coefficient under a temperature difference, thermoshape potential is induced. At maximum thermopower region, the thermoshape voltage is in the order of several millivolts per Kelvin for cold side temperature of 20K when temperature difference of 2K. Thermoshape junction constitutes not only a viable setup for the macroscopic manifestation and demonstration of quantum shape effects, but also their first possible device application.

%\chapter{Conclusion}
\chapter{Conclusion}

In this thesis, we introduced a new aspect of finite-size effects. We proposed the quantum shape effect which appears in strongly confined structures. We provided a thorough examination of the origins and the nature of quantum shape effects in the thermodynamics of non-interacting particles confined in impenetrable nanodomains. The main motivation was to explore this previously unnoticed effect and understand the fundamental physics behind it. We extended the quantum boundary layer methodology to capture the quantum shape effects in addition to the quantum size effects. We supported our findings both with numerical simulations and analytical calculations. Finally, we constructed quantum heat engines driven by quantum shape effects, in order to demonstrate possible thermodynamic applications of the effect.

Quantum shape effect is essentially a result of the wave nature of particles and the discrete energy spectrum caused by the confinement potential. To observe the effect, the conditions should be in favor of a comparable de Broglie wavelength with the domain sizes. Required conditions are not sufficient of course unless one does not create the conditions where boundaries of the system get really close to each other so that quantum boundary layers start to overlap. Core-shell nanoarchitectures are suitable candidates for the realization of nested domains. In recent years, research on core–shell semiconductor nanocrystals is also developed giving the precise control of size and shape of these structures. Other candidate nanoarchitectures for quantum shape effects are easily shapeable nanostructures like graphene sheets or nanoribbons. Controlling material properties by shape can open up a whole new direction in material science.

% Optical
Waveguides are the systems where the optical analogue of quantum shape effects can be observed. Optical versions of this effect can be exploited for instance in double-clad fibers where the pumping energy's absorption efficiency may be tuned by smooth shape variation\cite{dcfib,dcfib2}. One can also artificially create confinement conditions using optical traps where ultracold atoms or the system of interest is being confined\cite{trap1,trap2}.

There are some advantages that make quantum shape effects realizable in labs and some applications. For example, while most quantum systems can operate at cryogenic temperatures, quantum shape effects can appear even at room temperatures. It is not just a single and isolated effect, but can be demonstrated indirectly via many different physical mechanisms some of which are presented in the previous chapter. However, the effect is strong only in very small, extremely confined systems which may be challenging to manufacture smoothly with ease. The effect depends strongly on the disorders and imperfections on the boundaries of the material or inside the confinement domain. If the imperfections are small and random, which is mostly the case, we expect quantum shape effect to survive because the influence of disorders will be averaged out for all shape configurations and they will behave as fluctuations over the overall quantum shape effect behavior.

% Aşağıdakilerden de bahset burada: (sağlam referanslar ver bunlara)
%	Candidate nanoarchitectures for quantum shape effects
%Core-shell semiconductor nanocrystals
%Core-shell nanowires
%Quantum corrals
%Quantum shape effects in nanoelectronics

% FUTURE WORKS
Since the quantum shape effect is completely a new discovery, there are abundant things to study both from theoretical and application sides. It is a general, material-independent effect that can possibly be seen in many other exotic nanoscale systems like perhaps topological or superconductor materials\cite{supercondcore}. As a near future work, we are planning to investigate quantum shape effects in Bose-Einstein Condensates, where it may be possible for instance to control the condensation temperature by shape. Thermodynamic and transport properties of various Fermionic or Bosonic systems can be also explored under quantum shape effects. Topological thermodynamics in systems having more than one hole (note that nested domains investigated here have a single hole which is the core structure) can also be investigated.

Exploration of pure 1D systems is also of great theoretical importance as they constitute the simplest but, in many ways, exact systems that can be understood both physically and mathematically. We've already had some interesting results on such systems which may shed light onto various nanoscale thermodynamic and transport phenomena. A more clear and analytical understanding of energy moments would also help to clarify the notions of pressure, flux and directional quantities at quantum scale.

We restricted ourselves to quasistatic and time-independent processes in this thesis. Our future plan is to explore the quantum shape effects in interacting, time-dependent systems as well as with finite (constant, harmonic, quartic, etc.) confinement potentials. Opening electrical or magnetic fields, would possibly yield even more interesting direct or cross effects and may lead to further novel nanoscale devices. Study of the shape effects in quantum coherent systems is also a nice direction of research that would also help to the understanding of quantum thermodynamic systems.

%\appendix

\appendix
%\chapter{Supplementary Information About Calculations and Some Concepts}
\chapter{Appendix: Supplementary Information}

Here we give some additional details about the calculations done in the thesis. Theoretical works and analytical calculations are conducted by use of pen and lots of paper. A mathematical computation and symbolic manipulation software \textsc{wolfram mathematica} is extensively used for calculating algebraic equations, generating figures and post-processing data. A numerical finite element analysis software \textsc{comsol multiphysics} is used for calculating eigenvalues, eigenfunctions and other kind of numerical quantities, as well as for graphics and visual representations.

Also, in this Appendix A, we mention some attempts that we've done and approaches we've tried to implement for the improvement of the results and understanding of some parts of the thesis work. They don't necessarily contribute to the presented final results in the main parts of this thesis, but definitely increased our understanding and knowledge of the effect that we are dealing with.

\section{Details of Numerical Calculations}

% Finite Element Methods
We first created our confinement geometry with Dirichlet boundary conditions in \textsc{comsol} software and used \textit{Coefficient Form PDE} in the \textit{Mathematics} module for the solution of Schr\"{o}dinger equation for our arbitrary domain. The software uses finite element methods to numerically solve partial differential equations. There are three different finite element solution methods embedded which are MUMPS, PARDISO and SPOOLES. In terms of accuracy, we haven't noticed any detectable difference in our tests for the stationary Schr\"{o}dinger equation under the boundary conditions considered in this study. However, flexibilities and memory allocation properties of these methods are different. From our trials, we see that MUMPS is the fastest and the most flexible one in terms of allowing optional memory allocation. Hence, we chose MUMPS method.

% Meshing
We wanted to have a high numerical precision in our eigenvalues. Using the correct meshing is therefore a crucial issue in finite element method. The software has a bunch of mesh types and a wide range of mesh control parameters, including adaptive mesh capabilities (which is an important feature because of the local tiny perturbations that we consider in some cases). After our tests, we realized that free triangular mesh gives better accuracy than other types of meshes for our domains. Investigation of mesh errors is done by considering the number of mesh elements inside the minimum wavelength corresponding to the highest energy eigenvalue considered during the calculation of sums, so that increasing the number of mesh points won't cause any difference in the eigenvalue solutions. To this aim, we divided the wavelength corresponding to the largest eigenvalue that we truncate after, to the maximum size of our mesh element and found the number of mesh points inside the minimum wavelength. Then by increasing the number of mesh elements, we made a convergence analysis and determined the number of mesh elements accordingly. Around 70000 mesh elements are more than enough for the considered domains in the thesis. (Considered element size parameters of the meshing are: maximum element size: 0.1 nm, minimum element size: 0.000424 nm, maximum element growth rate: 1.1, curvature factor: 0.2, resolution of narrow regions: 1) An example of the meshing that we used is given in Fig. A.1.

\begin{figure}
\centering
\includegraphics[width=0.8\textwidth]{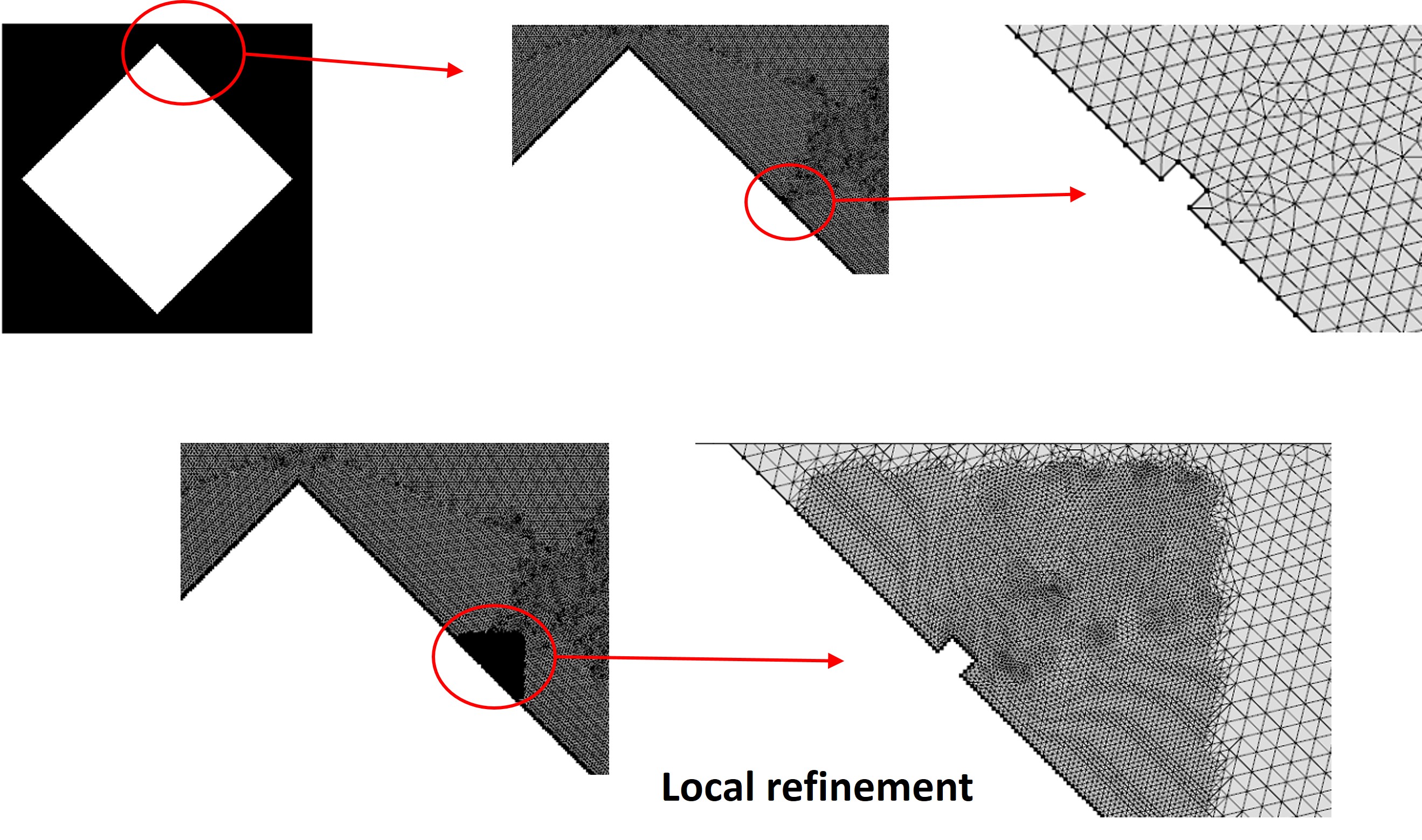}
\caption{Dense triangular meshing of our confinement domain in \textsc{comsol} environment. Local refinements are possible when necessary.}
\label{fig:cAf1}
\end{figure}

In addition to \textsc{comsol} software, for comparison we've made similar finite element calculations also in \textsc{mathematica} software using its built-in \textit{NDEigensystem} function. The results were the same, so we choose \textsc{comsol} for our eigenvalue calculations and \textsc{mathematica} for all other calculations and data processing.

% Truncation
Summations in statistical quantities are over all possible states which are infinitely many. No wonder, sums that we use are naturally convergent and one can truncate after some finite amount of terms. To determine the truncation point of eigenvalues, we made a truncation analysis. The proper number of eigenvalues depends on the particle's mass, temperature and the sizes of the domain. By choosing bare electron mass, room temperature and nanoscale domain sizes (taking transverse confinement parameter as unity), we determined the optimum number of eigenvalues for our arbitrary domain that we target to solve. We make sure that truncation errors are always negligible in all of our calculations.

% Sampling
We repeated the same solution procedure for each and every angular configuration that we wanted to consider. To span the angular ranges from $0^{\circ}$ to the symmetric periodicity angles, we used sampling. For the correct sensitivity of the sampling, we compared our results for $\Delta\theta=1^{\circ}$ and $\Delta\theta=0.25^{\circ}$ which we see no difference in the results. 

Numerical calculations are tested also with exactly solvable analytical models and it is ensured that all steps of numerical calculations are consistent and produce negligible total errors.

\section{Global and Local Boundary Perturbations}

Based on the physical quantity that is interested in, one can probe system's free energy response to the global or local perturbations on the boundaries, which are summarized in Fig. A.2. To calculate the torque (a global effect) exerted on inner square structure, we create small angular perturbations. We look to the difference of the unperturbed and perturbed free energies with respect to the angular perturbation, which gives the torque. Similarly, a linear perturbation on all boundaries of inner square is done to calculate the exerted pressure.

\begin{figure}
\centering
\includegraphics[width=0.9\textwidth]{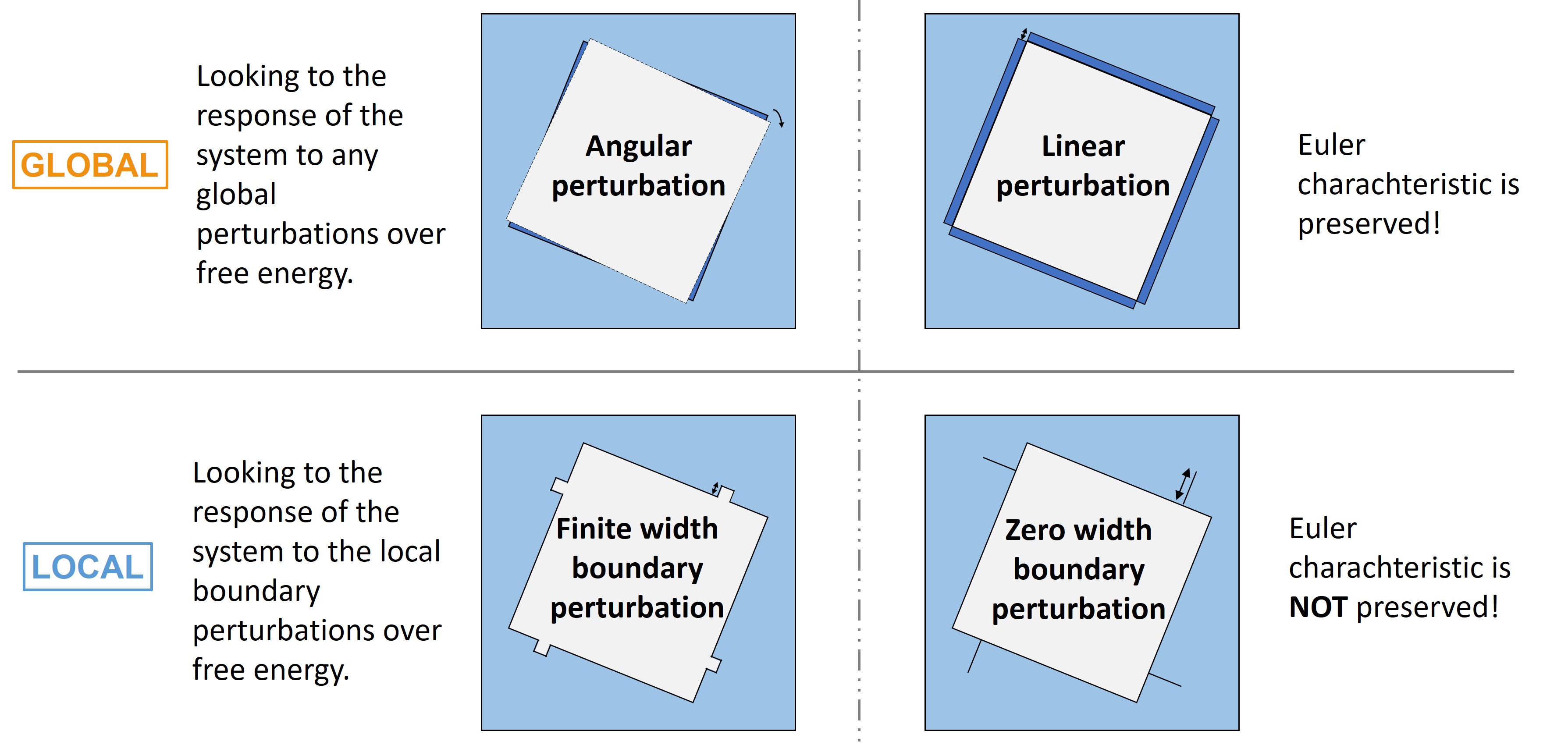}
\caption{Comparison of global and local boundary perturbation approaches. Keeping the axial symmetry is important.}
\label{fig:cAf2}
\end{figure}

As a justification of the existence of torque, we argue that pressure distribution should change non-uniformly along the walls of inner square. In order to find this pressure distribution, we implemented a local perturbation approach. We created tiny (having very small width and height) perturbations along the walls of inner square. There are many subtle points to consider when creating such local perturbations. How large can they be both in terms of width and height? Can and should they be infinitesimally thin? Should we create them inwards (removing a tiny part from the domain) or outwards (adding a tiny part to the domain)? We made extensive trials conceiving all of these along with many other considerations. There is no way to obtain results precisely and without affecting the system of course. By creating even tiny perturbations, we unavoidably change the actual area, periphery and number of vertices of our domain. This leads to some variation in eigenvalue spectrum which we wouldn't want to happen. In order to reduce this influence, we could create zero width perturbations (creating a perpendicular line boundary) instead of finite width ones. Hence, actual area stays the same whether it is perturbed or unperturbed, but just the periphery and vertices change. However, in spite of keeping the actual area constant during the zero width perturbations, the effective area which is felt by the system changes due to the quantum boundary layer that is generated by the perturbation even if it has zero thickness.

Another thing to notice is while global perturbations preserve the Euler characteristic (a topological invariant defined as the sum of the number of vertices and faces minus the edges of a shape) of the domain, local perturbations do not. Thus, not only shape but also size variables of the domain changes, when the Euler characteristic is not preserved. Tiny changes in area, periphery and number of vertices occur. Since the changes are tiny as the perturbation, lowest eigenvalues are almost not affected from these variations, but only higher eigenvalues of which the contributions to the global physical quantities are less. All in all, based on our tests, finite width perturbations give more accurate results than other local perturbations, whereas zero width perturbations give much more smoother graphs.

One other distinction of the perturbations is whether they are inwards or outwards, see Fig. A.3 for their comparison. Both of them affects the domain in a distinct way such as the sufficient mesh density differs. In the outward one, domain barely feels the existence of the perturbation, therefore much finer mesh is required than the inward perturbation. We have found an analytical formula representing the influence of finite inward (+) and outward (-) perturbations to the exerted pressure, which results to the following modified expression of pressure:
\begin{equation}
P_{\pm}=n_{cl}k_BT\left[1\pm\frac{1}{2}\frac{\Delta h}{L}+2\delta\left(\frac{1}{L\mp\Delta h}\right)\pm\left(2\delta\right)^2\left(\frac{1}{\left(L\mp\Delta h\right)^2}\mp\frac{1}{L^2}\right)\right]
\end{equation}
which is derived originally from Eq. (4.18). $\delta$ is the thickness of quantum boundary layer. The second term in the square bracket is the second order term in Taylor expansion, the third term is the quantum size effect correction and the last term represents the second order quantum size correction. The second and third term are more or less comparable to each other and their behavior is determined by the ratio of $\Delta h/L$ as well as temperature. As long as the perturbation depth $\Delta h$ is much smaller than $L$, errors coming from this finite difference operation are reasonably small. Both perturbation methods give similar results, and we used the inward one due to its mesh efficiency.

\begin{figure}
\centering
\includegraphics[width=0.9\textwidth]{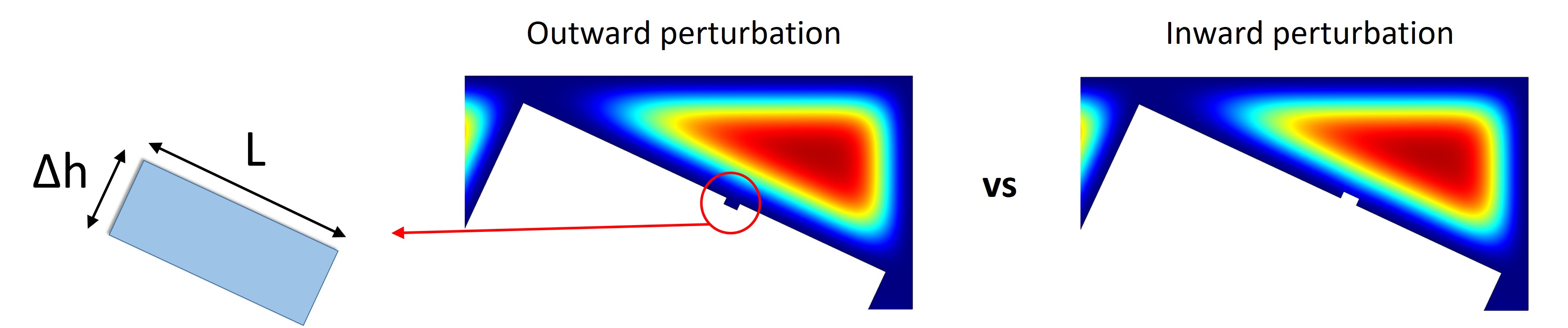}
\caption{Comparison of outward and inward local perturbations.}
\label{fig:cAf3}
\end{figure}

\section{Local Momentum Flux Approach}

There is one other approach we've tried for the calculation of torque, pressure and application points. It is a local approach again, but rather than making perturbations along the boundaries of the domain, we find the local quantum-mechanical momentum flux distributions inside the domain. We sought for a consistency between the pressure determined from free energy by global boundary perturbations and the local momentum flux of the particles. We exploited the equivalence of pressure to the momentum flux, which we showed it in Appendix A.6. Our purpose was to find the local positions near to boundaries of the inner square walls where local momentum flux equals to the pressure exerted to the wall. Thereby, we wanted to show the formation and existence of a pressure boundary layer. If we could generalize it to any domain like it has been done in quantum boundary layer, then it might be possible to obtain the local pressure at any position along the domain boundaries without doing extensive calculations.

\begin{figure}[!b]
\centering
\includegraphics[width=0.75\textwidth]{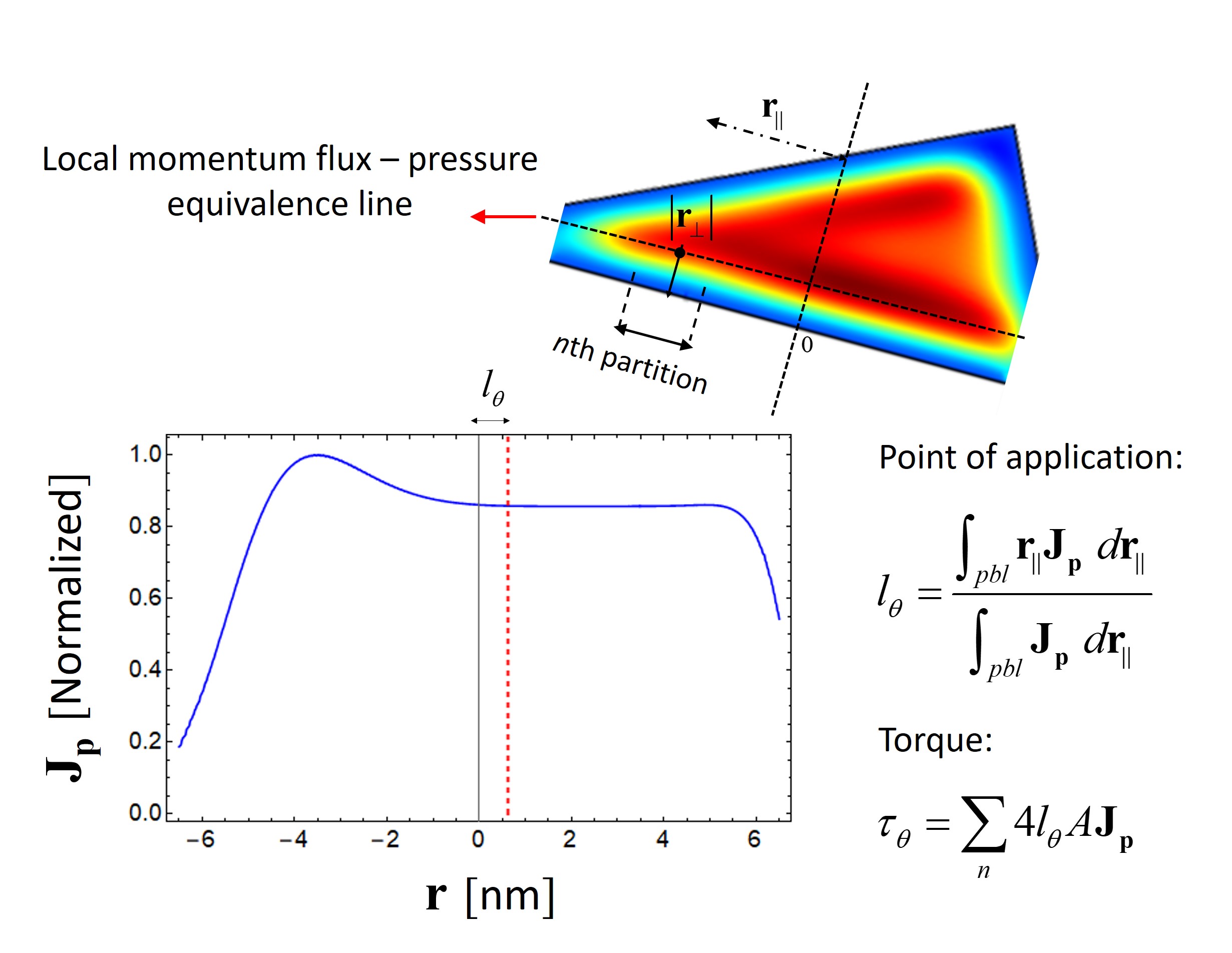}
\caption{Finding the local momentum flux - pressure equivalence line (pressure boundary layer) near to the boundary where pressure exerts. Point of applications can be found by integrating the local momentum flux along the pressure boundary layer that is parallel to the actual boundary.}
\label{fig:cAf4}
\end{figure}

Unlike the other approaches, eigenvalues are not sufficient to calculate the exact local properties, we need eigenfunctions as well. We numerically calculated eigenfunctions corresponding to the already obtained eigenvalues. Just like in the standard derivation of probability flux in quantum mechanics, one can obtain the momentum flux by using the Schr\"{o}dinger equation in conjunction with continuity equation as follows:
\begin{equation}
\frac{\partial \rho_p}{\partial t}+\vec{\mathbf{\nabla}}\cdot\mathbf{j_p}=0, \quad\quad \hat{H}\Psi=i\hbar\frac{\partial \Psi}{\partial t} \;\;\Rightarrow\;\; \mathbf{j_p}=m\left(\Psi^{*}\mathbf{\hat{v}}^2\Psi-\Psi\mathbf{\hat{v}}^2\Psi^{*}\right)=\mathbf{\hat{v}}^2\rho_p
\end{equation}
where $\rho_p=m\Psi^{*}\mathbf{\hat{v}}\Psi$ is momentum density, $\mathbf{\hat{v}}=-(i\hbar/m)\vec{\mathbf{\nabla}}$ is the velocity operator and $\mathbf{j_p}$ is momentum flux operator. Note that $\mathbf{j_p}$ contains the summation of both left and right fluxes described by the Wronskian $W(\Psi,\Psi^{*})$ and gives zero at equilibrium. However, we are interested in the flux towards the inner domain walls which is a one-sided flux. Total momentum flux in one direction can simply be found by using statistical mechanics,
\begin{equation}
\mathbf{J_p}=\sum_{\varepsilon}\mathbf{j_p}f=-N\frac{\hbar^2}{m}\frac{\sum_{\varepsilon}\exp(-\tilde{\varepsilon})\Psi^{*}\vec{\mathbf{\nabla}}^2\Psi}{\sum_{\varepsilon}\exp(-\tilde{\varepsilon})}=-N\frac{\hbar^2}{m}\left\langle \Psi^{*}\vec{\mathbf{\nabla}}^2\Psi\right\rangle_{ens}=Nm\left\langle \Psi^{*}\vec{\mathbf{\nabla}}^2\Psi\right\rangle_{ens}
\end{equation}
where $f$ is the relevant distribution function (here the Maxwell-Boltzmann is shown as an example) and $\mathit{ens}$ subscript with brackets denotes the ensemble average. This approach is known also as quantum hydrodynamics formalism\cite{qhydo}. Calculation of point of applications now can be done by integrating momentum flux times their application points along the wall and dividing it to the boundary layer integral of momentum flux. All of which can be obtained using eigenvalues and eigenfunctions.

In Fig. A.4, we show the line where local momentum flux equals to the non-uniform pressure exerted on a tiny partition of the wall. The local momentum flux along this equivalence line is plotted for the particular part of the domain that is shown in the Figure. A small peak can be seen near to the region where two boundaries approach themselves. This is an expected behavior of such system, as we will see the similar type of functional behaviors in our pure 1D analyses in Section A.4. Calculation of application points and torque are also given in Fig. A.4.

We have tried various numerical attempts to make the approach work, however in the end, we abandoned this approach as it gives very low numerical precision even though we obtained correct functional behaviors most of the time. Besides, there is no need to force in this direction, since the local pressure distribution that is calculated from the free energy by local perturbations provides quite reasonable and clean results. Nevertheless, we acquired some physical intuition about the results which strengthens our interpretations.

\section{Energy Moments and Density Distributions}

For the sake of understanding more of the local properties of our system, we investigated energy moments of the ensemble-averaged densities. Here as an example we used the Maxwell-Boltzmann distribution, but any other distribution can also be used. The general expression for various energy moment densities is written in its dimensionless (normalized to classical values) form
\begin{equation}
\tilde{n}_{\varepsilon}^{(m)}=\frac{n_{\varepsilon}^{(m)}}{n_{cl,\varepsilon}^{(m)}}=\frac{\sum_{\varepsilon}\tilde{\varepsilon}^{m/2}\exp(-\tilde{\varepsilon})\left|\Psi(\mathbf{r})^2\right|}{\frac{1}{V}\sum_{\varepsilon}\tilde{\varepsilon}^{m/2}\exp(-\tilde{\varepsilon})},
\end{equation}
where $m$ superscript denotes the order of the energy moment. For $m=0$, it gives the usual particle density whereas for $m=1$ and $m=2$, it gives the momentum and energy densities with position dependence. Therefore $\tilde{n}_{\varepsilon}^{(m)}(\mathbf{r})$ represents the particle, momentum and energy density distributions for a particular coordinate inside the domain so that $n_{\varepsilon}^{(0)}=n$, $n_{\varepsilon}^{(1)}=p$ and $n_{\varepsilon}^{(2)}=u$. Numerical simulation results of the energy densities for nested square domain is given in Fig. A.5.

\begin{figure}
\centering
\includegraphics[width=0.75\textwidth]{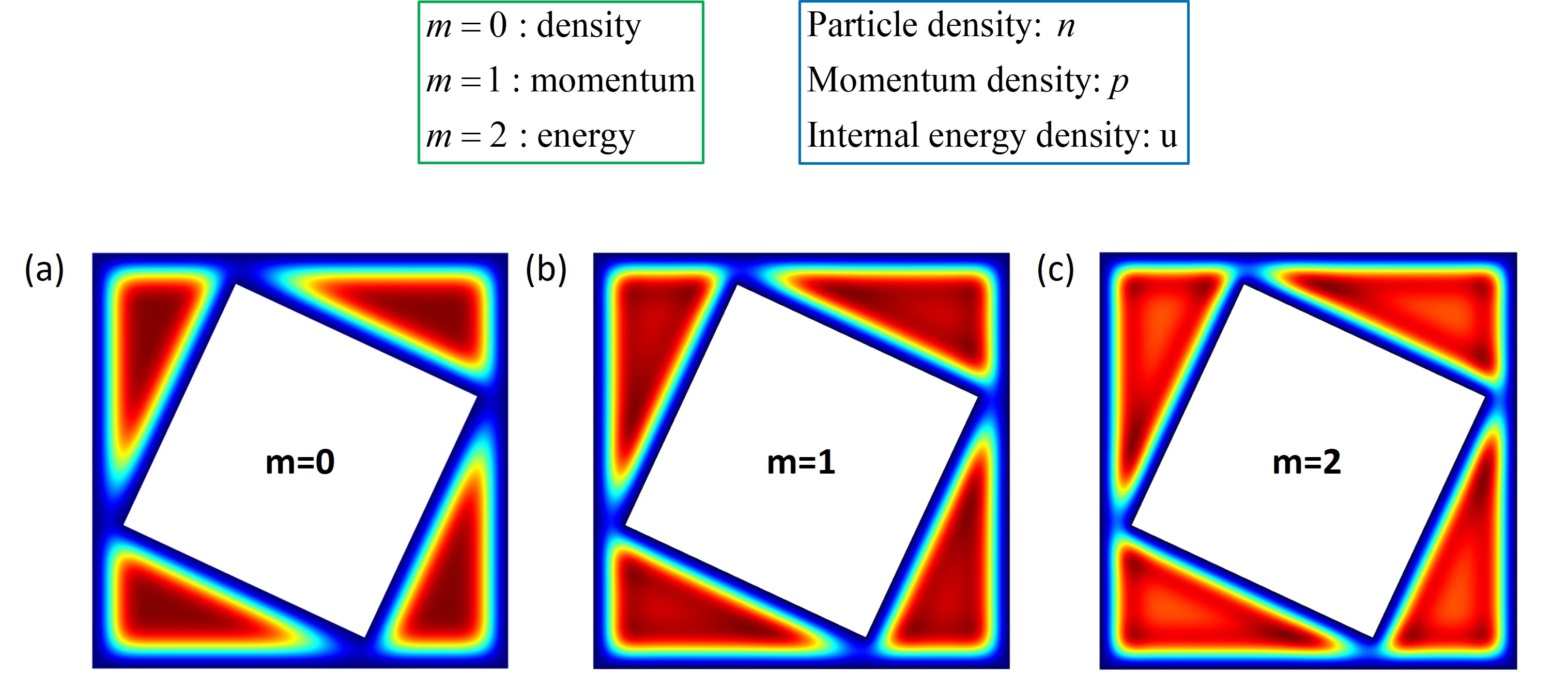}
\caption{Energy moment densities together with the corresponding quantities. (a) Particle (b) isotropic momentum and (c) energy density distributions inside the nested square domain for $\theta=25^{\circ}$ configuration.}
\label{fig:cAf5}
\end{figure}

Examination of different moments of energy may provide valuable information about the local properties of the system. Particle density distribution, as we have shown before, gives crucial information about the behavior of particles confined in such a domain shape, onto which the quantum boundary layer methodology is built. It also provides a tentative visual explanation about the existence of torque. Similarly, momentum distribution inside the domain, which is obtained when $m=1$ gives information about how the momentum of particles is distributed along the confinement domain. This justifies the non-uniform distribution of momentum and energy inside the domain. Note that despite intuitively correct, it does not always mean that the less particle localized in a region the less momentum flux. Proper examination can be done using momentum flux density distribution. Although momentum flux density is a tensorial quantity, in this section for simplicity we only plot isotropic momentum density, which is a vectorial quantity just like the energy flux density. In other words, by taking the square root of its square we turned it into a scalar quantity first, but since it is a flux density, it becomes vectorial. Energy density distribution ($m=2$) also gives useful information about why confinement energy is higher for some angular configurations and lower for some others. We compared particle density, isotropic momentum and energy densities for the nested square domain in Fig. A.5.

\begin{figure}
\centering
\includegraphics[width=0.85\textwidth]{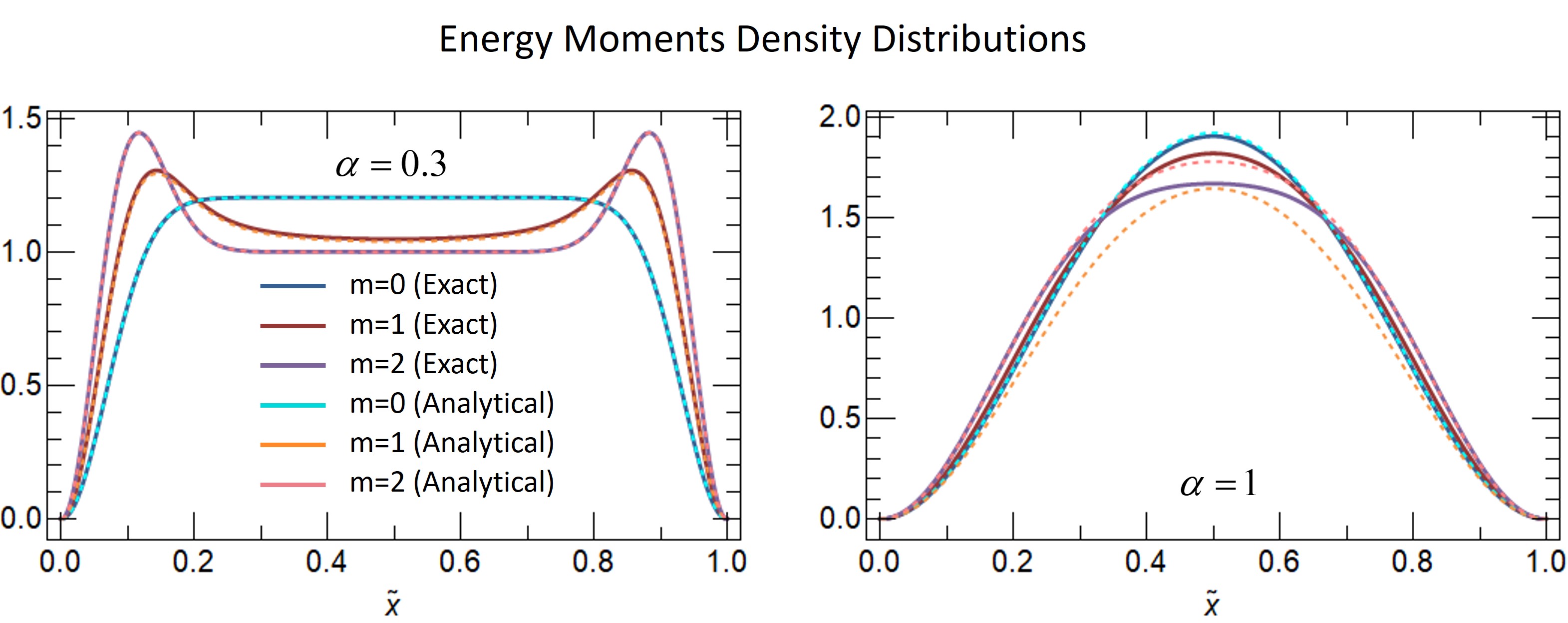}
\caption{Comparisons of exact and analytical expressions for two different confinement strengths (Left figure $\alpha=0.3$ and right figure $\alpha=1$) of density distributions of different moments. Solid curves represent the exact (numerical) calculations whereas the dashed curves represent the analytical calculations based on Eq. (A.5). Blue (cyan), red (orange) and purple (pink) curves represent the exact (analytical) results for $m=0$, $m=1$ and $m=2$ moments respectively.}
\label{fig:cAf6}
\end{figure}

Quantum boundary layer theory was constructed considering 1D domain and it has been proven to be successful on many different confined systems with higher dimensions. To this end, here we would like to investigate again 1D system and plot the distributions of energy moment densities in Fig. A.6 for confinement values of $\alpha=0.3$ and $\alpha=1$ corresponding to the moderate and strong confinements. Blue, red and purple curves represent the particle density, momentum density and the energy density respectively. The results are in coherence with the Fig. A.4 where we've found the momentum flux distribution from a completely different approach. Characteristic peaks of momentum and energy densities are present. One can find the corresponding fluxes by multiplying the expressions with the velocity of the particular direction. Momentum flux equals to the internal energy density in 1D case, since there is no off-diagonal terms for the possibility of velocity combinations in 1D.

Using Poisson summation formula, for 1D case, it is possible to obtain analytical expressions for density distributions for all moments. The expression we found for Maxwell-Boltzmann statistics is written as
\begin{equation}
\begin{split}
\tilde{n}_{\varepsilon}^{(m)} &=\frac{\sum_i \left(\alpha i\right)^m\exp\left(-(\alpha i)^2\right)\left|\Psi(\textbf{r})\right|^2}{\frac{1}{V}\sum_i \left(\alpha i\right)^m\exp\left(-(\alpha i)^2\right)}\\
&\approx\left[1- _1F_1\left(\frac{m+1}{2};\frac{1}{2};-\frac{\pi^2\tilde{x}^2}{\alpha^2}\right)\right]\left[1- _1F_1\left(\frac{m+1}{2};\frac{1}{2};-\frac{\pi^2(1-\tilde{x})^2}{\alpha^2}\right)\right]
\end{split}
\end{equation}
where $_1F_1(a;b;c)$ is Kummer confluent hypergeometric function. In the Fig. A.6, we compared the accuracies of the analytical results with the exact ones. As is seen from Fig. A.6, analytical and numerical results are in a very good agreement for $\alpha=0.3$, especially for $m=0$ and $m=1$ cases. Our analytical representation gives accurate results for all moments as long as the confinement is not very strong so that the conditions of bounded continuum is satisfied. It is also seen that for large confinement values like $\alpha=1$, the results of numerical and analytical calculations start to deviate. The larger the confinement, the higher the deviation. The analysis in Fig. A.6 explains also the energy moment densities that are compared in Fig. A.5. Higher the moment, closer the density near to the boundaries. 

\section{The First Order Quantum Boundary Layer Approach}

As we said before, in order for quantum shape effects to appear in confined systems, overlaps of quantum boundary layers are essential. The method suggested in this thesis is quite successful as it is shown by many examples. Nevertheless, a room for improvement to the methodology is also possible which we would like to mention here. The usual quantum boundary layer methodology is an approximate one as is known. It approximates the gradually changing density distribution with a step function which can only have values 1 or 0. Calculating overlaps with this method was possible and indeed proved to be accurate enough to represent quantum shape effects in most of the confined systems in this thesis. On the other hand, in reality overlaps do not start abruptly when the distance between boundaries is shorter than $2\delta$ as it can be clearly seen from Fig. 3.10 where there is a variation in effective area even before the critical value where the overlaps start according to the quantum boundary layer method. Evacuation of the region near to boundaries start not from $\delta$ or even $2\delta$, but visibly around $2.5\delta$, see Fig. A.7 black curve. We can make a better approximation to the exact density distribution of particles near to boundaries. Instead of the usual stepwise approximation, we can approximate the density with a ramp function, Fig. A.7. Let's call the stepwise one as $0$\textsuperscript{th} order and name the ramp one as the $1$\textsuperscript{st} order quantum boundary layer approaches. This should definitely give much more accurate results than the usual $0$\textsuperscript{th} order quantum boundary layer approach. The comparison of both approaches is given in Fig. A.7 where they are compared with the exact distribution.

\begin{figure}[h]
\centering
\includegraphics[width=0.65\textwidth]{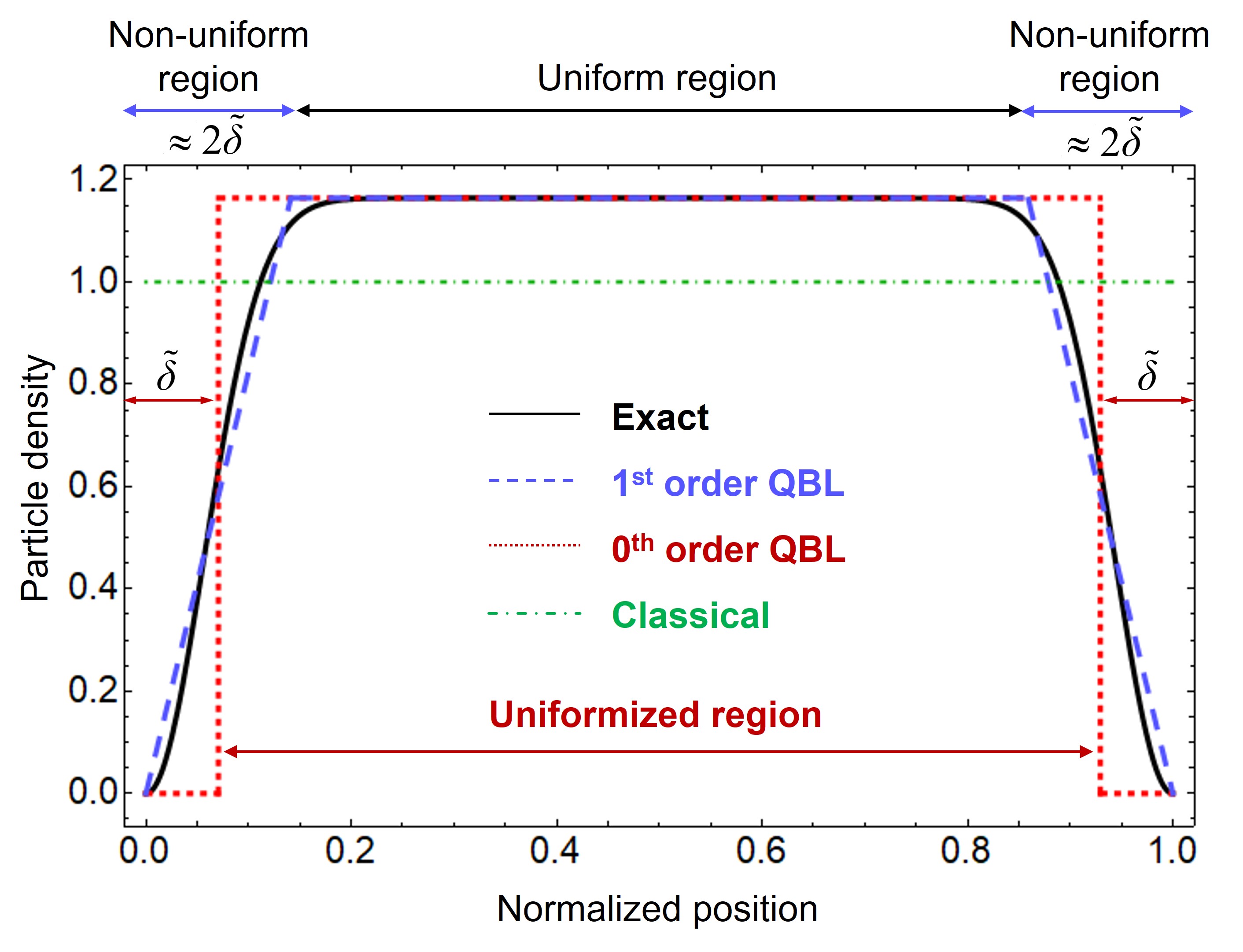}
\caption{Approximations of quantum boundary layer for density distributions. Ensemble-averaged quantum density distribution of particles confined in a 1D domain with length $L$. Solid black and dot-dashed green curves represent the exact and classical density distributions respectively. $0$\textsuperscript{th} order quantum boundary layer is presented by dotted red lines where the exact density distribution is approximated by empty regions with thickness $\tilde{\delta}$ near to boundaries and uniform density region in the remaining parts. $1$\textsuperscript{st} order quantum boundary layer denoted by dashed blue lines is the linear approximation to the non-uniform density region with thickness $2\tilde{\delta}$. Here $\tilde{n}=n/n_{cl}$, $\tilde{\delta}=\delta/L$ and $\tilde{x}=x/L$ where $x$ is the position.}
\label{fig:cAf7}
\end{figure}

\begin{figure}[h]
\centering
\includegraphics[width=0.95\textwidth]{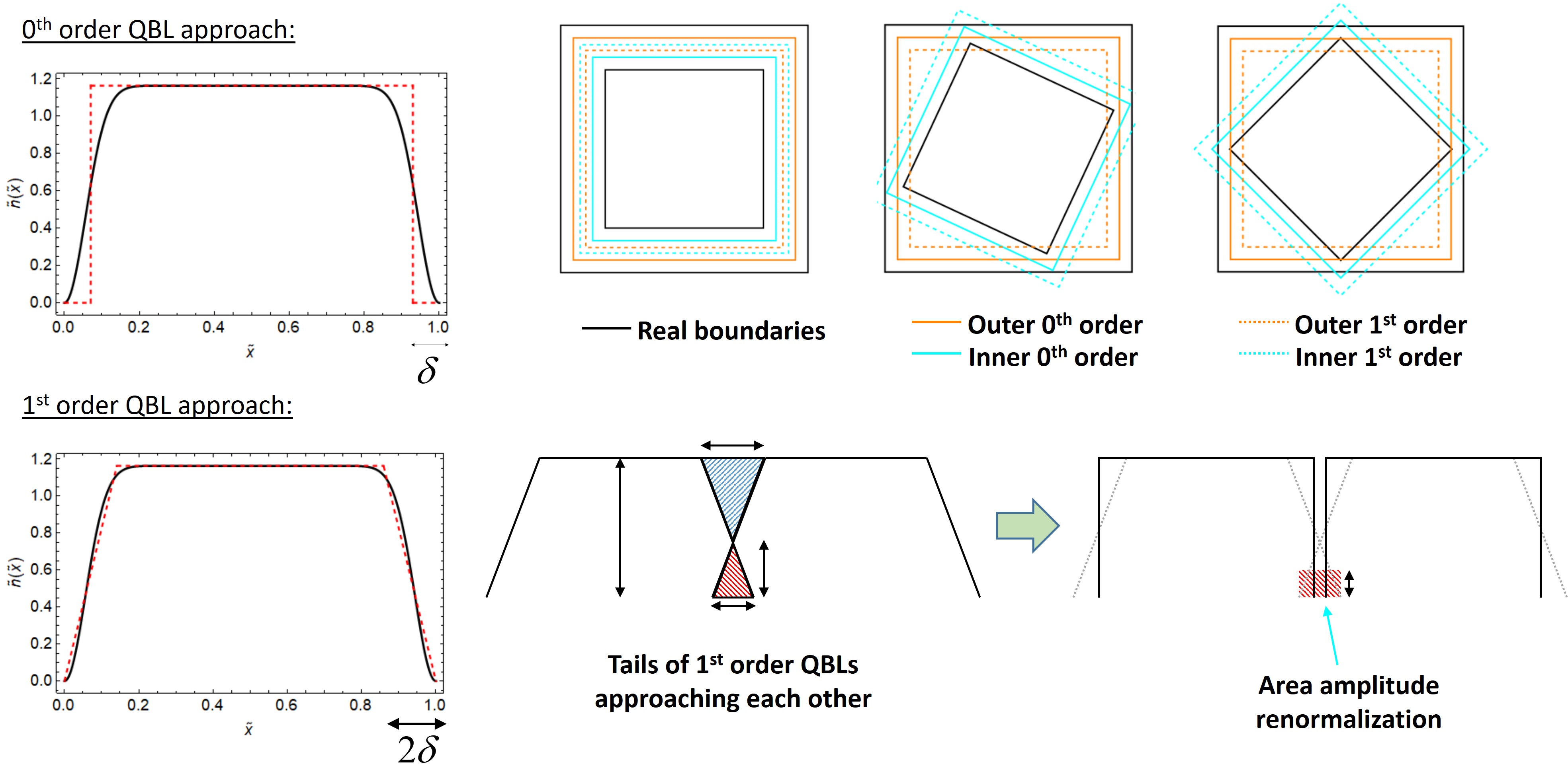}
\caption{Comparison of $0$\textsuperscript{th} and $1$\textsuperscript{st} order quantum boundary layer approaches. In the domain boundary analysis, real boundaries are shown by black color. The $0$\textsuperscript{th} order quantum boundary layers are represented by solid orange and cyan colors respectively, whereas corresponding $1$\textsuperscript{st} order ones are shown by dashed lines. Details of area amplitude renormalization is not given here.}
\label{fig:cAf8}
\end{figure}

\begin{figure}
\centering
\includegraphics[width=0.8\textwidth]{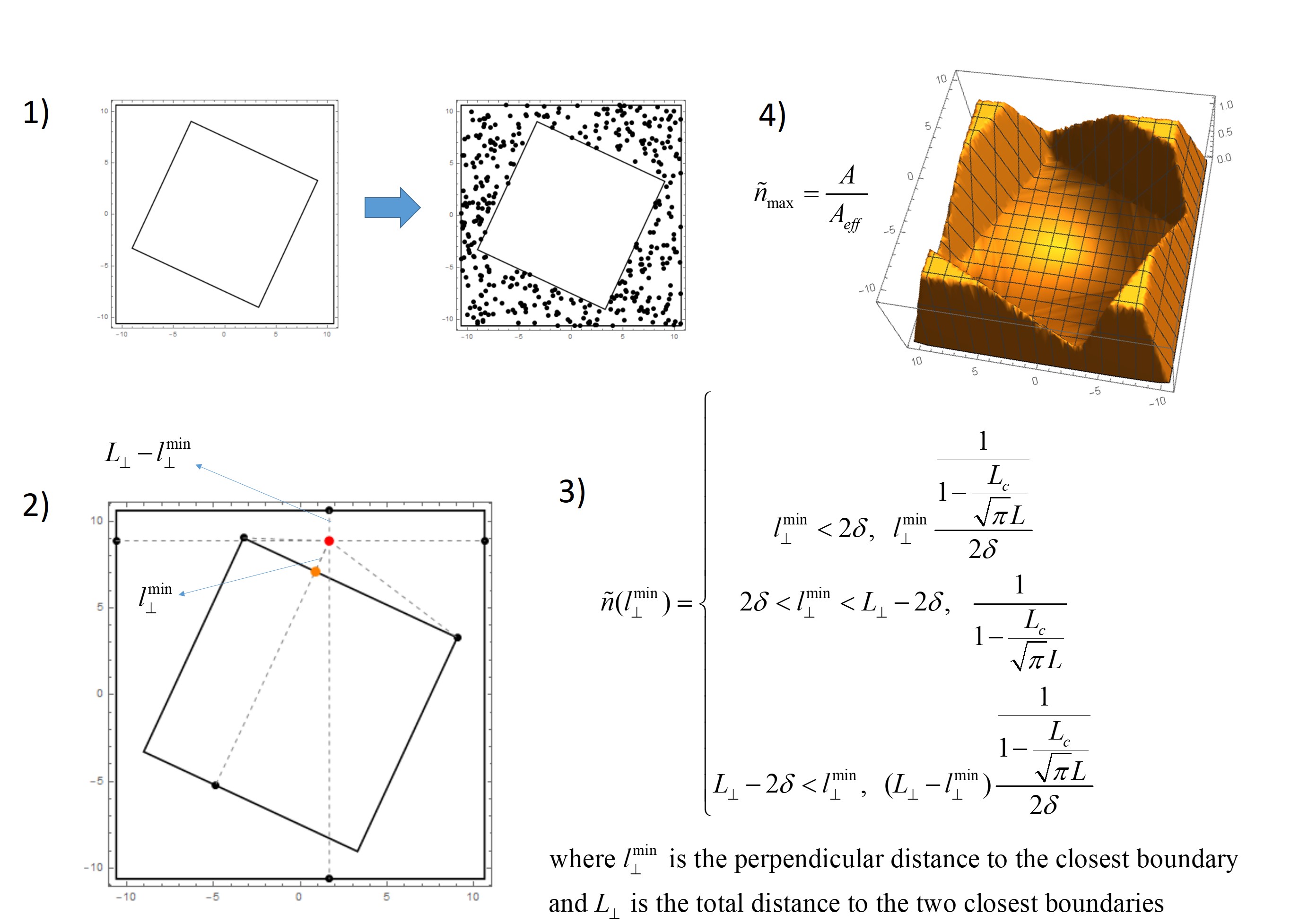}
\caption{Implementation of Monte Carlo method for the calculation of density distributions by analytical $1$\textsuperscript{st} order quantum boundary layer approach.}
\label{fig:cAf9}
\end{figure}

The shape dependence of partition function has a sigmoid type functional behavior which can be mimicked by $\tanh$ function. We have found hybrid ways to approximate the effective area by using the $1$\textsuperscript{st} order approach for less complicated geometries and $0$\textsuperscript{th} order for others. However, we need to establish an elegant methodology with a physical meaning rather than just doing some mathematical tricks.

We've done several attempts to construct an analytical or at least a numerical approach for the development of $1$\textsuperscript{st} order quantum boundary layer methodology. In this $1$\textsuperscript{st} order QBL method, the overlaps of the opposing boundaries start from $4\delta$ distance instead of $2\delta$. Therefore it is possible to capture even tiny details about the density distribution. Similar to the $0$\textsuperscript{th} order method, middle regions are equal to $1/(1-2\delta/L)$, but there is a gradual decrease near to the boundaries, rather than a sharp one. The main difficulty is unlike the $0$\textsuperscript{th} order method, in this one, overlaps would have a weight factor. Near boundary overlaps would have very different contribution to the effective density (or effective area) than the far boundary overlaps. This brings additional variables to the problem of overlap volume calculations and makes the analytical geometric calculations highly complicated due to sensitive positional dependence of the QBL. Nevertheless, we defined a new parameter called the area amplitude renormalization factor, as its name suggests, it renormalizes the contributions of the overlaps to the effective area. A rough picture of the methodology is presented in Fig. A.8. For the simplest cases of nested square domain (which are $0^{\circ}$ and $45^{\circ}$ degree configurations) we predicted the correct effective area with relative errors much less than $0.1\%$, which is remarkably accurate. However, implementing the same procedure is very difficult analytically for the degrees in between, let alone the arbitrary geometries.

In order to test whether our analytical approach could be reliable at all for any kind of domain, we decided to invoke numerical methods. We applied a Monte Carlo method to test the accuracy of the proposed methodology for arbitrary domains. The algorithm (see Fig. A.9) goes like this: (1) First, we draw our domain's boundaries and assign random points inside of our confinement domain. (2) Then for each point, we take the distance from the point to the normal (perpendicular) of the closest boundary. (3) We calculate the weight factor of that point according to the area amplitude renormalization factor of $1$\textsuperscript{st} order approach and assign that value for that particular point. (4) We repeat this procedure for each and every point, thereby generating the domain's density distribution as a whole. The approach gives the correct functional behavior with some fluctuations and with moderate errors overall. Of course, there are many non-trivial things to be considered, like how we calculate perpendicular distances from and to the angled boundaries. Numerical errors are also concern here. 

% Exact QBL
Instead of the $1$\textsuperscript{st} order approach, another possible improvement of QBL could be to go into the heart of the problem of determining the QBL thickness. Instead of using the first two terms of PSF, we may use elliptic theta functions as full solutions for the partition function under infinite summation form and then calculate QBL thickness from the full solutions, while keep staying within $0$\textsuperscript{th} order approach. How does QBL thickness modify in strongly confined regions? Should we stick with the overlap approach or can we come up with a better, even more accurate one? We don't know yet. The improvements on QBL methodology is still ongoing.

\section{Pressure - Momentum Flux Equivalence}

In this section, we derive and show the mathematical equivalence of pressure and momentum flux in rectangular confinement domains for Maxwell-Boltzmann, Fermi-Dirac and Bose-Einstein statistics. We start with the derivations in Maxwell-Boltzmann statistics.

\subsection{Equivalence for Maxwell-Boltzmann statistics}

Thermodynamic pressure, Helmholtz free energy (for large $N$) and partition function in Maxwell-Boltzmann statistics are given respectively as
\begin{equation}
P=-\left(\frac{\partial F}{\partial V}\right),
\end{equation}
\begin{equation}
F=-Nk_BT\left(\ln\zeta-\ln N+1\right),
\end{equation}
\begin{equation}
\zeta=\sum_{\{i_1,i_2,i_3\}}\exp(-\tilde{\varepsilon}),
\end{equation}
where $\tilde{\varepsilon}=\varepsilon/k_BT$ are dimensionless energy eigenvalues of a simple 3D rectangular domain. When we plug partition function into the free energy and then the resulting expression into the pressure, we get the pressure by following steps:
\begin{equation}
\begin{split}
P& =-\frac{1}{L_2L_3}\frac{\partial}{\partial L_1}\left(-Nk_BT\ln\zeta\right)=\frac{Nk_BT}{L_2L_3}\frac{\partial\alpha_1}{\partial L_1}\frac{\partial}{\partial\alpha_1}\ln\zeta \\
&=\frac{Nk_BT}{L_2L_3}\left(-\frac{\alpha_1}{L_1}\right)\frac{1}{\zeta}\frac{\partial\zeta}{\partial\alpha_1}=-n_{cl}k_BT\frac{\alpha_1}{\zeta}\sum\left(-2\alpha_1i_1^2\exp(-\tilde{\varepsilon})\right) \\
&=2n_{cl}k_BT\frac{\sum(\alpha_1i_1)^2\exp(-\tilde{\varepsilon})}{\sum\exp(-\tilde{\varepsilon})}=2n_{cl}k_BT\frac{\sum(\alpha_1i_1)^2f}{\sum f}.
\end{split}
\end{equation}
Pressure exerted to the $L_2\times L_3$ side (the area) of a rectangular box is then obtained as
\begin{equation}
P=2n_{cl}k_BT\frac{\sum(\alpha_1i_1)^2f}{\sum f}.
\end{equation}

To derive the pressure, we relied on thermodynamics equations. Now let's derive the momentum flux by considering the transport equations. From the transport theory, scalar momentum flux, translational kinetic energy and distribution function are given respectively as
\begin{equation}
J_p=\frac{1}{V}m\sum v_1^2 f,
\end{equation}
\begin{equation}
\frac{1}{2}mv_1^2=k_BT(\alpha_1 i_1)^2,
\end{equation}
\begin{equation}
f=\exp(\Lambda)\exp(-\tilde{\varepsilon}).
\end{equation}
Combining these equations gives the momentum flux
\begin{equation}
J_p=\frac{1}{V}2k_BT\exp(\Lambda)\sum\left(\alpha_1 i_1\right)^2\exp(-\tilde{\varepsilon}).
\end{equation}
From the number of particles expression
\begin{equation}
N=\sum\exp(\Lambda)\exp(-\tilde{\varepsilon}),
\end{equation}
fugacity is written as
\begin{equation}
\exp(\Lambda)=\frac{n_{cl}V}{\sum\exp(\tilde{\varepsilon})}.
\end{equation}
When we plug the fugacity into Eq. (A.14), we get momentum flux
\begin{equation}
J_p=\frac{1}{V}2k_BT\frac{n_{cl}V}{\sum\exp(\tilde{\varepsilon})}\sum\left(\alpha_1 i_1\right)^2\exp(-\tilde{\varepsilon}).
\end{equation}
By simple reordering, it becomes
\begin{equation}
J_p=2n_{cl}k_BT\frac{\sum(\alpha_1 i_1)^2\exp(\tilde{\varepsilon})}{\sum\exp(\tilde{\varepsilon})}=2n_{cl}k_BT\frac{\sum(\alpha_1 i_1)^2 f}{\sum f}.
\end{equation}
So it is shown that pressure, Eq. (A.10), is equal to momentum flux, Eq. (A.18), in Maxwell-Boltzmann statistics.

\subsection{Equivalence for Fermi-Dirac and Bose-Einstein statistics}

The derivation in Fermi-Dirac and Bose-Einstein statistics are done together in this section. Thermodynamic pressure, Helmholtz free energy and partition function are given respectively as
\begin{equation}
P=-\left(\frac{\partial F}{\partial V}\right),
\end{equation}
\begin{equation}
F=N\mu-k_BTZ,
\end{equation}
\begin{equation}
Z_{\substack{\text{FD}\\ \text{BE}}}=\sum_{\{i_1,i_2,i_3\}}\ln\left[1\pm\exp(\Lambda-\tilde{\varepsilon})\right],
\end{equation}
where the upper and lower signs correspond to Fermi-Dirac and Bose-Einstein statistics respectively in all equations that they appear. Plugging the free energy into pressure gives
\begin{equation}
P=-\frac{k_BT}{L_2L_3}\frac{\partial}{\partial L_1}(N\Lambda-Z)=-\frac{Nk_BT}{L_2L_3}\left(\frac{\partial\Lambda}{\partial L_1}-\frac{1}{N}\frac{\partial Z}{\partial L_1}\right).
\end{equation}
There are derivatives in Eq. (A.22) which need to be found analytically. Derivative of dimensionless chemical potential with respect to $L_1$ can be found by taking the advantage of the fact that number of particles does not depend on the changes in $L_1$. From the derivative of $N$ with respect to $L_1$ we get
\begin{equation}
\frac{\partial N}{\partial L_1}=0 \rightarrow\frac{\partial\Lambda}{\partial L_1}=-\frac{2}{L_1}\frac{\sum(\alpha_1 i_1)^2 f(1\mp f)}{\sum f(1\mp f)}.
\end{equation}
Also, the derivative of partition function with respect to $L_1$ can be found directly as
\begin{equation}
\frac{\partial Z}{\partial L_1}=-\frac{2}{L_1}\sum f\frac{\sum(\alpha_1 i_1)^2 f(1\mp f)}{\sum f(1\mp f)}+\frac{2}{L_1}\sum(\alpha_1 i_1)^2 f.
\end{equation}
By plugging these derivatives into Eq. (A.22), pressure becomes
\begin{equation}
P=2n_{cl}k_BT\frac{\sum(\alpha_1i_1)^2f}{\sum f}.
\end{equation}
This expression has exactly the same form of the one derived in Maxwell-Boltzmann statistics in the previous section. The only difference comes from the distribution functions. 

The distribution function in Fermi-Dirac and Bose-Einstein statistics is given as
\begin{equation}
f_{\substack{\text{FD}\\ \text{BE}}}=\frac{1}{1\pm\exp(-\Lambda+\tilde{\varepsilon})}.
\end{equation}
Combining Eq. (A.11) and Eq. (A.12) with Eq. (A.26) gives the momentum flux as
\begin{equation}
J_p=\frac{1}{V}2k_BT\sum\frac{(\alpha_1 i_1)^2}{1\pm\exp(-\Lambda+\tilde{\varepsilon})}.
\end{equation}
From the number of particles $N=\sum f$, and the expression for density
\begin{equation}
\frac{1}{V}=\frac{n_{cl}}{N},
\end{equation}
momentum flux then becomes
\begin{equation}
J_p=2n_{cl}k_BT\frac{\sum(\alpha_1i_1)^2f}{\sum f}.
\end{equation}
which is the same as the expression Eq. (A.18). Thus, it is seen here that pressure and momentum flux are mathematically equivalent independent of statistics. 

\bibliography{phdref}
\bibliographystyle{unsrt}
\end{document}